\documentclass[11pt]{article}
\usepackage{amssymb,amsmath,amsfonts}
\usepackage{graphicx,color,enumitem}
\usepackage{amsthm} 
\usepackage{bm}
\usepackage[numbers,sort&compress]{natbib}
\usepackage{soul}
\usepackage{comment}
\usepackage{geometry}
\usepackage{stmaryrd}
\usepackage{pifont}   
\usepackage{tikz}
\usetikzlibrary{decorations.pathreplacing}
\usetikzlibrary{patterns}
\usepackage{mathtools}
\usepackage[ruled,vlined]{algorithm2e}
 \mathtoolsset{showonlyrefs}
\usepackage[colorinlistoftodos, textwidth=4cm, shadow]{todonotes}
\RequirePackage[colorlinks=true,citecolor=blue,urlcolor=blue]{hyperref}

\usepackage{subcaption}

\usepackage{array,multirow,makecell}
\setcellgapes{1pt}
\makegapedcells
\newcolumntype{R}[1]{>{\raggedleft\arraybackslash }b{#1}}
\newcolumntype{L}[1]{>{\raggedright\arraybackslash }b{#1}}
\newcolumntype{C}[1]{>{\centering\arraybackslash }b{#1}}

\def \mra{\mathrm{a}} 
 
\def \mrs{\mathrm{s}} 
\def \mrf{\mathrm{f}} 
\def \mrb{\mathrm{b}} 
 
\def\boM{{\boldsymbol M}}

\usepackage{bbm}

\renewcommand{\d}{{\rm d}}

\newcommand{\E}{{\mathbb E}}
\newcommand{\F}{{\mathbb F}}

\renewcommand{\P}{{\mathbb P}}

\newcommand{\R}{{\mathbb R}}
\renewcommand{\S}{{\mathbb S}}

\newcommand{\Acal}{{\mathcal A}}

\newcommand{\Lcal}{{\mathcal L}}

\newcommand{\Mcal}{{\mathcal M}}

\newcommand{\Tcal}{{\mathcal T}}

\newcommand{\Ac}{{\mathcal A}}

\newcommand{\Zcal}{{\mathcal Z}}

\newcommand{\vertiii}[1]{{\left\vert\kern-0.25ex\left\vert\kern-0.25ex\left\vert #1 
    \right\vert\kern-0.25ex\right\vert\kern-0.25ex\right\vert}}

\newcommand{\Fc}{{\mathcal F}}

\newcommand{\Tc}{{\mathcal T}}

\newcommand{\blue}{\color{blue}}
\definecolor{darkgreen}{rgb}{0,0.7,0}

\newcommand{\iii}{{\vert\kern-0.25ex\vert\kern-0.25ex\vert}}
\newcommand{\bec}[1]{\begin{equation} \begin{cases} #1\end{cases} \end{equation}}
\newcommand{\bes}[1]{\begin{equation} \begin{split} #1\end{split} \end{equation}}

\definecolor{blue0}{RGB}{0,77,153} 
\definecolor{red0}{RGB}{179,0,77} 
\definecolor{green0}{RGB}{134,219,76} 
\definecolor{gray0}{RGB}{84,97,110}

\newtheorem{theorem}{Theorem}

\newtheorem{remark}[theorem]{Remark}
\theoremstyle{definition}
\numberwithin{equation}{section}
\numberwithin{theorem}{section}

\begin{document}

\title{Differential learning methods for solving fully nonlinear PDEs}

\author{William LEFEBVRE\thanks{BNP Paribas Global Markets, Universit\'e Paris Cité and Sorbonne Universit\'e, Laboratoire de Probabilit\'es, Statistique et Mod\'elisation (LPSM, UMR CNRS 8001), Building Sophie Germain, Avenue de France, 75013 Paris, \texttt{wlefebvre at lpsm.paris}} \quad Gr\'egoire LOEPER\thanks{BNP Paribas Global Markets, School of Mathematics, Monash University, Clayton Campus, VIC, 3800, Australia, \texttt{gregoire.loeper at monash.edu}} \quad Huy\^en PHAM \thanks{Universit\'e Paris Cité and Sorbonne Universit\'e, Laboratoire de Probabilit\'es, Statistique et Mod\'elisation (LPSM, UMR CNRS 8001), Building Sophie Germain, Avenue de France, 75013 Paris, \texttt{pham at lpsm.paris}}
}


\maketitle

\begin{abstract}
We propose machine learning methods for solving fully nonlinear partial differential equations (PDEs) with convex Hamiltonian. 
Our algorithms are conducted in two steps. First the PDE is rewritten in its dual stochastic control representation form, and the corresponding optimal feedback control is estimated using a neural network. Next, three different methods are presented to approximate the associated value function, i.e., the solution of the initial PDE, on the entire space-time domain of interest.  
The proposed deep learning algorithms rely on various loss functions obtained either from regression or pathwise versions of the martingale representation and its differential relation, and compute simultaneously the solution and its derivatives. Compared to existing methods, the addition of a differential loss function associated to the gradient, and augmented training sets with Malliavin derivatives of the forward process, 
yields a better estimation of the PDE's solution derivatives, in particular of the second derivative, which is usually difficult to approximate. 
Furthermore, we leverage our methods to design algorithms for solving families of PDEs when varying terminal condition 
(e.g. option payoff in the context of mathematical finance) by means of the class of DeepOnet neural networks aiming to approximate functional operators.  
Numerical tests illustrate the accuracy of our methods on the resolution of a fully nonlinear PDE associated to the pricing of options  with linear market impact, and on the Merton portfolio selection problem. 
\end{abstract}


\vspace{5mm}


\newpage

\section{Introduction}

This paper is devoted to the resolution of fully nonlinear partial differential equations (PDEs) of the form
\bec{ \label{PDEH} 
    \partial_t u +  H(x,D_x u, D_x ^2 u) &= \; 0, \quad \quad  (t,x) \in  [0,T)\times\R^d, \\
    \hspace{2.5cm}  u(T,x) &= \; g(x),  \quad  x \in  \R^d, 
}
where the Hamiltonian  $H : \R^d\times \R^d \times \R^{d\times d} \rightarrow \R\cup \{\infty\}$ is a lower semi-continuous convex function with respect to the two last arguments $(z,\gamma)$, 
and $g$ a measurable function on $\R^d$. The numerical resolution of this class of PDE is a notorious challenging problem, and it is especially difficult to obtain a good approximation of the second spatial derivative $D_x^2 u$ of the solution in this fully nonlinear context. In the last years, significant progress has been achieved towards these challenges with several numerical methods using techniques from deep learning, see the recent surveys by \cite{becetal20} and \cite{gerphawar21}: 
A first class of approximation algorithms, called  {\it Physics Informed Neural Network}  (PINNs) \cite{raissi2019physics}, also known as {\it Deep Galerkin method} (DGM) \cite{sirignano2018dgm}, directly approximates the solution to the PDE by a neural network, and its partial derivatives by automatic differentiation, by minimizing the loss function arising from the residual of the PDE evaluated on a  random grid in the space-time domain. A second class of algorithms relies on the backward stochastic differential representation of 
the PDE in the semi-linear case by minimizing either a global loss function (see \cite{Ehanjen17}, and extensions in 
\cite{BEJ19}, \cite{penetal20},  \cite{nusric21}), or sequence of loss functions from backward recursion (see \cite{hure2020deep}, and variations-extensions in \cite{pham2021neural}, \cite{BBCJN19}, \cite{gerphawar21}).


In this article, we consider numerical methods for fully nonlinear PDEs 
based on machine learning techniques that are conducted in two steps. 
The starting point of our approach is to rewrite the PDE \eqref{PDEH} with convex Hamiltonian in its stochastic control representation form following the duality arguments of \cite{soner2013dual}. An approximation of the associated optimal feedback control  is then obtained using a neural network by a global optimization, as described in \cite{HanE16} and \cite{gobmun05}. 
Based on this control approximation, two main approaches using neural networks are then developed in order to approximate the value function, hence the solution of the initial PDE on the space-time domain, which is formulated as a conditional expectation with respect to the approximate optimal state process.

The first one, called {\it Differential regression learning}, is inspired by \cite{huge2020differential}. In this paper, the authors compute conditional expectations of an option payoff in the spirit of the Longstaff-Schwartz method \cite{longstaff2001valuing}, by parametrizing it with a neural network and performing the regression simultaneously on the value and on the derivative of this neural network. The addition of a regression loss on the derivative, where the derivative of the network, computed by automatic differentiation, is regressed against the pathwise derivative of the conditional expectation integrand, improves the estimation of the first derivative of the conditional expectation and empirically speeds up the training by allowing to train the network on smaller batches. 
We adapt this method to our context. Indeed, having approximated the optimal control of the stochastic control problem associated to the PDE, the associated value function can be expressed as the conditional expectation of a running payoff of optimally controlled state trajectories, while its gradient is represented also as a conditional expectation formula by differentiation of the payoff.  
This representation formulae provide two loss functions that will be minimized alternately in order to learn by neural network approximation both the solution of the PDE and its gradient.

The second approach, called Pathwise learning, is inspired by \cite{vidales2018unbiased}, where the authors compute the conditional expectation of a payoff by using the Feynman-Kac formula to derive a pathwise control variate corresponding to the hedging strategy. In their work, the derivative of the conditional expectation value, present in the hedging integral, is approximated by a neural network and optimized so as to minimize the variance of the conditional expectation estimator. This approach is analogous to the one derived in \cite{potters2001hedged}, with the addition of machine learning techniques. In our case, given the approximation of the stochastic control associated to the PDE, a martingale representation of the payoff is derived on optimally controlled trajectories. Our first pathwise method, called {\it Pathwise martingale learning}, uses this martingale representation to train the value and the first derivative of a neural network. This method is also in the spirit of the deep BSDE method of \cite{becetal20}, but the minimization of our loss function provides directly an approximation of the solution and its gradient on the space-time domain. 
Our second method, called {\it Pathwise differential learning}, considers furthermore the derivative of this martingale representation, computed by automatic differentiation, which gives another loss function to be minimized in order to train neural networks for approximating the value function and its first and second derivatives. Such differential representation has been also considered in the recent paper \cite{negyesi21} for designing a deep learning scheme with one-step loss functions as in the deep backward approach in \cite{hure2020deep} for solving forward backward SDEs with new estimation and error control  of the $Z$ process.  
Actually, the addition of this derivative loss function permits a better approximation of the terminal condition of the PDE and improves the overall approximation of the PDE solution's value and derivatives on the entire domain.  

Finally, we leverage our deep learning algorithms for solving families of PDEs when varying the terminal condition. 
In other words, the input is a function $g_K$ with parameter $K$, and the output is the solution to the PDE with terminal condition $g_K$. 
This is performed by means of the class of DeepOnet neural networks aiming to approximate functional operators.  These networks, introduced in \cite{lu2019deeponet}, rely on a universal approximation theorem for operators \cite{chen1995universal} stating that a neural network with a single hidden layer can approximate accurately any nonlinear continuous operator. The DeepOnet realizes this theorem in practice and can be used to learn the mapping between the terminal function of a PDE and its solution.

The outline of the paper is organized as follows. In Section \ref{sec:duality_stochastic_control_representation}, we present the problem, recall the dual stochastic control representation of fully nonlinear PDEs and outlines the different methods. In Section \ref{sec:differential_regression_learning}, we present the theory of the Differential regression learning method, give the expression of the losses used to train the neural network and present the advantages of adding a loss to train the first derivative of the neural network. In Section \ref{sec:pathwise_learing}, the Pathwise and Pathwise differential methods are developed and the expressions of the losses used to train the neural network are given. In Section \ref{sec:numerical_results}, the implementation details and pseudo-codes of the different algorithms are presented along with validation tests and numerical results of the three methods on the Merton porfolio selection problem and on the Black-Scholes with linear market impact PDE. Finally, Section \ref{sec:resolution_any_terminal_condition} presents a method to solve nonlinear parabolic PDEs with parametric terminal condition $g_K$ for parameter values $K$ in a compact set. The codes of our algorithms are available on  \url{https://colab.research.google.com/drive/1xyE1U3SqN4Hjia2d3pOsCXCCDGWRUqsH?usp=sharing}.


\vspace{1mm}

\noindent {\bf Notations.} We end this introduction with some notations that will be used in the sequel of the paper. 
The scalar product between two vectors $b$ and $z$ is denoted by $b \cdot z$, and $| \cdot |$ is  the Euclidian norm.  
Given two matrices $A$ $=$ $(A_{ij})$ and $B$ $=$ $(B_{ij})$, we denote by $A:B$ $=$ ${\rm Tr}(A^\top B)$ $=$ $\sum_{i,j} A_{ij} B_{ij}$ its inner product, and by $|A|$ the Frobenius norm of $A$. 
Here $\top$ is the transpose matrice operator.  
$\S^d$ is the set of $d\times d$ symmetric matrices with real coefficients equipped with the partial order: $\gamma_1$ $\leq$ $\gamma_2$ iff $\gamma_2-\gamma_1$ 
$\in$ $\S_+^d$, the set of positive semidefinite matrices in $\S^d$.

Let $\boM$ $=$ $(\boM_{i_1i_2i_3})$ $\in$ $\R^{d_1\times d_2\times d_3}$ be  a tensor of order $3$. For $p$ $=$ $1,2,3$, the $p$-mode product of  $\boM$ with a vector $b$ $=$ $(b_i)$ $\in$ $\R^{d_p}$, is denoted by $\boM\bullet_p b$, and it is a tensor of order $2$, i.e. a matrix 
defined elementwise as 
\bes{
\big( \boM\bullet_1 b)_{i_2i_3} & = \; \sum_{i_1=1}^{d_1} M_{i_1i_2i_3} b_{i_1}, \;    \big( \boM\bullet_2 b)_{i_1i_3} \;  = \; \sum_{i_2=1}^{d_2}  M_{i_1i_2i_3} b_{i_2}, \;   \big( \boM\bullet_3 b)_{i_1i_2} \;  = \; \sum_{i_3=1}^{d_3} M_{i_1i_2i_3} b_{i_3} 
}

 The $p$-mode product of a $3$-th order tensor $\boldsymbol{M}$ $\in$ $\R^{d_1\times d_2\times d_3}$ with a matrix $B$  $=$ $(B_{ij})$ $\in$ $\R^{d_p\times d}$, also denoted by $\boldsymbol{M}\bullet_p B$,  is a $3$-th order 
 tensor 
 defined elementwise as
\bes{
\big(\boldsymbol{M}\bullet_1 B \big)_{\ell i_2i_3} & \; = \sum_{i_1 =1}^{d_1} M_{i_1i_2i_3} B_{i_1\ell}, \quad \big(\boldsymbol{M}\bullet_2 B \big)_{i_1\ell i_3}  \; = \sum_{i_2 =1}^{d_2} M_{i_1i_2i_3} B_{i_2\ell} \\
\big( \boM\bullet_3 B)_{i_1i_2\ell} & = \; \sum_{i_3=1}^{d_3} M_{i_1i_2i_3} B_{i_3\ell}. 
}

Finally, the tensor contraction (or partial trace) of a $3$-th order tensor $\boldsymbol{M}$ $\in$ $\R^{d_1\times d_2\times d_3}$ whose dimensions $d_p$ and $d_q$ are equal is denoted as $\mathrm{Tr}_{p,q} \boldsymbol{M}$. This tensor contraction is a tensor of order 1, i.e. a vector, defined elementwise as 
\bes{
    \big(\mathrm{Tr}_{1,2} \boldsymbol{M} \big)_{i_3} = \; \sum_{\ell=1}^{d_1} M_{\ell \ell i_3}, \;     \big(\mathrm{Tr}_{1,3} \boldsymbol{M} \big)_{i_2} = \; \sum_{\ell=1}^{d_1} M_{\ell i_2 \ell}, \;     \big(\mathrm{Tr}_{2,3} \boldsymbol{M} \big)_{i_1} = \; \sum_{\ell=1}^{d_2} M_{i_1 \ell \ell}.
}

\section{Dual stochastic control representation of fully nonlinear PDE}
\label{sec:duality_stochastic_control_representation}


We consider a fully nonlinear partial differential equation (PDE) of parabolic type: 
\bec{
\label{eq:pde}
    \partial_t u +  H(x,D_x u, D_x ^2 u) &= \; 0, \quad \quad  (t,x) \in  [0,T)\times\R^d, \\
    \hspace{2.5cm}  u(T,x) &= \; g(x),  \quad  x \in  \R^d, 
}    
where the Hamiltonian  $H : \R^d\times \R^d \times \R^{d\times d} \rightarrow \R\cup \{\infty\}$ is a lower semi-continuous convex function w.r.t  the two last arguments $(z,\gamma)$, 
and $g$ a measurable function on $\R^d$. As it is usual, we assume that $H(x,z,\gamma)$ $=$ $H(x,z,\gamma^\top)$, 
and that $\gamma$ $\in$ $\S^d$ $\mapsto$ 
$H(x,z,\gamma)$ is nondecreasing, 
 
Without loss of generality, we may then assume that $H$ is in a Bellman form: 
\begin{align}  \label{Hbel} 
H(x,z,\gamma) &= \; \sup_{a \in A} \big[ \mrb(x,a).z + \frac{1}{2} \sigma\sigma^\top(x,a)  : \gamma + f(x,a) \big],  \;\;  (x,z,\gamma) \in \R^d\times\R^d\times\S^d, 
\end{align} 
for some measurable functions $b$ $:$ $\R^d\times A$ $\rightarrow$ $\R^d$, $\sigma$  $:$ $\R^d\times A$ $\rightarrow$ $\R^{d\times m}$, $f$ $:$ $\R^d\times A$ $\rightarrow$ $\R$, and with $A$ 
some subset of $\R^q$.  Indeed, such form may arise directly from the dynamic programming equation of a stochastic control problem. Otherwise, it can be written in this formulation by following the duality argument as in \cite{soner2013dual}.  We introduce the concave conjugate of the function $H(x,z,\gamma)$ w.r.t. the last two variables, i.e. 
\bes{
\label{eq:multid_dual_H}
 \mrf(x,b,c) &:= \;   \underset{z \in \R^d, \gamma \in \R^{d\times d}}{\inf}\ \big[ H(x,z,\gamma) -  b\cdot z -  \frac{1}{2} c : \gamma  \big],\quad x \in \R^d, b \in \R^d, c \in \R^{d\times d}, 
}
and notice that $\mrf(x,b,c)$ $=$ $\mrf(x,b,c^\top)$ as $H(x,z,\gamma)$ $=$ $H(x,z,\gamma^\top)$. 
By the Fenchel-Moreau duality relation, we then get 
\bes{
H(x,z,\gamma) & = \; \underset{b \in \R^d, c \in \S^{d}}{\sup}\ \big[ b\cdot z +  \frac{1}{2} c : \gamma  + \mrf(x,b,c)   \big] \\
&= \; \underset{(b,c)\in D_f}{\sup}\ \big[ b\cdot z +  \frac{1}{2} c : \gamma  + \mrf(x,b,c)  \big],  \quad \mbox{ for } x \in \R^d, z \in \R^d, \gamma \in \S^d,
}
where $D_f$ $:=$ $\{(b,c) \in \R^d\times\S^d: \mrf(x,b,c) > -  \infty\}$ $\subset$ $\R^d\times\S_+^d$ by the nondecreasing monotonicity of $\gamma$ $\mapsto$ $H(x,z,\gamma)$.  
By assuming that $H$ is uniformly continuous in $x$, we notice that the domain $D_f$ does not depend on $x$. 
Since for any $c$ $\in$ $\S_+^d$, there exists a unique  $\mrs$ $\in$ $\S_+^d$ s.t. $c$ $=$ $\mrs^2$, the above duality relation is in the Bellman form \eqref{Hbel} with $a$ $=$ $(b,\mrs)$ $\in$ 
$A$ $=$ $\{(b,\mrs) \in \R^d\times\S_+^d: \mrf(x,b,\mrs^2) > - \infty\}$, $\mrb(x,a)$ $=$ $b$, $\sigma(x,a)$ $=$ $\mrs$, $f(x,a)$ $=$ $\mrf(x,b,\mrs^2)$.

It is well-known that the solution to the PDE \eqref{eq:pde} with an Hamiltonian $H$ as in \eqref{Hbel} admits the stochastic representation:
\bes{
\label{eq:var_form}
    u(t,x) & = \;  \underset{\alpha \in {\cal A} }{\sup}\ \E\Big[ g(X_T^{t,x,\alpha}) +  
    \int_t^T f(X_s^{t,x,\alpha},\alpha_s) \mathrm{d}s \Big], \quad (t,x) \in [0,T] \times \R^d, 
}
where $X$ $=$ $X^{t,x,\alpha}$ is solution to the stochastic differential equation 
\bes{
\label{eq:diffusion}
    \mathrm{d}X_s \; = \;  \mrb(X_s,\alpha_s)  \mathrm{d}s + \sigma(X_s,\alpha_s)  \mathrm{d}W_s, \quad t \leq s \leq T, \;\;\;  
    X_t = x,
}
on a filtered  probability space  $(\Omega,\Fc,\F=(\Fc_t)_{0\leq t\leq T},\P)$ along with a $m$-dimensional Brow\-nian motion $W$, and  the control $\alpha$ $\in$ ${\cal A}$ is a pair of $\F$-progressively measurable processes valued in $A$,   
satisfying  suitable integrability conditions for ensuring under some  Lipschitz assumptions on the coefficients $\mrb$, $\sigma$ that the SDE \eqref{eq:diffusion} admits a unique strong solution. 

\vspace{1mm} 
 
Problem \eqref{eq:var_form} is a standard stochastic control problem with controlled Markov state process $X$ governed by \eqref{eq:diffusion}, and it is well-known that when it exists 
the optimal control $\hat\alpha$ $\in$ $\Ac$ is in closed-loop (or feedback) form, i.e.
\bes{
\hat\alpha_s  &=\; \hat\mra(s,\hat X_s^{t,x}), \quad  t\leq s \leq T, \; (t,x) \in [0,T]\times\R^d, 
}
for some measurable function $\hat\mra$ $:$ $[0,T]\times\R^d$ $\rightarrow$ $A$ $\subset$ $\R^q$, where 
$\hat X$ is the state process controlled by $\hat\alpha$. From the time-consistency of the stochastic control problem \eqref{eq:var_form}, we notice that this feedback form $\hat\mra$ does not depend on the starting point  $(t,x)$ $\in$ $[0,T]\times\R^d$ of the value function. 
Furthermore,  the value function $u$ is given by the conditional expectation 
\bes{ \label{ulin} 
u(t,x) &= \; \E \Big[ g(\hat X_T) +  \int_t^T f(\hat X_s,\hat\mra(s,\hat X_s))  \d s \big| \hat X_t = x \Big]. 
}

Our resolution method for approximating a solution to \eqref{eq:pde} on the whole domain $[0,T]\times\R^d$ 
is performed in two steps: 
\begin{itemize}
\item[1.]  First, following the deep learning approach in \cite{HanE16}, we shall  use neural networks functions  $\mra_\theta$ $:$ $[0,T]\times\R^d$ $\rightarrow$ $A$ $\subset$ $\R^q$, to approximate the optimal feedback control $\hat b$, by maximizing over parameters $\theta$ the objective function
\bes{ \label{defJtheta}
J(\theta) &= \; \E \Big[ g(X_T^\theta) + \int_0^T f(X_t^\theta,\mra_\theta(t,X_t^\theta))  \d t \Big], 
}
where $X^\theta$ solves
\bes{
dX_t^\theta &= \; \mrb(X_t^\theta,\mra_\theta(t,X_t^\theta)) \d t + \sigma(X_t^\theta,\mra_\theta(t,X_t^\theta))  \d W_t, \quad 0 \leq t \leq T,
}
with initial condition $X_0^\theta$ distributed acccording to some law $\mu_0$ on $\R^d$.   
We denote by $\theta^*$ the ``optimal parameter" that maximizes $J(\theta)$, and set $\mra^*$ $=$ $\mra_{\theta^*}$.  
We denote by $X^*$ $=$ $X^{\theta^*}$ an approximation of the optimal state process $\hat X$. For the numerical implementation, we discretize in time the process $X^\theta$ and the integral over $f$ in \eqref{defJtheta}, and apply a stochastic gradient ascent algorithm based on samples of $X^\theta$. 
The pseudo-code is presented in Algorithm \ref{algo:scheme_control_global_method}. 
\item[2.] Once we get an approximation of the optimal feedback control, we could in principle compute $u(t,x)$ from the Feynman-Kac representation \eqref{ulin} by Monte-Carlo simulations of $X^*$. However, with the purpose of solving the PDE \eqref{eq:pde} on the whole domain, this has to be performed for every point $(t,x)$ $\in$ $[0,T]\times\R^d$, which is not feasible in practice.  Instead, we apply 
three types of differential learning  
methods for approximating simultaneously the value function $u$, as well as its derivative:  
(i) the first one, called {\it differential regression learning}, is directly inspired from the original approach  
in \cite{huge2020differential}, and 
gives an approximation of $u$ and its first derivative $D_x u$ from the minimization of two loss functions 
based on least-square regressions,
(ii) the second one in the spirit of \cite{potters2001hedged}, \cite{vidales2018unbiased}, which uses a contingent claim hedging strategy as a Monte Carlo control variate,  
approximates the value function and its first derivative from the minimization of a single loss function based on the martingale representation in \eqref{ulin}, and is refe\-red to as {\it pathwise martingale learning} method. 
(iii) the third one, called {\it pathwise differential learning}, provides in addition an accurate approximation of the second derivative $D_x^2 u$ of $u$. 
We develop these three methods and present their pseudo-codes in the next sections. 
\end{itemize}  
Notice that since the  neural network $\mra^*$ is by nature a suboptimal feedback policy, the approximation  computed in the second step provides a lower bound for the value function $u$ solution to the PDE.

\section{Differential regression  learning}
\label{sec:differential_regression_learning}

From the conditional expectation representation \eqref{ulin}, and its fundamental characterization property as an 
$L^2$-regression, we have  
\bes{
u(t,\hat X_t) &= \; {\rm arg}\min_{v_t}  \E \big| \hat Y_T^t - v_t(\hat X_t) \big|^2 (\hat X_t), \quad \mbox{ for all } t \in [0,T],
}
where the  target payoff  is 
\begin{align}  \label{defYhat} 
\hat Y_T^t &=\;  g(\hat X_T) + \int_t^T f(\hat X_s,\hat\mra(s,\hat X_s)) \d s, \quad t \in [0,T], 
\end{align}
and the argmin is taken over measurable real-valued functions $v_t$ on $\R^d$ s.t. $v_t(\hat X_t)$ is square-integrable.  

This suggests to use a class of neural networks (NN) functions $\vartheta^\eta$ on $[0,T]\times\R^d$, with parameters $\eta$, for approximating the value function $u$, and a loss function
\begin{align}
\hat L_{val}(\eta) &= \; \E \Big[ \int_0^T \big| \hat Y_T^t - \vartheta^\eta(t,\hat X_t) \big|^2 \d t \Big] \nonumber \\
& \simeq \; \E \Big[ \int_0^T \big| Y_T^{*,t}  - \vartheta^\eta(t,X_t^*) \big|^2 \d t \Big] \; = : L^*_{val}(\eta), \label{defLval} 
\end{align} 
where 
\begin{align}\label{defYstar} 
Y_T^{*,t} &= \; g(X_T^*) +  \int_t^T f(X_s^*,\mra^*(s,X_{s}^*)) \d s, \quad t \in [0,T].   
\end{align} 
As pointed out in \cite{huge2020differential}, the training of the loss function $L_{val}^*$ in \eqref{defLval} 
would require a vast number of samples (often of order millions) to learn accurate approximation of the value function, and is furthermore 
prone to overfitting. Indeed, by training a neural network to minimize $L_{val}^*$, we would obtain a function which interpolates the random points generated during training. This comes with two shortcomings. First, a large number of training samples is needed to get satisfactory values of the solution and a good enough generalisation to untrained domains. Second, the functions obtained by this method are usually noisy. If we are interested in the derivatives of the PDE solution, as it is the case in finance for example, where \textit{greeks} are computed in order to hedge contingent claims, the solution computed might not be accurate enough. Some standard methods, such as Ridge and Lasso penalisations allow to reduce overfitting but come at the cost of adding bias and an arbitrary penalty and do not ensure that the derivative of the network will be a good approximation of the derivative of the PDE solution. To circumvent these issues, and following the idea in \cite{huge2020differential}, we propose to consider furthermore the learning of the derivative of the value function. This method relies on pathwise differentiation of the target payoff $\hat Y_T^t$ for deriving the gradient of the value function (see Chapter 7 in \cite{glasserman2013monte}):
\begin{align} \label{repDxv} 
D_x  u(t, \hat X_t) &= \; \E \Big[ 
\hat Z_T^t
\big| \Fc_t \Big], \quad t \in [0,T],  
\end{align} 
where $\hat Z_T^t$ $=$ $D_{\hat X_t} \hat Y_T^t$. 
This suggests to complete the learning of the value function together with its derivative by considering furthermore the loss function
\begin{align}
\hat L_{der}(\eta) &= \; \E \Big[  \int_0^T \big|  \hat Z_T^t - D_x \vartheta^\eta(t,\hat X_t) \big|^2 \d t  \Big] \nonumber \\
& \simeq \; \E \Big[ \int_0^T    \big| Z_T^{*,t}  - D_x \vartheta^\eta(t,X_{t}^*) \big|^2 \d t  \Big]  \; =: \; L^*_{der}(\eta), \label{defLder} 
\end{align}
where $Z_T^{*,t}$ $=$ $D_{X_{t}^*} Y_T^{*,t}$ valued in $\R^d$, 
is obtained by automatic differentiation as
\begin{align} \label{derZ} 
Z_T^{*,t} &= \;  \big(D_{X_{t}}X_T\big)^\top  D_x g(X_T)  
+ \int_t^T  \big(D_{X_{t}}X_s\big)^\top  D_x f^{\mra^*}(s,X_s) \d s, \quad t \in [0,T],   
\end{align} 
where we denote by $f^{\mra^*}(t,x)$ $=$ $f(x,\mra^*(t,x))$, and assuming that $g$ and $f$ are continuously differentiable.  Notice that $D_x f^{\mra^*}$ $=$ $D_x f$ $+$ $(D_x\mra^*)^\top D_a f$, where  
the derivatives $D_x \mra^*$ of the approximate optimal feedback control in the class of neural networks can be efficiently computed by automatic differentiation.
Here, to alleviate notations, we have dropped the superscript $*$ for the state $X$ $=$ $X^*$. Actually, when $g$ and $f$ are only piecewise-differentiable, the above relation still holds when the 
marginal law of  $X_s$ is absolutely continuous with respect to Lebesgue measure on $\R^d$, which is satisfied under nondegeneracy conditions on the diffusion coefficients (see Theorem 2.3.2 in \cite{nua95}). 
We recall that the flow derivative of the 
optimal state process, 
valued in $\R^{d\times d}$, is solution to  the  SDE (see e.g.  \cite{protter2005stochastic}) 
\begin{align} \label{derivflow} 
D_{X_t} X_s &=   I_d + \int_t^s D_x \mrb^{\mra^*}(r,X_r) D_{X_t} X_r \d r + D_x \sigma_j^{\mra^*}(r,X_r) D_{X_t} X_r \d W^j_r,   \;   t \leq s \leq T, 
\end{align}
where we denote by $\mrb^{\mra^*}(t,x)$ $=$ $\mrb(x,\mra^*(t,x))$, $\sigma^{\mra^*}(t,x)$ $=$ $\sigma(x,\mra^*(t,x))$, and  use the Einstein summation convention over the repeated index $j$ $=$ $1,\ldots,d$, with $\sigma_j^{\mra^*}$ (resp. $\sigma_j$)  the $j$-th column of the matrix $\sigma^{\mra^*}$ (resp. $\sigma$).  
Notice that $D_x\mrb^{\mra^*}$ $=$ $D_x\mrb$ $+$ $D_a\mrb D_x \mra^*$, and $D_x\sigma_j^{\mra^*}$ $=$ $D_x\sigma_j$ $+$ $D_a\sigma_j D_x \mra^*$. 

\begin{remark}
We have an alternative representation \eqref{repDxv}  for the gradient of $v$, which avoids smoothness assumptions on the coefficients. It is expressed with $\hat Z_T^t$ given by  (see \cite{mazha02}): 
\begin{align} \label{Zmalliavin} 
\hat Z_T^t &= \; g(\hat X_T) \hat H_T^t + \int_t^T f(\hat X_s,\hat\mra(s,\hat X_s)) \hat H_s^t \d s, \quad t \in [0,T],
\end{align} 
with the so-called Malliavin weights $\hat H_s^t$, $t\leq s$, given by
\bes{
H_s^t &= \; \frac{1}{s-t} \int_t^s \sigma^{-1}(t,\hat X_t)^\top \sigma^{-1}(r,\hat X_r) D_{\hat X_t}  \hat X_r  \sigma(t,\hat X_t) \d W_r, 
}
where $\sigma^{-1}$ $=$ $\sigma^\top(\sigma\sigma^\top)^{-1}$ is the right-inverse of the matrix $\sigma$ assumed to be of full rank. Therefore, in the loss function $L_{der}^*$, instead of  $Z_T^{*,t}$ as in \eqref{derZ}, we can use alternately $Z_T^{*,t}$ as in \eqref{Zmalliavin}, with $\hat X$ approximated by $X^*$, and 
$\hat\mra$ approximated by $\mra^*$. 
\end{remark}

\begin{figure}[h]
    \centering
    \begin{subfigure}{.49\linewidth}
        \centering
        \includegraphics[height=5.5cm]{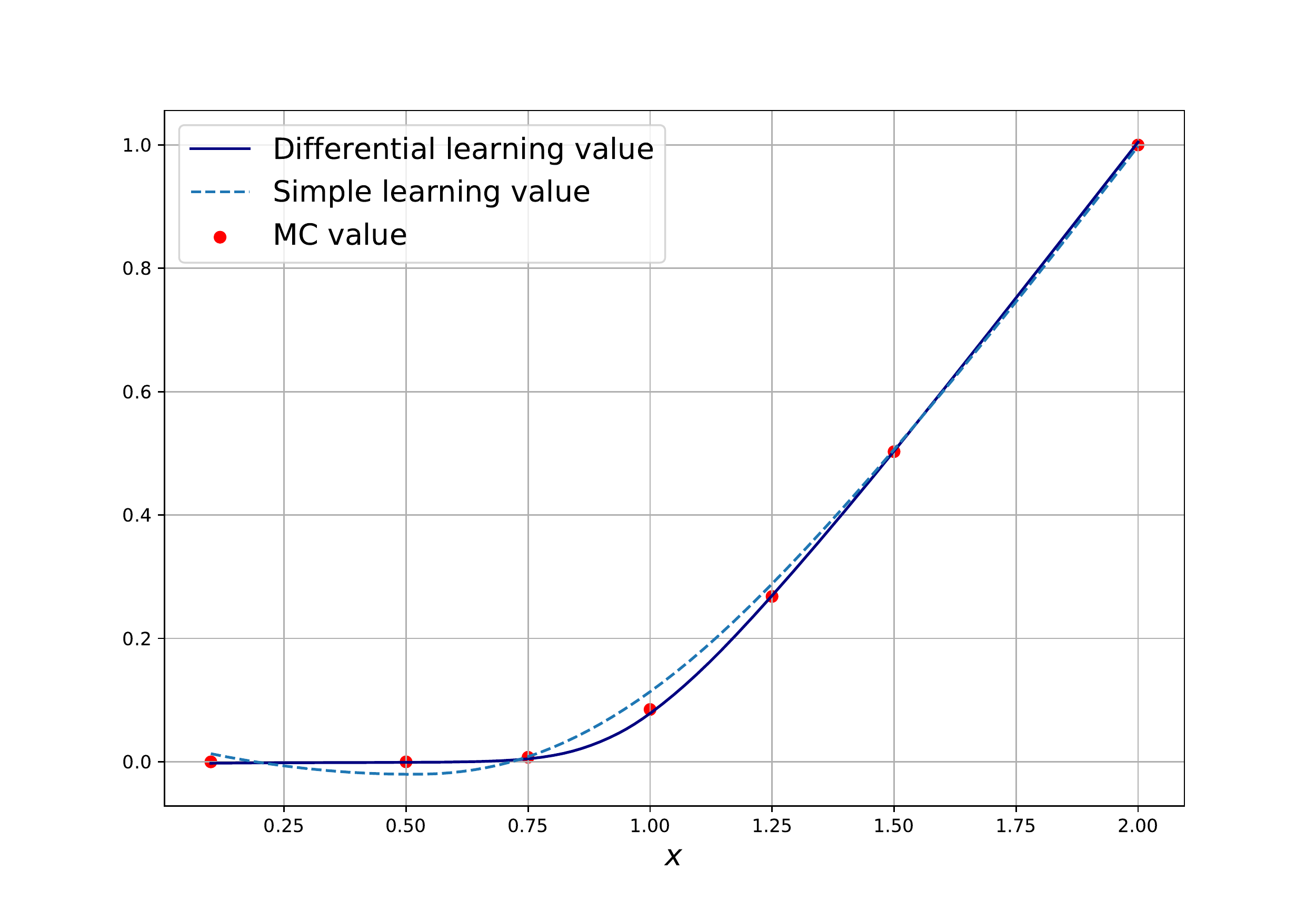}
    \end{subfigure}
    \begin{subfigure}{.49\linewidth}
        \centering
        \includegraphics[height=5.5cm]{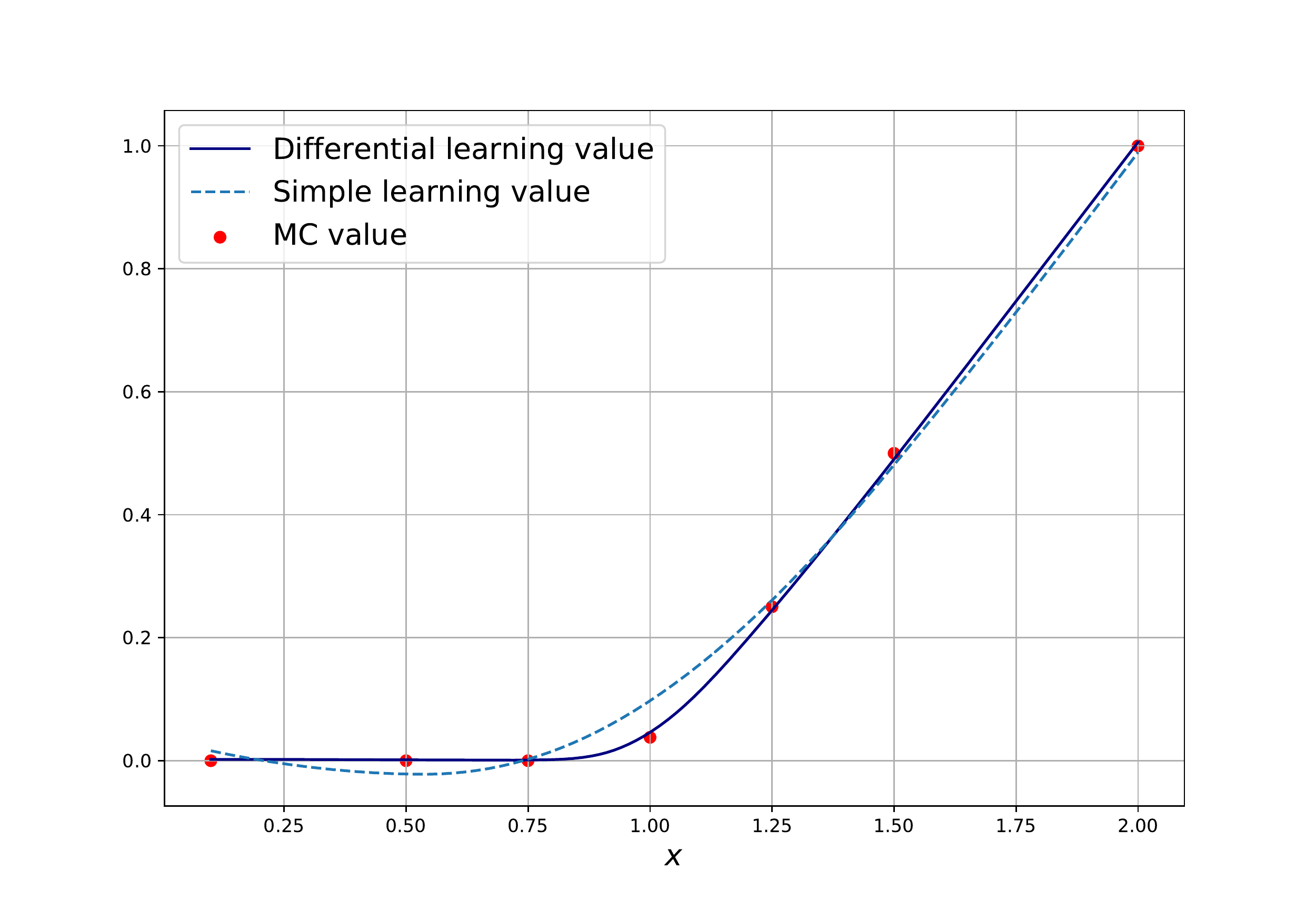}
    \end{subfigure}
    \begin{subfigure}{.49\linewidth}
        \centering
        \includegraphics[height=5.5cm]{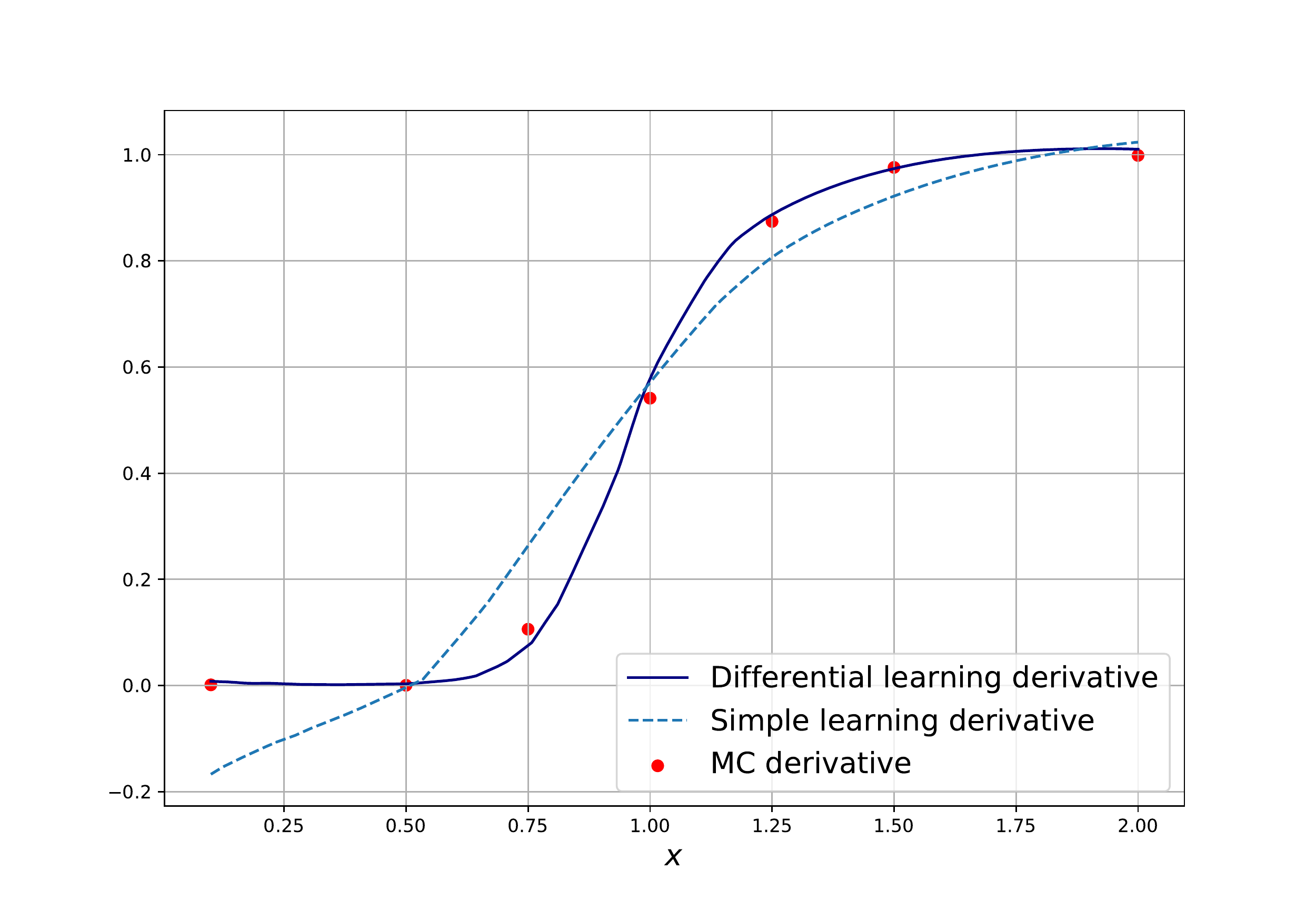}
        \caption[short]{$t=0.5$}
    \end{subfigure}
    \begin{subfigure}{.49\linewidth}
        \centering
        \includegraphics[height=5.5cm]{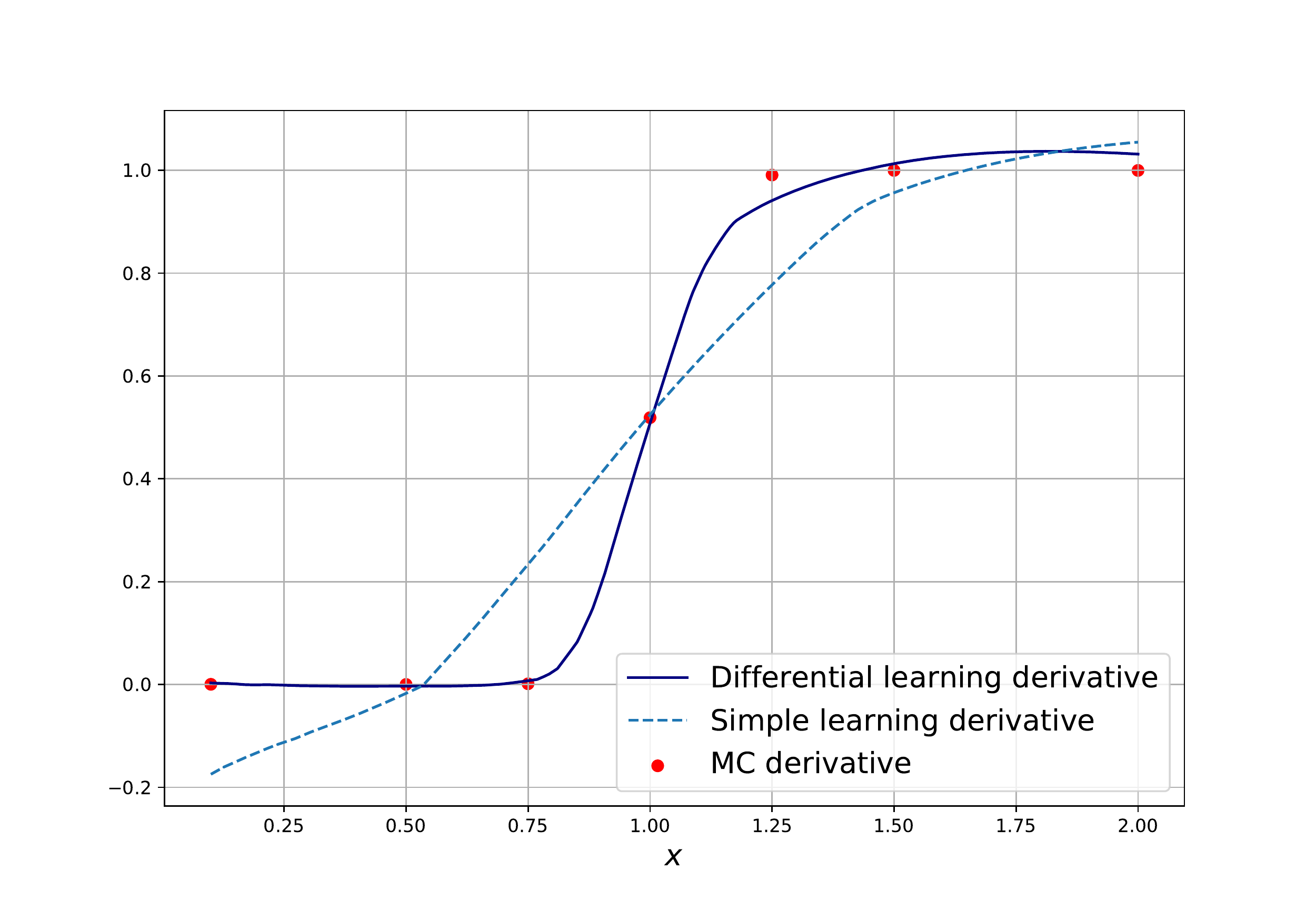}
        \caption[short]{$t=0.9$}
    \end{subfigure}
    \caption{
    \label{fig:comparison_diff_simple_learning_call}
    Value function values (first line) and derivatives (second line) obtained by \textit{Differential Learning} (navy curve) (Algorithm \ref{algo:scheme_value_differential_learning}), \textit{Simple learning} (blue dashed curve) and Monte Carlo (red dots) plotted as functions of $x$, for fixed values of $t$.
    }
\end{figure}

To illustrate the interest of learning the derivative of the value function, we plot in Figure \ref{fig:comparison_diff_simple_learning_call} the results obtained by learning the value and the derivative (\textit{Differential regression learning}) or by learning just the value (\textit{Simple learning}) of the call option price with market impact (see the application  presented in Section \ref{sec:BS_linear_market_impact}). 

As a reference, we compute the option price on chosen points $(t,x)\in \R_+ \times \R$ by Monte-Carlo, as explained in section \ref{sec:validation_tests}.

In the differential regression learning method, we use a combination of the loss $L^*_{val}$ and $L^*_{der}$ in the training of the neural networks for approximating the value function and its derivative.   We train alternately the value and the derivative of the network by taking a gradient step to minimize $L^*_{val}$ every even number of epochs and a gradient step to minimize $L^*_{der}$ every odd number of epochs. An alternative method would be to minimize a convex combination of $L^*_{val}$ and $L^*_{der}$ with weights $w_{val}$ and $w_{der} = 1 - w_{val}$. These weights could be chosen by performing a grid search or a random search, such as advocated in \cite{bergstra2012random} for the choice of neural network hyperparameters. Some theoretical arguments could also be derived in order to choose these weights, as it has been done in \cite{van2021optimally} to optimally choose the weights of the losses associated to different constraints of a PDE when using Physics Informed Neural Networks \cite{raissi2019physics}. Our approach proves to be effective, as shown by the numerical results in Sections \ref{sec:Merton} and \ref{sec:BS_linear_market_impact}, and avoids the need to chose a value for this additional hyperparameter.  The algorithmic implementation and the pseudo-codes are described in Section \ref{section:numerical_differential_learning}.

\section{Pathwise learning}
\label{sec:pathwise_learing}

\subsection{Pathwise martingale learning}

This approach is based on the martingale representation related to relation \eqref{ulin}, which leads by It\^o's formula to  the equation:
\begin{align}\label{marIto} 
\hat Y_T^t  &= \;  u(t,\hat X_t)  + \int_t^T  \big( D_x u(s,\hat X_s) \big)^\top \sigma^{\hat\mra}(s,\hat X_s) \d W_s, \quad t \in [0,T], 
\end{align} 
where we recall that $\hat Y_T^t$ is given in \eqref{defYhat}, and denote $\sigma^{\hat\mra}(t,x)$ $=$ $\sigma(x,\hat\mra(t,x))$.  
This suggests to use a class of neural networks (NN) functions $\vartheta^\eta$ on $[0,T]\times\R^d$, with parameters $\eta$, for approximating the value function $u$, and a loss function
\bes{
\hat L_{mar}(\eta) & = \; \E \Big[ \int_0^T \big| \hat Y_T^t - \vartheta^\eta(t,\hat X_t) -  \int_t^T  \big( D_x \vartheta^\eta(s,\hat X_s) \big)^\top \sigma^{\hat\mra}(s,\hat X_s) \d W_s \big|^2  \d t \Big]  \\
& \simeq \;  \E \Big[ \int_0^T \big| Y_T^{*,t}  - \vartheta^\eta(t,X_t^*) -  \int_t^T  \big( D_x \vartheta^\eta(s,X_s^*) \big)^\top \sigma^{\mra^*}(s,X_s^*) \d W_s \big|^2  \d t \Big] \\
& = : \; L^*_{mar}(\eta). 
}
Alternately, we can use two classes of neural networks: one $\vartheta^\eta$ from $[0,T]\times\R^d$ into $\R$, with parameters $\eta$,  for the approximation of $u$, and a second one $\Zcal^\delta$ from  $[0,T]\times\R^d$ into $\R^d$, with parameters $\delta$, for the approximation of $D_x u$. 
We then  consider a loss function
\bes{
\tilde L^*_{mar}(\eta,\delta) & := \;   \E \Big[ \int_0^T \big| Y_T^{*,t}  - \vartheta^\eta(t,X_t^*) -  \int_t^T   \Zcal^\delta(s,X_s^*)^\top \sigma^{\mra^*}(s,X_s^*) \d W_s \big|^2  \d t \Big].  
}
Notice that compared to the deep BSDE approach in \cite{HJE17}, which considers a loss function from the misfit between the l.h.s (the target)  and r.h.s.  of  \eqref{marIto} at time $0$, namely
\bes{
\tilde L^*_{DBSDE}(y_0,\delta) & := \;   \E \Big[  \big| Y_T^{*,0}  - y_0 -  \int_0^T   \Zcal^\delta(s,X_s^*)^\top \sigma^{\mra^*}(s,X_s^*) \d W_s \big|^2  \Big],  
}
our loss functions $L^*_{mar}$ or $\tilde L^*_{mar}$  take into account the misfit between the l.h.s and r.h.s.  of  \eqref{marIto} at any time $t$ $\in$ $[0,T]$, since 
our goal is to approximate the solution $u$ (and its derivative) on the whole domain $[0,T]\times\R^d$ (and not only at time $t$ $=$ $0$).

\subsection{Pathwise differential learning}\label{sec:pathwise_differential_learning}

We can further compute the pathwise derivative in the martingale representation relation \eqref{marIto} in order to obtain  a second estimator linking the first and second derivatives of $u(t,x)$.  
Indeed, by \cite{elkquepen97}, we have 
\bes{
D_{\hat X_t} \hat Y_T^t &= \;  D_x u(t,\hat X_t) 
+ \int_t^T \Big( \big[ D_x \sigma^{\hat\mra}(s,\hat X_s) \bullet_3 D_{\hat X_t} \hat X_s  \big] \bullet_1 D_x u(s,\hat X_s) \\
&  \hspace{3.5cm} + \; \sigma^{\hat\mra}(s,\hat X_s)^\top D_{x}^2 u(s,\hat X_s) D_{\hat X_t} \hat X_s \Big)^\top \d W_s, \quad t \in [0,T].  
}

This suggests to use a class of neural networks (NN) functions $\vartheta^\eta$ on $[0,T]\times\R^d$, with parameters $\eta$, for approximating the value function $u$, and a loss function 
\bes{
L^*_{dermar}(\eta) &= \; \E \Big[ \int_0^T \Big| Z_T^{*,t} -  D_x \vartheta^\eta(t,X_t) 
-  \int_t^T \Big( \big[ D_x \sigma^{\mra^*}(s,X_s) \bullet_3 D_{X_t}  X_s  \big] \bullet_1 D_x \vartheta^\eta(s,X_s) \\
&  \hspace{3cm} + \; \sigma^{\mra^*}(s,X_s)^\top D_{x}^2 \vartheta^\eta(s,X_s) D_{X_t} X_s \Big)^\top \d W_s \Big|^2 \d t \Big],  
}
where we recall that $Z_T^{*,t}$ is given in \eqref{derZ}, and we omit the superscript $*$  in the approximation of  the optimal state process $X$ $=$ $X^*$. 
Alternatively, we can use three classes of neural networks: one $\vartheta^\eta$ from $[0,T]\times\R^d$ into $\R$, with parameters $\eta$,  for the approximation of $u$,  a second one $\Zcal^\delta$ from  $[0,T]\times\R^d$ into $\R^d$, with parameters $\delta$, for the approximation of $D_x u$, and a third one 
$\Gamma^\epsilon$ from  $[0,T]\times\R^d$ into $\S^{d} $, with parameters $\epsilon$, for the approximation of $D_{xx} u$, and consider a loss function 
\bes{
\tilde L^*_{dermar}(\delta,\epsilon) &= \; \E \Big[ \int_0^T \Big| Z_T^{*,t} -  \Zcal^\delta(t,X_t) -  \int_t^T \Big( \big[ D_x \sigma^{\mra^*}(s,X_s) \bullet_3 D_{X_t}  X_s  \big] \bullet_1  \Zcal^\delta(s,X_s) \\
&  \hspace{3cm} + \; \sigma^{\mra^*}(s,X_s)^\top \Gamma^\epsilon(s,X_s) D_{X_t} X_s \Big)^\top \d W_s \Big|^2 \d t \Big].  
}
As with the Differential regression learning method presented in Section \ref{sec:differential_regression_learning}, the neural network can be trained either by minimising a convex combination of the losses $L_{mar}^*$ and $L_{dermar}^*$ or by minimising these losses individually. During our numerical experiment, we minimised these two losses individually, but contrary to the algorithm used in the Differential regression learning, for each epoch, a gradient step was made to minimise $L_{mar}^*$ and then another one was made to minimize $L_{dermar}^*$. 

\begin{remark}
This optimisation scheme was found to be more effective when optimising the neural network parameters for the Pathwise differential learning. For the Differential regression learning method, making a gradient step on only one of these two losses at each epoch gave better results. The difference between these two optimization schemes lies in the fact that if both the losses $L_{mar}^*$ and $L_{dermar}^*$ are optimised during an epoch, these two losses are computed using the "old" network weights $\eta$, then these weights are modified two times, first by making a gradient step to minimize $L_{mar}^*$, and then by taking another gradient step to minimize $L_{dermar}^*$, as written below
\bes{
\textrm{One epoch:}\\
\eta^{'} &\leftarrow \eta - \nabla_{\eta} L_{mar}^* (\eta),\\
\eta^{''} &\leftarrow \eta^{'} - \nabla_{\eta} L_{dermar}^* (\eta).
}
When the network is optimised by minimizing alternatively one of these two losses at each epoch, as in the Differential regression method, one of the losses, say $L_{mar}^*$, is computed using the "old" weights $\eta$. A gradient step is then made to minimize this loss and obtain new network weights $\eta^{'}$, which are then used in the next epoch to compute the loss $L_{mar}^*$ and make the next gradient step, as written below
\bes{
\textrm{One epoch:}\\
\eta^{'} &\leftarrow \eta - \nabla_{\eta} L_{mar}^* (\eta),\\
\textrm{Next epoch:}\\
\eta^{''} &\leftarrow \eta^{'} - \nabla_{\eta} L_{dermar}^* (\eta^{'}).
}
\end{remark}

\section{Numerical results}
\label{sec:numerical_results}

In this section, we detail  the implementation of the methods presented above.  Our algorithms are discretised in time  for the training of the processes and for the integrals that appear in the loss functions, and which are approximated by Riemann sums.  In the sequel, we are then given a mesh grid 
$\Tc_N$ $=$ $\{0=t_0 < t_1 < \ldots < t_N= T\}$ of $[0,T]$ with $\Delta t_n$ $=$ $t_{n+1}-t_n$, for $n$ $=$ $0,\ldots,N-1$.  

Our codes are  written in Python and we use the Tensorflow library to implement the neural networks and compute the derivatives present in our calculations by auto-differentiation (AAD). 

Finally, we illustrate our results with some examples of applications in finance.

\subsection{Approximation of the optimal control}
\label{sec:numerical_approximation optimal control}

As a first step, we consider a neural network $\mra_\theta$ from $[0,T]\times\R^d$ into $A$ $\subset$ $\R^q$ for the approximation of the feedback control, and the associated discretised state process
\bes{
X^\theta_{t_{n+1}} & = \;  X^\theta_{t_{n}}  +  \mrb(X^\theta_{t_{n}},\mra_\theta(t_n,X^\theta_{t_{n}})) \Delta t  +  \sigma(X^\theta_{t_{n}},\mra_\theta(t_n,X^\theta_{t_{n}})) \Delta W_{t_n}, \quad n=0,\ldots,N-1,
} 
starting from $X_0$ $\sim$ $\mu_0$ (probability distribution on $\R^d$), and where $\Delta W_{t_n}$ $=$ $W_{t_{n+1}}-W_{t_n}$. 
As in \cite{hure2021deep}, to constrain the output of the neural network $a_\theta$ to be in the control space $A$, we define a custom activation function $\sigma_A$ for the output layer of the network. This activation function $\sigma_A$ is chosen depending on the form of the control space $A$. When $A=\R^q$, $\sigma_A$ is equal to the identity function. When the control space if of the form $A= \prod_{i=1}^q [a_i, \infty)$, one can take the component-wise ReLU activation function (possibly shifted and scaled); when $A=\prod_{i=1}^q [a_i, b_i]$, for $a_i \leq b_i$, $i=1,...,q$, one can take the component-wise sigmoid activation function (possibly shifted and scaled).
For the numerical experiments presented here, we used the ELU (Exponential Linear Unit) activation function, defined as
\bes{
    ELU(x) = 
    \begin{cases}
        \hspace{0.5cm} x & x>0\\
        \alpha (e^x -1) & x \leq 0,
    \end{cases}
}
with parameter $\alpha$ for the hidden layers.

\begin{figure}[h!]
    \centering
    \includegraphics[width=0.6\linewidth]{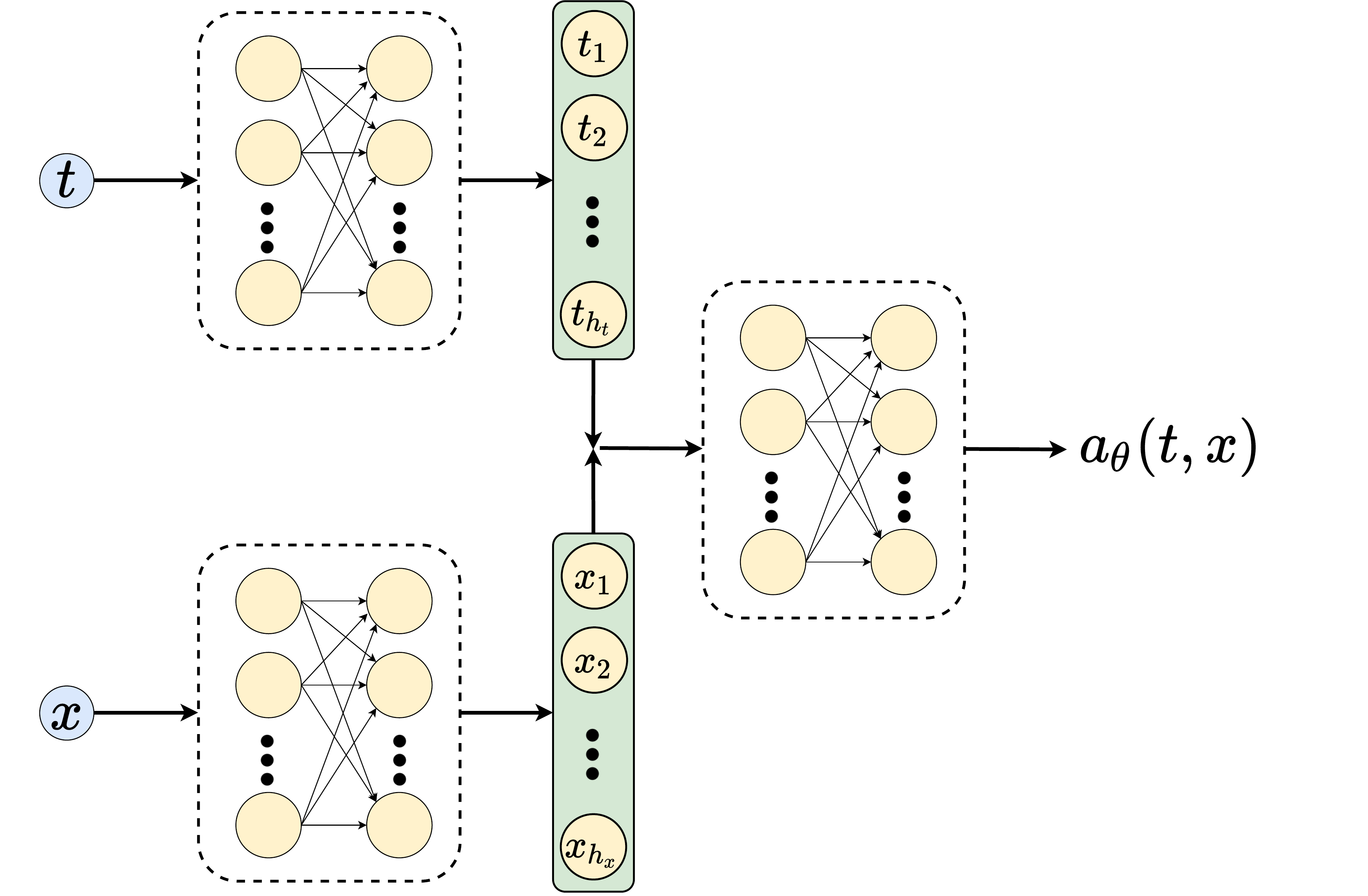}
\caption{Structure of the neural network used to approximate the optimal control and value function.}
\label{fig:structure_network}
\end{figure}

The structure of the neural networks used to approximate the feedback control is represented in  Figure \ref{fig:structure_network}. It is composed of two dense feed-forward sub-networks composed of two layers of $n$ neurons, taking respectively the time $t$ and the state $x$ as input. The outputs of these two sub-network, which are in $\R^{n}$, are then concatenated and inputed in a third dense network composed of two layers of $n$ neurons and a last layer of $q$ neurons which outputs the approximation of the control in $\R^q$. This structure adds more flexibility compared to the network structure usually implemented, where the time and state variables are directly concatenated and passed through a dense feed-forward network. It allows to use different activation functions in each sub-network and adapts well to situations where the network is used to approximate a function which has very different behaviors in its time and state variables. The same structure is used for the neural networks used to approximate the value function by Differential Regression learning or Pathwise learning. In the applications presented in this article, we used neural networks with $n=50$ neurons per layer and $q=1$.

For the training of the neural network control $\mra_\theta$, we use a batch of $M$ independent trajectories $\{x_{t_n}^{m,\theta}, t_n \in \Tc_N\}$, $m$ $=$ $1,\ldots,M$,  of $\{X_{t_n}^\theta, t_n \in \Tc_N\}$, and apply 
a stochastic gradient ascent method to the empirical gain function:
\bes{
J_M(\theta) &= \; \frac{1}{M} \sum_{m=1}^M \Big[ g(x_T^{m,\theta}) +  \sum_{n=0}^{N-1} f\big(x_{t_n}^{m,\theta},\mra_\theta(t_n,x_{t_n}^{m,\theta})\big) \Delta t_n \Big].  
}
The pseudo-code is described in  Algorithm \ref{algo:scheme_control_global_method}.  
The output of this algorithm yields a parameter $\theta^*$, and so 
an approximation of the optimal feedback control with  $\mra^*$ $=$ $\mra_{\theta^*}$, and of the associated optimal state process with $X^*$ $=$ $X^{\theta^*}$. 
 In the sequel, to alleviate notations, we shall omit the superscript $*$, and simply denote $\mra$ and $X$. 

\begin{algorithm}
\scriptsize
\SetAlgoLined
\KwResult{A set of optimized parameters $\theta^*$;}
 Initialize the learning rate $l$ and the neural network $\mra_\theta$\; 
 Generate an $\R^{N+1}$-valued time grid $0=t_0 < t_1 < ... < t_N = T$ with time steps $(\Delta t_n)_{n=0,...,N-1}$\;
 Generate a batch of $M$ starting points $X_0$ $\sim$ $\mu_0$ and Brownian increments $(\Delta W_{t_n})_{n=0,...,N-1}$ in $R^d$\;
 \For{each batch element $m$}{
        Compute the trajectory $(x_{t_n}^{m, \theta})_{n=0,...,N}$ through the scheme
        \bes{
            x_{t_{n+1}}^{m, \theta} & = \;  x_{t_{n}}^{m, \theta}  +  \mrb(x_{t_{n}}^{m, \theta},\mra_\theta(t_n,x_{t_{n}}^{m, \theta})) \Delta t_n  +  \sigma(x_{t_{n}}^{m, \theta},\mra_\theta(t_n,x_{t_{n}}^{m, \theta})) \Delta w_{t_n}^m,
        }
        from the generated starting point $x_{t_0}^m$ and Brownian increments $(\Delta w_{t_n}^m)_{n=0,...,N-1}$\; 
    } 
 \For{each epoch}{
    Compute the batch loss 
        \bes{
        J_M(\theta) &= \; \frac{1}{M} \sum_{m=1}^M \Big[ g(x_T^{m,\theta}) +  \sum_{n=0}^{N-1} f\big(x_{t_n}^{m,\theta},\mra_\theta(t_n,x_{t_n}^{m,\theta})\big) \Delta t_n \Big]
        }
    Compute the gradients $\nabla_{\theta} J_M(\theta)$\;
    Update $\theta \leftarrow \theta - l \nabla_{\theta} J_M(\theta)$\;
}
\textbf{Return:} The set of  optimized parameters $\theta^*$\;
\caption{Deep learning scheme to solve the stochastic control problem \eqref{eq:var_form}}
\label{algo:scheme_control_global_method}
\end{algorithm}

\subsection{Differential regression  learning algorithm}
\label{section:numerical_differential_learning}

We consider a neural network $\vartheta^\eta$ from $[0,T]\times\R^d$ into $\R$ for the approximation of the value function.  The derivatives $D_x g$, $D_x f$,  $D_x\mrb$, $D_x \sigma$, $D_x\mra$, and 
$D_x \vartheta^\eta$  that appear in the differential regression learning methods are computed straightforwardly by auto-differentiation.  Concerning the flow derivative of the approximate optimal state process, it is computed by time discretization of 
\eqref{derivflow}, which can be efficiently obtained by storing  the one-step derivatives: 
\begin{eqnarray*}
D_{X_{t_n}} X_{t_{n+1}} &=& I_d + D_x b^{\mra^*}(t_n,X_{t_n}) \Delta t_n +  \sum_{j=1}^d D_x \sigma_j^{\mra^*}(t_n,X_{t_n}) \Delta W^j_{t_n}, \quad n=0,\ldots,N-1, 
\end{eqnarray*}
and then use the chain rule
\begin{eqnarray*}
D_{X_{t_n}} X_{t_{p}} &=&
D_{X_{t_n}} X_{t_{n+1}} \cdots D_{X_{t_{p-1}}} X_{t_p}, \quad  \mbox{ for } n <  p \in \llbracket 0, N\rrbracket 
\end{eqnarray*}
 
The target payoff and its derivative are then computed as 
\bes{
Y_T^{t_n} &= \; g(X_T) +  \sum_{p=n}^{N-1} f^{\mra^*}(t_p,X_{t_p}) \Delta t_p,  \quad n=0,\ldots,N,  \\
Z_T^{t_n} &= \;  \big(D_{X_{t_n}}X_T\big)^\top  D_x g(X_T)   \; + \;  \sum_{p=n}^{N-1}   \big(D_{X_{t_n}}X_{t_p}\big)^\top  D_x f^{\mra^*}(t_p,X_{t_p})  \Delta t_p,
}
with the convention that the above sum over $p$ is zero when $n$ $=$ $N$.

For the training of the neural network $\vartheta^\eta$, we use a batch  of $M$ independent samples $(x_{t_n}^m,y_T^{m,t_n}, z_T^{m,t_n})$, $m$ $=$ $1,\ldots,M$,  of $(X_{t_n},Y_T^{t_n},Z_T^{t_n})$, $n$ $=$ 
$0,\ldots,N$, and apply stochastic gradient descent for the minimization of the mean squared error functions
\bes{\label{eq:MSE_differential_learning}
MSE_{val}(\eta) &= \; \frac{1}{M} \sum_{m=1}^M \sum_{n=0}^{N-1} \big| y_T^{m,t_n} - \vartheta^\eta(t_n,x_{t_n}^m) \big|^2 \Delta t_n  \\
MSE_{der}(\eta) & = \; \frac{1}{M} \sum_{m=1}^M \sum_{n=0}^{N-1} \frac{1}{\| z_T^{t_n} \|^2}  \big| z_T^{m,t_n} - D_x \vartheta^\eta(t_n,x_{t_n}^m) \big|^2 \Delta t_n. 
}

Here, as in \cite{huge2020differential} (Appendix 2), we normalize the derivative loss by the $L_2$ norm of the target derivative computed along the batch dimension $\| z_T^{t_n} \|^2 := \big( \sum_{m=1}^M \sqrt{z_T^{t_n}}\big)^2$.

The pseudo-code is described in  Algorithm \ref{algo:scheme_value_differential_learning}.

\begin{remark}
Notice that in the above  expressions of the mean squared errors, the neural network is trained from the time $t_0 = 0$ up to time $t_{N-1} < T$ and is thus not trained on the terminal condition of the PDE \eqref{eq:pde}. This choice is justified by the fact that the neural network is already implicitly trained to fit the terminal condition at time $T$ since the terminal function $g$ appears in the losses. Furthermore, we observed that the regularity of the terminal function affects the performance of the neural network. If the terminal function is at least of class $C^1$, no problem arises as the regularity of the solution to the  PDE \eqref{eq:pde} is the same on the domain $[0,T) \times \R^d$ and on the terminal domain $\{T\} \times \R^d$. If we use a neural network $\vartheta^\eta$ of regularity $C^1$, we will then be able to approximate the PDE solution on the entire domain $[0,T] \times \R^d$.
However, if the terminal condition's regularity is less than $C^1$, it will be difficult for the neural network to approximate the PDE solution on the entire domain $[0,T] \times \R^d$. Indeed, the solution of parabolic PDEs is often smoother than its terminal (or initial) condition, thus the neural network will have to approximate a function that has a continuous first derivative on the domain $[0,T) \times \R^d$ and a discontinuous first derivative on $\{T\} \times \R^d$. If the neural network used is not $C^1$, which is the case for a network with ReLU activation functions for example, the network will give a good approximation of the terminal condition but will give a worst fit of the solution on the domain $[0,T) \times \R^d$, and particularly of its derivatives. On the contrary, if the neural network used is $C^1$, which is the case when the ELU activation function is used, the network will give a good approximation of the solution on $[0,T) \times \R^d$ but will give a worse approximation of the terminal condition. The difficulty thus comes from the fact that we try to obtain a solution on the entire domain of the PDE. As the terminal condition is known and the quantity of interest is the solution of the PDE \eqref{eq:pde} on the domain $[0,T) \times \R^d$, we choose to use a $C^1$ neural network trained on this domain.
\end{remark}

\begin{algorithm}[H] 
\scriptsize
\SetAlgoLined
\KwResult{A set of optimized parameters $\eta^*$;}
 Initialize the learning rate $l$, the neural networks $\vartheta^\eta$\; 
 Generate an $\R^{N+1}$-valued time grid $0=t_0 < t_1 < ... < t_N = T$ with time steps $(\Delta t_n)_{n=0,...,N-1}$\;
 Generate a batch of $M$ starting points $X_0$ $\sim$ $\mu_0$ and Brownian increments $(\Delta W_{t_n})_{n=0,...,N}$ in $R^d$\;
 \For{each batch element $m$}{
    Compute the trajectory $(x_{t_n}^{m})_{n=0,...,N}$ through the scheme
        \bes{
            x_{t_{n+1}}^{m} & = \;  x_{t_{n}}^{m}  +  \mrb^{\mra^*}(t_n,x_{t_{n}}^{m}) \Delta t_n  +  \sigma^{\mra^*}(t_n,x_{t_{n}}^{m}) \Delta w_{t_n}^m,
        }
        from the generated starting point $x_{t_0}^m$, Brownian increments $(\Delta w_{t_n}^m)_{n=0,...,N-1}$ and previously trained control $a=a_{\theta^*}$\;
    Compute the value and derivative targets $(y_T^{m, t_n})_{n=0,...,N}$ and $(z_T^{m, t_n})_{n=0,...,N}$\;
}
 \For{each epoch}{
    \If{Epoch number is even}{
        Compute, for every batch element $m$, the integral $\sum_{n=0}^{N-1} \big| y_T^{m,t_n} - \vartheta^\eta(t_n,x_{t_n}^m) \big|^2 \Delta t_n$\;
        Compute the batch loss $MSE_{val}(\eta)$\;
        Compute the gradient $\nabla_{\eta} MSE_{val}(\eta)$\;
        Update $\eta \leftarrow \eta - l \nabla_{\eta} MSE_{val}(\eta)$\;
    }
    \Else{
        Compute, for every batch element $m$, the integral $\sum_{n=0}^{N-1} \big| z_T^{m,t_n} - D_x \vartheta^\eta(t_n,x_{t_n}^m) \big|^2 \Delta t_n$.\;
        Compute the batch loss $MSE_{der}(\eta)$\;
        Compute the gradient $\nabla_{\eta} MSE_{der}(\eta)$\;
        Update $\eta \leftarrow \eta - l \nabla_{\eta} MSE_{der}(\eta)$\;
    }
}
\textbf{Return:} The set of  optimized parameters  $\eta^*$\;
\caption{Deep learning scheme for Differential Regression learning}
\label{algo:scheme_value_differential_learning}
\end{algorithm}

\subsection{Pathwise learning algorithms}
\label{section:numerical_pathwise_learning}

We consider a neural network $\vartheta^\eta$ from $[0,T]\times\R^d$ into $\R$  for the approximation of the value function. For the training of this neural network, we use a batch  of $M$ independent samples $(x_{t_n}^m,y_T^{m,t_n},\Delta w_{t_n}^m)$, $m$ $=$ $1,\ldots,M$,  of $(X_{t_n},Y_T^{t_n},\Delta W_{t_n})$, $n$ $=$ 
$0,\ldots,N$, and apply stochastic gradient descent for the minimization of  the mean squared error  function
\bes{ \label{MSEmar} 
MSE_{mar}(\eta) &= \; \frac{1}{M} \sum_{m=1}^M \sum_{n=0}^{N-1} \Big| y_T^{m,t_n} - \vartheta^\eta(t_n,x_{t_n}^m) \\
& \hspace{3cm}  - \;  \sum_{p=n}^{N-1}  \big( D_x \vartheta^\eta(t_p,x_{t_p}^m) \big)^\top \sigma(x_{t_p}^m,\mra^*(t_p,x_{t_p}^m)) \Delta w_{t_p}^m   \Big|^2 \Delta t_n,  
}

The pseudo-code for the pathwise martingale learning is described in  Algorithm \ref{algo:scheme_value_pathwise_martingale_learning}.   

\vspace{1mm}

\begin{algorithm}[H] 
\scriptsize
\SetAlgoLined
\KwResult{A set of optimized parameters $\eta^*$;}
 Initialize the learning rate $l$, the neural networks $\vartheta^\eta$\; 
 Generate an $\R^{N+1}$-valued time grid $0=t_0 < t_1 < ... < t_N = T$ with time steps $(\Delta t_n)_{n=0,...,N-1}$\;
 Generate a batch of $M$ starting points $X_0$ $\sim$ $\mu_0$ and Brownian increments $(\Delta W_{t_n})_{n=0,...,N}$ in $R^d$\;
 \For{each batch element $m$}{
    Compute the trajectory $(x_{t_n}^{m})_{n=0,...,N}$ through the scheme
        \bes{
            x_{t_{n+1}}^{m} & = \;  x_{t_{n}}^{m}  +  \mrb^{\mra^*}(t_n,x_{t_{n}}^{m}) \Delta t_n  +  \sigma^{\mra^*}(t_n,x_{t_{n}}^{m}) \Delta w_{t_n}^m,
        }
        from the generated starting point $x_{t_0}^m$, Brownian increments $(\Delta w_{t_n}^m)_{n=0,...,N-1}$ and previously trained control $a=a_{\theta^*}$\;
    Compute the value target $(y_T^{m, t_n})_{n=0,...,N}$\;
}
 \For{each epoch}{
    Compute, for every batch element $m$, the integral $\sum_{n=0}^{N-1} \Big| y_T^{m,t_n} - \vartheta^\eta(t_n,x_{t_n}^m)  - \;  \sum_{p=n}^{N-1}  \big( D_x \vartheta^\eta(t_p,x_{t_p}^m) \big)^\top \sigma^{\mra^*}(t_p,x_{t_p}^m) \Delta w_{t_p}^m   \Big|^2 \Delta t_n$\;
    Compute the batch loss $MSE_{mar}(\eta)$\;
    Compute the gradient $\nabla_{\eta} MSE_{mar}(\eta)$\;
    Update $\eta \leftarrow \eta - l \nabla_{\eta} MSE_{mar}(\eta)$\;
}
\textbf{Return:} The set of  optimized parameters  $\eta^*$\;
\caption{Deep learning scheme for Pathwise martingale learning with 1 NN}
\label{algo:scheme_value_pathwise_martingale_learning}
\end{algorithm}

\vspace{1mm}

Alternately, we can use another neural network $\Zcal^\delta$ from $[0,T]\times\R^d$ into $\R^d$  
for the approximation of the gradient of the solution, and then use the mean squared error function
\bes{
\tilde MSE_{mar}(\eta,\delta) &= \; \frac{1}{M} \sum_{m=1}^M \sum_{n=0}^{N-1} \Big| y_T^{m,t_n} - \vartheta^\eta(t_n,x_{t_n}^m) \\
& \hspace{3cm}  - \;  \sum_{p=n}^{N-1}  \big( \Zcal^\delta(t_p,x_{t_p}^m) \big)^\top \sigma^{\mra^*}(x_{t_p}^m) \Delta w_{t_p}^m   \Big|^2 \Delta t_n.   
}

The pseudo-code for the pathwise martingale learning with two neural networks is described in  Algorithm \ref{algo:scheme_value_pathwise_martingale_learningbis} in Appendix \ref{appendix:algorithms_multiple_NN}.   

\vspace{1mm}

\vspace{2mm}

For the differential version of this algorithm, we also use a neural network $\vartheta^\eta$ from $[0,T]\times\R^d$ into $\R$ for the approximation of the value function
that we train by using the same batch of $M$ independent samples and applying stochastic gradient descent for the minimization of both  the mean squared error functions defined in \eqref{MSEmar} and  
the following one: 

\bes{
MSE_{dermar}(\eta) &= \; \frac{1}{M} \sum_{m=1}^M \sum_{n=0}^{N-1} \Big| z_T^{m,t_n} -  D_x \vartheta^\eta(t_n,x_{t_n}^m) \\
& \hspace{2cm} - \;   \sum_{p=n}^{N-1} \Big( \big[ D_x \sigma^{\mra^*}(t_p,x_{t_p}^m) \bullet_3 D_{x_{t_n}}  x_{t_p}^m  \big] \bullet_1 D_x \vartheta^\eta(t_p,x_{t_p}^m) \\
&  \hspace{2cm} + \; \sigma^{\mra^*}(t_p,x_{t_p}^m)^\top D_{xx} \vartheta^\eta(t_p,x_{t_p}^m) D_{x_{t_n}} x_{t_p}^m \Big)^\top \Delta w_{t_p}^m \Big|^2 \Delta t_n.   
}

\vspace{1mm}

The pseudo-code is described in Algorithm \ref{algo:scheme_value_pathwise_differential_learning}. 

\vspace{1mm}

\begin{algorithm}[H] 
\scriptsize
\SetAlgoLined
\KwResult{A set of optimized parameters $\eta^*$;}
 Initialize the learning rate $l$, the neural networks $\vartheta^\eta$\; 
 Generate an $\R^{N+1}$-valued time grid $0=t_0 < t_1 < ... < t_N = T$ with time steps $(\Delta t_n)_{n=0,...,N-1}$\;
 Generate a batch of $M$ starting points $X_0$ $\sim$ $\mu_0$ and Brownian increments $(\Delta W_{t_n})_{n=0,...,N}$ in $R^d$\;
 \For{each batch element $m$}{
    Compute the trajectory $(x_{t_n}^{m})_{n=0,...,N}$ through the scheme
        \bes{
            x_{t_{n+1}}^{m} & = \;  x_{t_{n}}^{m}  +  \mrb^{\mra^*}(t_n,x_{t_{n}}^{m}) \Delta t_n  +  \sigma^{\mra^*}(t_n,x_{t_{n}}^{m}) \Delta w_{t_n}^m,
        }
        from the generated starting point $x_{t_0}^m$, Brownian increments $(\Delta w_{t_n}^m)_{n=0,...,N-1}$ and previously trained control $a=a_{\theta^*}$\;
    Compute the value and derivative targets $(y_T^{m, t_n})_{n=0,...,N}$ and $(z_T^{m, t_n})_{n=0,...,N}$\;
}
\For{each epoch}{
    Compute, for every batch element $m$, the integral $\sum_{n=0}^{N-1} \Big| y_T^{m,t_n} - \vartheta^\eta(t_n,x_{t_n}^m) -  \sum_{p=n}^{N-1}  \big( D_x \vartheta^\eta(t_p,x_{t_p}^m) \big)^\top \sigma^{\mra^*}(t_p,x_{t_p}^m) \Delta w_{t_p}^m   \Big|^2 \Delta t_n$\;
    Compute the batch loss $MSE_{mar}(\eta)$\;
    Compute the gradient $\nabla_{\eta} MSE_{mar}(\eta)$\;
    Update $\eta \leftarrow \eta - l \nabla_{\eta} MSE_{mar}(\eta)$\;
    Compute, for every batch element $m$, the integral $\sum_{n=0}^{N-1} \Big| z_T^{m,t_n} -  D_x \vartheta^\eta(t_n,x_{t_n}^m) -  \sum_{p=n}^{N-1} \Big( \big[ D_x \sigma^{\mra^*}(t_p,x_{t_p}^m) \bullet_3 D_{x_{t_n}}  x_{t_p}^m \big] \bullet_1 D_x \vartheta^\eta(t_p,x_{t_p}^m)  $ \\
    $+ \sigma^{\mra^*}(t_p,x_{t_p}^m)^\top D_{xx} \vartheta^\eta(t_p,x_{t_p}^m) D_{x_{t_n}} x_{t_p}^m \Big)^\top \Delta w_{t_p}^m \Big|^2 \Delta t_n$\;
    Compute the batch loss $MSE_{dermar}(\eta)$\;
    Compute the gradient $\nabla_{\eta} MSE_{dermar}(\eta)$\;
    Update $\eta \leftarrow \eta - l \nabla_{\eta} MSE_{dermar}(\eta)$\;
}
\textbf{Return:} The set of  optimized parameters  $\eta^*$\;
\caption{Deep learning scheme for Pathwise differential learning with 1 NN }
\label{algo:scheme_value_pathwise_differential_learning}
\end{algorithm}

\vspace{2mm}

Alternately, in addition to the NN $\vartheta^\eta$ for $u$,  we can use neural networks $\Zcal^\delta$ and $\Gamma^\epsilon$ for the gradient and the Hessian that we train with the loss function
\bes{
\tilde MSE_{dermar}(\delta, \epsilon) &= \; \frac{1}{M} \sum_{m=1}^M \sum_{n=0}^{N-1} \Big| z_T^{m,t_n} -  \Zcal^\delta(t_n,x_{t_n}^m) \\
& \hspace{2.5cm} -  \sum_{p=n}^{N-1} \Big( \big[ D_x \sigma^{\mra^*}(t_p,x_{t_p}^m) \bullet_3 D_{x_{t_n}}  x_{t_p}^m  \big] \bullet_1 \Zcal^\delta(t_p,x_{t_p}^m) \\
&  \hspace{2cm} + \; \sigma^{\mra^*}(t_p,x_{t_p}^m)^\top \Gamma^\epsilon(t_p,x_{t_p}^m) D_{x_{t_n}} x_{t_p}^m \Big)^\top \Delta w_{t_p}^m \Big|^2 \Delta t_n,  
}
where $D_{x_{t_n}} x_{t_p}^m$ denotes the flow derivative of the approximate optimal state process at time $t_p$ w.r.t the state at time $t_n$ along the path $m$ of the batch.

\vspace{1mm}

The pseudo-code for this version using three neural networks is described in Algorithm \ref{algo:scheme_value_pathwise_differential_learning3NN} in Appendix \ref{appendix:algorithms_multiple_NN}.

\vspace{1mm}

\subsection{Validation tests}\label{sec:validation_tests}

The deep learning methods described above provide an approximation $\vartheta$ of the solution $u$ to the PDE, which relies up-front on the approximation $\mra^\theta$ of the optimal control $\hat\mra$ arising from the dual stochastic control representation. 

We can test and validate the convergence and accuracies of these approximations as follows. 
On the one hand, as in the DGM  and PINN methods, see  \cite{sirignano2018dgm} and  \cite{raissi2019physics}, we can compute the losss $\Lcal_{res}$ and $\Lcal_{term}$, associated respectively to the residual and to the terminal condition of the partial differential equation \eqref{eq:pde}

\bes{ \label{residual} 
    \Lcal_{res} := \frac{1}{|\Tcal||\chi|} \sum_{t\in\Tcal, x\in\chi} \Big| \partial_t \vartheta^\eta +  H(x,D_x \vartheta^\eta, D_x ^2 \vartheta^\eta) \Big|^2,
}
\bes{ \label{terminal_condition} 
    \Lcal_{term} := \frac{1}{|\chi|} \sum_{x\in\chi} \Big| \vartheta^\eta (T,x) - g(x) \Big|^2,
}
where the time grid $\Tcal$ is composed of times that were not used during the network training, so as to verify the generalisation of the solution obtained, and $\chi$ is a bounded space grid in $\R^d$. 
As the state diffusion $(X_t^\theta)_{0\leq t\leq T}$ is not bounded, the bounds  of the space grid are fixed arbitrarily depending on the domain of interest. Alternatively, these bounds could be chosen based on the distribution of the values attained during the computation of the diffusion scheme of $(X_t^\theta)_{0\leq t\leq T}$.

On the other hand, by noting that the optimal control should satisfy the optimality condition 
\bes{
    D_a \mrb(x,\hat \mra). D_x u(t,x) + \frac{1}{2} \mathrm{Tr}_{1,2} \big( D_a \sigma\sigma^\top(x,\hat \mra) \bullet_2 D_{xx}u(t,x) \big) + D_a f(x,\hat \mra) = 0,
}
we can check the accuracy  the approximation $\mra_\theta$ of the optimal control by computing the following loss
\begin{align}
    \Lcal_{optim} &= \; \frac{1}{|\Tcal||\chi|} \sum_{t\in\Tcal, x\in \chi} \Big| D_a \mrb(x, \mra_\theta) D_x \vartheta^\eta (t,x) + \frac{1}{2} \mathrm{Tr}_{1,2} \big(D_a \sigma\sigma^\top(x, \mra_\theta) \big) \bullet_2 D_{xx} \vartheta^\eta (t,x) \big) \\
    & \hspace{5cm} + \;  D_a f(x, \mra_\theta) \Big|^2,
\end{align} 
on the same grid $\Tcal \times \chi$ as before.

Another validation method, which is more graphical, consists in approximating numerically the optimal control as described in Section \ref{sec:numerical_approximation optimal control} and then computing the value function and its first derivative for some chosen points $(t,x)$ by Monte Carlo simulations:  
\bes{
\vartheta_{MC}(t,x)  = \; \frac{1}{M}\sum_{m=1}^M\Big[ g(x_{t_N}^{m,t,x}) +  \sum_{p=n}^{N-1} f(x_{t_p}^{m,t,x},\mra_\theta(t_p, x_{t_p}^{m,t,x})) \Delta t_p \Big],
}
\bes{
D_x \vartheta_{MC}(t,x)  &= \; \frac{1}{M}\sum_{m=1}^M\Big[ \big( D_x x_{t_N}^{m,t,x} \big)^\top D_x g(x_{t_N}^{m,t,x}) \\
 & \hspace{1.5cm} + \;  \sum_{p=n}^{N-1} \Big( \big( D_x x_{t_N}^{m,t,x} \big)^\top D_x f(x_{t_p}^{m,x},\mra_\theta(t_p, x_{t_p}^{m,t,x})) \\
 & \hspace{1.5cm} + \; \big( D_x \mra_\theta(t_p, x_{t_p}^{m,t,x}) D_x x_{t_p}^{m,t,x} \big)^\top D_a f(x_{t_p}^{m,t,x},\mra_\theta(t_p, x_{t_p}^{m,t,x})) \Big) \Delta t_p \Big] ,\\
}
with $t = t_n$. We then plot these Monte Carlo points alongside the value functions obtained by using neural networks to check that the machine learning methods described in the previous sections are able to approximate the value function corresponding to the approximated optimal control $\mra_\theta$.

\subsection{Example of Merton portfolio selection} \label{sec:Merton}

We consider the Bellman equation:
\begin{equation}
\left\{
\begin{array}{rcl} 
\partial_t u + \sup_{a \in \R} \big[ a x b D_ x u +  \frac{1}{2} a^2 x^2 \sigma^2  D_{x}^2 u \big] & = &  0, \quad \quad (t,x) \in [0,T) \times (0,\infty), \\
 \hspace{3cm}    u(T,x) &= & g(x), \quad x \in (0,\infty), 
\end{array}
\right.
\end{equation}
which arises from the Merton portfolio selection problem where an agent invests a proportion $\alpha$ $=$ $(\alpha_t)_t$ of her wealth $X$ $=$ $X^\alpha$ in a stock following a Black-Scholes model with rate of return 
$b$ $\in$ $\R$, and constant volatility $\sigma$ $>$ $0$. The controlled wealth dynamics is then governed by 
\bes{
dX_t &=\;  X_t \alpha_t  b \d t + X_t \alpha_t \sigma \d W_t, 
}
and the goal of the investor is to maximize over $\alpha$ her expected terminal wealth $\E[ g(X_T)]$, with $g$ some utility function, i.e. concave and nondecreasing, on $(0,\infty)$.

When the utility function $g$ is of power type, i.e. $g(x)$ $=$ $x^\gamma/\gamma$, for some $\gamma$ $<$ $1$, $\gamma$ $\neq$ $0$,  it is well-known that the optimal control is constant equal to 
\bes{
\hat\mra &= \; \frac{b}{\sigma^2(1-\gamma)}, 
}
while the value function is explicitly given by
\bes{
u(t,x) &= \; e^{\rho(T-t)} g(x), \quad \mbox{ with  }  \quad \rho = \frac{b^2}{2\sigma^2} \frac{\gamma}{1-\gamma}.  
}
These closed-form expressions serve as benchmarks for comparing our results computed by the differential learning algorithms.

\vspace{1mm}
In order to check that the value function approximation obtained is a lower bound of the true one, we compute in Table \ref{tableMerton_diffClosedMC} the difference between the closed form value function and the estimation of the value function on points $(t,x)$ obtained by computing the expectation \eqref{ulin} by Monte Carlo on $1e^6$ trajectories controlled by the Deep Learning approximation of the optimal control.We compute the value functions on a grid $t \in \{0, 0.5, 0.9\}$, $x\in \{1e^{-2}, 0.5, 0.75, 1, 1.25, 1.5, 2\}$ with parameters $b=0.2$, $\sigma=0.2$ and power utility with exponent $\gamma=0.5$ and present in the table the difference between the closed form value and the Monte Carlo approximation. For clarity of presentation we present the results averaged over $t$ in this table.

\begin{scriptsize}
\begin{table}[h]
\centering
\begin{tabular}{|c|c|c|c|c|c|c|c|}
\hline  & $x=1 e^{-2}$ & $x=0.5$ &  $x=0.75$ &  $x=1$ &  $x=1.25$ &  $x=1.5$ & $x=2$ \\
\hline  Difference & & & & & & & \\
closed - MC & $2.699e^{-4}$ & $1.811e^{-3}$ & $2.208e^{-3}$ & $2.543e^{-3}$ & $2.839e^{-3}$ & $3.106e^{-3}$ & $3.582e^{-3}$\\
\hline
\end{tabular}
\caption{Difference between the closed form value function and the value computed by Monte Carlo on $1e^6$ trajectories on points $x\in \{1e^{-2}, 0.5, 0.75, 1, 1.25, 1.5, 2\}$ and averaged over times $t \in \{0, 0.5, 0.9\}$ for the Merton problem with parameters $b = 0.2$, $\sigma = 0.2$ and power utility with exponent $\gamma=0.5$.} 
\label{tableMerton_diffClosedMC} 
\end{table}
\end{scriptsize}

We compute in Table \ref{tableMerton}, the residual losses defined in  \eqref{residual} for the NN $\vartheta^\eta$ obtained by the various deep learning methods: the differential learning scheme (Algorithm \ref{algo:scheme_value_differential_learning}), the pathwise martingale learning with $1$ NN (Algorithm \ref{algo:scheme_value_pathwise_martingale_learning}) and the pathwise differential learning with $1$ NN (Algorithm \ref{algo:scheme_value_pathwise_differential_learning}). Since two gradient steps are performed during each training epoch of the Pathwise differential learning method, we indicate the training time for 500 epochs for this method whereas the training time of the two other methods is indicated for 1000 epochs. For each of these algorithms, 8192 starting points and Brownian trajectories are used in order to train the neural networks. We also provide the training time for each of these algorithms. On this table we see that the Pathwise differential learning method is the slowest to train but yields the best results in terms of residual and terminal losses. We see that for all three methods the difference between the residual loss only and the sum of residual and terminal loss is small, meaning that with all three methods the neural network managed to fit the terminal condition of the PDE during the training.

\begin{scriptsize}
\begin{table}[h]
\centering
\begin{tabular}{|c|c|c|c|}
\hline  & Diff. regr. learning & Path. 1NN &  Path. diff. 1NN \\
\hline  Residual loss  & $1.538e^{-1}$ & $1.752e^{-1}$ & $7.872e^{-2}$ \\
\hline  Residual loss  & & & \\
+ terminal loss & $1.548e^{-1}$ & $1.758e^{-1}$ & $7.894e^{-2}$ \\
\hline  Training time & 274s & 297s & 525s \\
 &  1000 epochs & 1000 epochs & 500 epochs  \\
\hline
\end{tabular}
\caption{Residual and boundary losses computed on a $102$x$102$ time and space grid with $t\in[0,0.9]$ and $x\in[1e^{-2}, 2]$ for the Merton problem with parameters $b = 0.2$, $\sigma = 0.2$ and power utility with exponent $\gamma=0.5$.} 
\label{tableMerton} 
\end{table}
\end{scriptsize}

\vspace{1mm}

\begin{figure}[htp]
    \centering
    \begin{subfigure}{.32\linewidth}
        \centering
        \includegraphics[height=3.75cm]{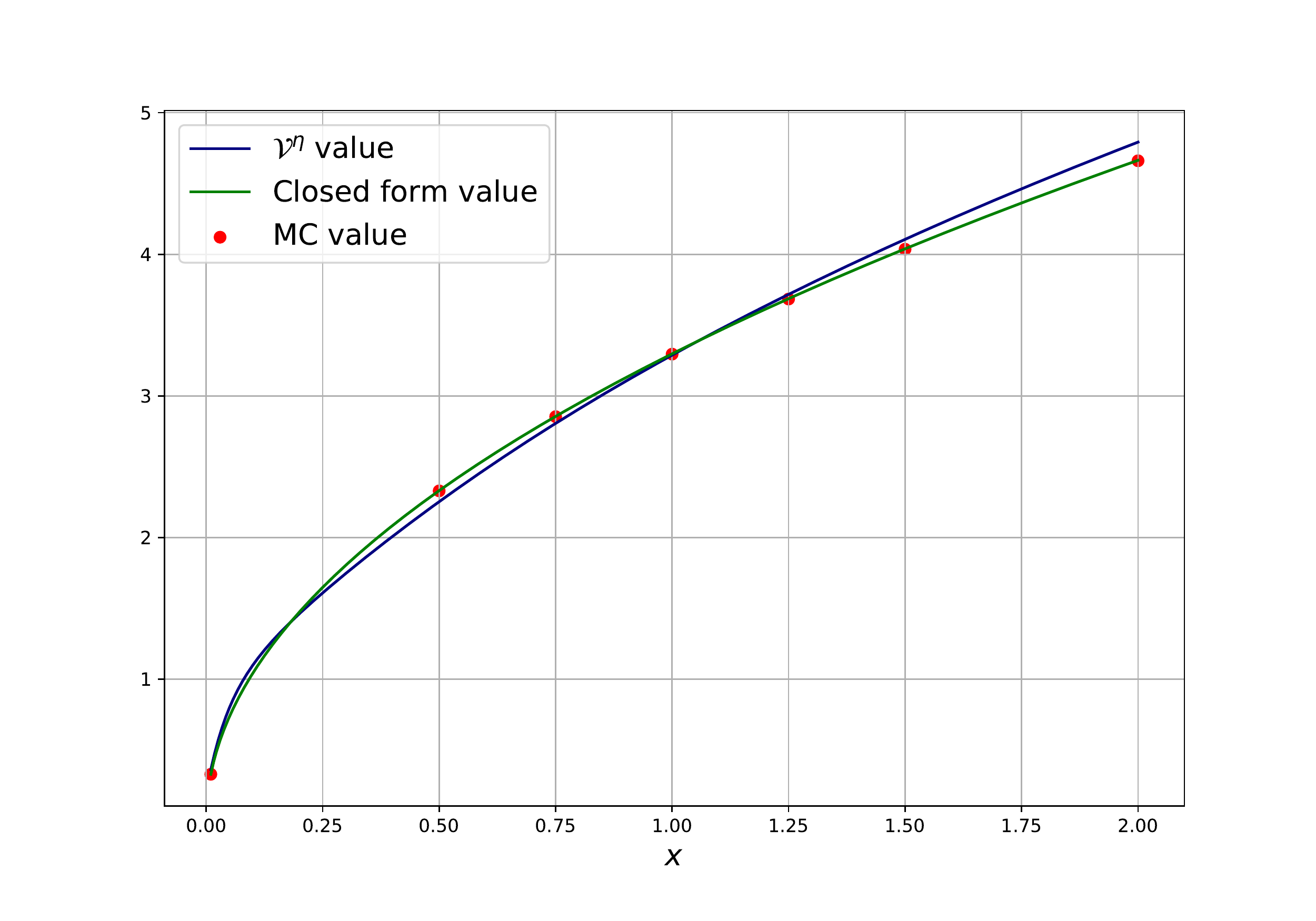} 
    \end{subfigure}
    \begin{subfigure}{.32\linewidth}
        \centering
        \includegraphics[height=3.75cm]{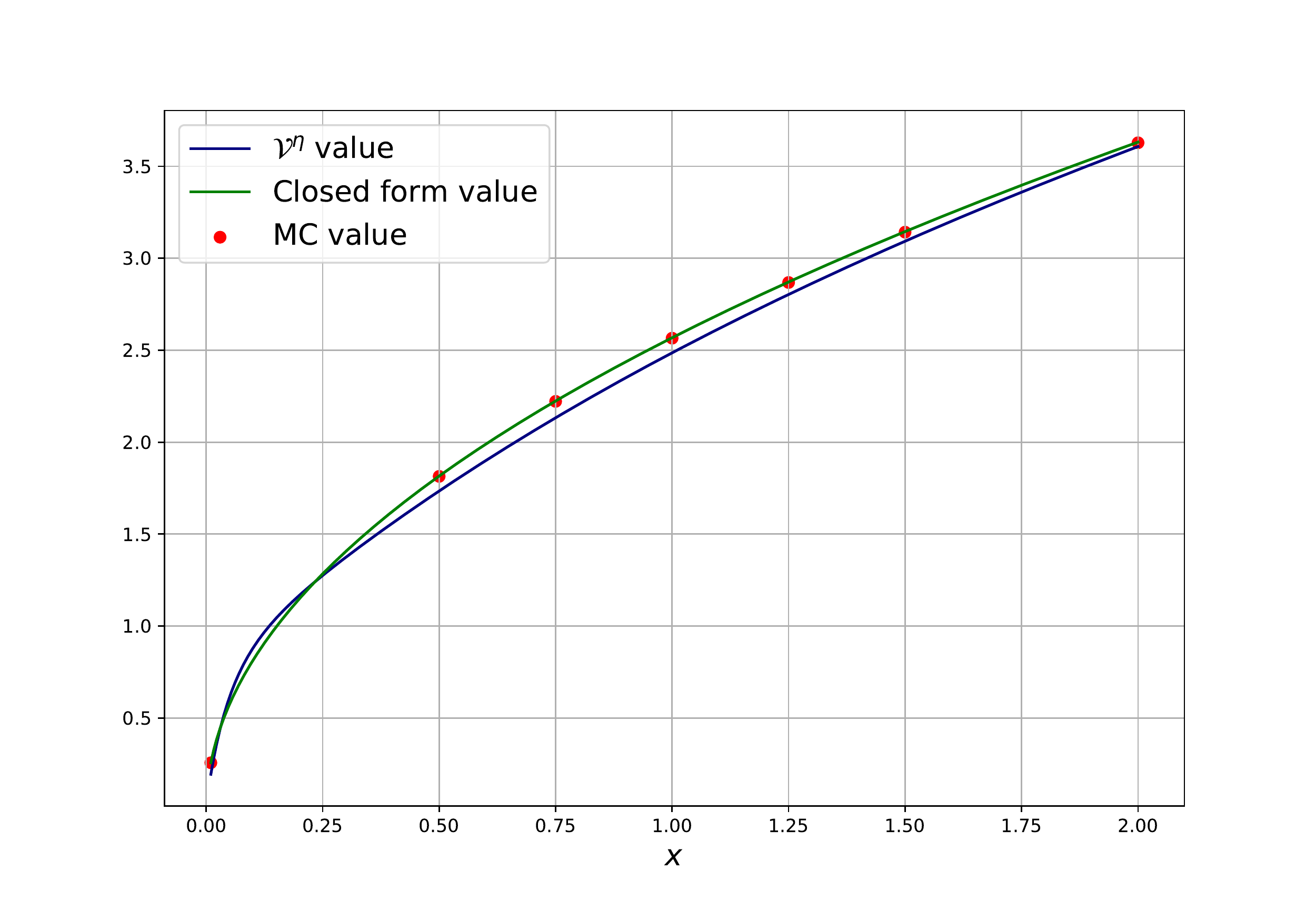}
    \end{subfigure}
    \begin{subfigure}{.32\linewidth}
        \centering
        \includegraphics[height=3.75cm]{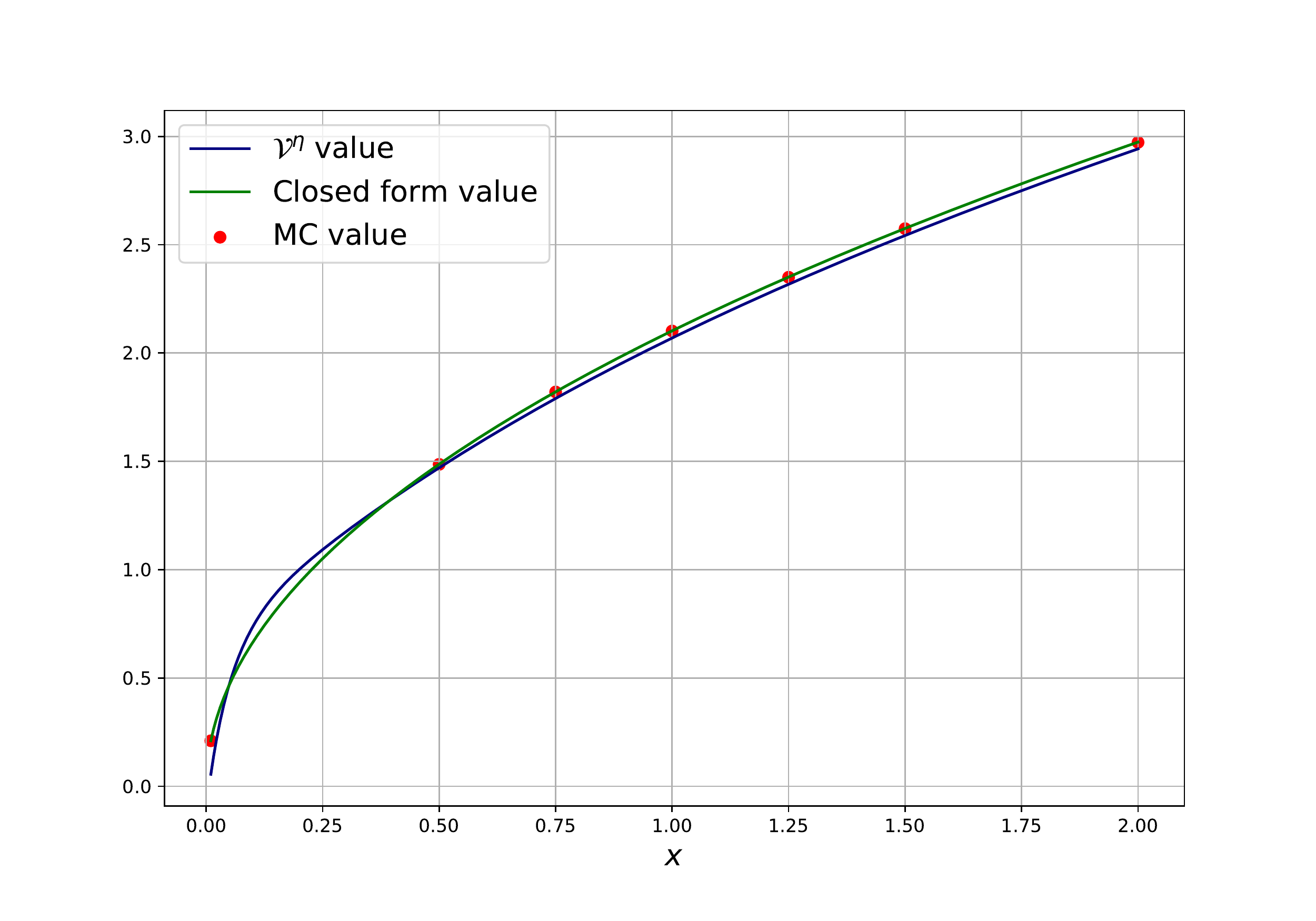}
    \end{subfigure}
    \begin{subfigure}{.32\linewidth}
        \centering
        \includegraphics[height=3.75cm]{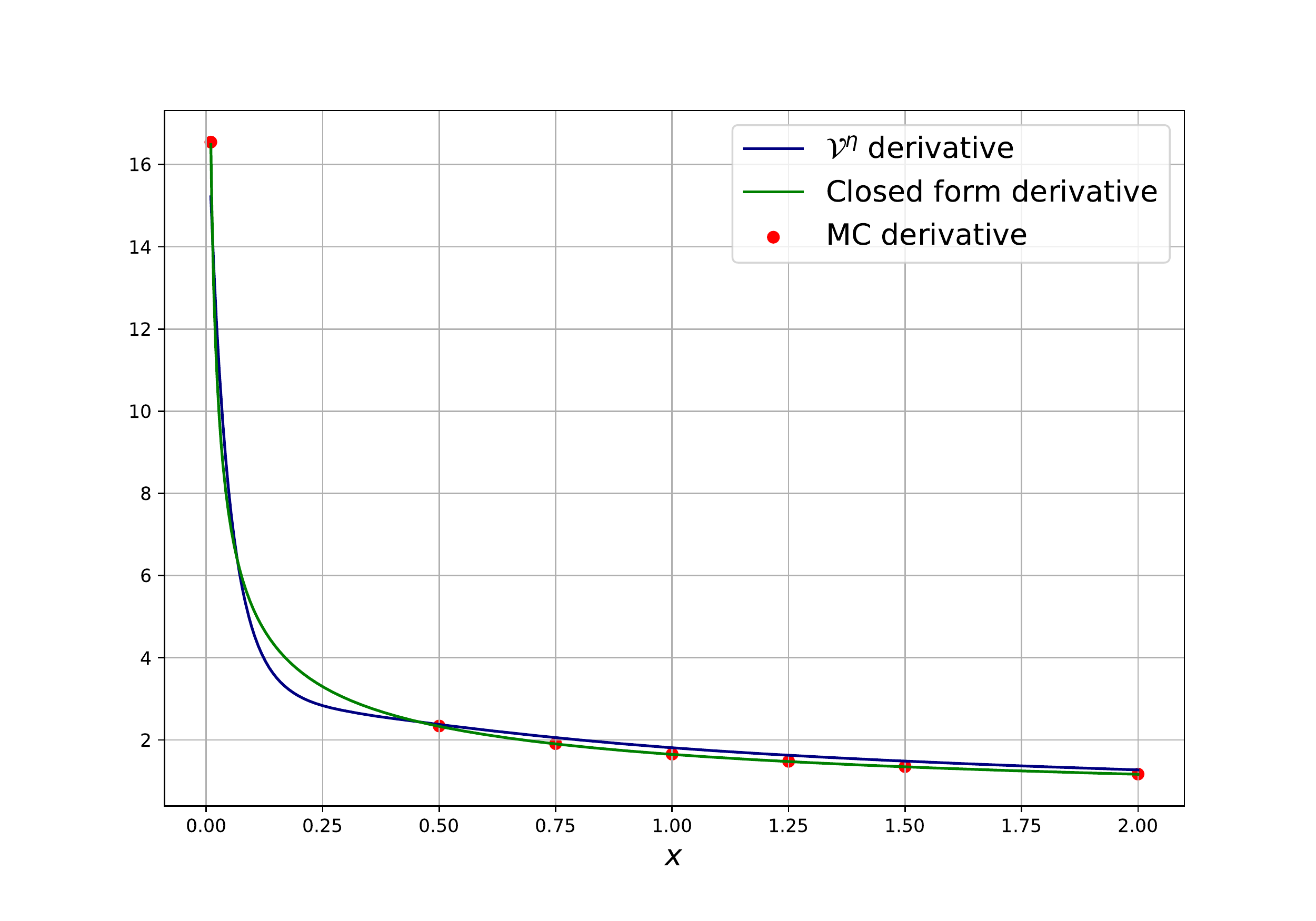} 
    \end{subfigure}
    \begin{subfigure}{.32\linewidth}
        \centering
        \includegraphics[height=3.75cm]{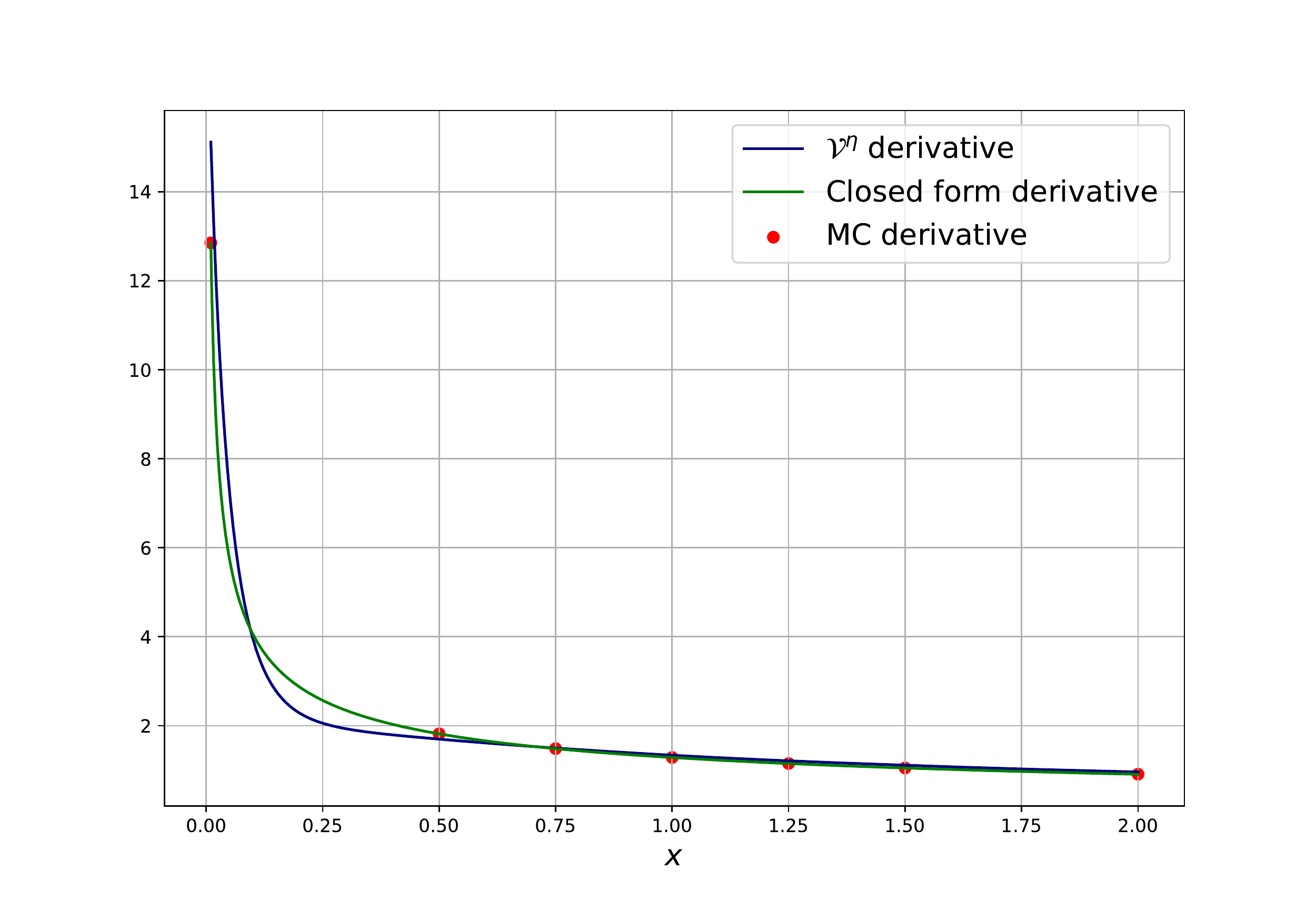}
    \end{subfigure}
    \begin{subfigure}{.32\linewidth}
        \centering
        \includegraphics[height=3.75cm]{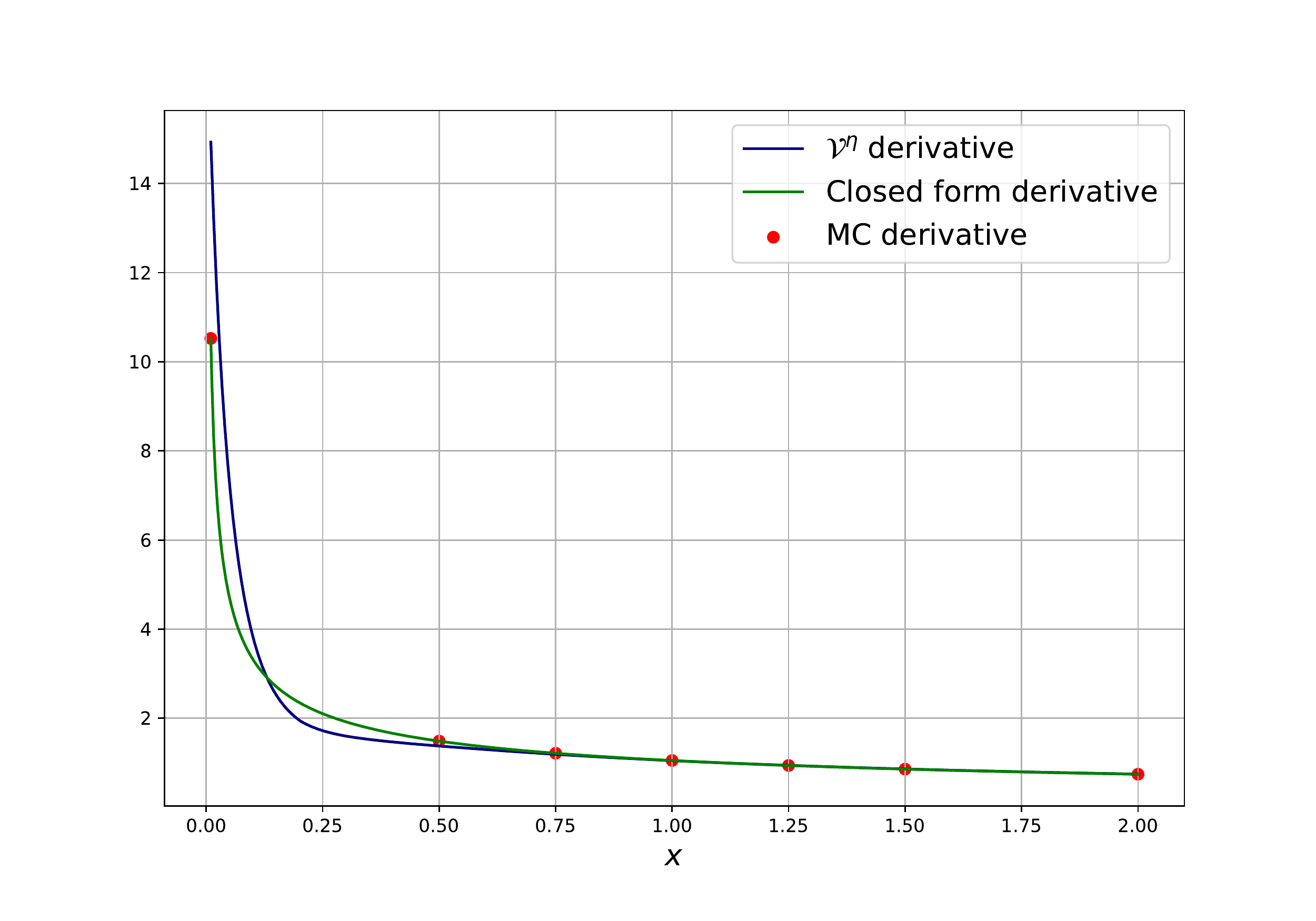}
    \end{subfigure}
    \begin{subfigure}{.32\linewidth}
        \centering
        \includegraphics[height=3.75cm]{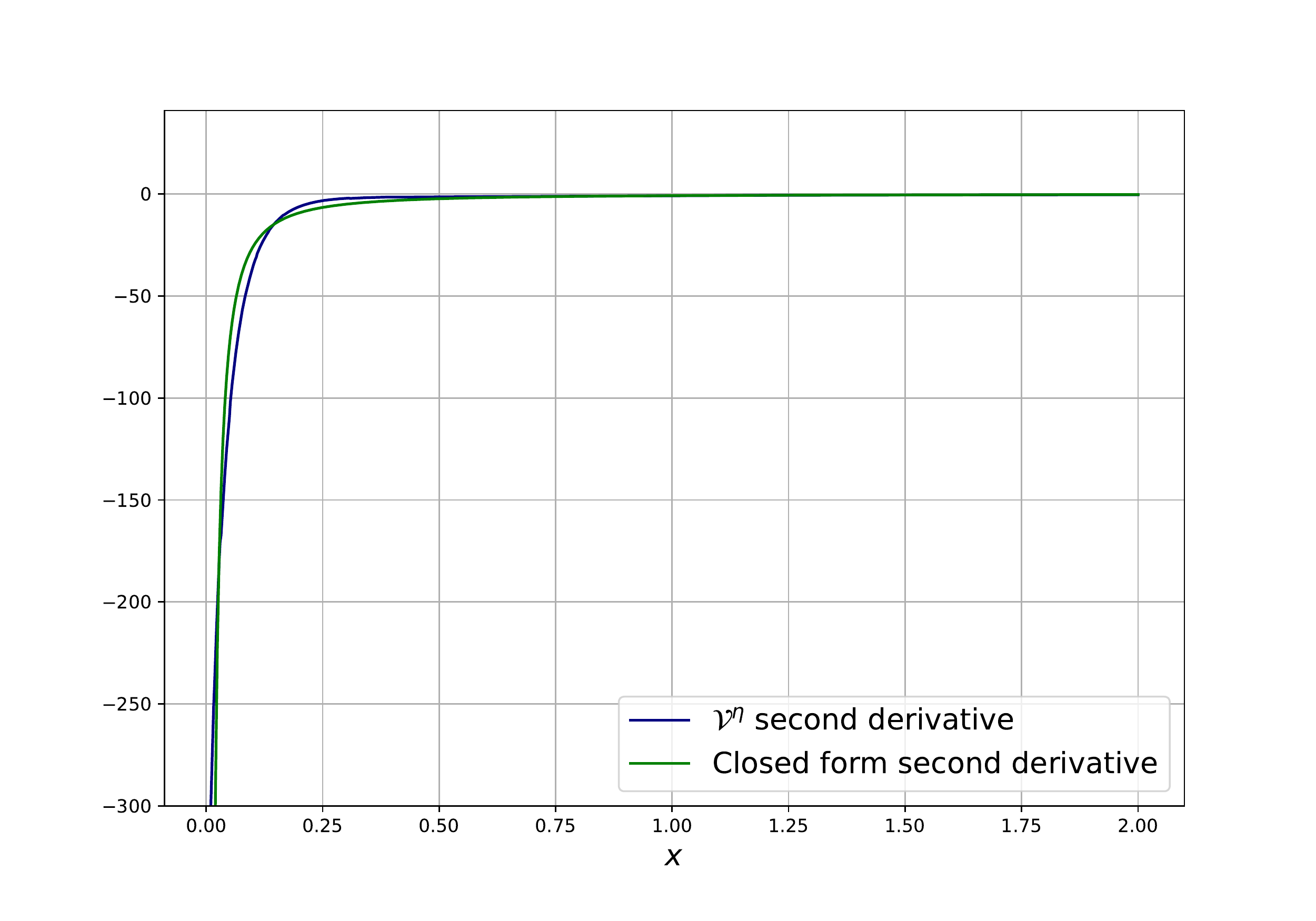} 
        \caption[short]{$t=0$}
    \end{subfigure}
    \begin{subfigure}{.32\linewidth}
        \centering
        \includegraphics[height=3.75cm]{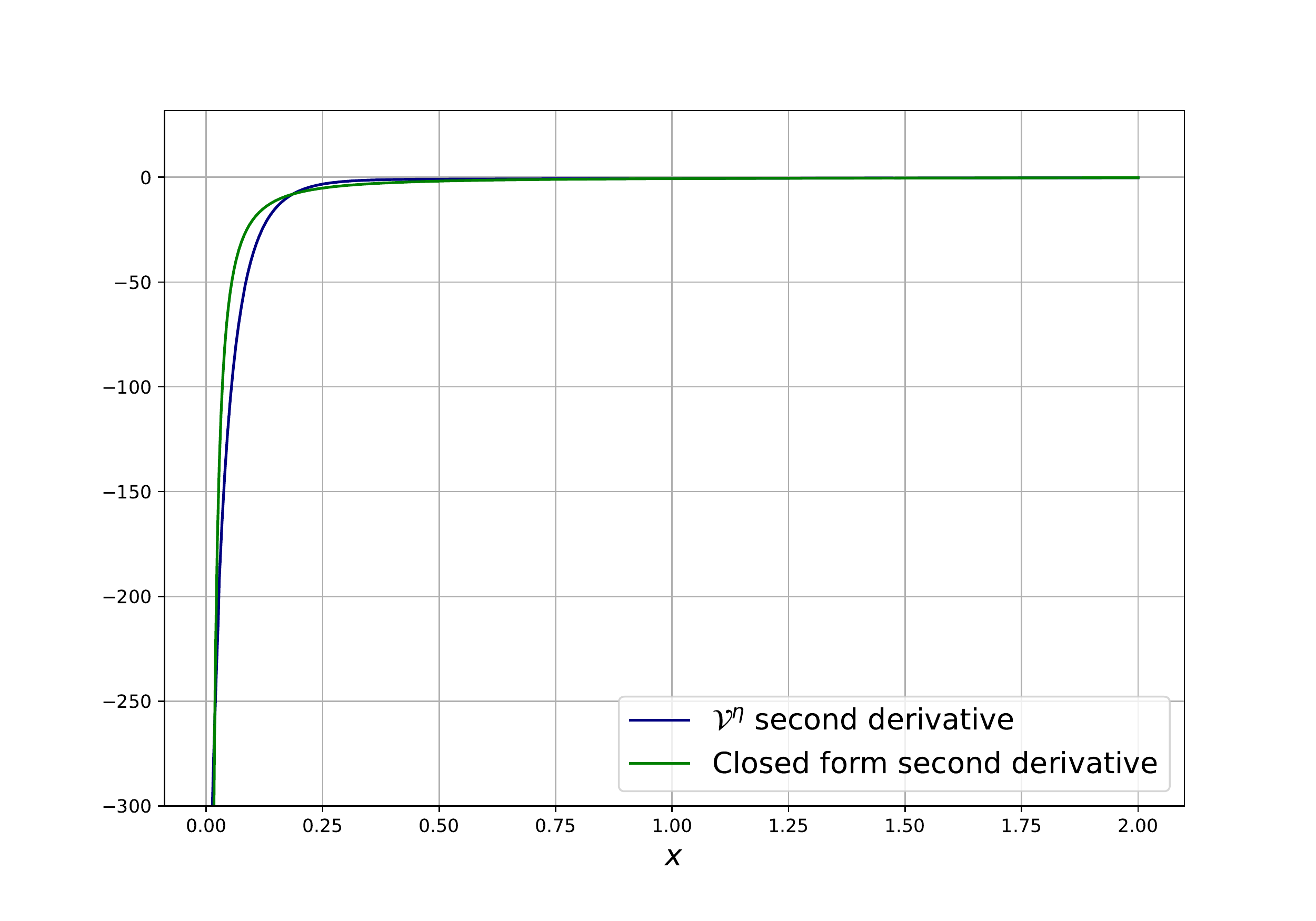}
        \caption[short]{$t=0.5$}
    \end{subfigure}
    \begin{subfigure}{.32\linewidth}
        \centering
        \includegraphics[height=3.75cm]{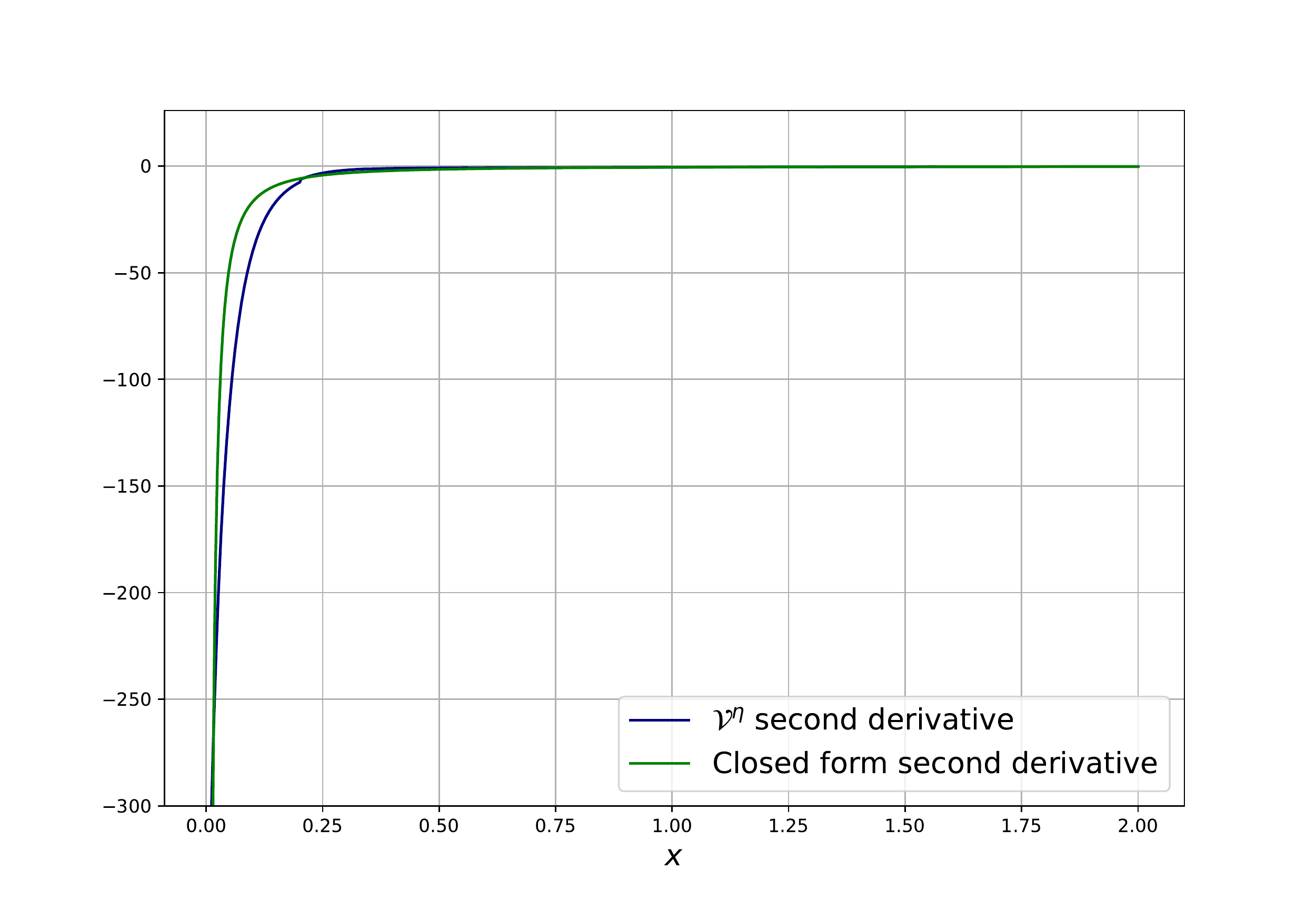}
        \caption[short]{$t=0.9$}
    \end{subfigure}
    \caption{
    \label{fig:value_differential_learning_merton_gamma_05}
    Value function $\vartheta^\eta$ and its first and second derivative obtained by Differential regression Learning (Algorithm \ref{algo:scheme_value_differential_learning}) for the Merton problem with parameters $b = 0.2$, $\sigma = 0.2$ and power utility with exponent $\gamma=0.5$, plotted as functions of $x$, for fixed values of $t$.
    }
\end{figure}

\begin{figure}[htp]
    \centering
    \begin{subfigure}{.32\linewidth}
        \centering
        \includegraphics[height=3.75cm]{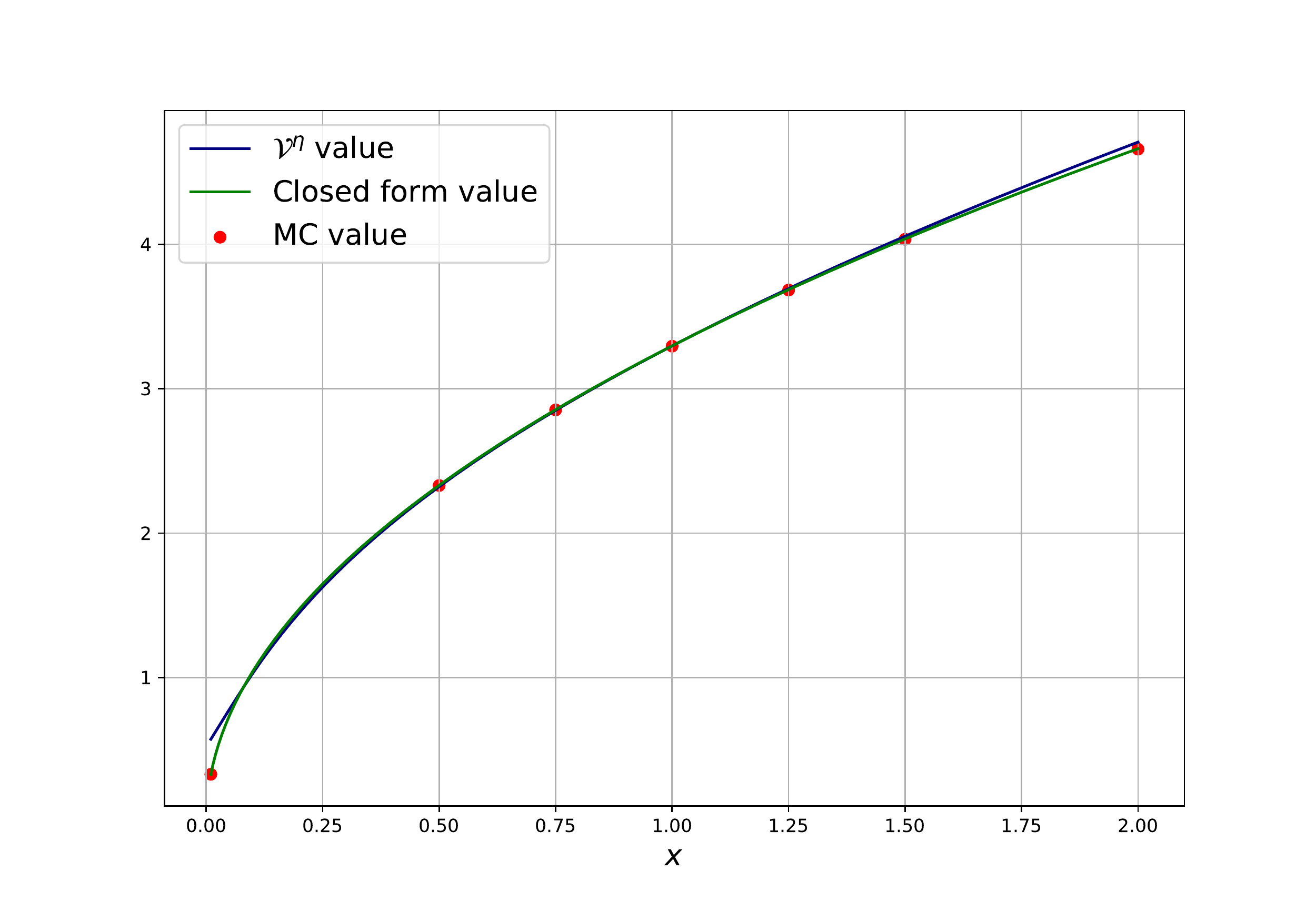} 
    \end{subfigure}
    \begin{subfigure}{.32\linewidth}
        \centering
        \includegraphics[height=3.75cm]{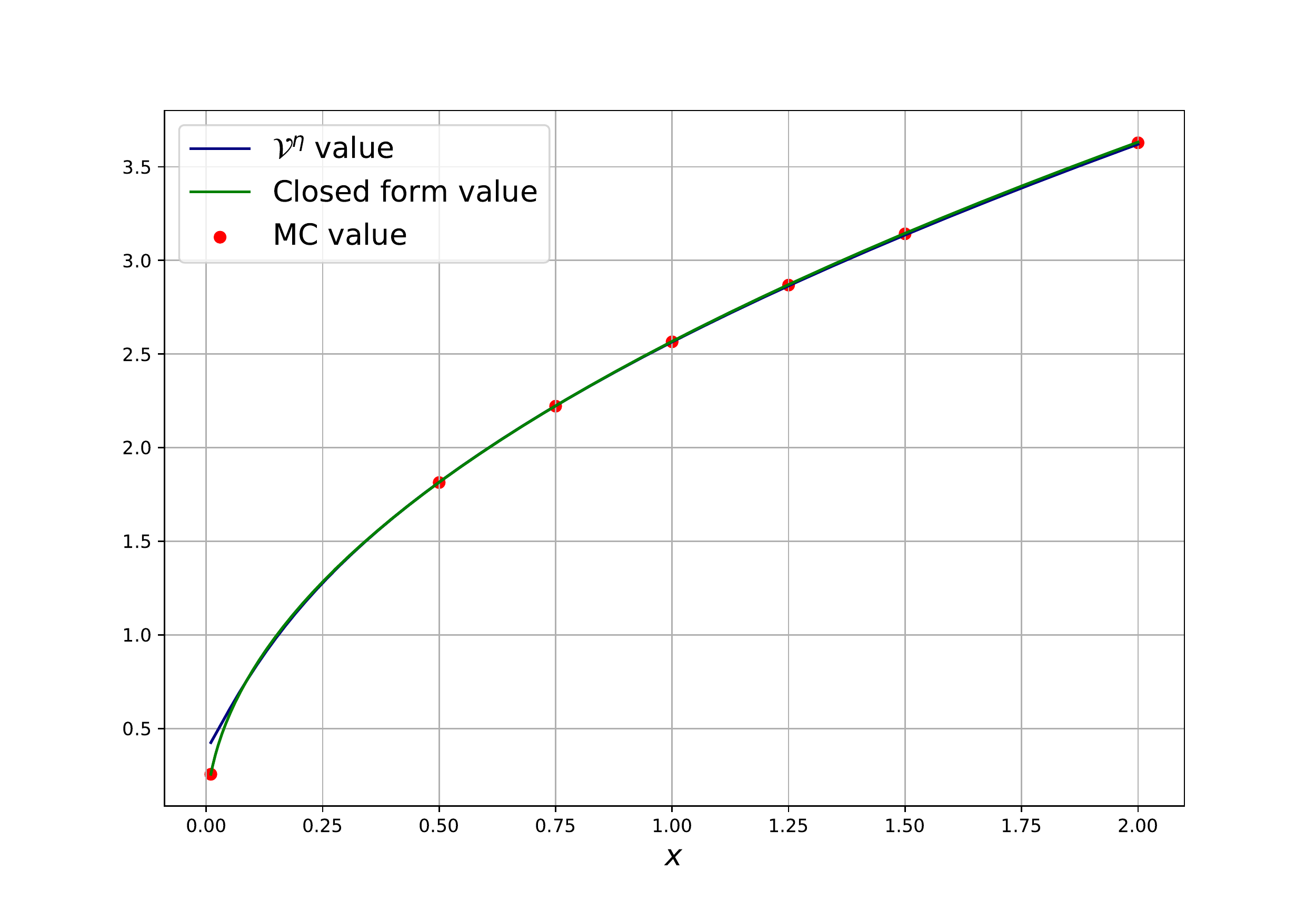}
    \end{subfigure}
    \begin{subfigure}{.32\linewidth}
        \centering
        \includegraphics[height=3.75cm]{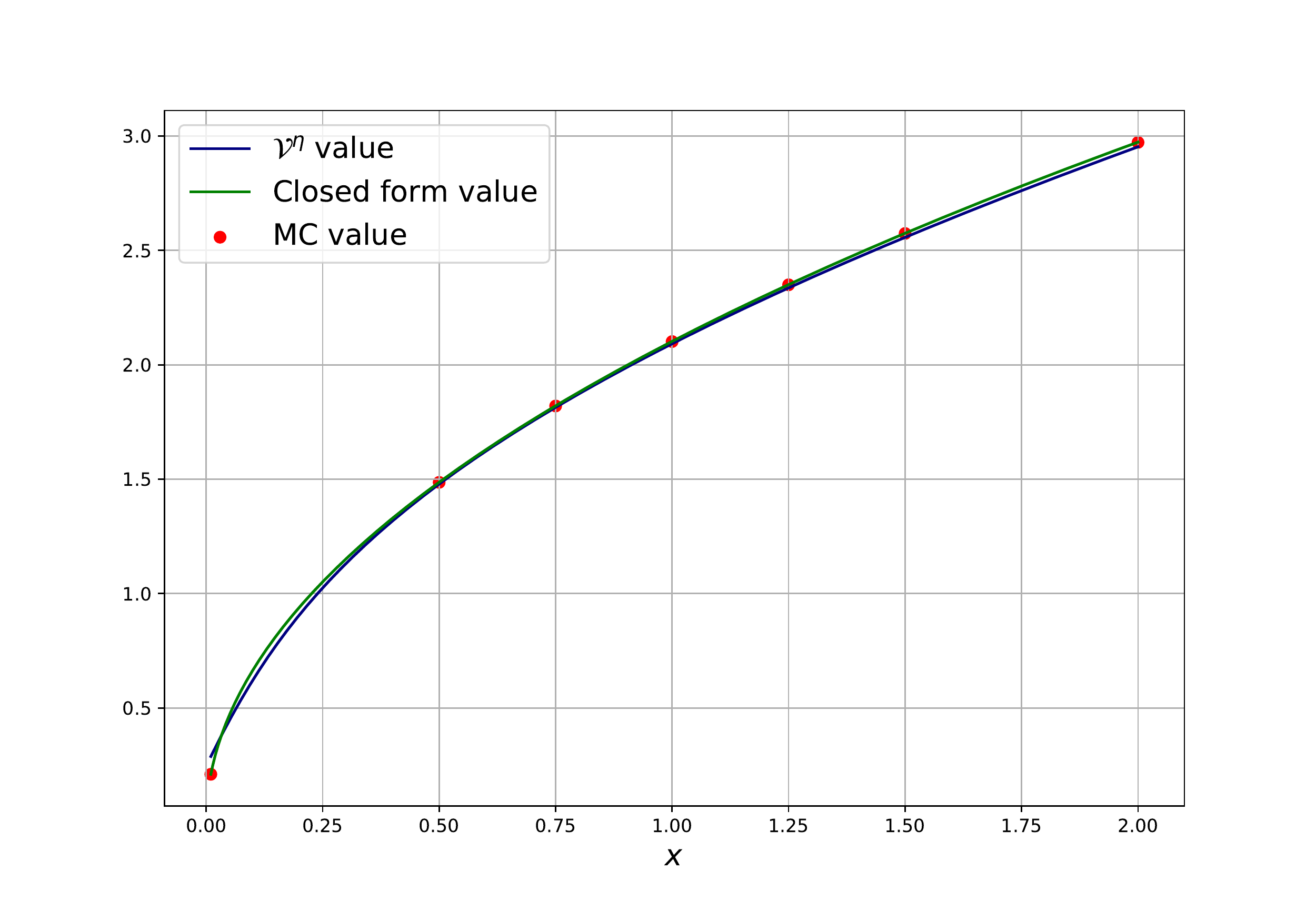}
    \end{subfigure}
    \begin{subfigure}{.32\linewidth}
        \centering
        \includegraphics[height=3.75cm]{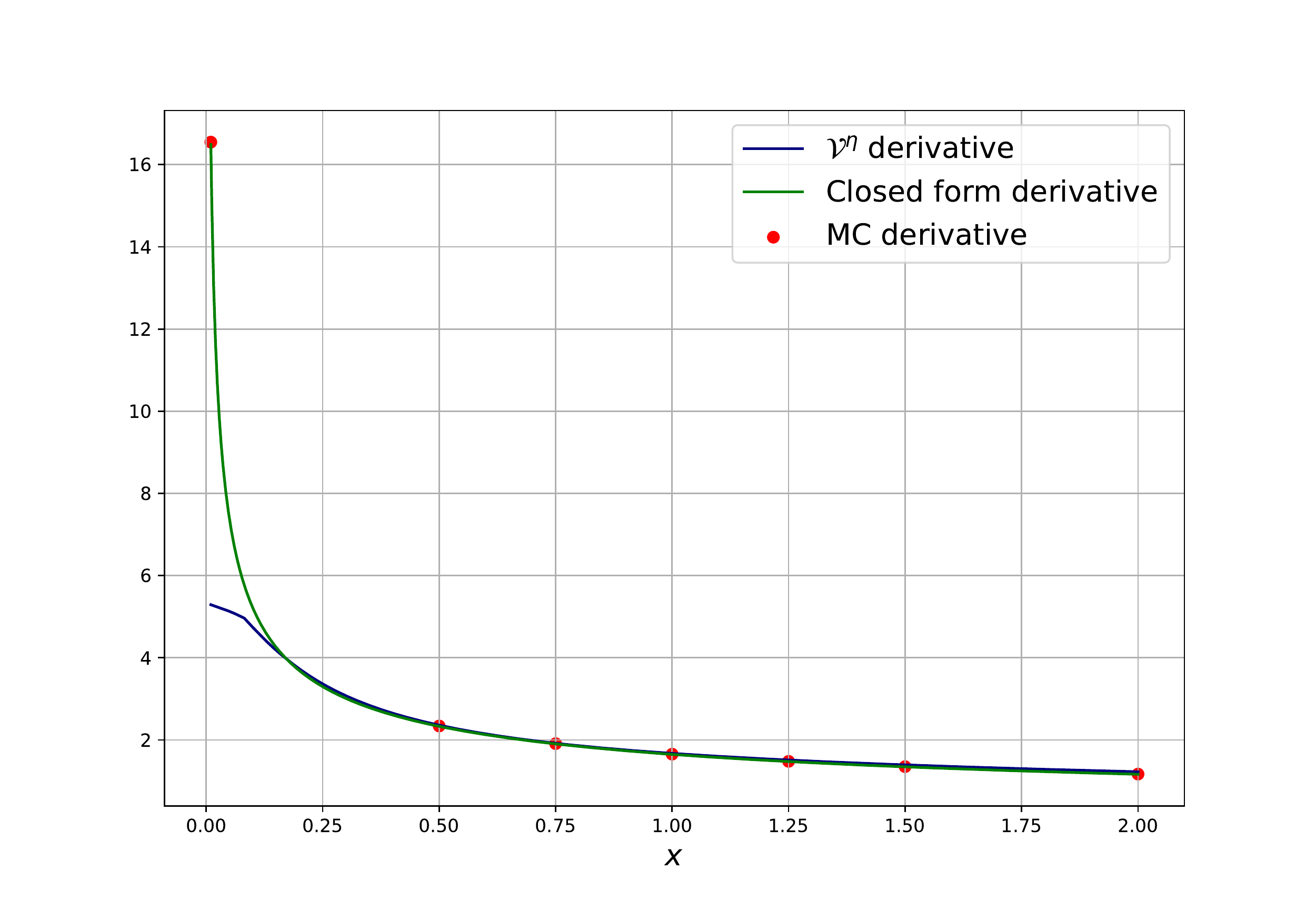} 
    \end{subfigure}
    \begin{subfigure}{.32\linewidth}
        \centering
        \includegraphics[height=3.75cm]{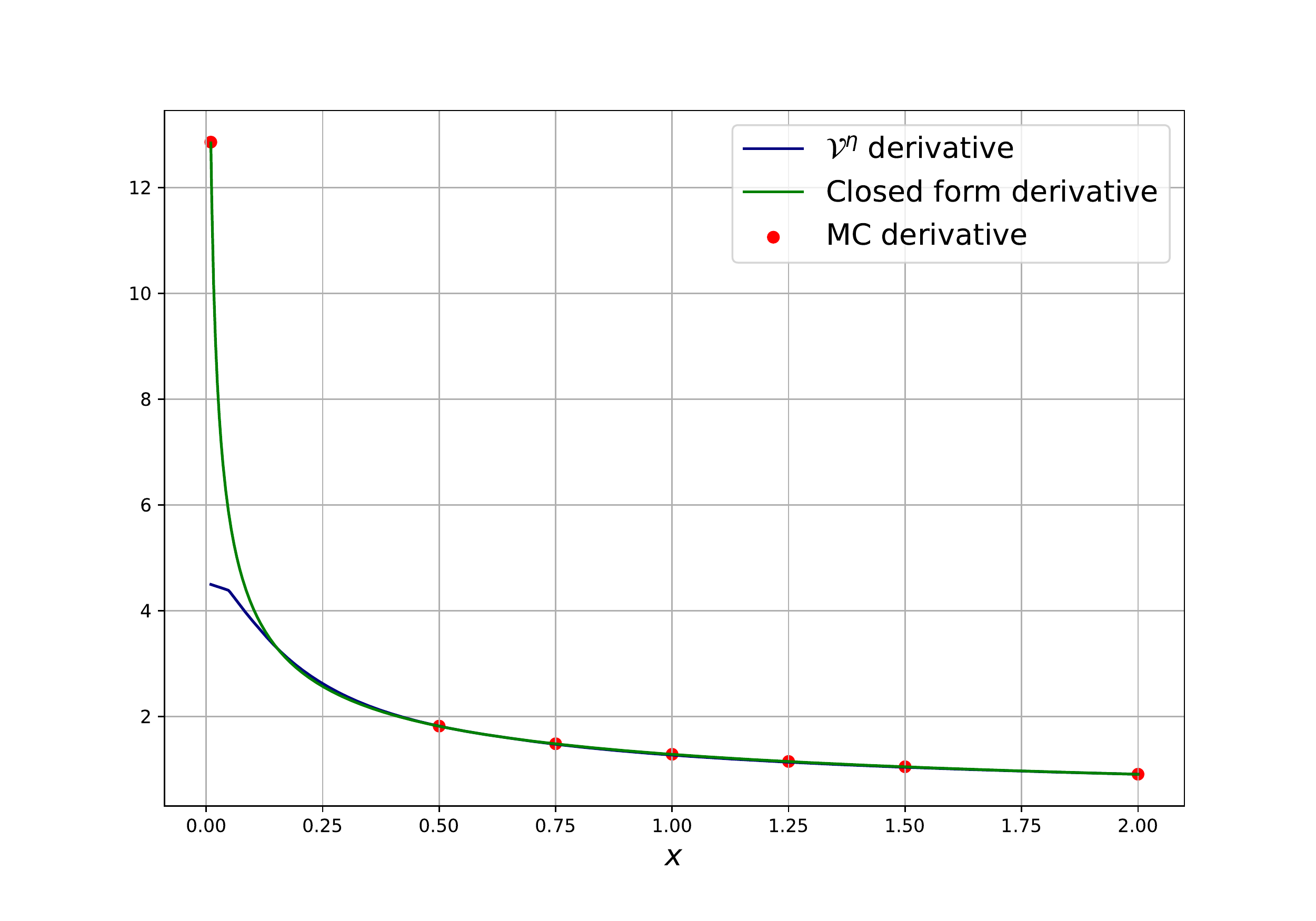}
    \end{subfigure}
    \begin{subfigure}{.32\linewidth}
        \centering
        \includegraphics[height=3.75cm]{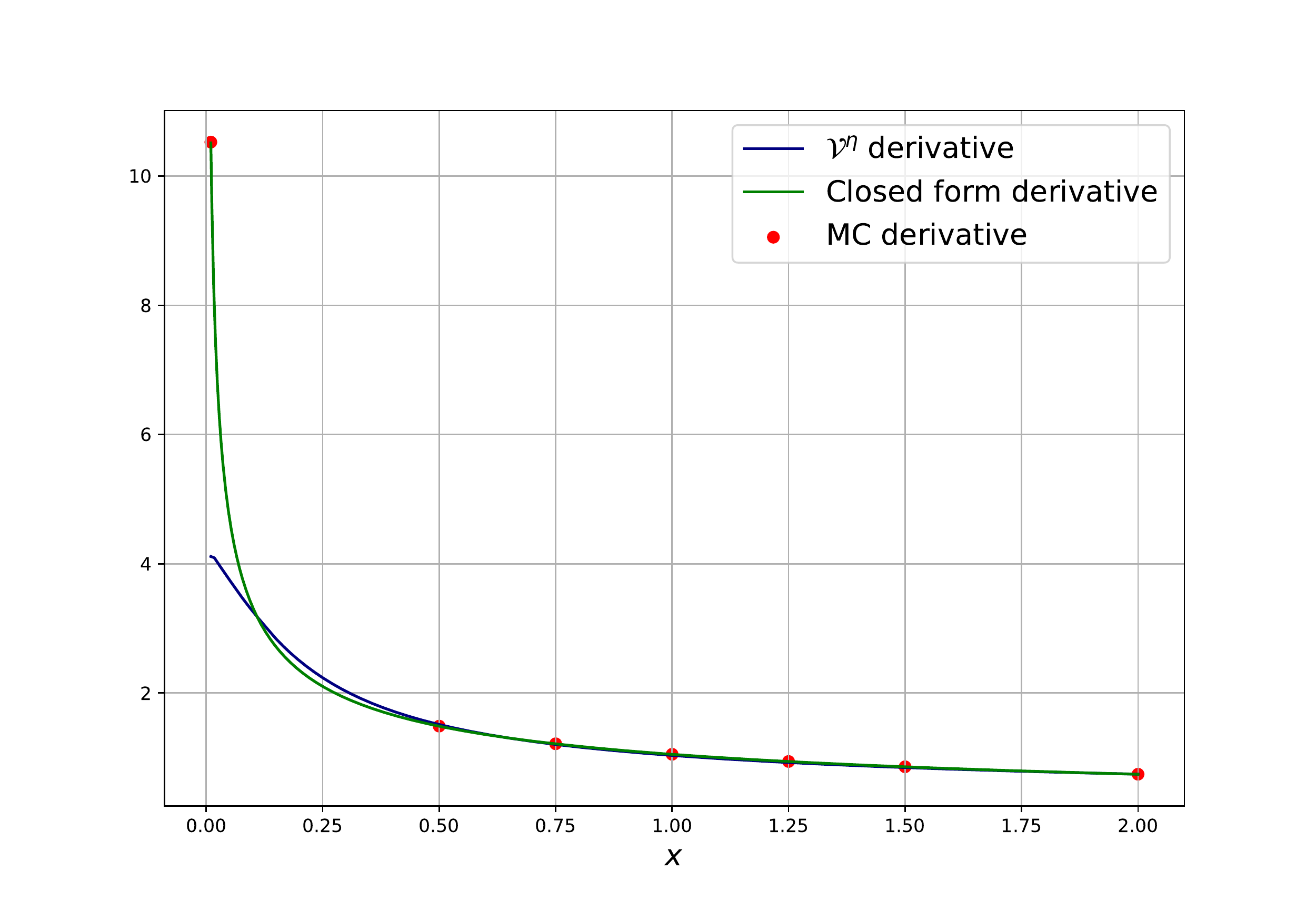}
    \end{subfigure}
    \begin{subfigure}{.32\linewidth}
        \centering
        \includegraphics[height=3.75cm]{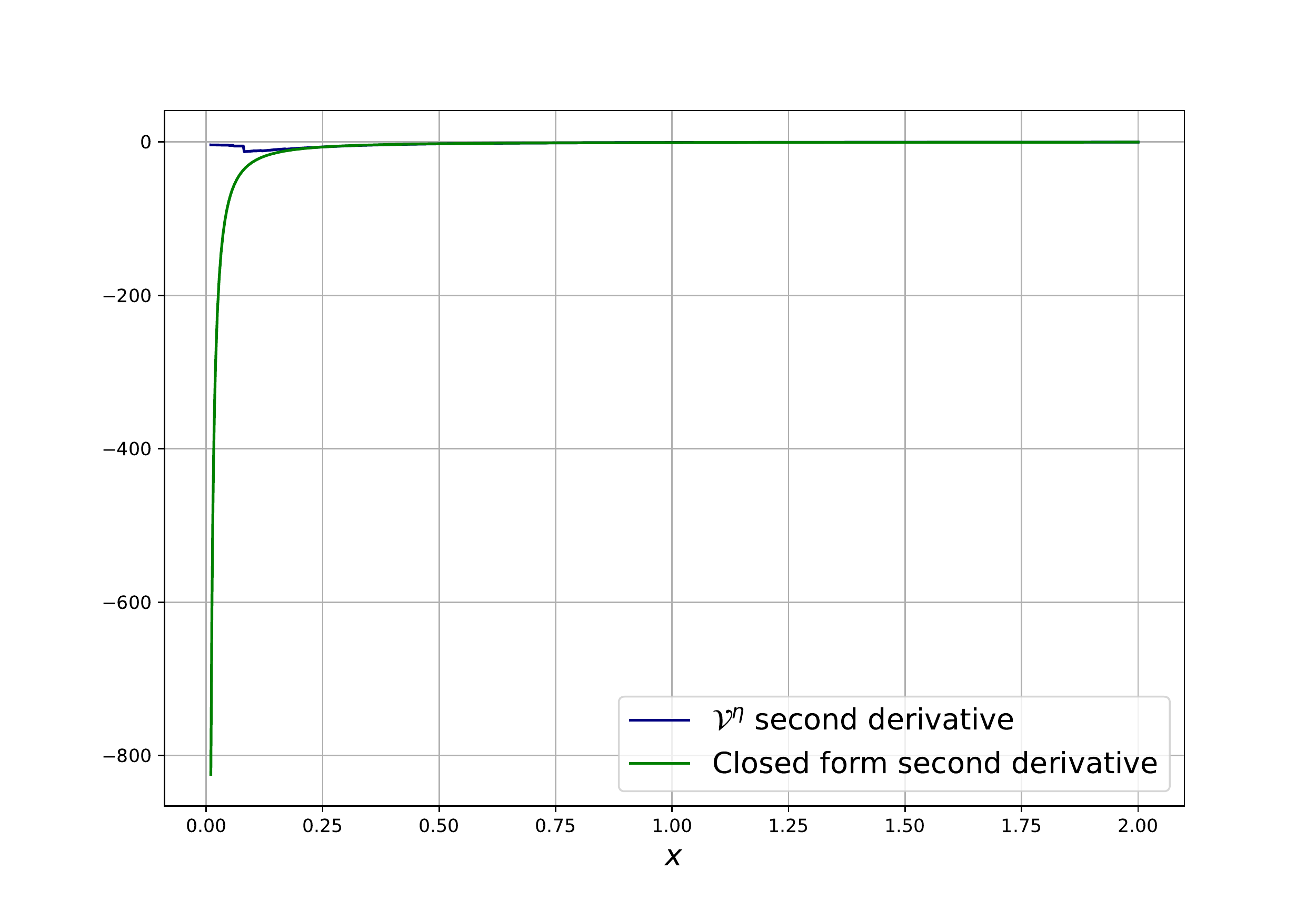} 
        \caption[short]{$t=0$}
    \end{subfigure}
    \begin{subfigure}{.32\linewidth}
        \centering
        \includegraphics[height=3.75cm]{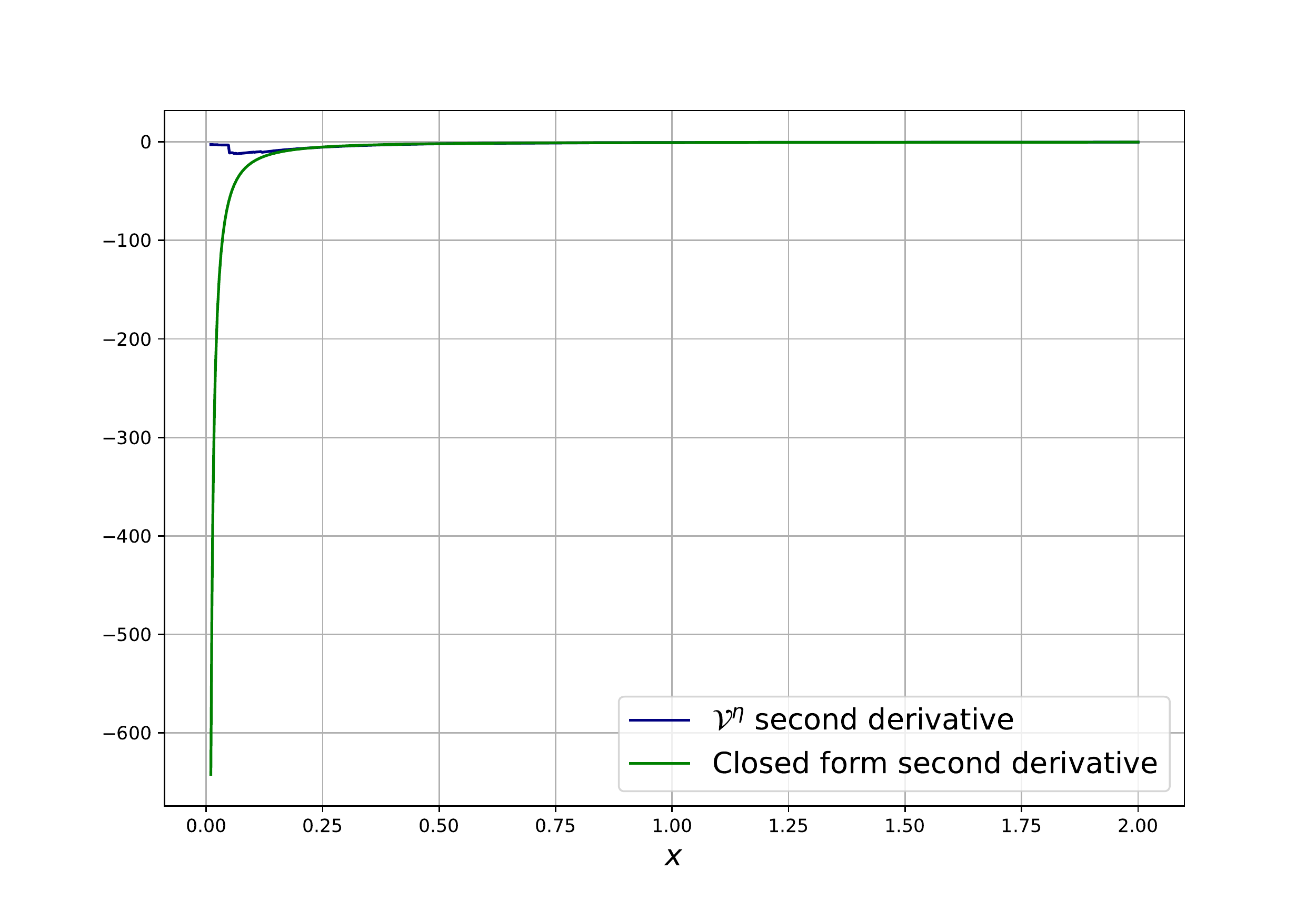}
        \caption[short]{$t=0.5$}
    \end{subfigure}
    \begin{subfigure}{.32\linewidth}
        \centering
        \includegraphics[height=3.75cm]{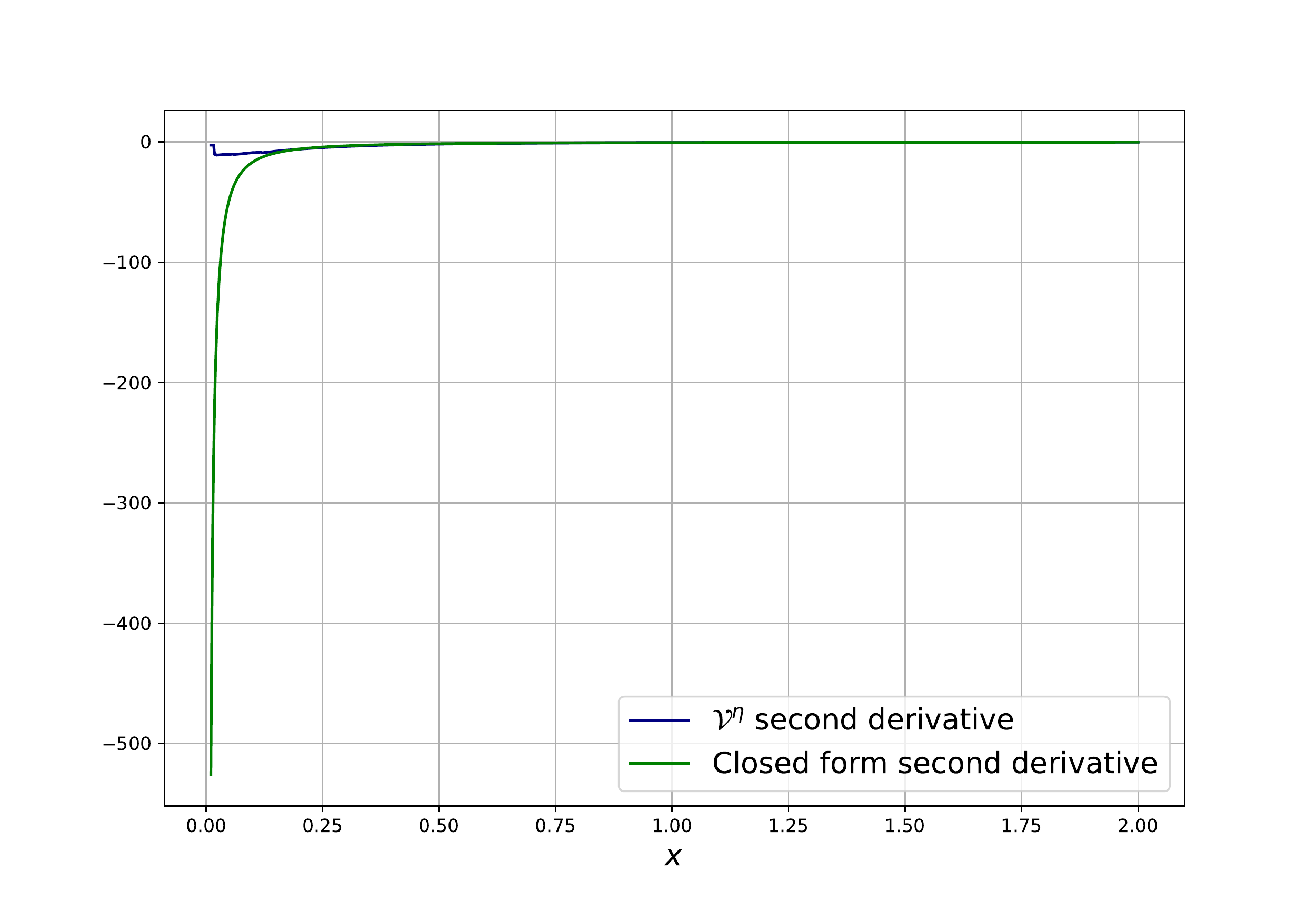}
        \caption[short]{$t=0.9$}
    \end{subfigure}
    \caption{
    \label{fig:value_pathwise_learning_merton_gamma_05}
    Value function $\vartheta^\eta$ and its first and second derivative obtained by Pathwise learning (Algorithm \ref{algo:scheme_value_pathwise_martingale_learning}) for the Merton problem with parameters $b = 0.2$, $\sigma = 0.2$ and power utility with exponent $\gamma=0.5$, plotted as functions of $x$, for fixed values of $t$.
    }
\end{figure}

\begin{figure}[htp]
    \centering
    \begin{subfigure}{.32\linewidth}
        \centering
        \includegraphics[height=3.75cm]{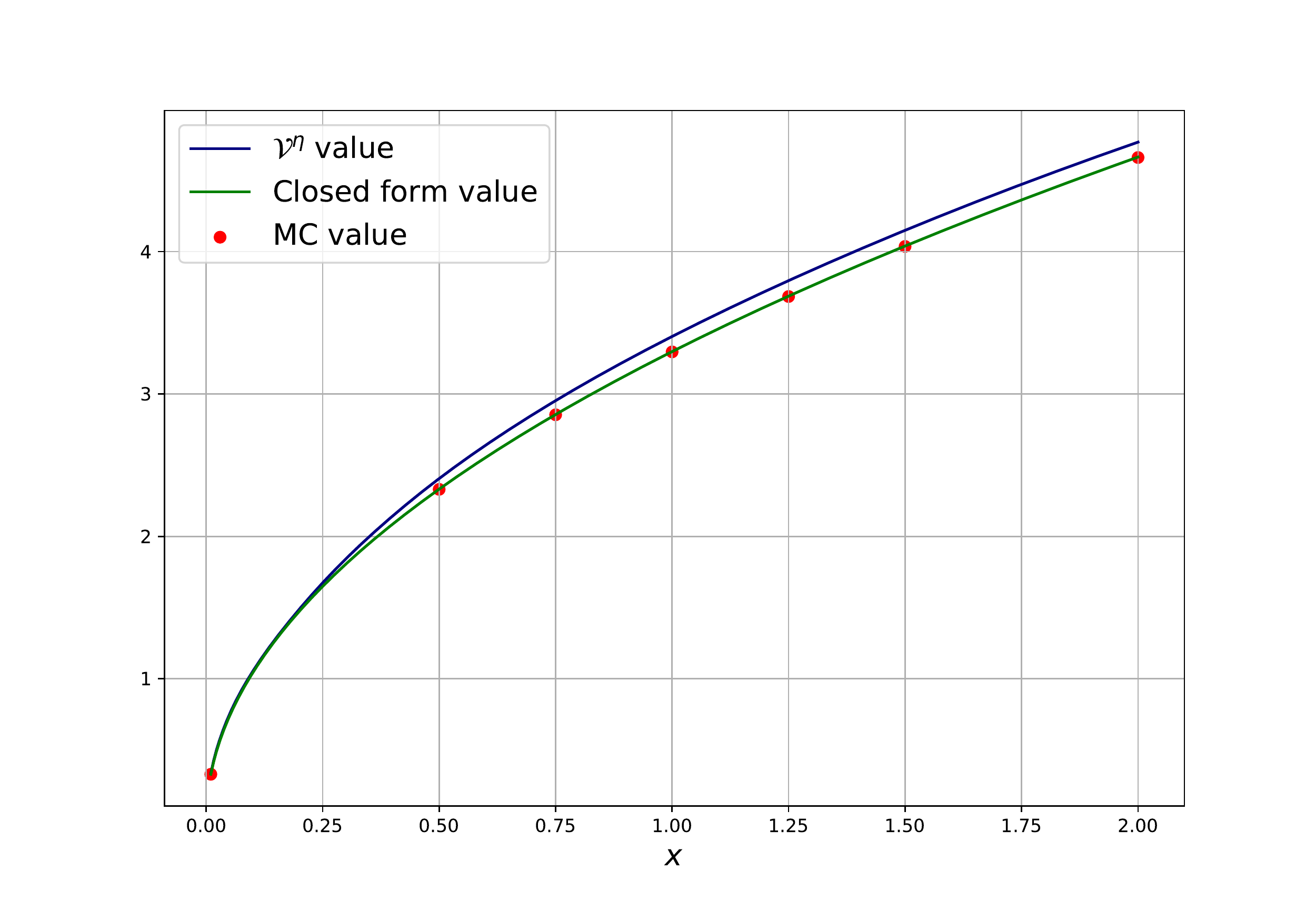} 
    \end{subfigure}
    \begin{subfigure}{.32\linewidth}
        \centering
        \includegraphics[height=3.75cm]{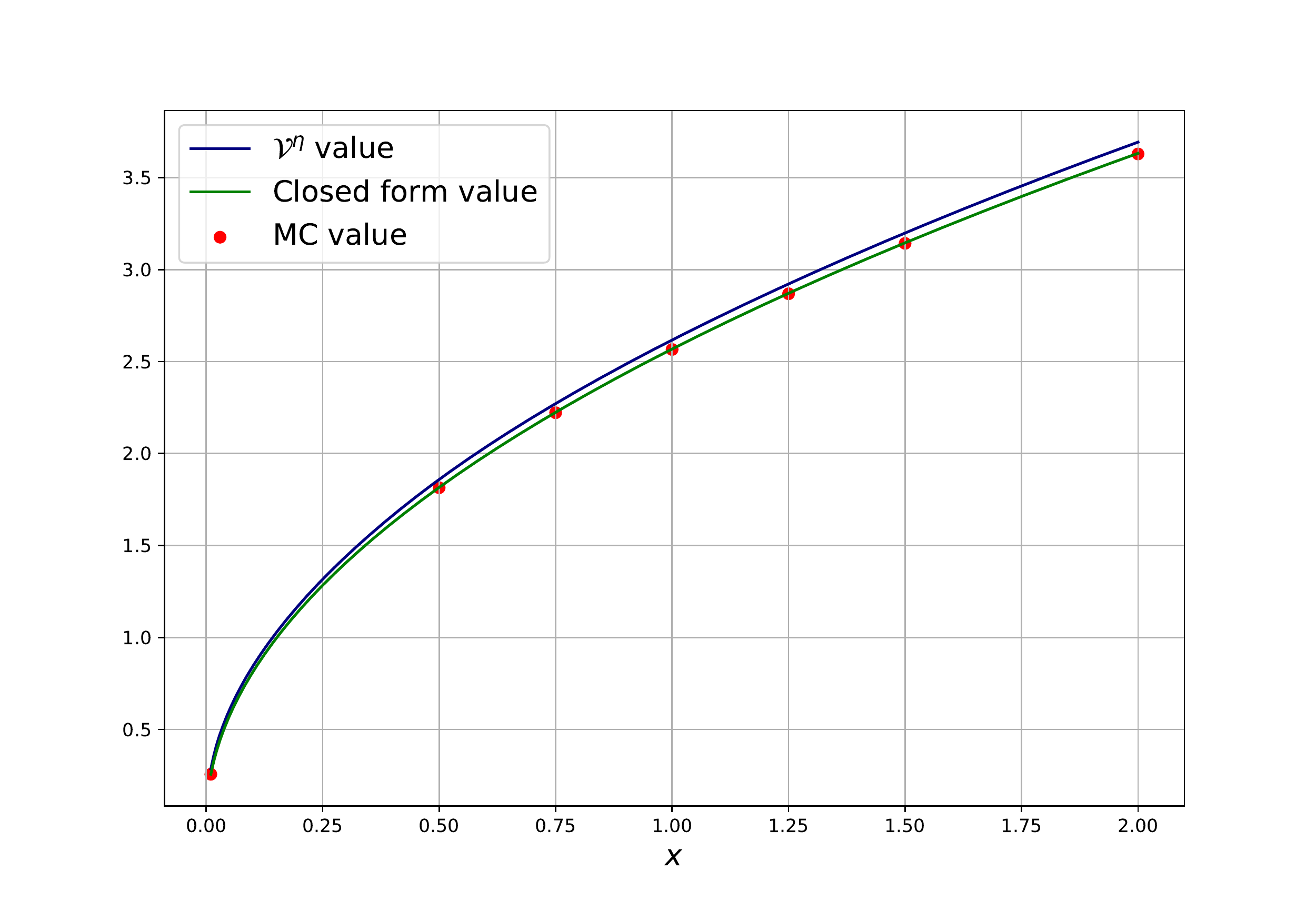}
    \end{subfigure}
    \begin{subfigure}{.32\linewidth}
        \centering
        \includegraphics[height=3.75cm]{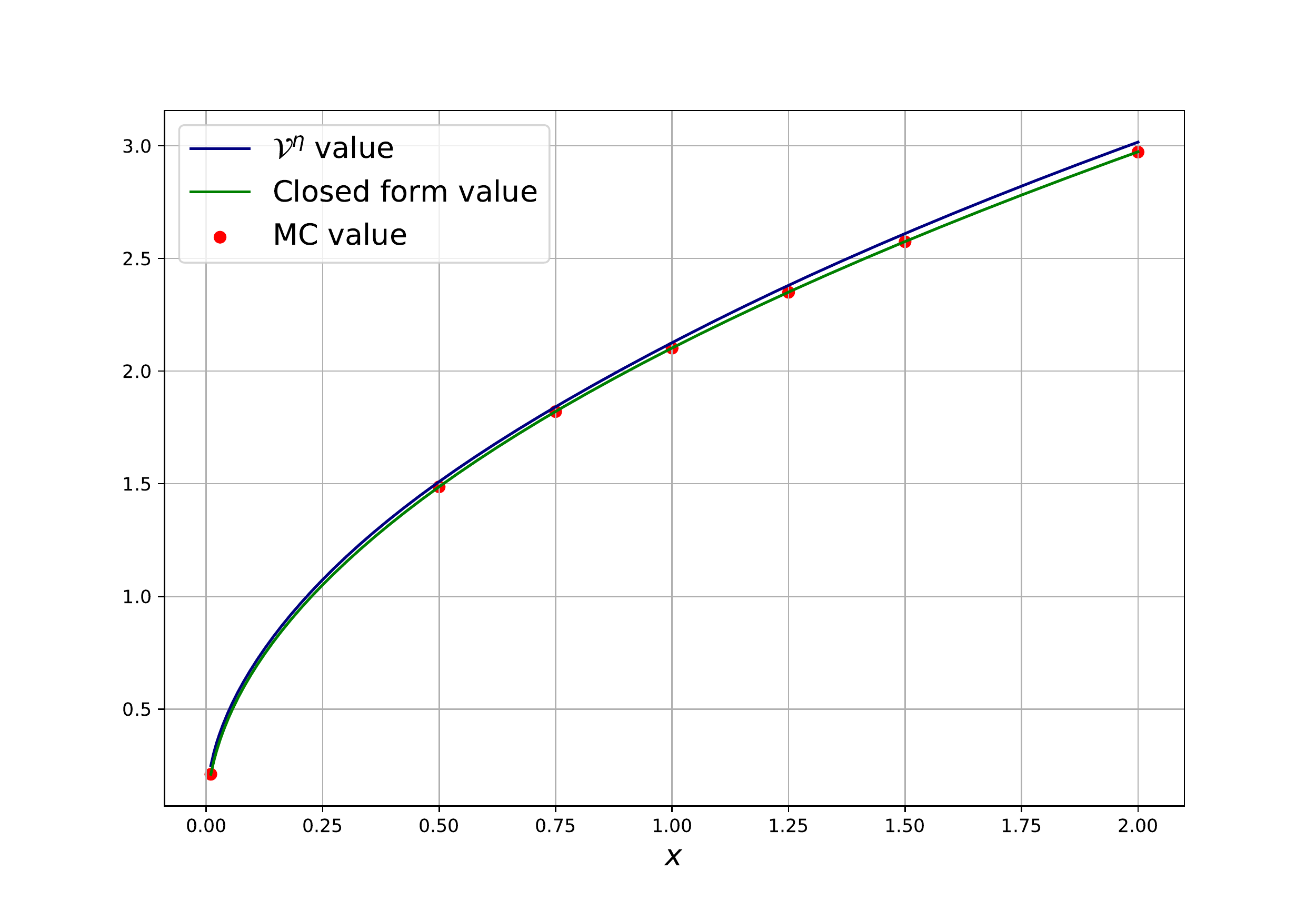}
    \end{subfigure}
    \begin{subfigure}{.32\linewidth}
        \centering
        \includegraphics[height=3.75cm]{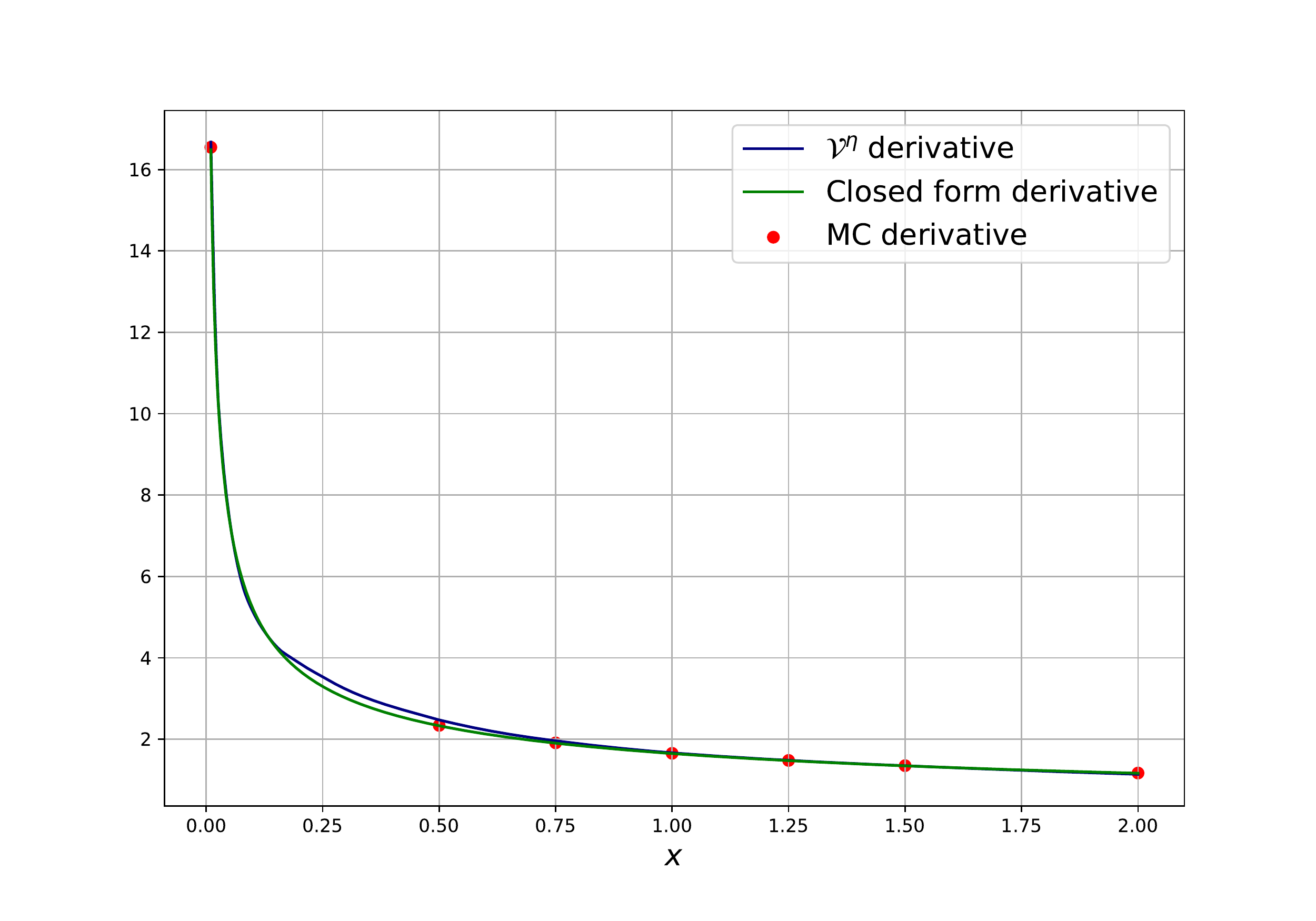} 
    \end{subfigure}
    \begin{subfigure}{.32\linewidth}
        \centering
        \includegraphics[height=3.75cm]{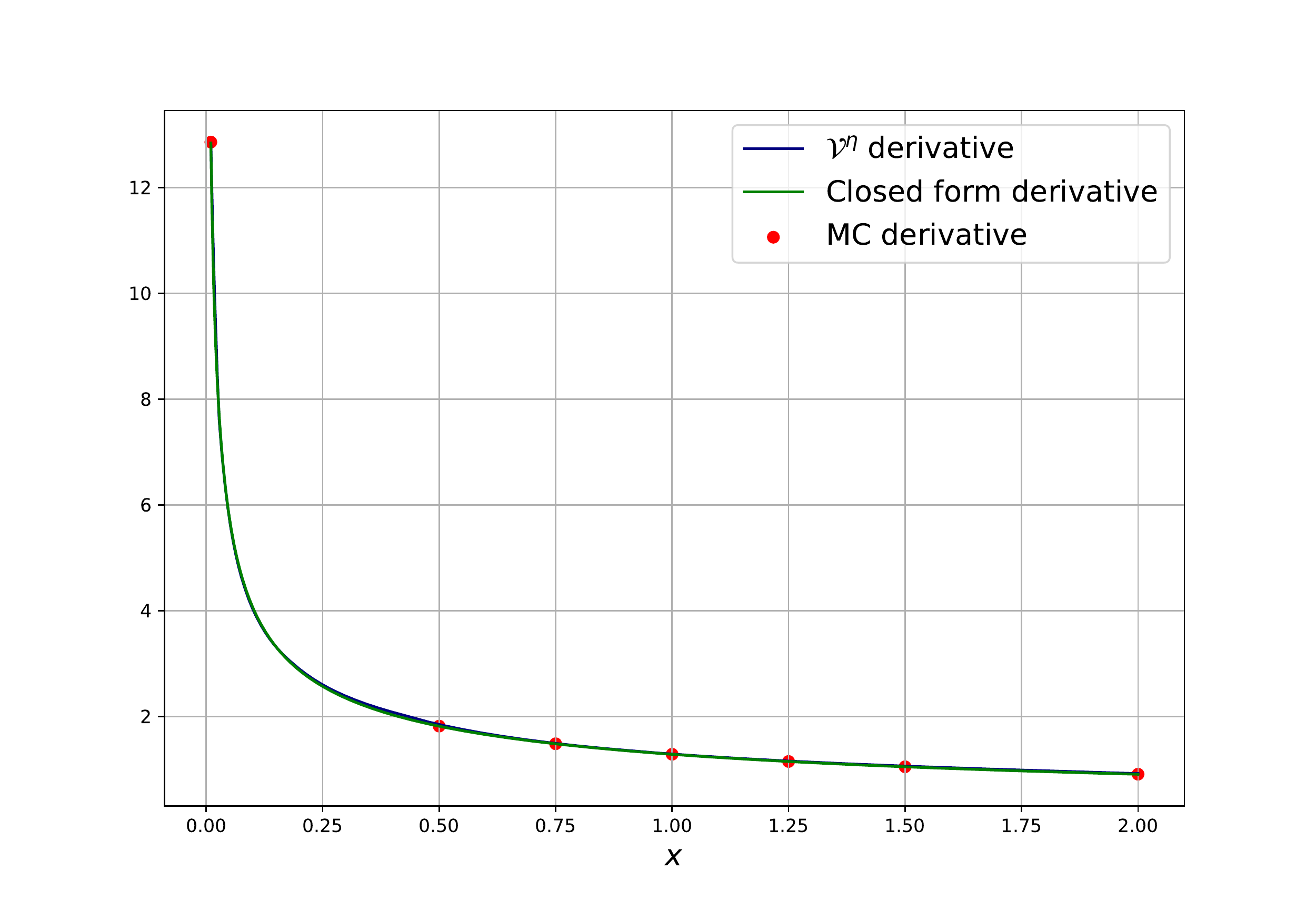}
    \end{subfigure}
    \begin{subfigure}{.32\linewidth}
        \centering
        \includegraphics[height=3.75cm]{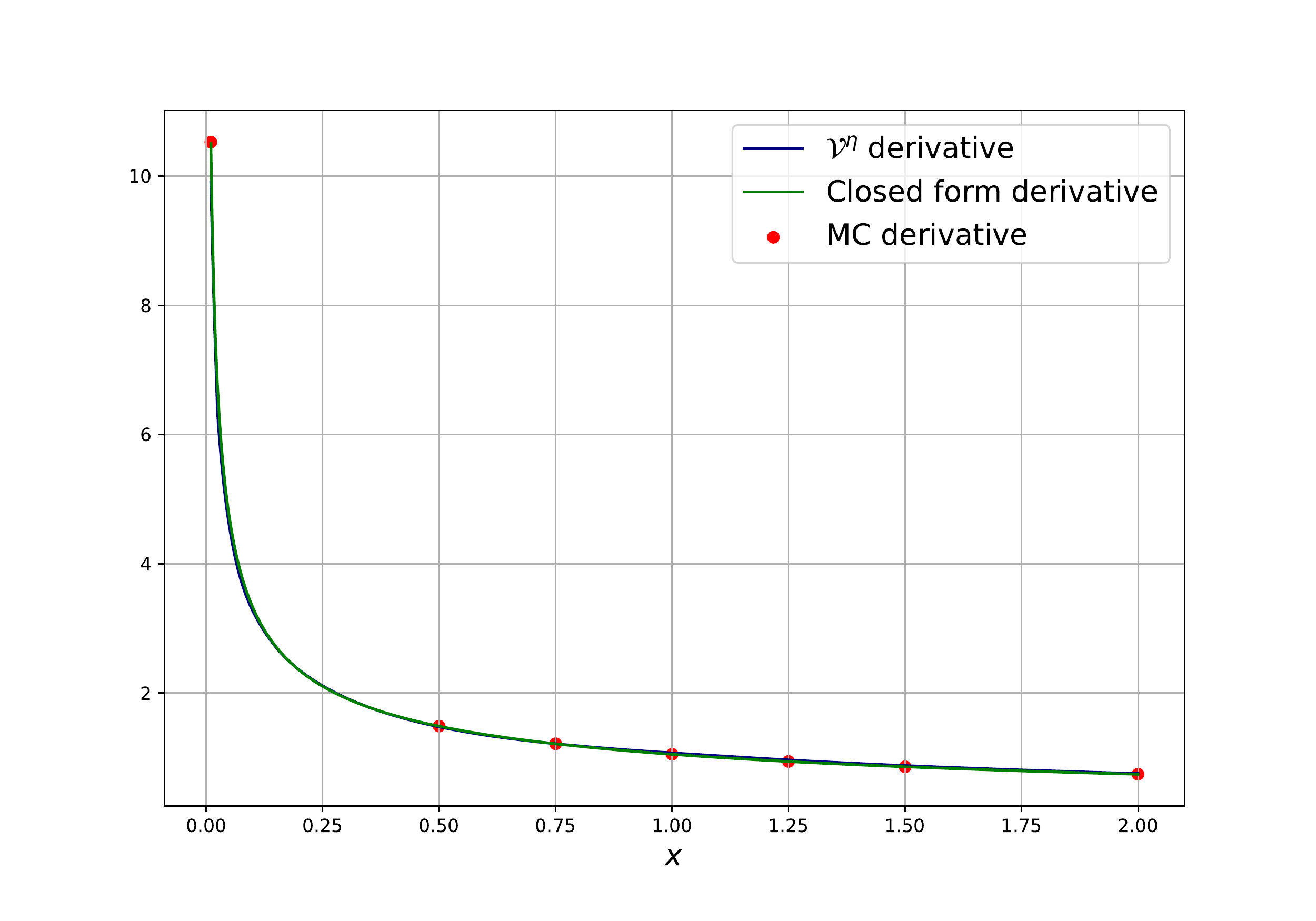}
    \end{subfigure}
    \begin{subfigure}{.32\linewidth}
        \centering
        \includegraphics[height=3.75cm]{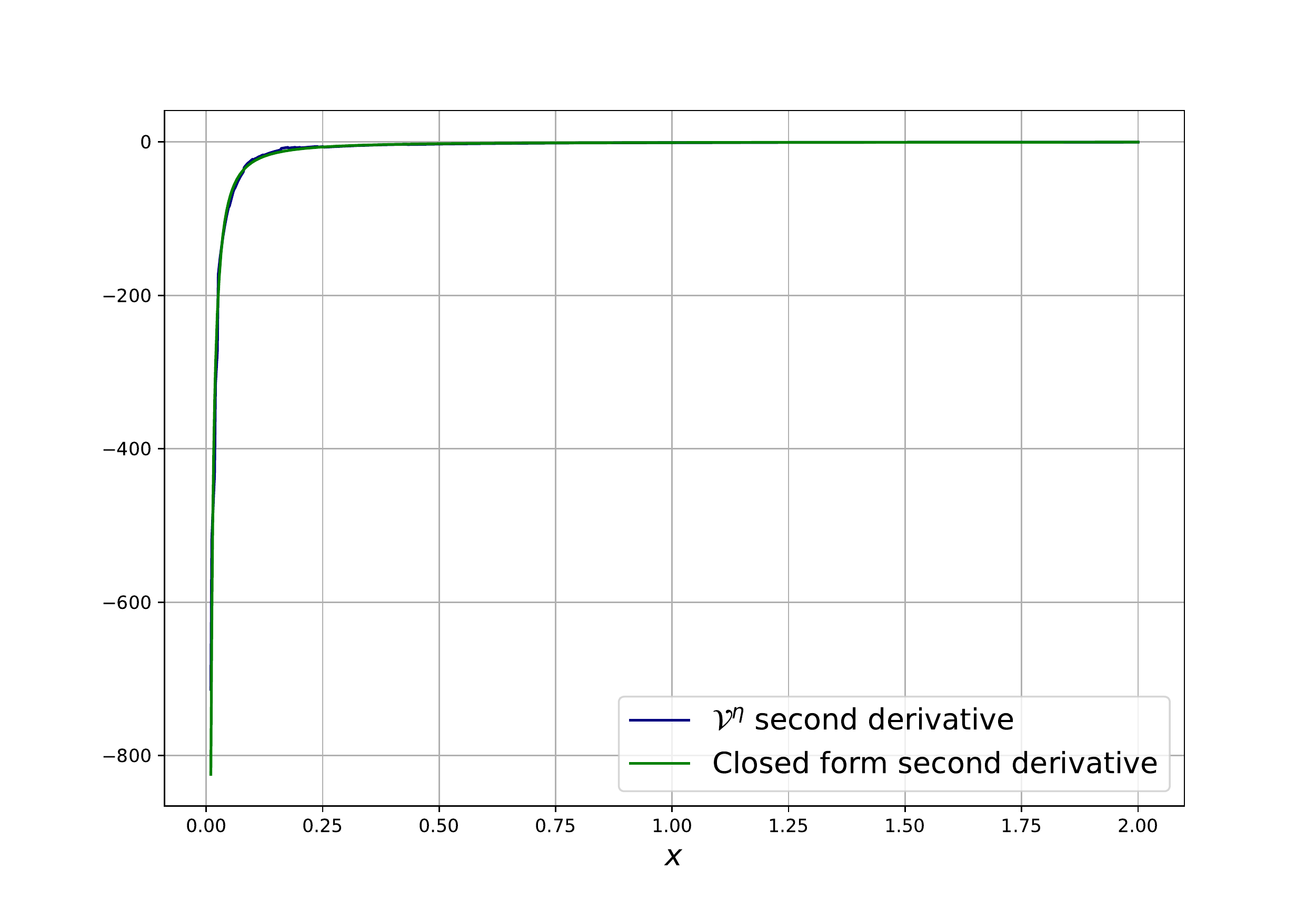} 
        \caption[short]{$t=0$}
    \end{subfigure}
    \begin{subfigure}{.32\linewidth}
        \centering
        \includegraphics[height=3.75cm]{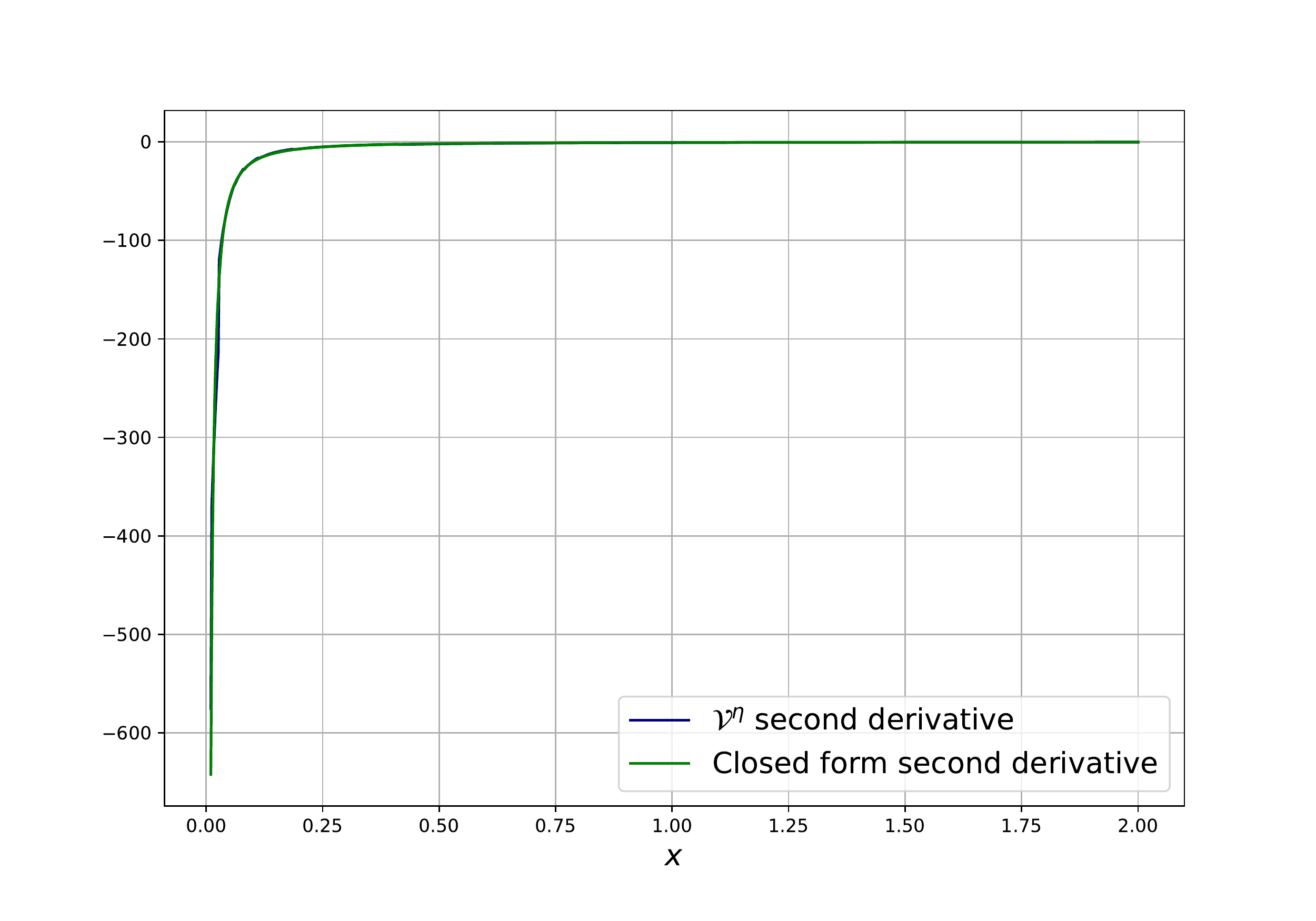}
        \caption[short]{$t=0.5$}
    \end{subfigure}
    \begin{subfigure}{.32\linewidth}
        \centering
        \includegraphics[height=3.75cm]{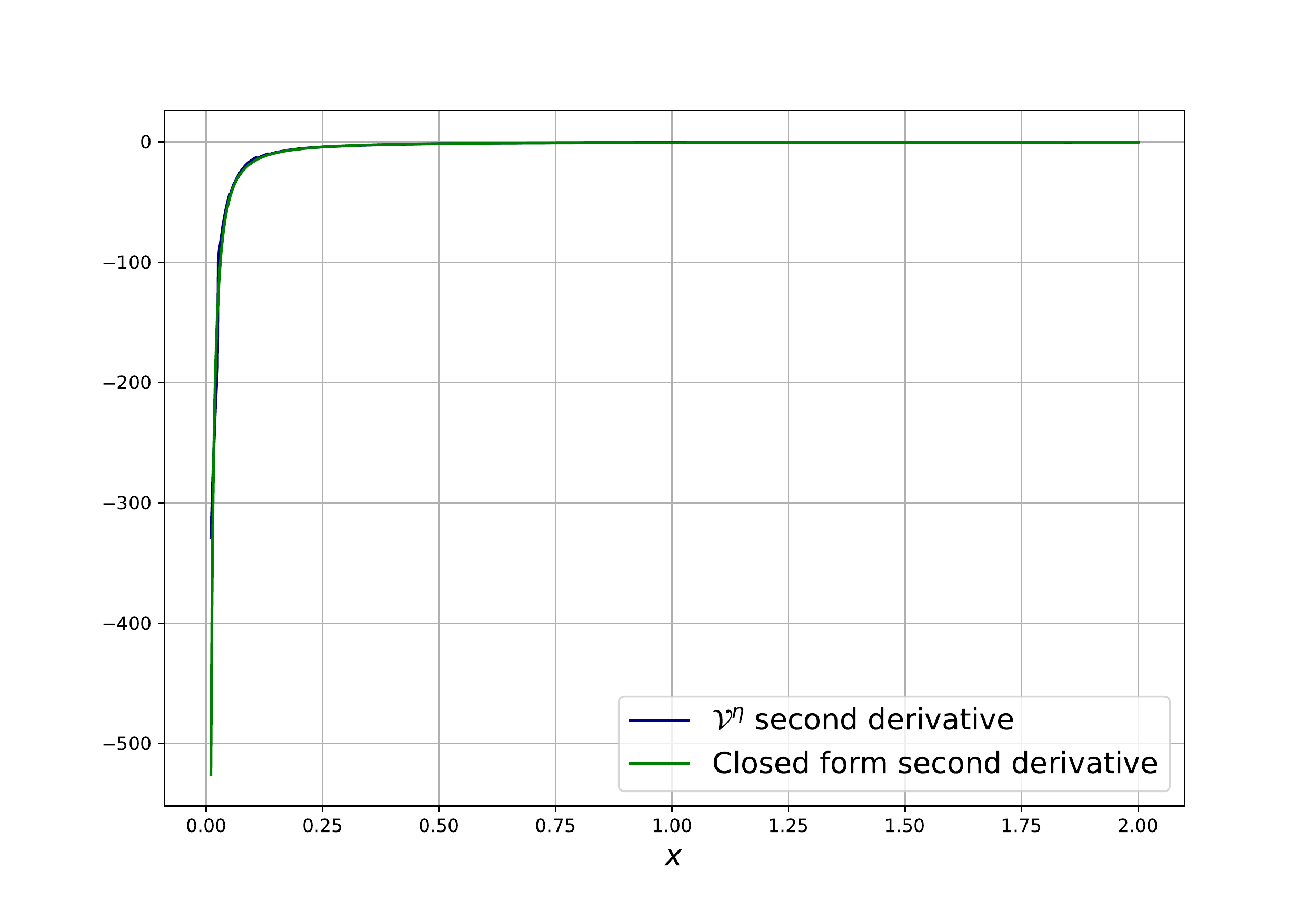}
        \caption[short]{$t=0.9$}
    \end{subfigure}
    \caption{
    \label{fig:value_pathwise_differential_learning_merton_gamma_05}
    Value function $\vartheta^\eta$ and its first and second derivative obtained by Pathwise differential learning (Algorithm \ref{algo:scheme_value_pathwise_differential_learning}) for the Merton problem with parameters $b = 0.2$, $\sigma = 0.2$ and power utility with exponent $\gamma=0.5$, plotted as functions of $x$, for fixed values of $t$.
    }
\end{figure}

We plot the value function $\vartheta^\eta(t,x)$ and its derivatives $\partial_x \vartheta^\eta (t,x)$ and $\partial_{xx} v_\eta (t,x)$ for fixed values $t=0$, $t=0.5$, $t=0.9$, and for parameter values $b=0.2$, $\sigma = 0.2$, $\gamma = 0.5$, and compare it with the closed-form solution of the problem. Figure \ref{fig:value_differential_learning_merton_gamma_05} corresponds to the Differential regression learning method, Figure \ref{fig:value_pathwise_learning_merton_gamma_05} corresponds to the pathwise martingale learning  while Figure \ref{fig:value_pathwise_differential_learning_merton_gamma_05} corresponds to the pathwise differential learning method. These graphs are coherent with the results presented in Table \ref{tableMerton}. We see that the Pathwise differential provides the best fit, in particular for the first and second derivatives of the solution, followed by the Differential regression method. The Pathwise method provides a good fit for the value of the PDE solution but does not manage to fit very well its first and second derivatives for small values of $x$.

\subsection{Example of the Black-Scholes model with linear market impact}\label{sec:BS_linear_market_impact}

We consider the option pricing problem with linear market impact as studied in \cite{loeper2018option}, which leads to a nonlinear Black Scholes (BS) equation in the form \eqref{eq:pde} with $g$ the option payoff and an Hamiltonian $H$ given on $(0,\infty)\times\R$  by 
\bes{ \label{Hpricing} 
H(x,\gamma) &= \; 
\left\{
\begin{array}{cc}
\frac{1}{2} \sigma^2 \frac{x^2 \gamma}{1-\lambda x^2\gamma}, & \mbox{ if } \lambda x^2 \gamma < 1 \\
\infty, &  \mbox{ otherwise.} 
\end{array}
\right.
}
where $\sigma$ $>$ $0$ is the volatility in the BS model, and $\lambda$ is a nonnegative constant related to the linear market impact. 
Notice that $H$ can be written in Bellman form as
\bes{
H(x,\gamma) &= \; \sup_{a\geq 0}\big[ \frac{1}{2}ax^2\gamma - \frac{1}{2\lambda}(\sqrt{a} - \sigma)^2 \big],
}
which corresponds to 
the dual stochastic control representation of the option price $u$ as 
\bes{ \label{dualpricing} 
u(t,x) &= \; \sup_\alpha \E \Big[ g(X_T^{t,x,\alpha}) - \frac{1}{2\lambda} \int_t^T \big( \sqrt{\alpha_s} - \sigma \big)^2 
\mathrm{d}s \Big],
}
where $X$ $=$ $X^{t,x,\alpha}$ is governed by the controlled dynamics
\bes{
\mathrm{d} X_s &= \; X_s \sqrt{\alpha_s} \mathrm{d}W_s, \quad t \leq s \leq T, \; X_t = x > 0,  
}
with a control process $\alpha$ valued in $\R_+$. 

We shall apply the various differential learning methods to this problem for two examples of option payoff.

\subsubsection{Closed-form solution for a logarithmic terminal cost}

We first consider the toy example where the option payoff $g$ is logarithmic: $g(x)$ $=$ $\ln x$. Indeed, in this case, we can check that  
the solution to the pricing PDE \eqref{eq:pde} with $H$ as in \eqref{Hpricing} is given in closed-form by  
\bes{
    u(t,x)= \ln(x) - \frac{\sigma^2}{2(1+\lambda)}(T-t),
}
while  the optimal control  to the dual stochastic control representation \eqref{dualpricing} is constant equal to 
\bes{
\hat\mra(t,x) &= \; \Big(\frac{\sigma}{1+\lambda}\Big)^2.
}

In order to check that the value function approximation obtained is a lower bound of the true one, we compute in Table \ref{tablelog_diffClosedMC} the difference between the closed form value function and the estimation of the value function on points $(t,x)$ obtained by computing the expectation \eqref{ulin} by Monte Carlo on $1e^6$ trajectories controlled by the Deep Learning approximation of the optimal control. We compute the value functions on a grid $t \in \{0, 0.5, 0.9\}$, $x\in \{1e^{-2}, 0.5, 0.75, 1, 1.25, 1.5, 2\}$ with parameter $\sigma=0.3$ and linear market impact factor $\lambda = 5e^{-3}$ and present in the table the difference between the closed form value and the Monte Carlo approximation. For clarity of presentation we present the results averaged over $t$ in this table.

\begin{scriptsize}
\begin{table}[h]
\centering
\begin{tabular}{|c|c|c|c|c|c|c|c|}
\hline  & $x=1 e^{-2}$ & $x=0.5$ &  $x=0.75$ &  $x=1$ &  $x=1.25$ &  $x=1.5$ & $x=2$ \\
\hline  Difference & & & & & & & \\
closed - MC & $2.673e^{-4}$ & $2.042e^{-4}$ & $2.058e^{-4}$ & $2.066e^{-4}$ & $2.099e^{-4}$ & $2.147e^{-4}$ & $2.300e^{-4}$\\
\hline
\end{tabular}
\caption{Difference between the closed form value function and the value computed by Monte Carlo on $1e^6$ trajectories on points $x\in \{1e^{-2}, 0.5, 0.75, 1, 1.25, 1.5, 2\}$ and averaged over times $t \in \{0, 0.5, 0.9\}$ for Black-Scholes problem with linear market impact factor $\lambda=5e^{-3}$ and parameter $\sigma = 0.3$.} 
\label{tablelog_diffClosedMC} 
\end{table}
\end{scriptsize}

In Table \ref{tablelog}, we compute the residual losses  defined in  \eqref{residual} for the NN $\vartheta^\eta$ obtained by the various deep learning methods: the differential learning scheme (Algorithm \ref{algo:scheme_value_differential_learning}), 
the pathwise martingale learning with $1$ NN (Algorithm \ref{algo:scheme_value_pathwise_martingale_learning}), the pathwise differential learning with $1$ NN (Algorithm \ref{algo:scheme_value_pathwise_differential_learning}). We also provide the training time for 500 epochs for each of these algorithms. On this table, we see that the Pathwise learning methods yield better results than the Differential regression learning methods. The difference between the residual loss only and the sum of the residual and terminal loss is small in all three methods, meaning that both the PDE solution's derivatives and terminal condition have been learned by the neural network.

\begin{scriptsize}
\begin{table}[h]
\centering
\begin{tabular}{|c|c|c|c|}
\hline  & Diff. regr. learning & Path. 1NN &  Path. diff. 1NN \\
\hline  Residual loss  & $2.046e^{-3}$ & $3.484e^{-4}$ & $6.644e^{-4}$ \\
\hline  Residual loss  & & & \\
+ terminal loss & $2.179e^{-3}$ & $3.864e^{-4}$ & $6.758e^{-4}$ \\
\hline  Training time & & & \\
(500 epochs)  &  163s & 130s & 525s  \\
\hline
\end{tabular}
\caption{Residual and boundary losses computed on a $102$x$102$ time and space grid with $t\in[0,0.9]$ and $x\in[0.1, 2]$ for a terminal logarithmic payoff, with parameter $\sigma = 0.3$ and linear market impact factor $\lambda = 5e^{-3}$.} 
\label{tablelog} 
\end{table}
\end{scriptsize}

\begin{figure}[htp]
    \centering
    \begin{subfigure}{.32\linewidth}
        \centering
        \includegraphics[height=3.75cm]{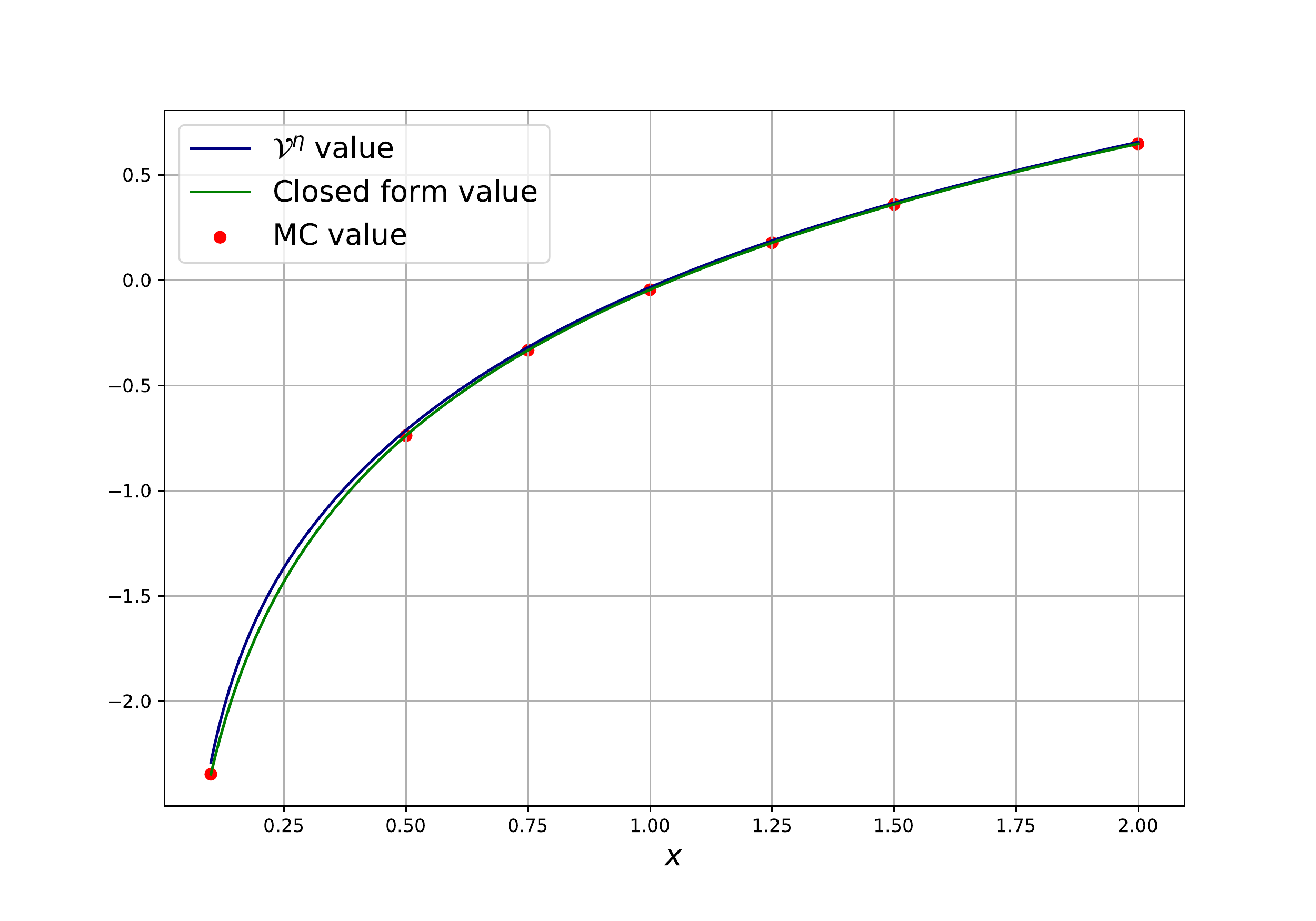} 
    \end{subfigure}
    \begin{subfigure}{.32\linewidth}
        \centering
        \includegraphics[height=3.75cm]{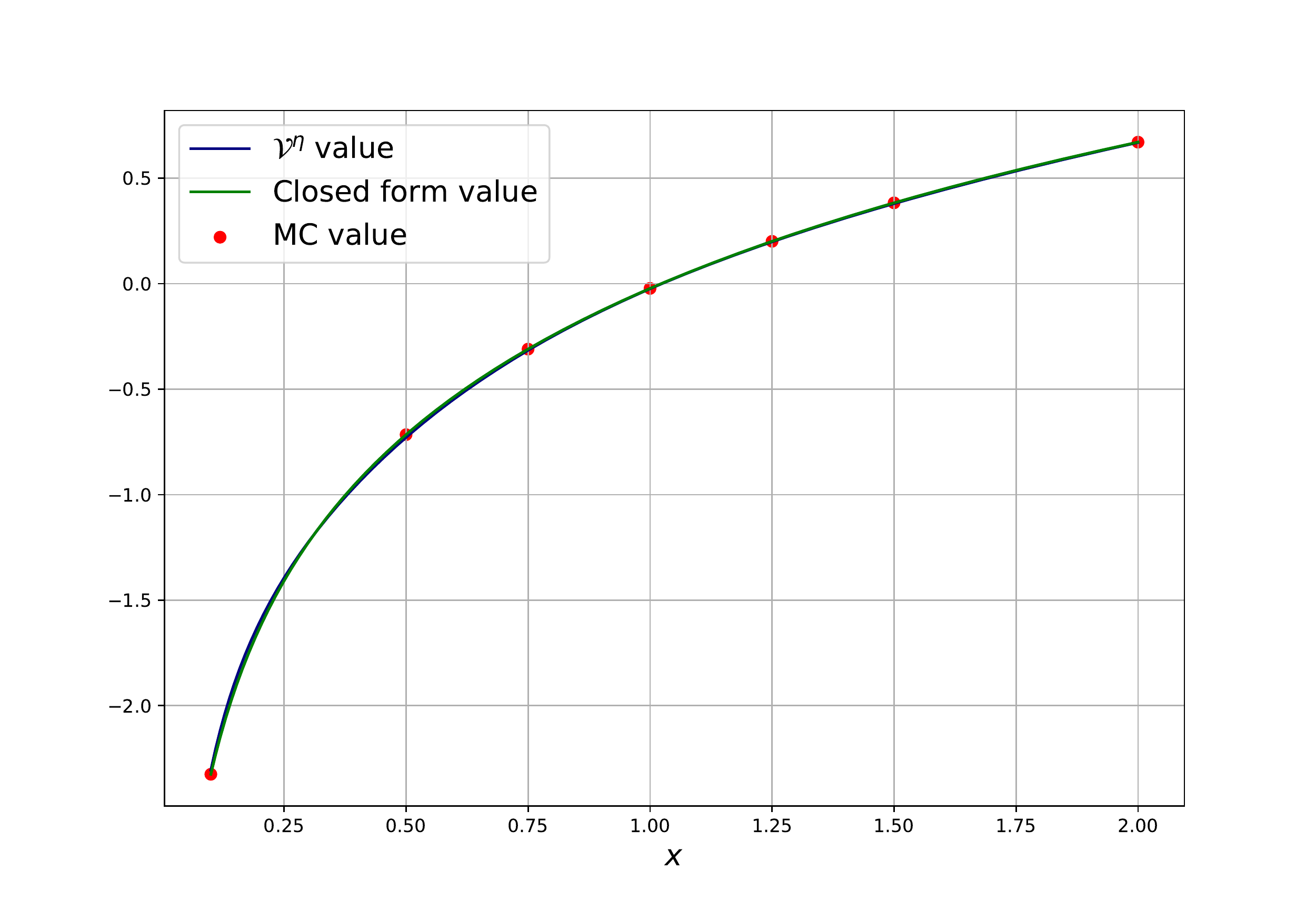}
    \end{subfigure}
    \begin{subfigure}{.32\linewidth}
        \centering
        \includegraphics[height=3.75cm]{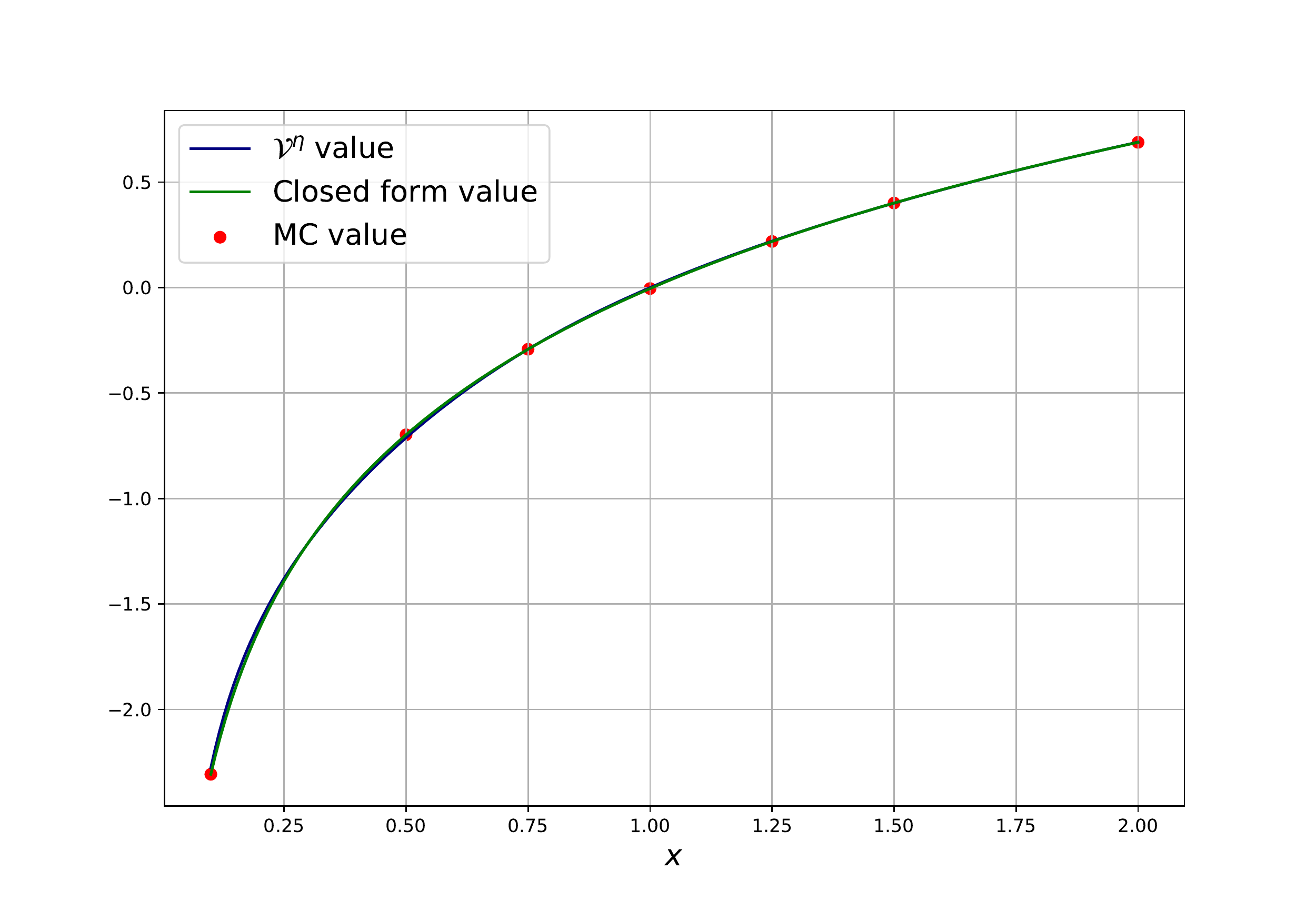}
    \end{subfigure}
    \begin{subfigure}{.32\linewidth}
        \centering
        \includegraphics[height=3.75cm]{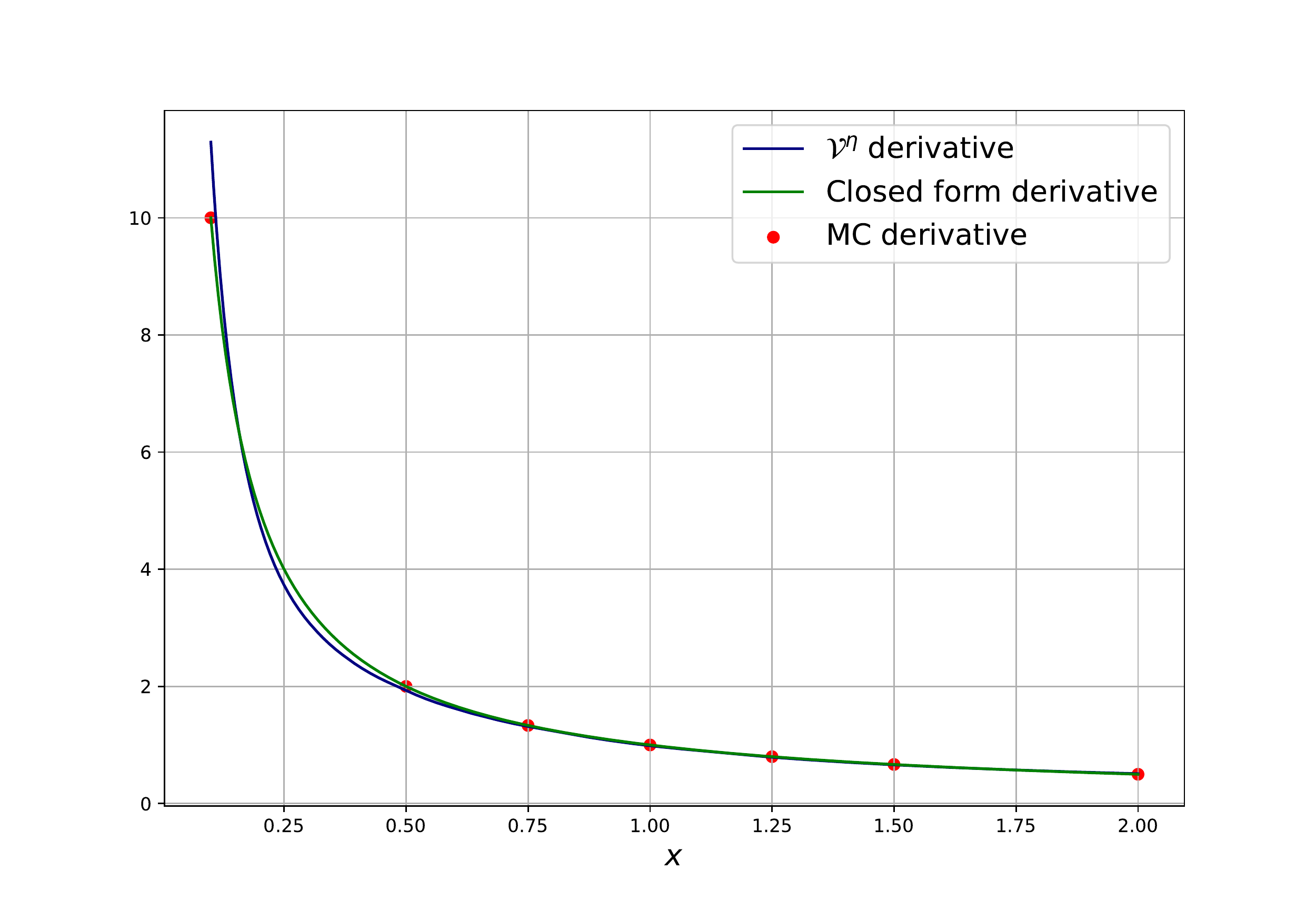} 
    \end{subfigure}
    \begin{subfigure}{.32\linewidth}
        \centering
        \includegraphics[height=3.75cm]{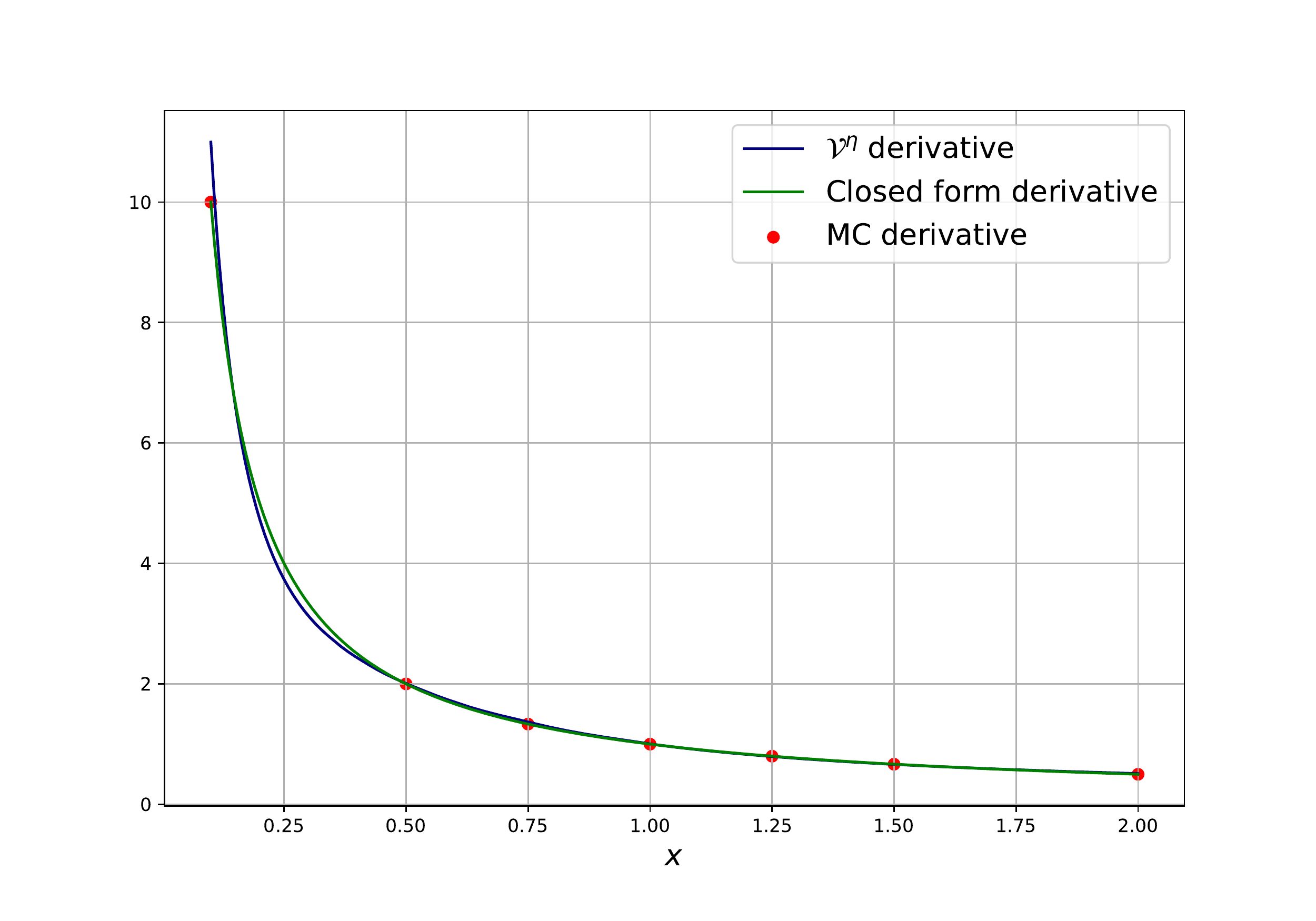}
    \end{subfigure}
    \begin{subfigure}{.32\linewidth}
        \centering
        \includegraphics[height=3.75cm]{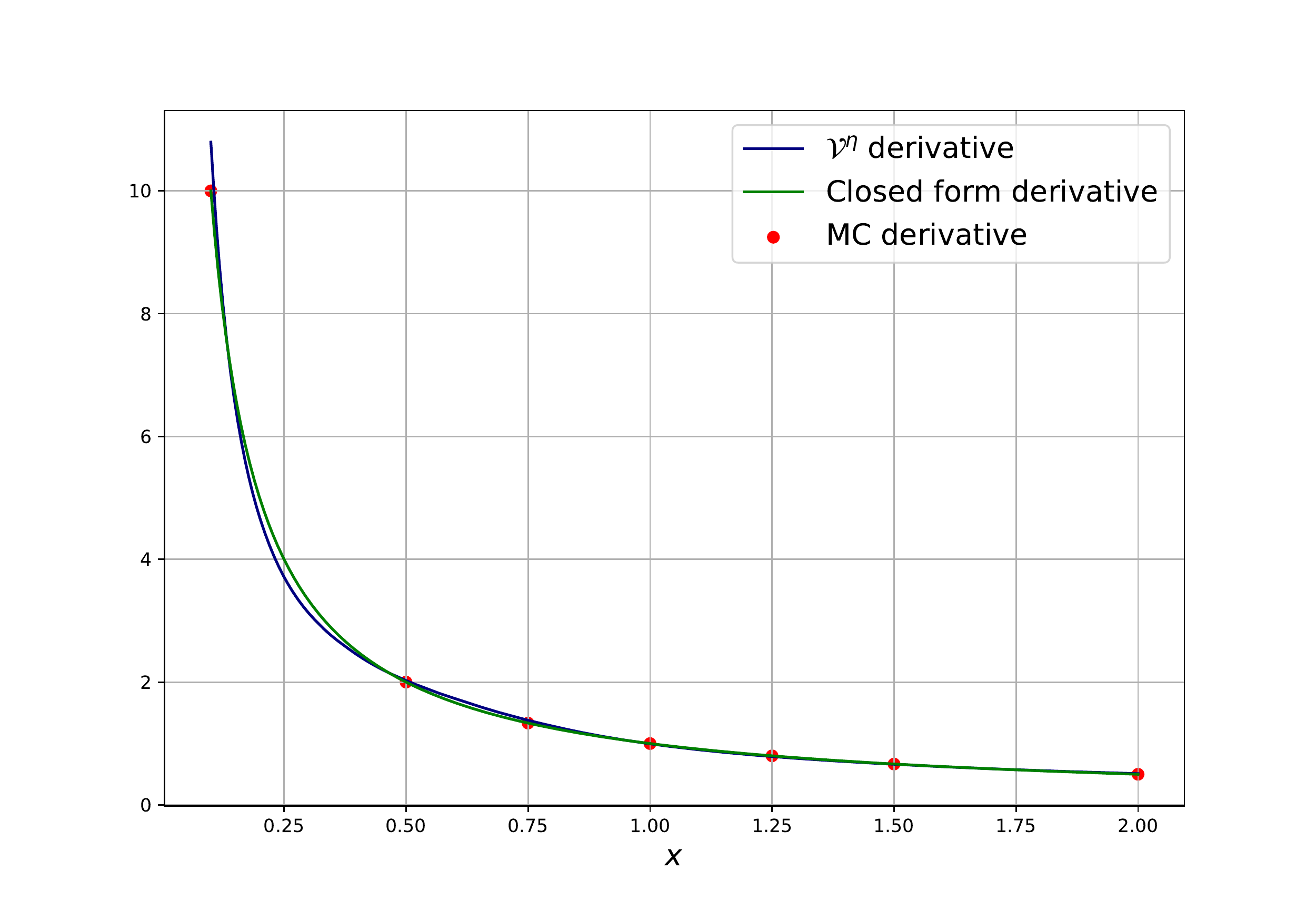}
    \end{subfigure}
    \begin{subfigure}{.32\linewidth}
        \centering
        \includegraphics[height=3.75cm]{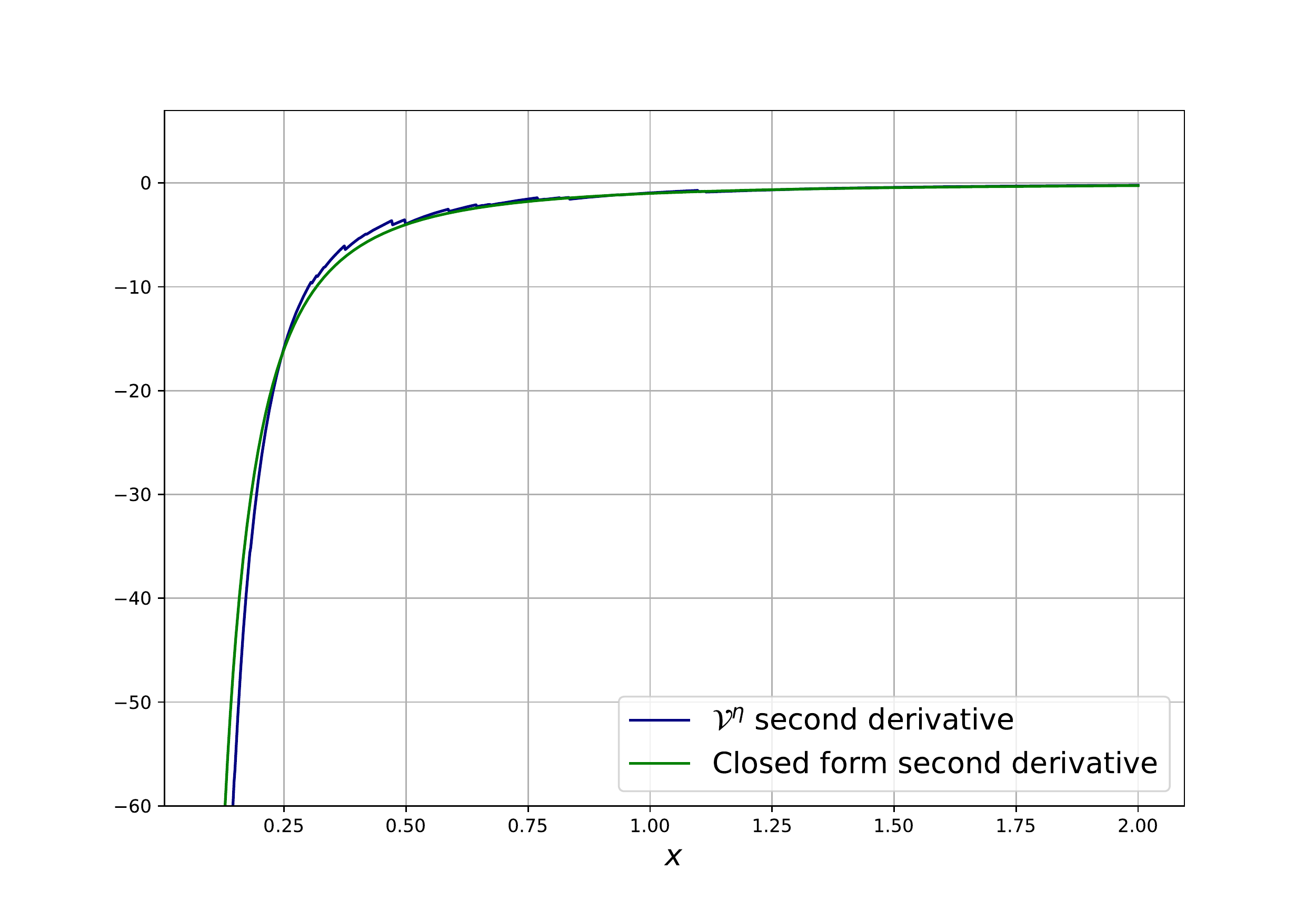} 
        \caption[short]{$t=0$}
    \end{subfigure}
    \begin{subfigure}{.32\linewidth}
        \centering
        \includegraphics[height=3.75cm]{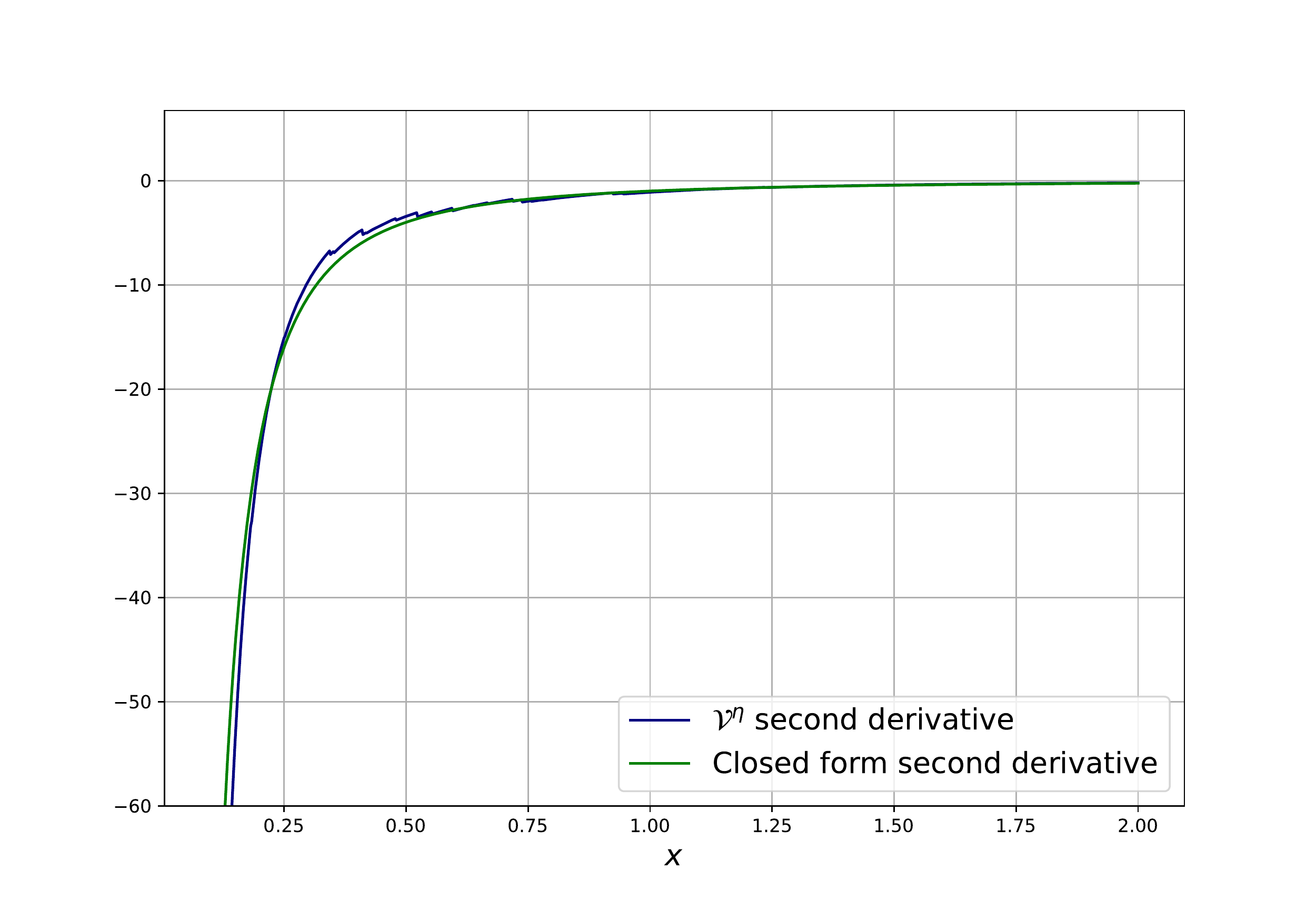}
        \caption[short]{$t=0.5$}
    \end{subfigure}
    \begin{subfigure}{.32\linewidth}
        \centering
        \includegraphics[height=3.75cm]{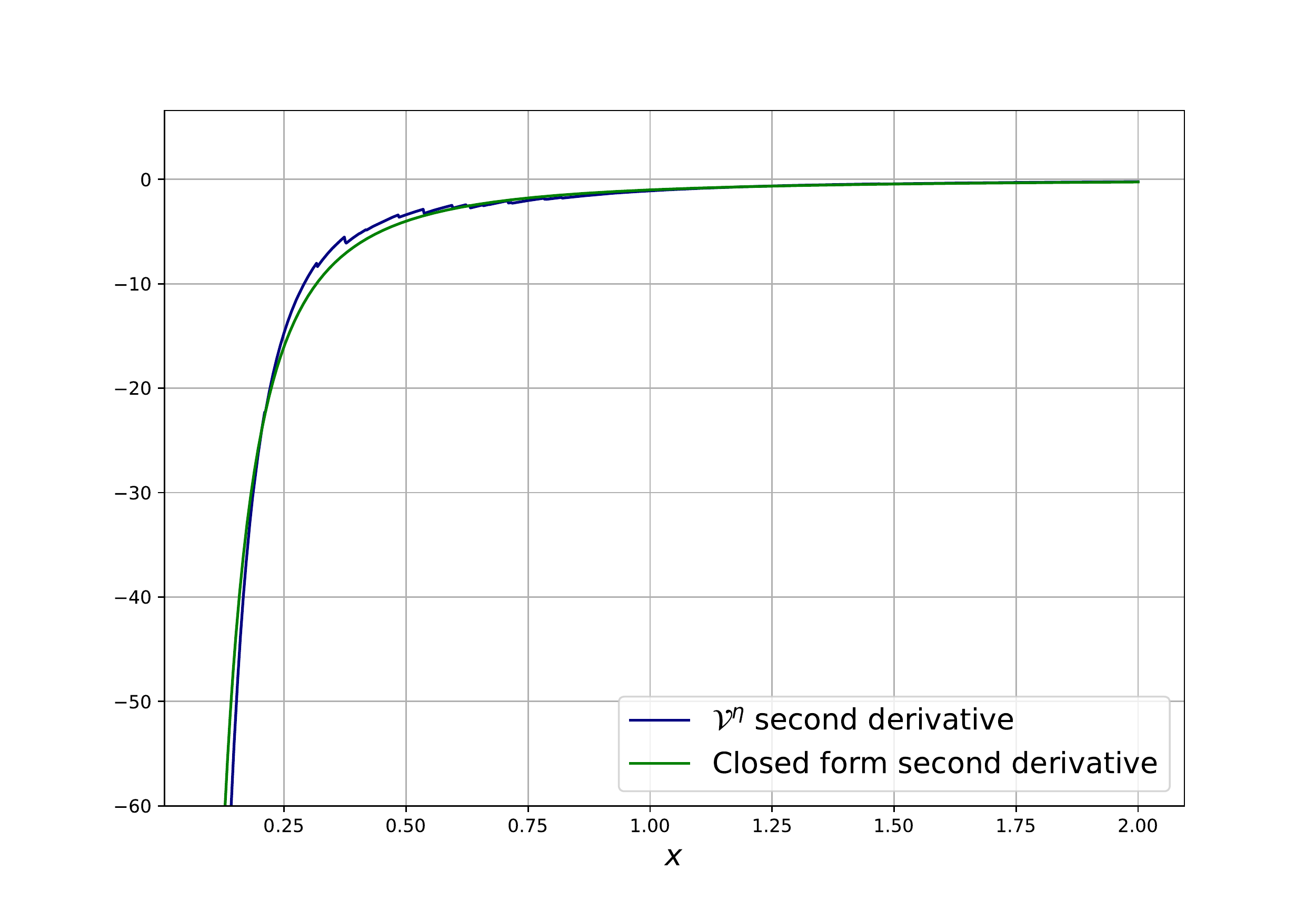}
        \caption[short]{$t=0.9$}
    \end{subfigure}
    \caption{
    \label{fig:value_differential_learning_log}
    Value function $\vartheta^\eta$ and its first and second derivative obtained by Differential Regression Learning (Algorithm \ref{algo:scheme_value_differential_learning}) for a logarithmic option payoff, with parameter $\sigma = 0.3$ and linear market impact factor $\lambda = 5e^{-3}$, plotted as functions of $x$, for fixed values of $t$.
    }
\end{figure}

\begin{figure}[htp]
    \centering
    \begin{subfigure}{.32\linewidth}
        \centering
        \includegraphics[height=3.75cm]{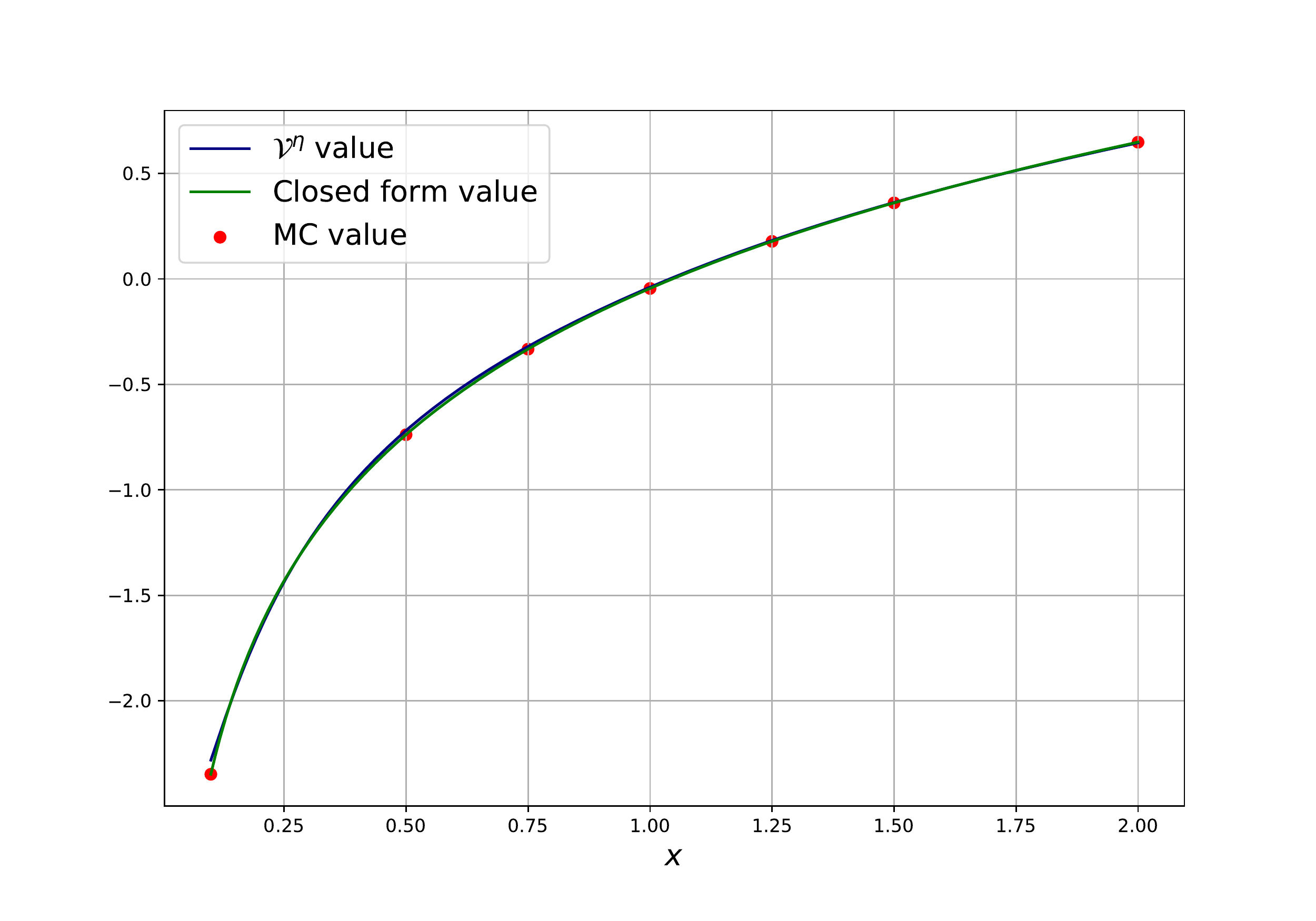} 
    \end{subfigure}
    \begin{subfigure}{.32\linewidth}
        \centering
        \includegraphics[height=3.75cm]{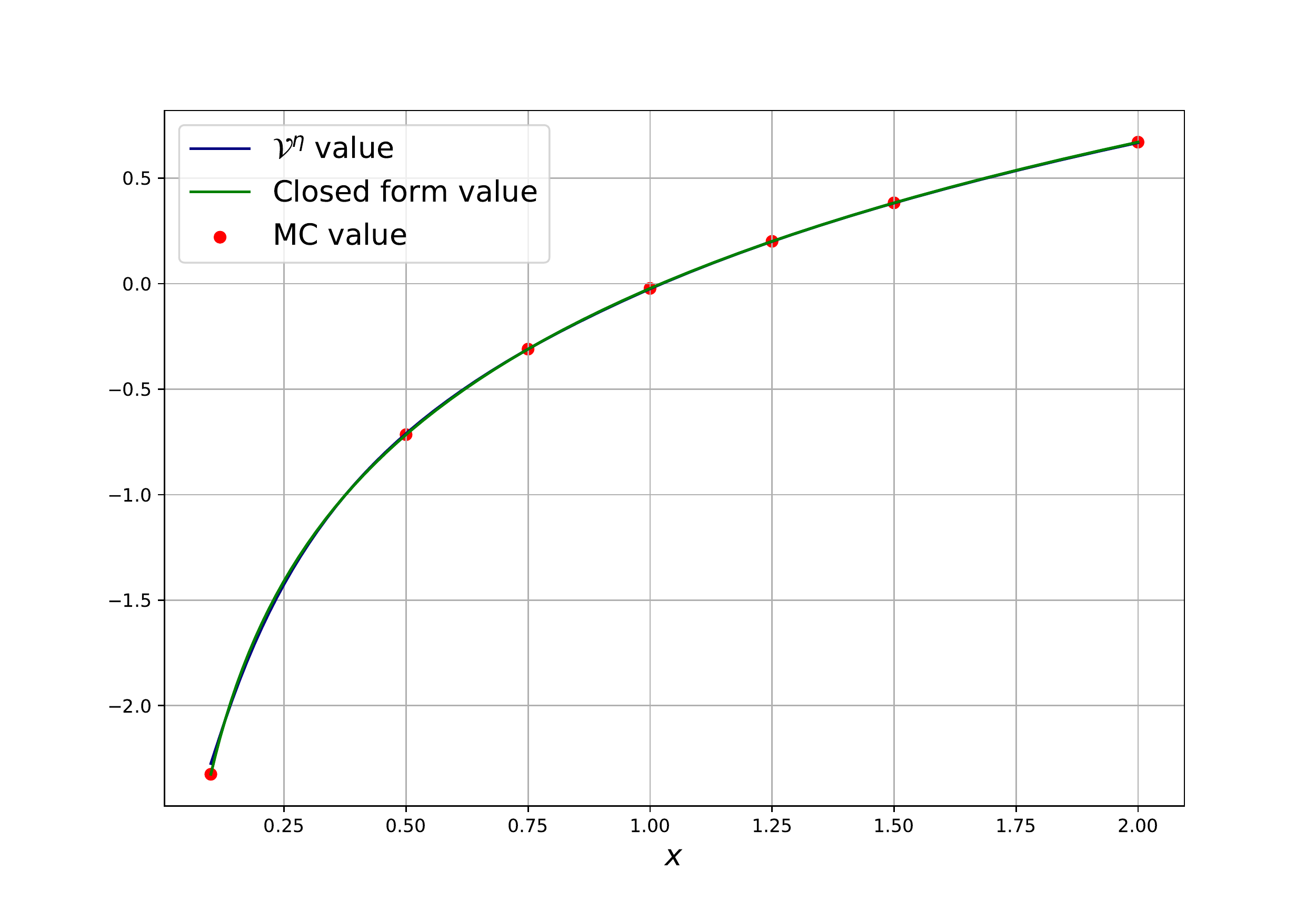}
    \end{subfigure}
    \begin{subfigure}{.32\linewidth}
        \centering
        \includegraphics[height=3.75cm]{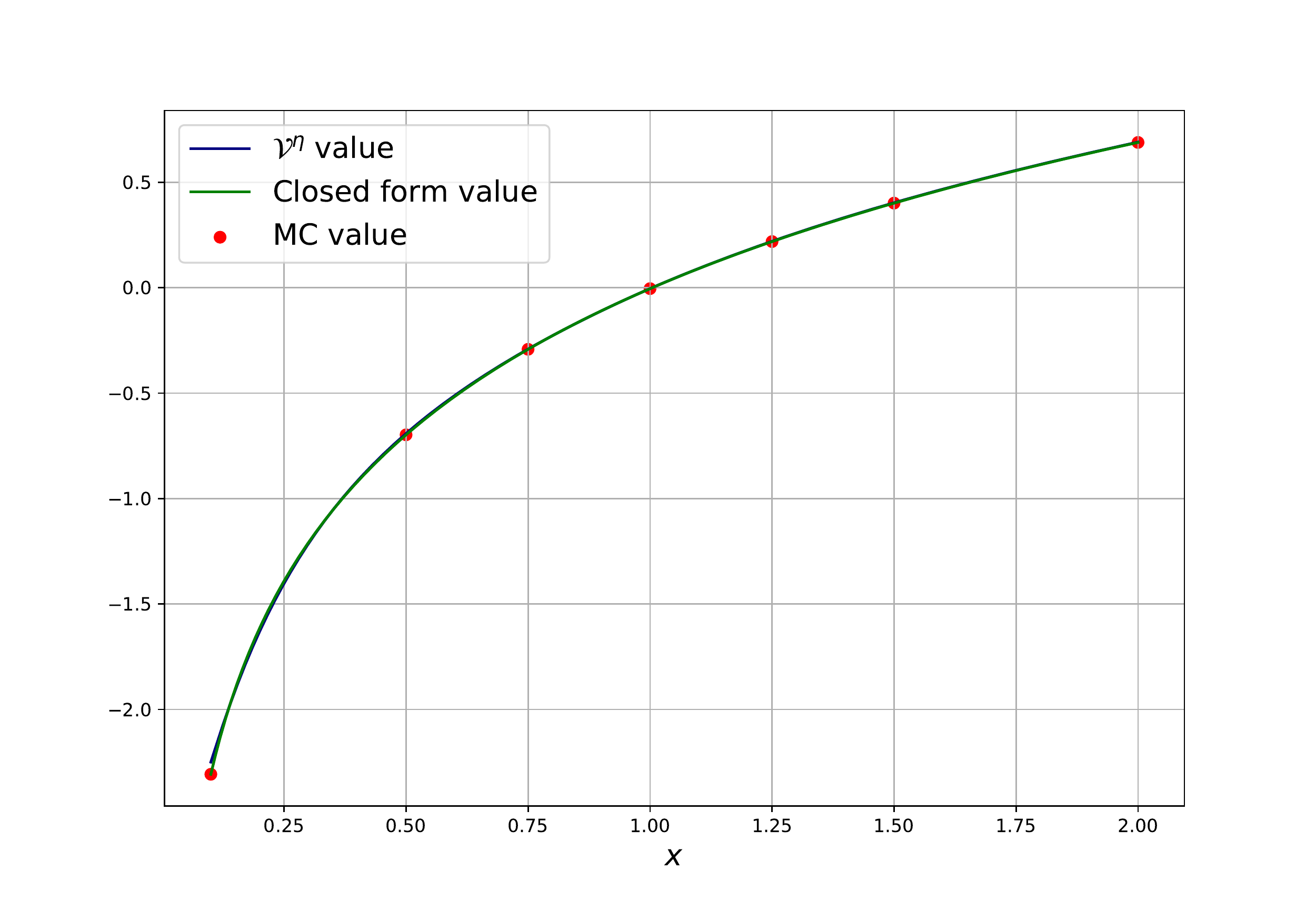}
    \end{subfigure}
    \begin{subfigure}{.32\linewidth}
        \centering
        \includegraphics[height=3.75cm]{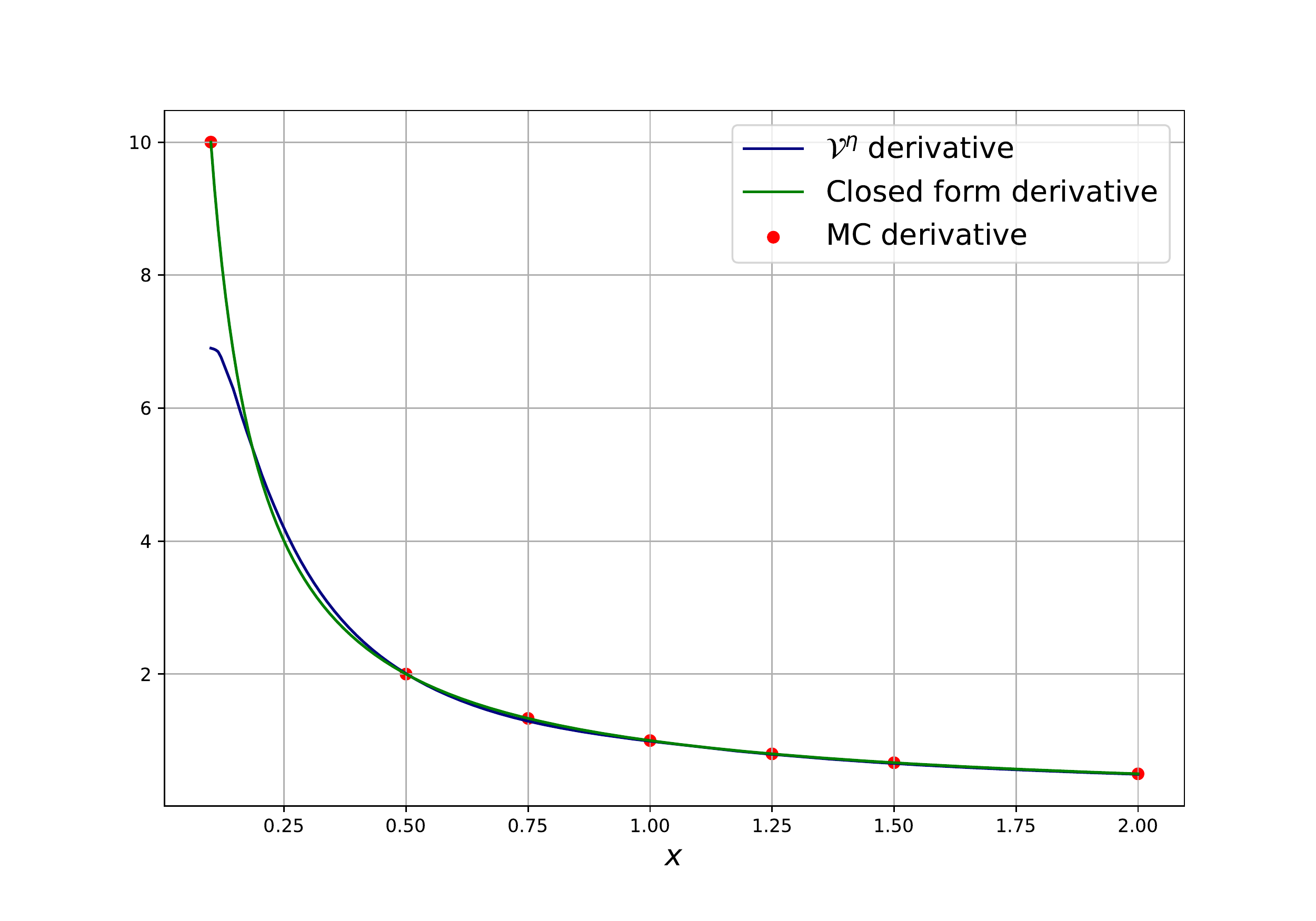} 
    \end{subfigure}
    \begin{subfigure}{.32\linewidth}
        \centering
        \includegraphics[height=3.75cm]{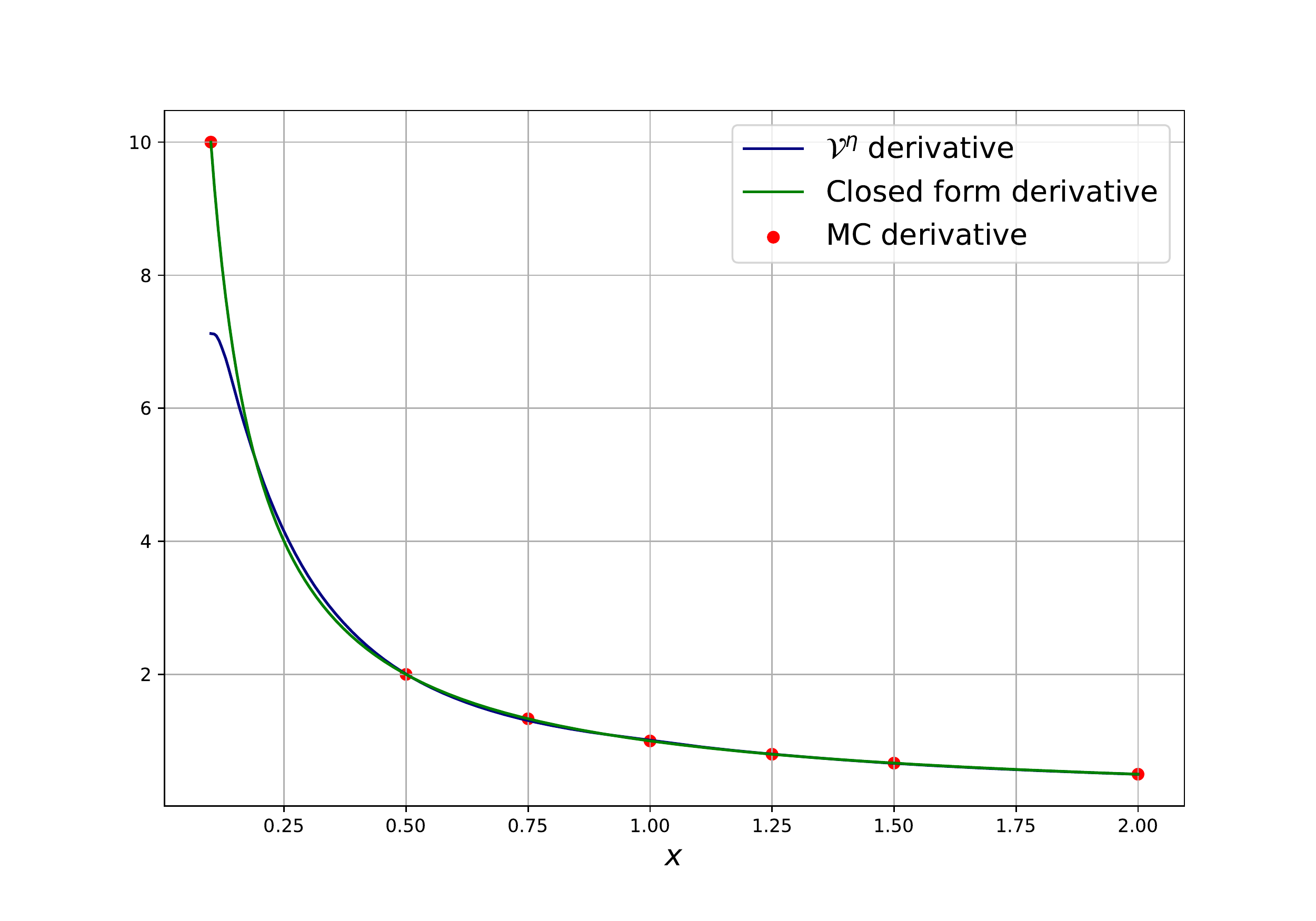}
    \end{subfigure}
    \begin{subfigure}{.32\linewidth}
        \centering
        \includegraphics[height=3.75cm]{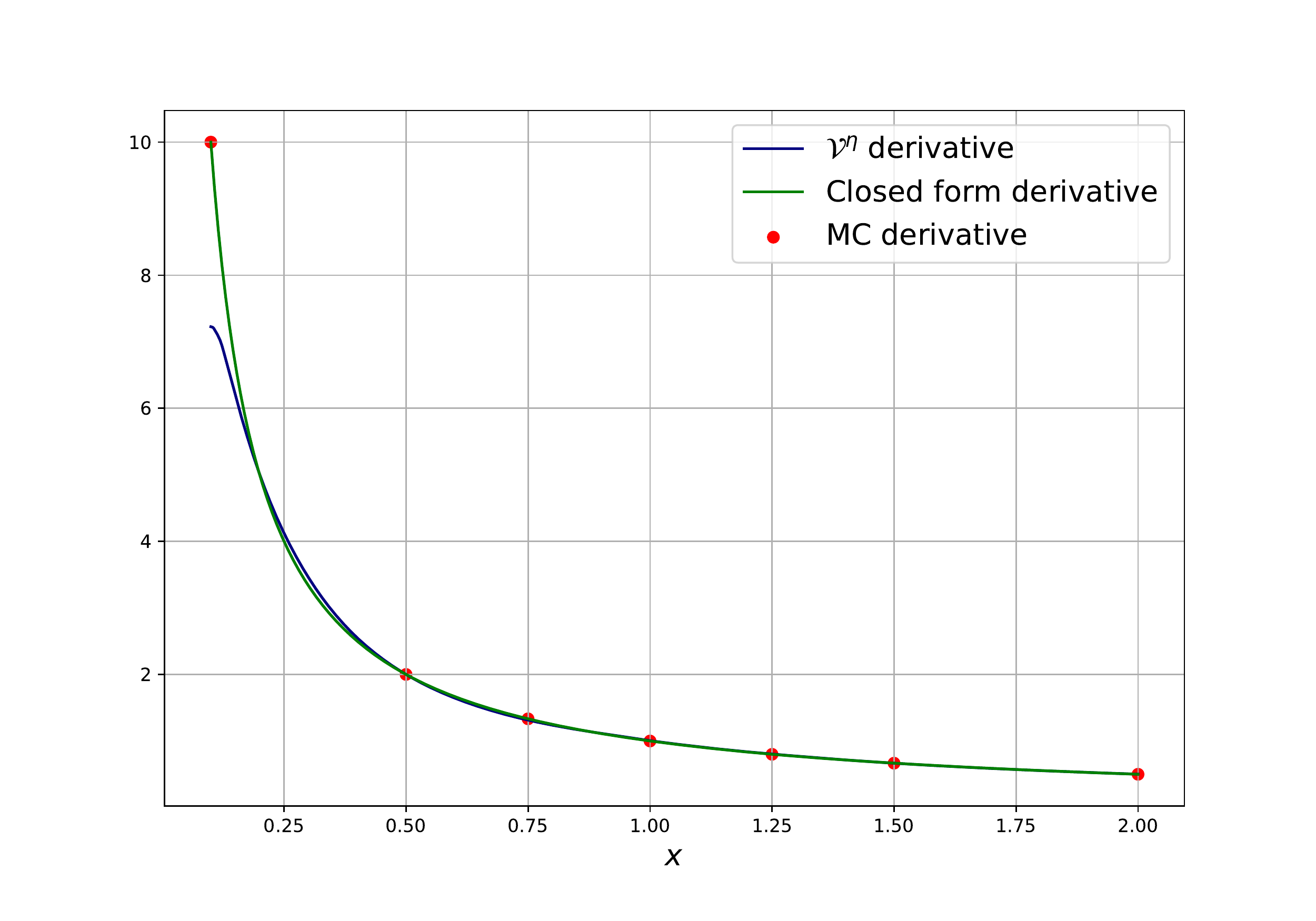}
    \end{subfigure}
    \begin{subfigure}{.32\linewidth}
        \centering
        \includegraphics[height=3.75cm]{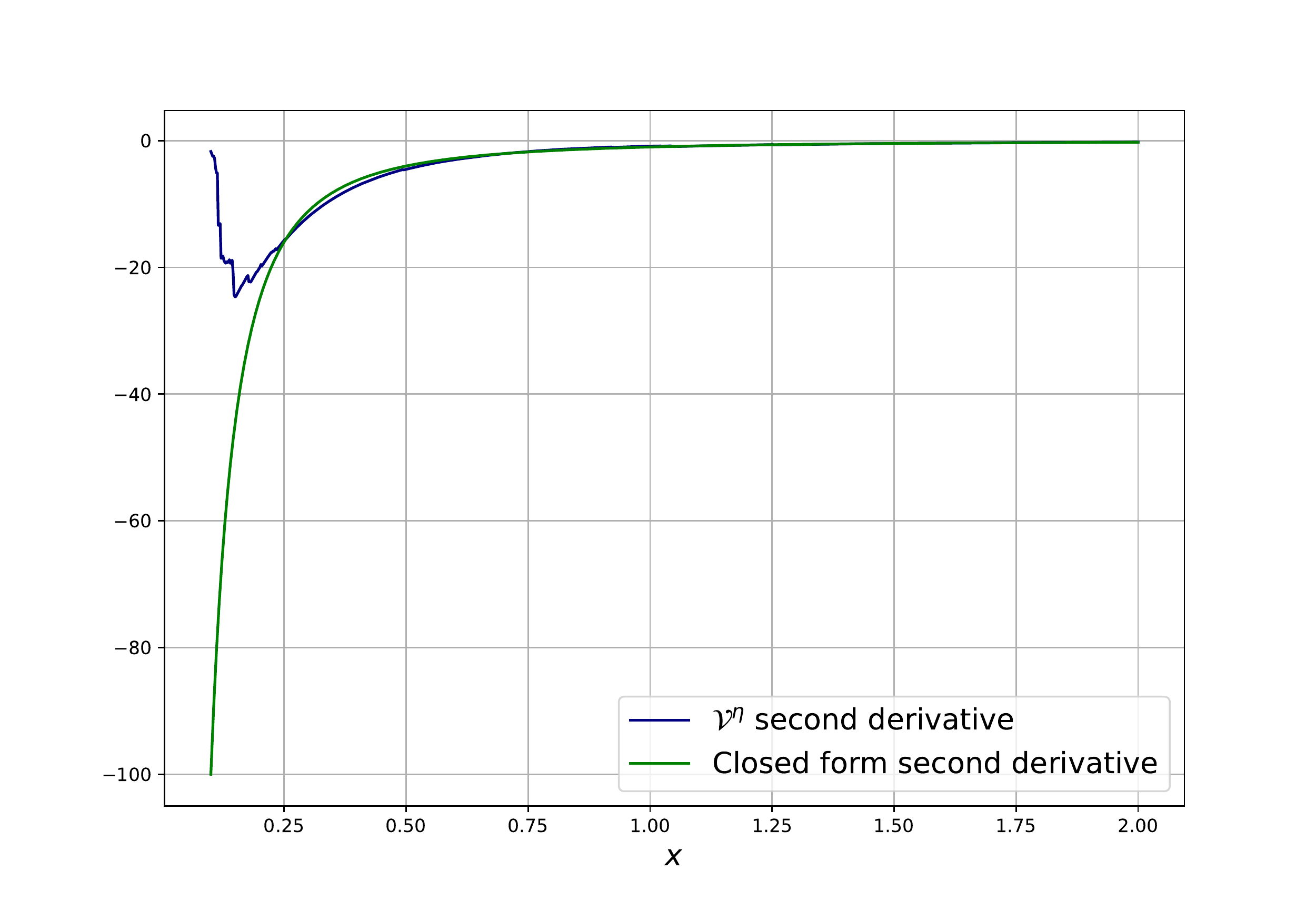} 
        \caption[short]{$t=0$}
    \end{subfigure}
    \begin{subfigure}{.32\linewidth}
        \centering
        \includegraphics[height=3.75cm]{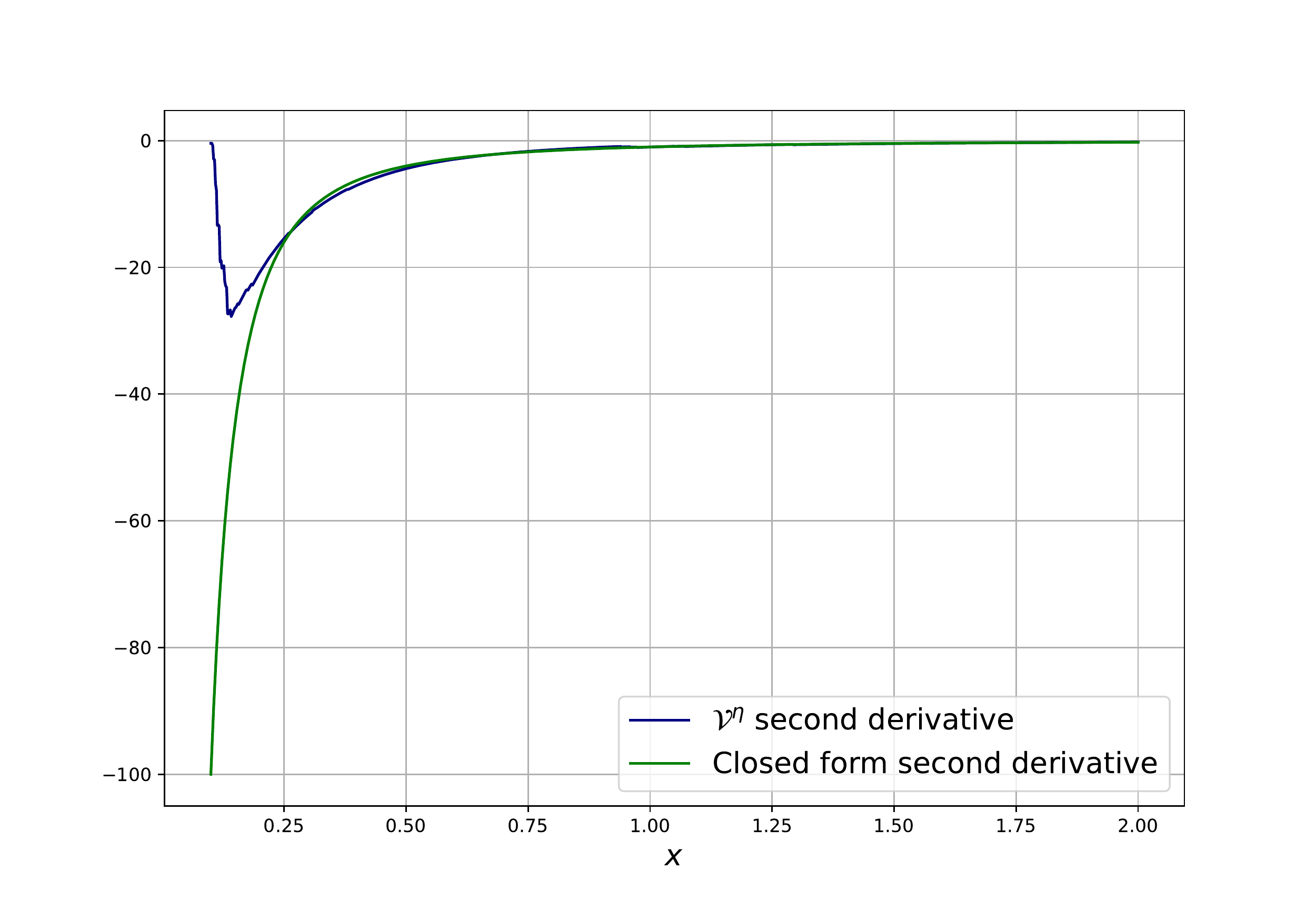}
        \caption[short]{$t=0.5$}
    \end{subfigure}
    \begin{subfigure}{.32\linewidth}
        \centering
        \includegraphics[height=3.75cm]{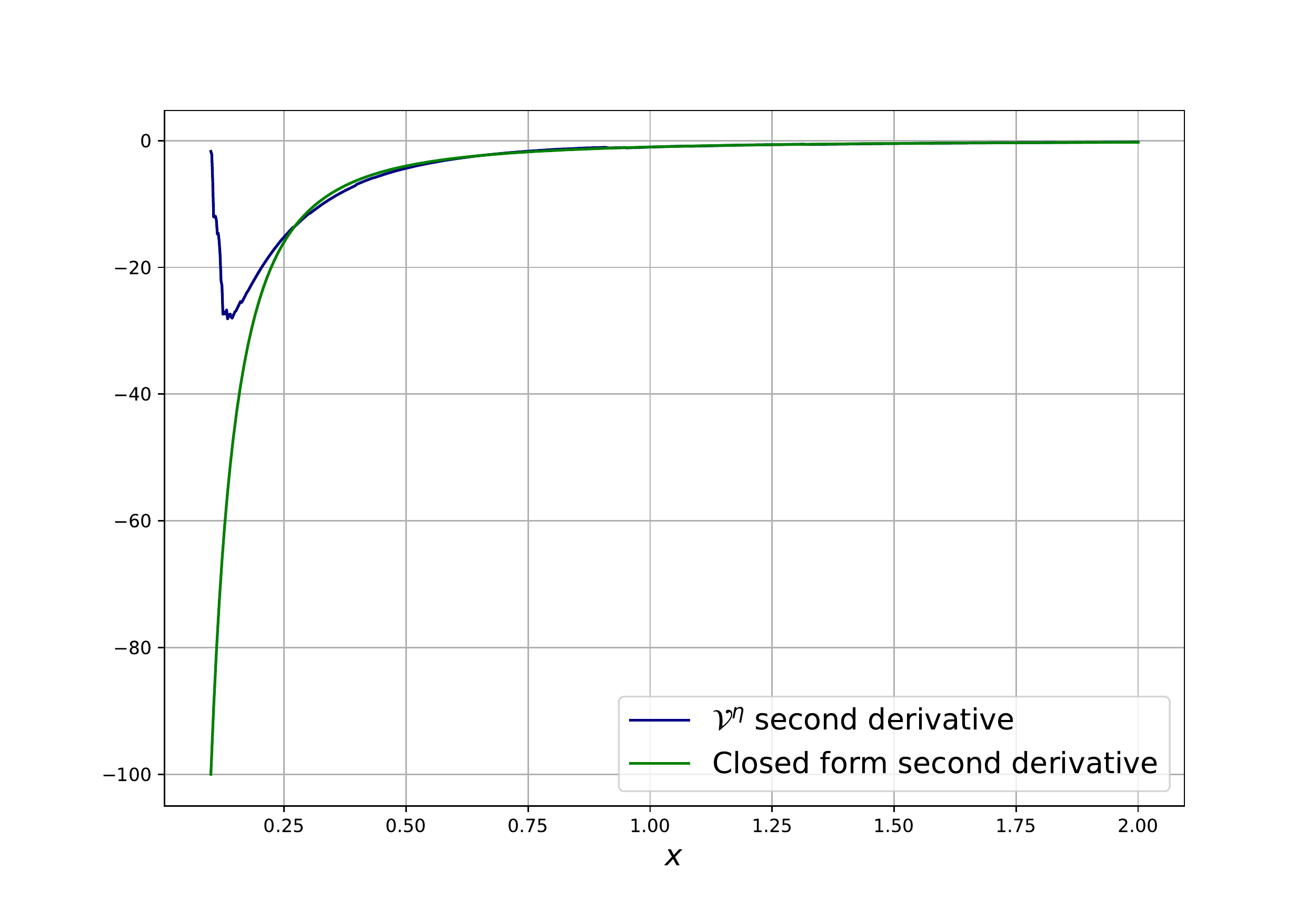}
        \caption[short]{$t=0.9$}
    \end{subfigure}
    \caption{
    \label{fig:value_pathwise_learning_log}
    Value function $\vartheta^\eta$ and its first and second derivative obtained by Pathwise Learning (Algorithm \ref{algo:scheme_value_pathwise_martingale_learning}) for a logarithmic option payoff, with parameter $\sigma = 0.3$ and linear market impact factor $\lambda = 5e^{-3}$, plotted as functions of $x$, for fixed values of $t$.
    }
\end{figure}

\begin{figure}[htp]
    \centering
    \begin{subfigure}{.32\linewidth}
        \centering
        \includegraphics[height=3.75cm]{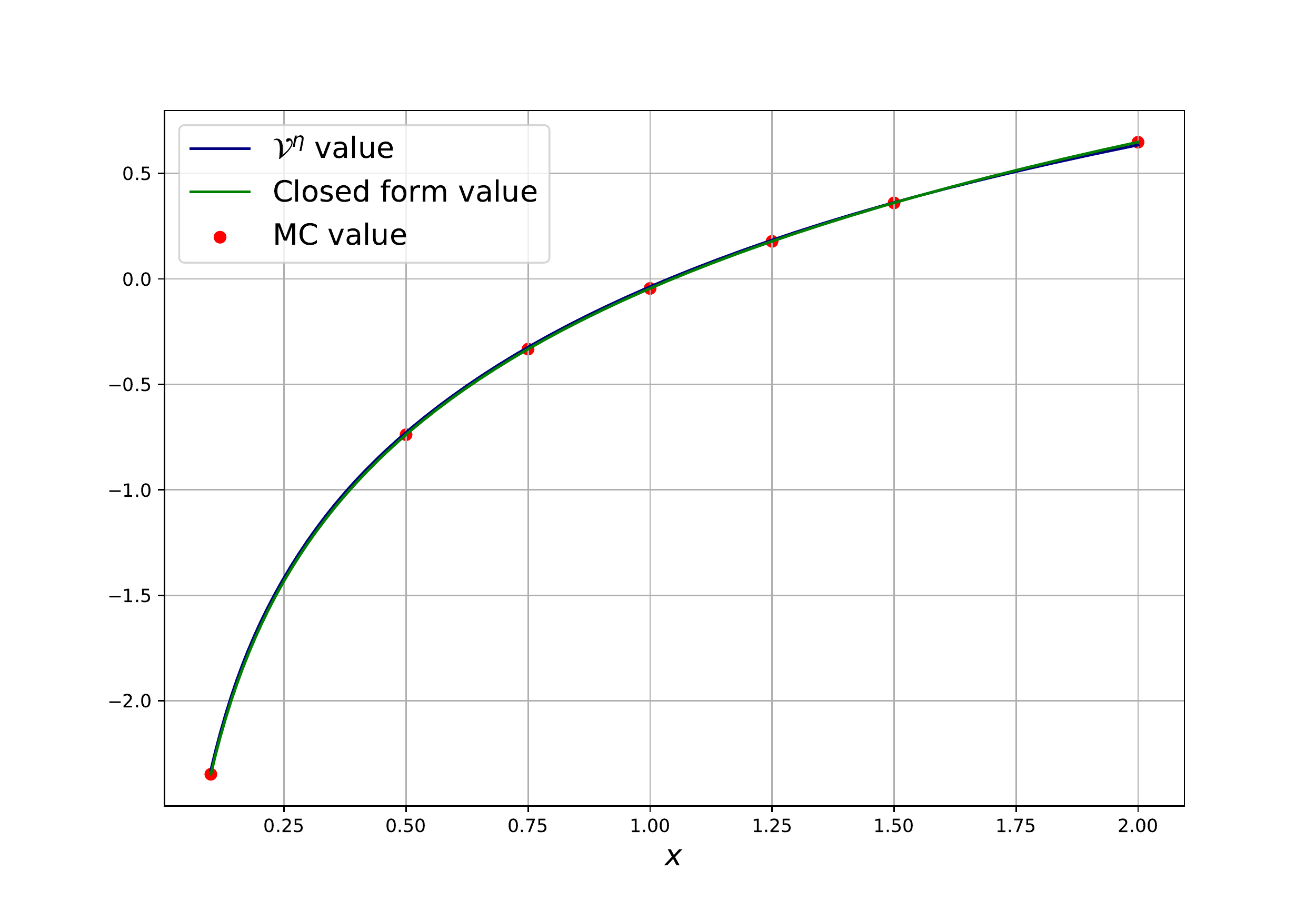} 
    \end{subfigure}
    \begin{subfigure}{.32\linewidth}
        \centering
        \includegraphics[height=3.75cm]{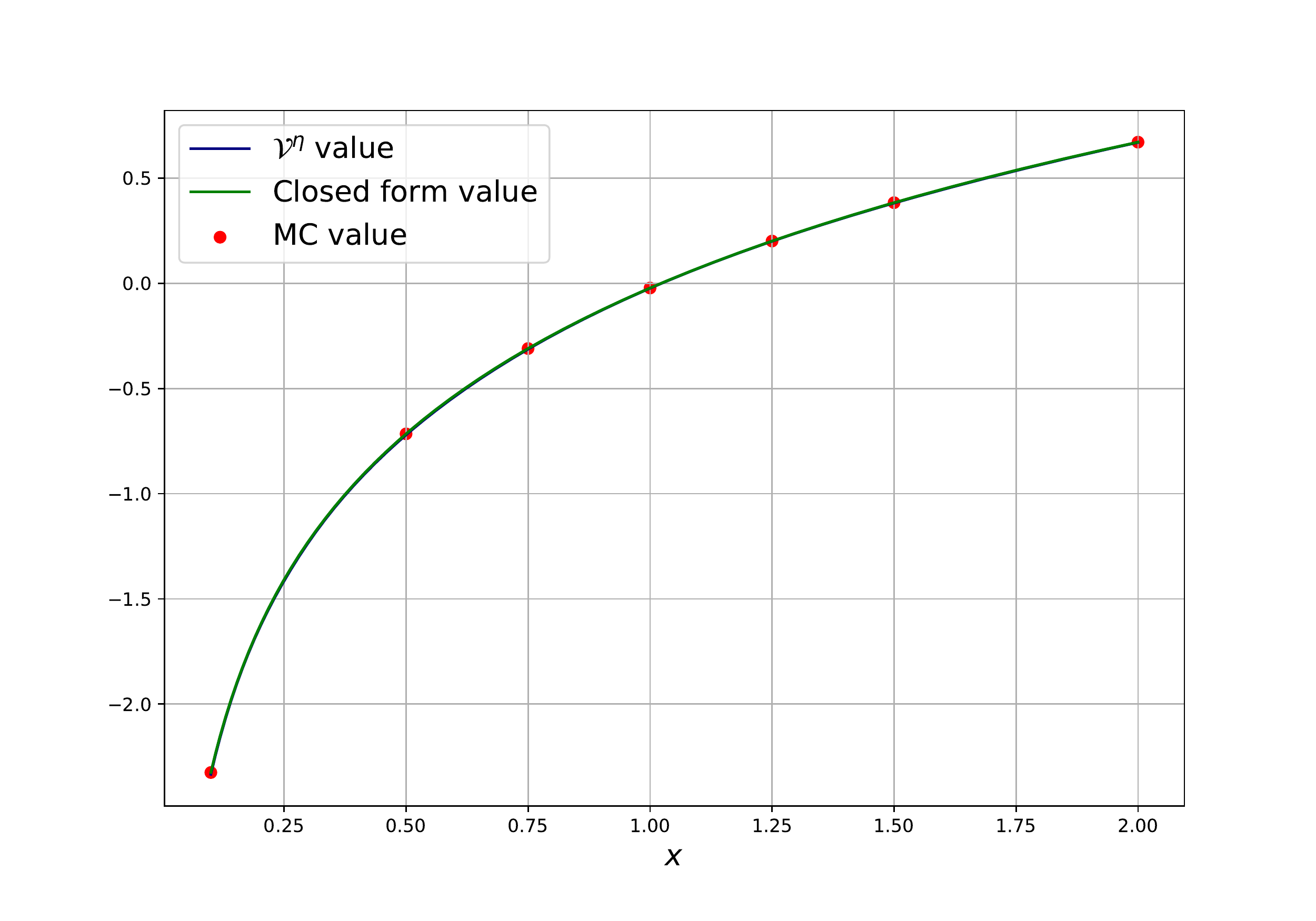}
    \end{subfigure}
    \begin{subfigure}{.32\linewidth}
        \centering
        \includegraphics[height=3.75cm]{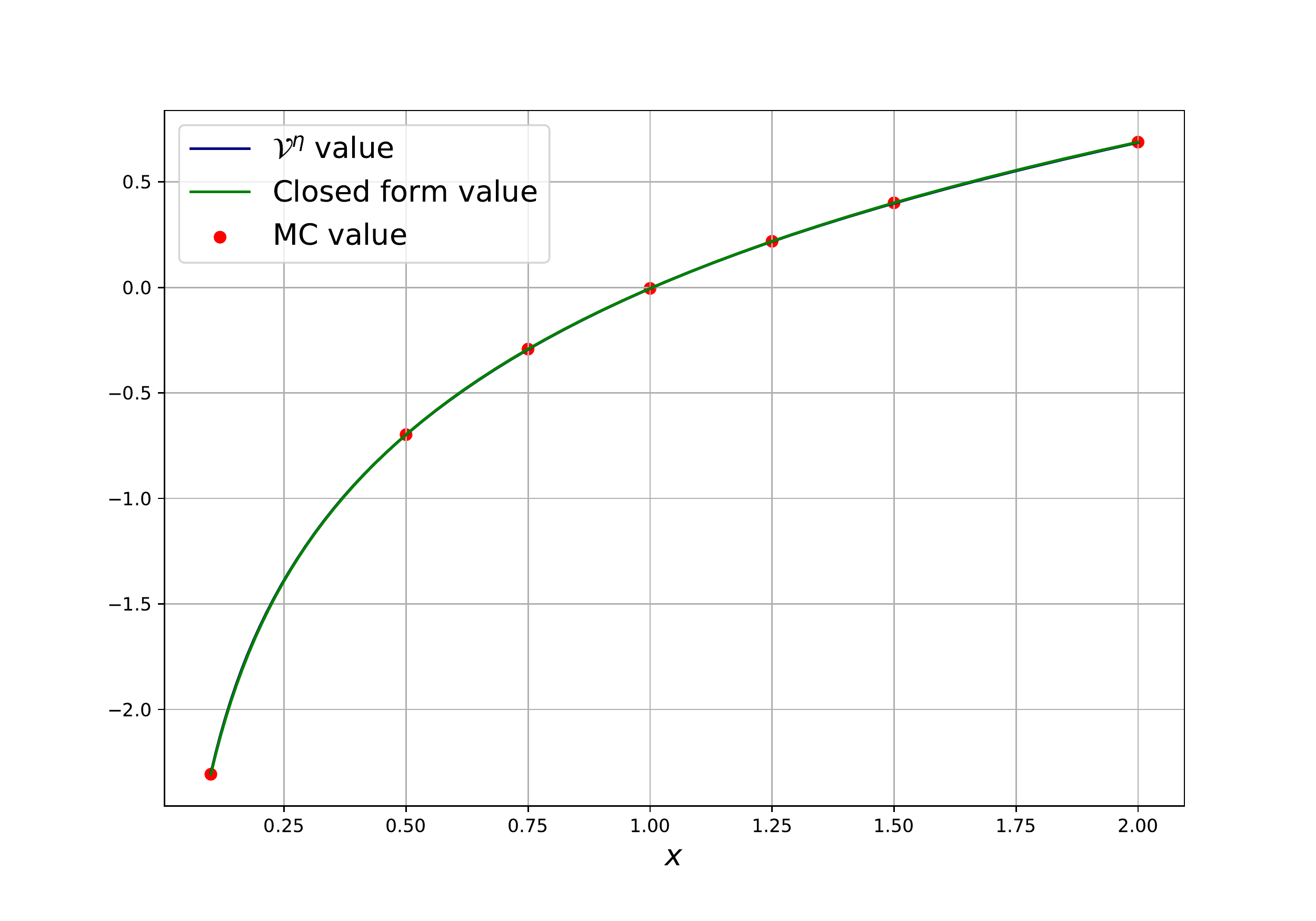}
    \end{subfigure}
    \begin{subfigure}{.32\linewidth}
        \centering
        \includegraphics[height=3.75cm]{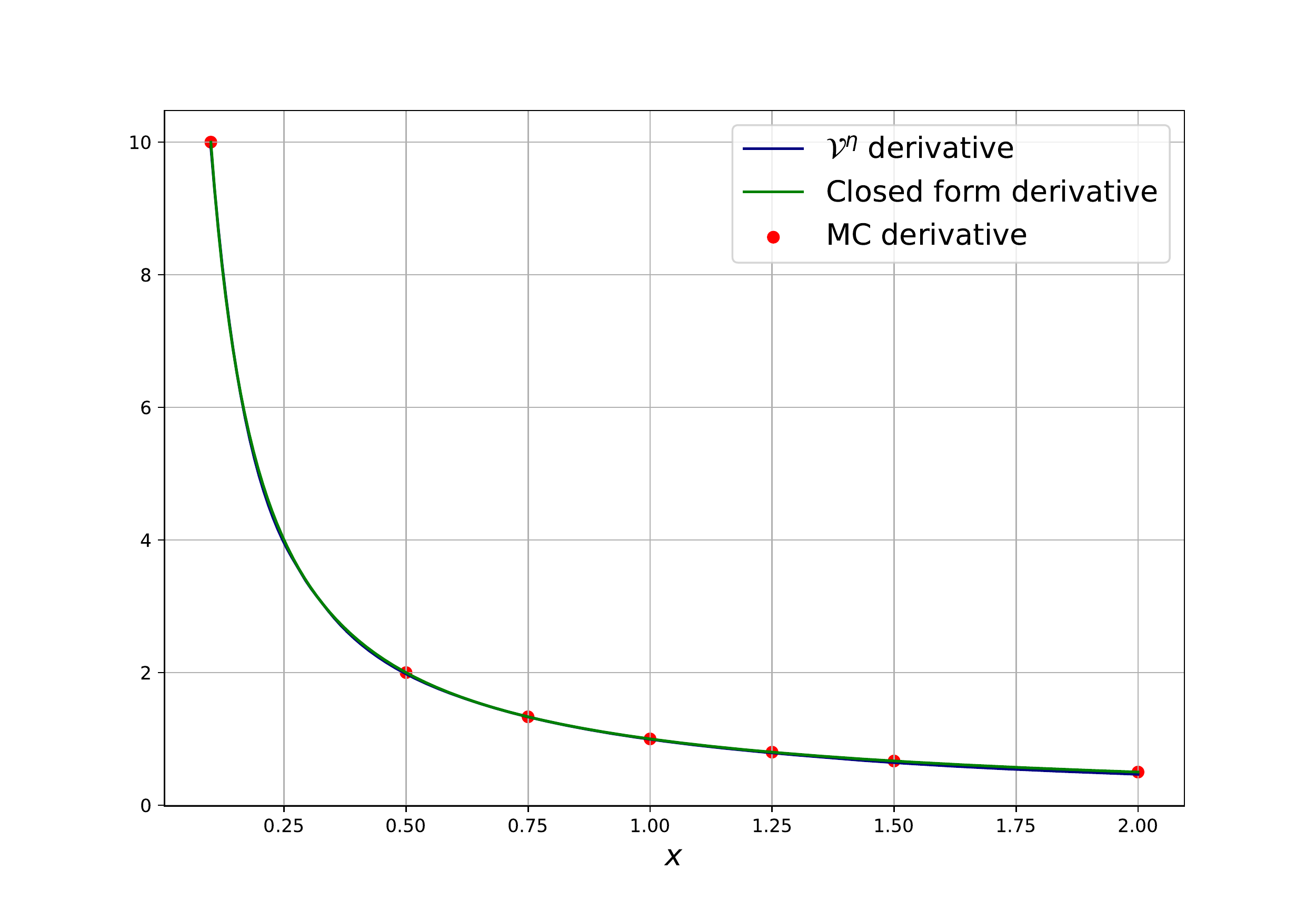} 
    \end{subfigure}
    \begin{subfigure}{.32\linewidth}
        \centering
        \includegraphics[height=3.75cm]{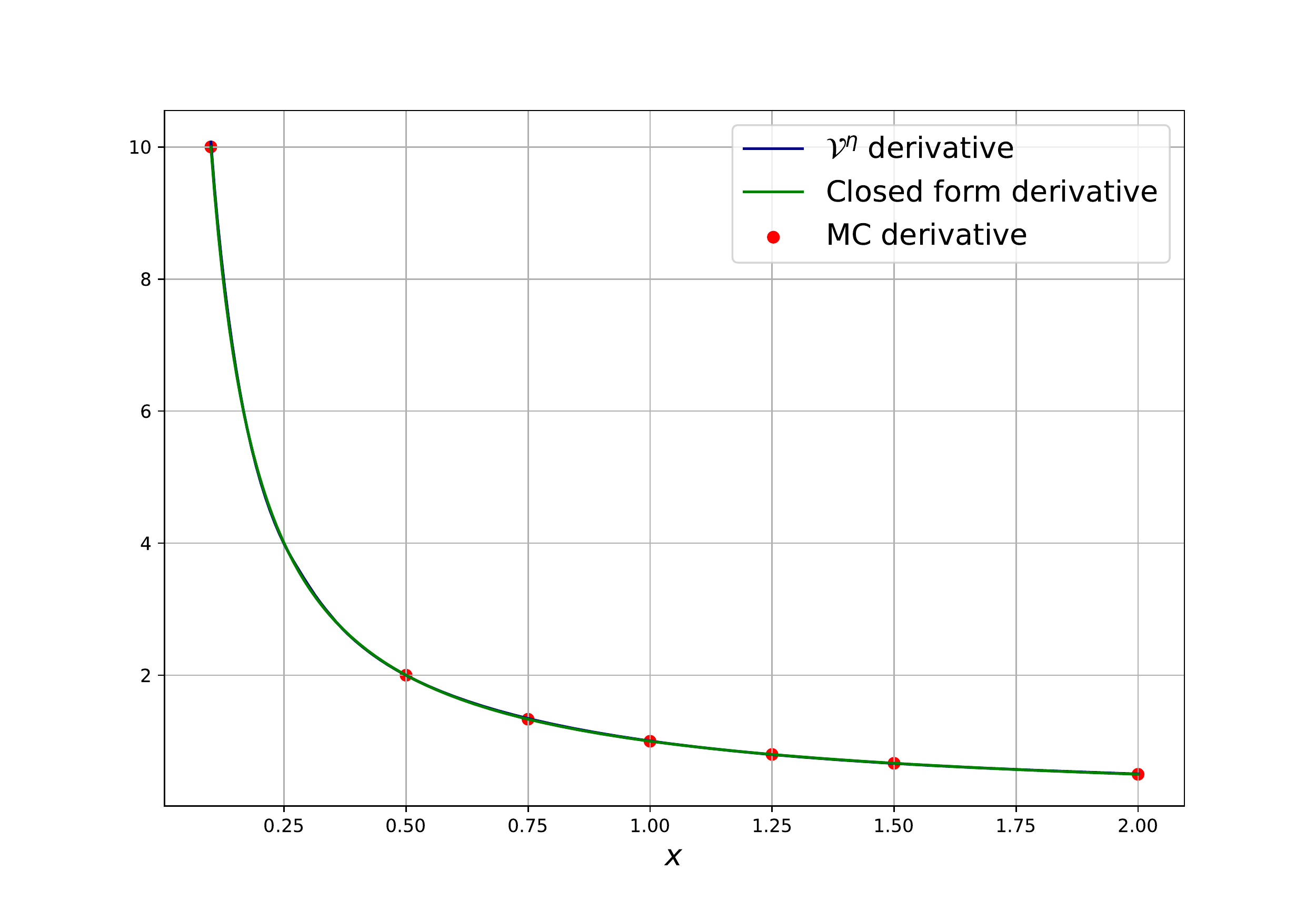}
    \end{subfigure}
    \begin{subfigure}{.32\linewidth}
        \centering
        \includegraphics[height=3.75cm]{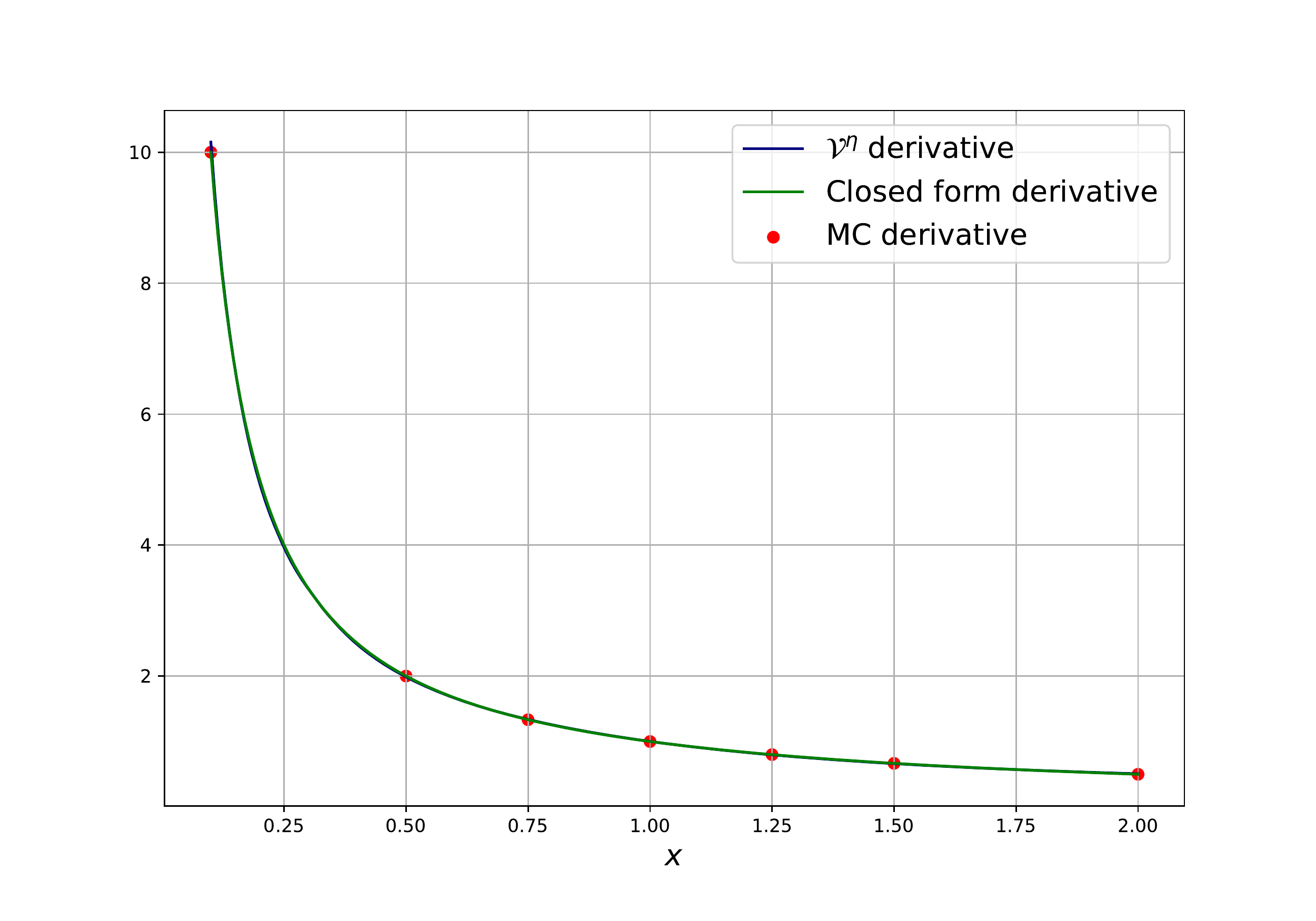}
    \end{subfigure}
    \begin{subfigure}{.32\linewidth}
        \centering
        \includegraphics[height=3.75cm]{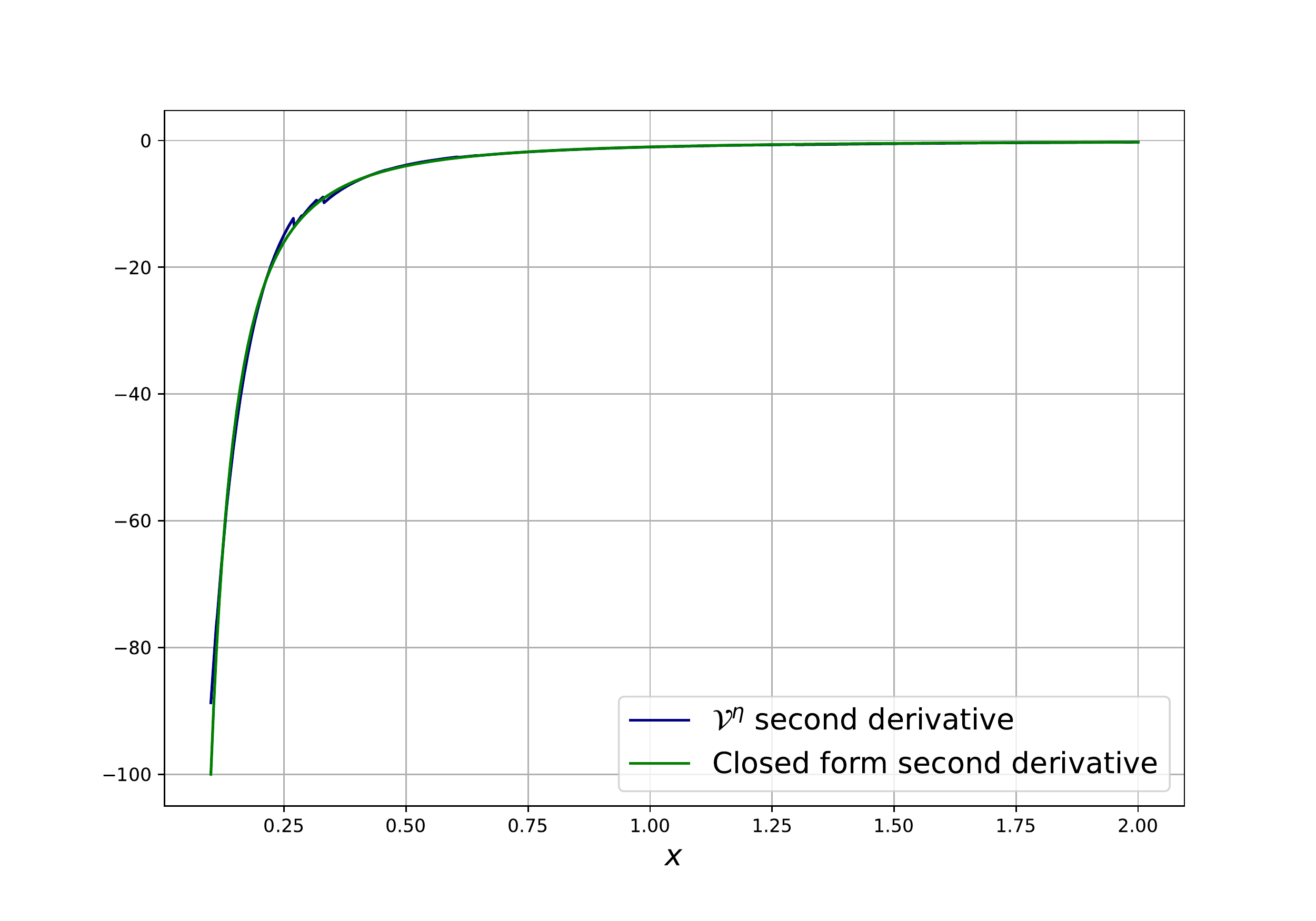} 
        \caption[short]{$t=0$}
    \end{subfigure}
    \begin{subfigure}{.32\linewidth}
        \centering
        \includegraphics[height=3.75cm]{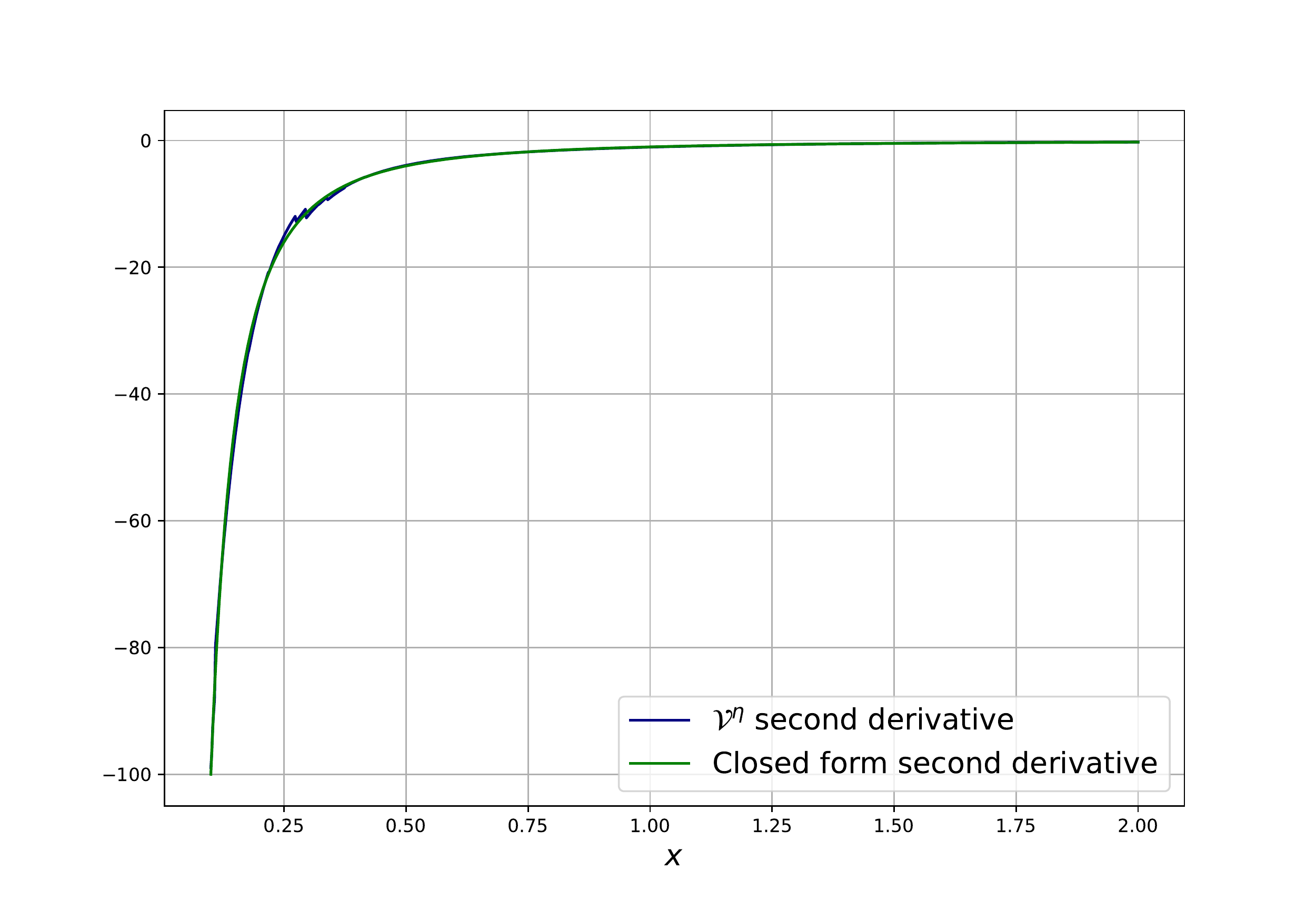}
        \caption[short]{$t=0.5$}
    \end{subfigure}
    \begin{subfigure}{.32\linewidth}
        \centering
        \includegraphics[height=3.75cm]{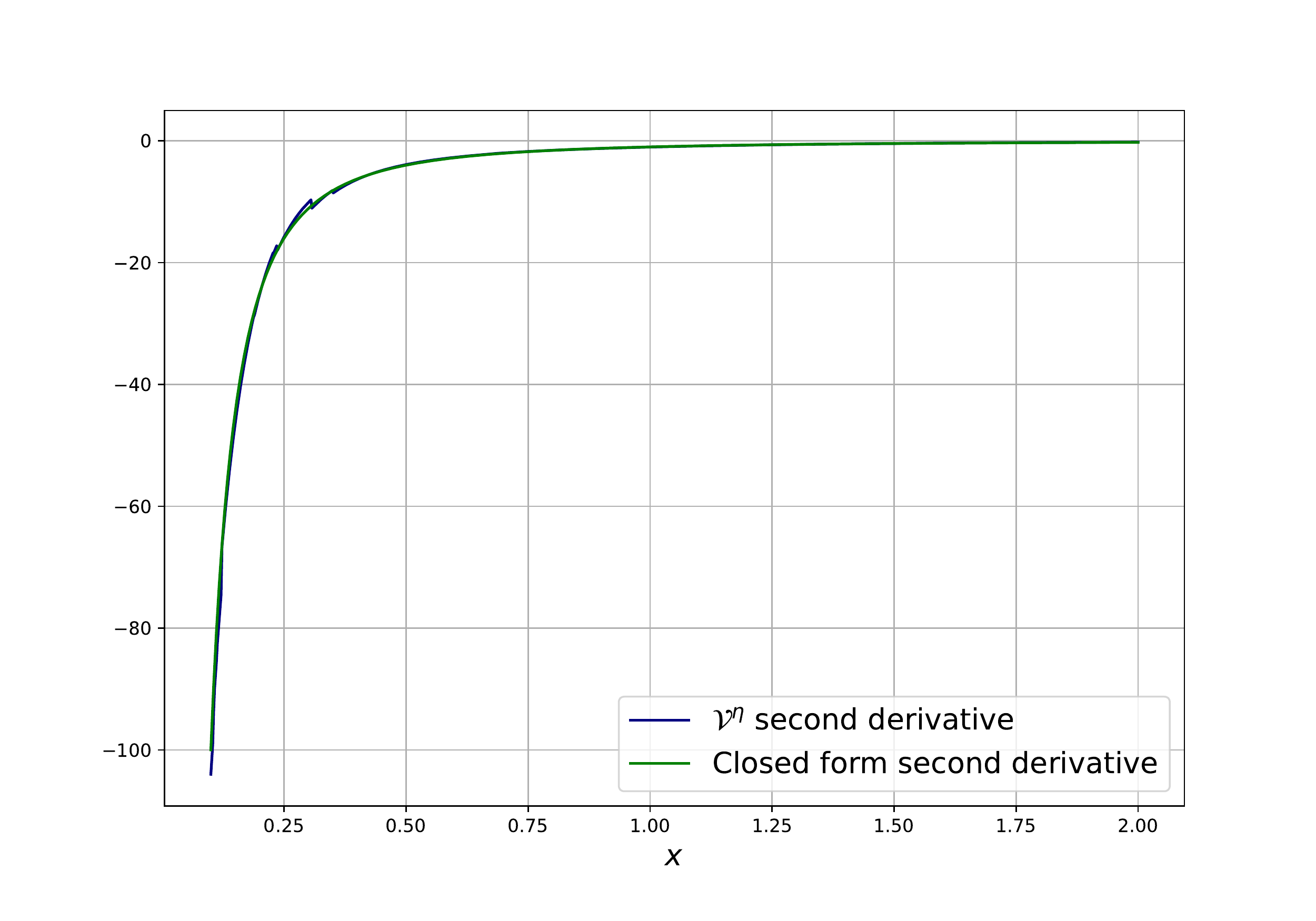}
        \caption[short]{$t=0.9$}
    \end{subfigure}
    \caption{
    \label{fig:value_pathwise_differential_learning_log}
    Value function $\vartheta^\eta$ and its first and second derivative obtained by Pathwise Differential Learning (Algorithm \ref{algo:scheme_value_pathwise_differential_learning}) for a logarithmic option payoff, with parameter $\sigma = 0.3$ and linear market impact factor $\lambda = 5e^{-3}$, plotted as functions of $x$, for fixed values of $t$.
    }
\end{figure}

We plot the value function $\vartheta^\eta(t,x)$ and its derivatives $\partial_x \vartheta^\eta (t,x)$ and $\partial_{xx} v_\eta (t,x)$ for fixed values $t=0$, $t=0.5$, $t=0.9$, parameter $\sigma = 0.3$ and linear market impact factor $\lambda = 5e^{-3}$, and compare it with the closed-form solution of the problem. Figure \ref{fig:value_differential_learning_log} corresponds to the Differential regression learning method, Figure \ref{fig:value_pathwise_learning_log} corresponds to the pathwise martingale learning  while Figure \ref{fig:value_pathwise_differential_learning_log} corresponds to the pathwise differential learning method. On these graphs, we can see that the three methods, as well as the Monte-Carlo values $\vartheta_{MC}$ and $D_x \vartheta_{Mc}$,  yield good approximations of the PDE solution and its derivatives.  
Notice however that the pathwise learning (see Figure \ref{fig:value_pathwise_learning_log}) does not provide a good approximation of the second derivative on the boundary points of the grid, namely the points that were not explored by the simulations, but when combining with the differential learning (see Figure \ref{fig:value_pathwise_differential_learning_log}), it greatly improves the approximation of the second derivative.

\subsubsection{Call-option type terminal condition}

We now consider a usual call option payoff with strike $K$ $=$ $1$, hence a function $g$ equal to: $g(x)$ $=$ $\max(x-1,0)$.

 Again, we compute in Table \ref{tablecall}, the residual losses defined in  \eqref{residual} for the NN $\vartheta^\eta$ obtained by the various deep learning methods: the differential learning scheme (Algorithm \ref{algo:scheme_value_differential_learning}), 
the pathwise martingale learning with $1$ NN (Algorithm \ref{algo:scheme_value_pathwise_martingale_learning}) and the pathwise differential learning with $1$ NN (Algorithm \ref{algo:scheme_value_pathwise_differential_learning}). We also provide the training time for each of these algorithms. On this table, we see that the Differential regression learning and the Pathwise differential methods yield better results than the "simple" Pathwise method. Despite having the lowest residual loss, the Pathwise method  gives the biggest terminal loss. This shows that, while the time and second space derivatives of the neural network give a low residual loss corresponding to the Hamiltonian \eqref{Hpricing}, the network does not manage to fit the terminal function. This phenomenon is also present, to a lesser extent, in the approximation given by the Differential regression learning method. Out of the three methods, the Pathwise differential yields the smallest residual and terminal losses.

\begin{scriptsize}
\begin{table}[h]
\centering
\begin{tabular}{|c|c|c|c|}
\hline  & Diff. regr. learning & Path. 1NN & Path. diff. 1NN \\
\hline  Residual loss  & $ 2.998 e^{-4}$ & $ 2.756 e^{-4}$ & $2.283  e^{-4}$ \\
\hline  Residual loss  & & & \\
+ terminal loss & $ 3.972 e^{-4}$ & $ 1.004 e^{-3}$ & $ 2.538 e^{-4}$ \\
\hline  Training time & 262s & 299s & 584s \\
 &  1000 epochs & 1000 epochs & 500 epochs  \\
\hline
\end{tabular}
\caption{Residual and boundary losses computed on a $102$x$102$ time and space grid with $t\in[0,0.9]$ and $x\in[0.1, 2]$ for a terminal call-option payoff $g(x)$ $=$ $\max(x-1,0)$, with parameter $\sigma = 0.3$ and linear market impact factor $\lambda = 5e^{-3}$.} 
\label{tablecall} 
\end{table}
\end{scriptsize}
 
\begin{figure}[htp]
    \centering
    \begin{subfigure}{.31\linewidth}
        \centering
        \includegraphics[height=3.67cm]{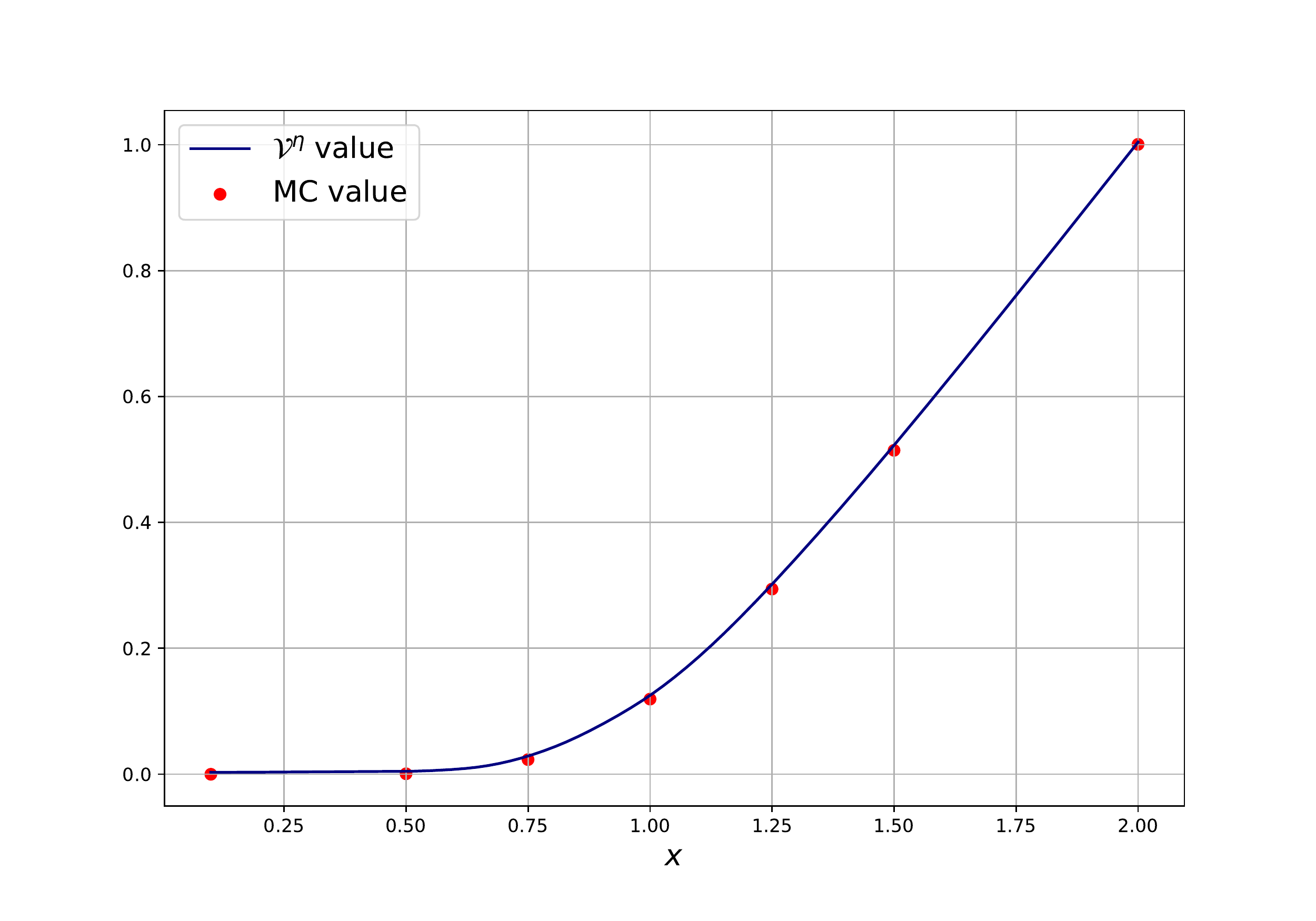} 
    \end{subfigure}
    \begin{subfigure}{.31\linewidth}
        \centering
        \includegraphics[height=3.67cm]{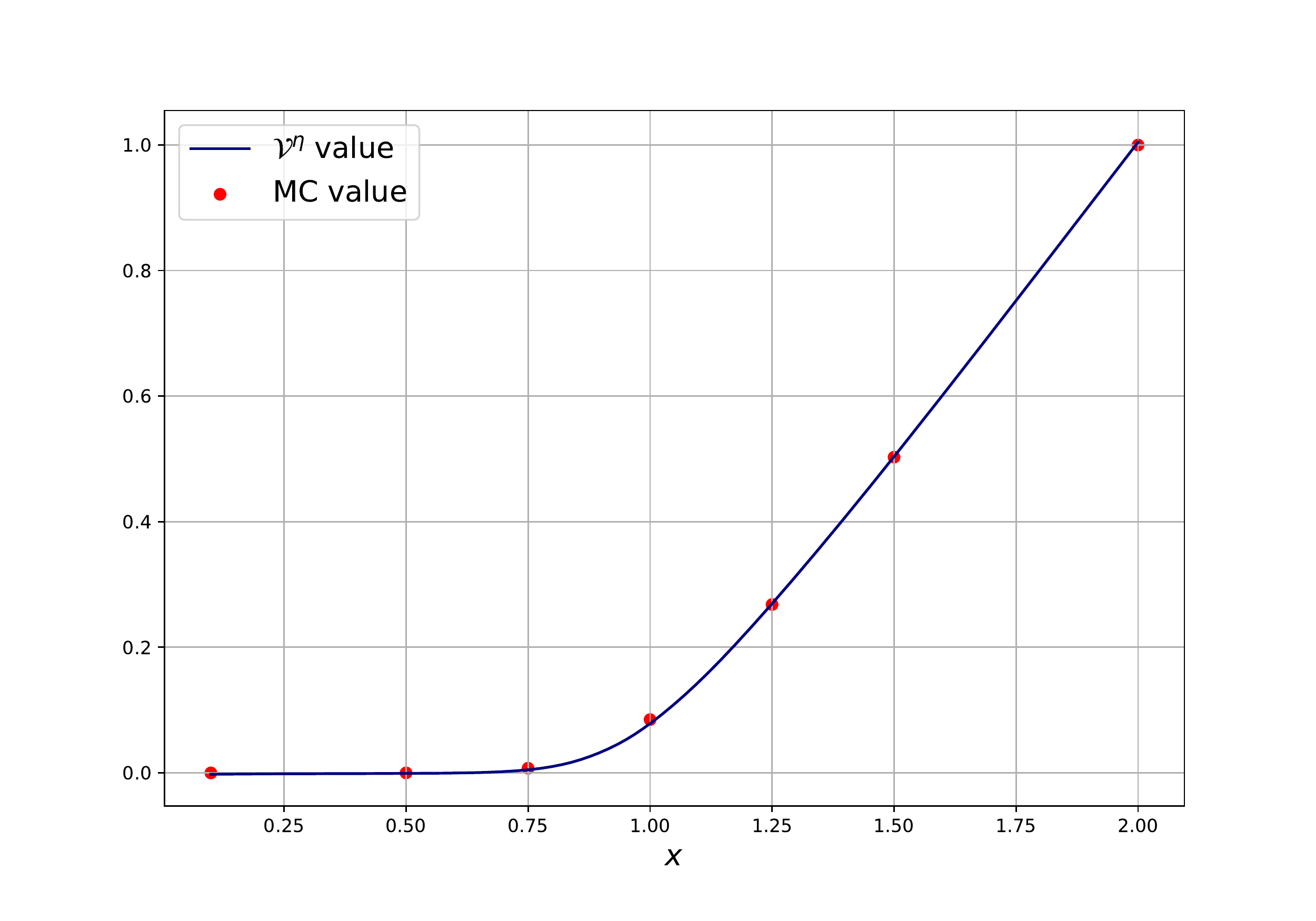}
    \end{subfigure}
    \begin{subfigure}{.31\linewidth}
        \centering
        \includegraphics[height=3.67cm]{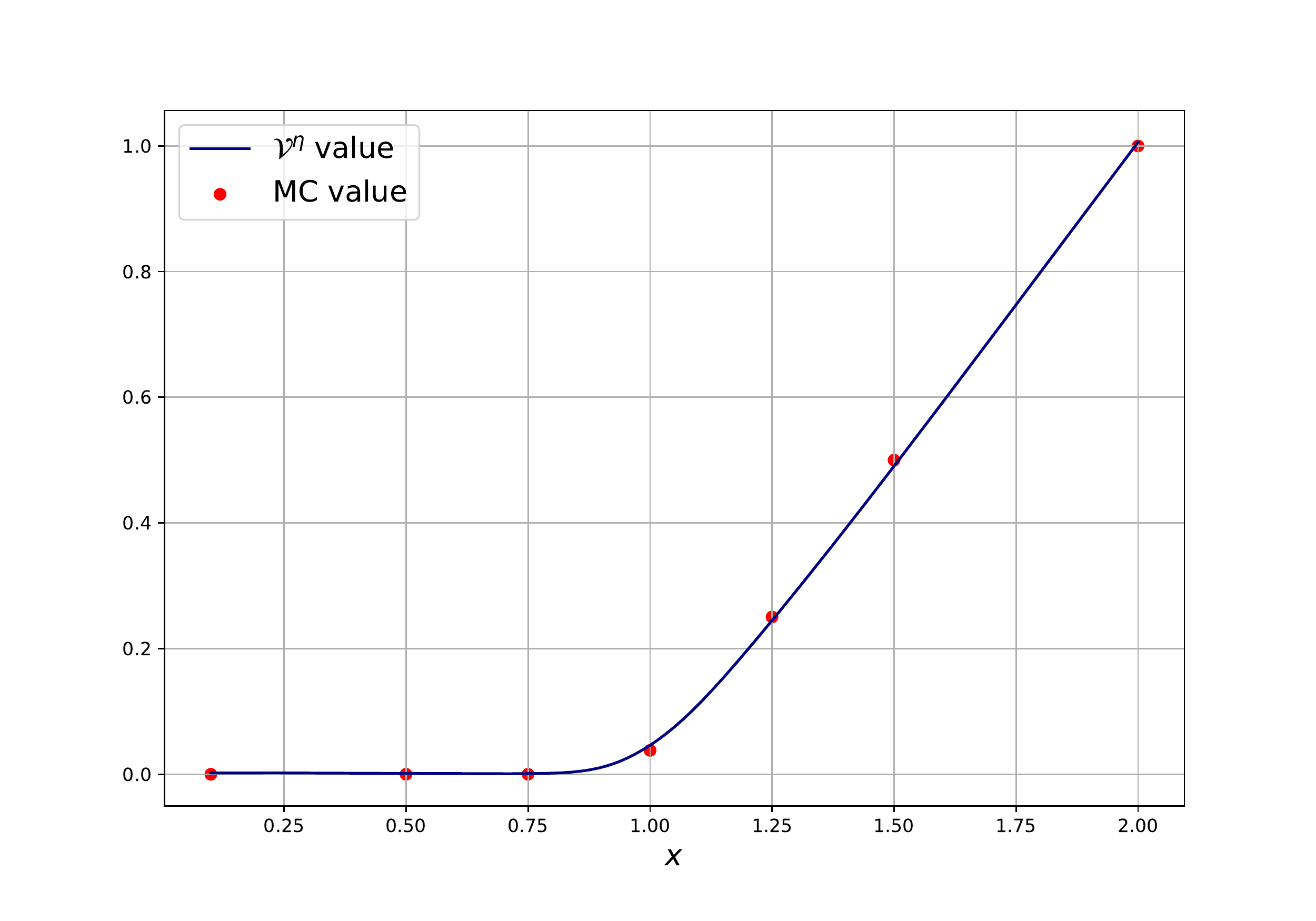}
    \end{subfigure}
    \begin{subfigure}{.31\linewidth}
        \centering
        \includegraphics[height=3.67cm]{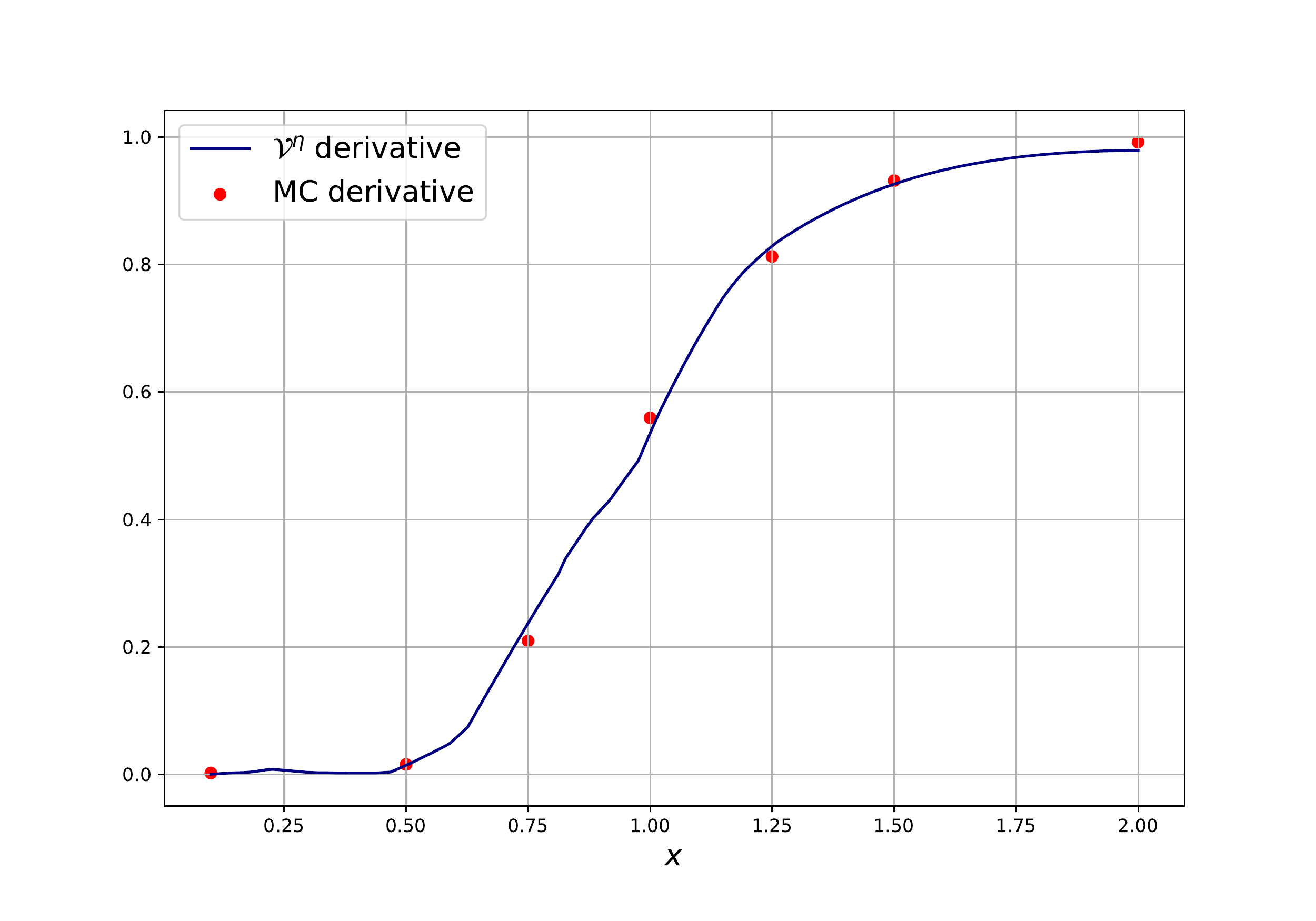} 
        \caption[short]{$t=0$}
    \end{subfigure}
    \begin{subfigure}{.31\linewidth}
        \centering
        \includegraphics[height=3.67cm]{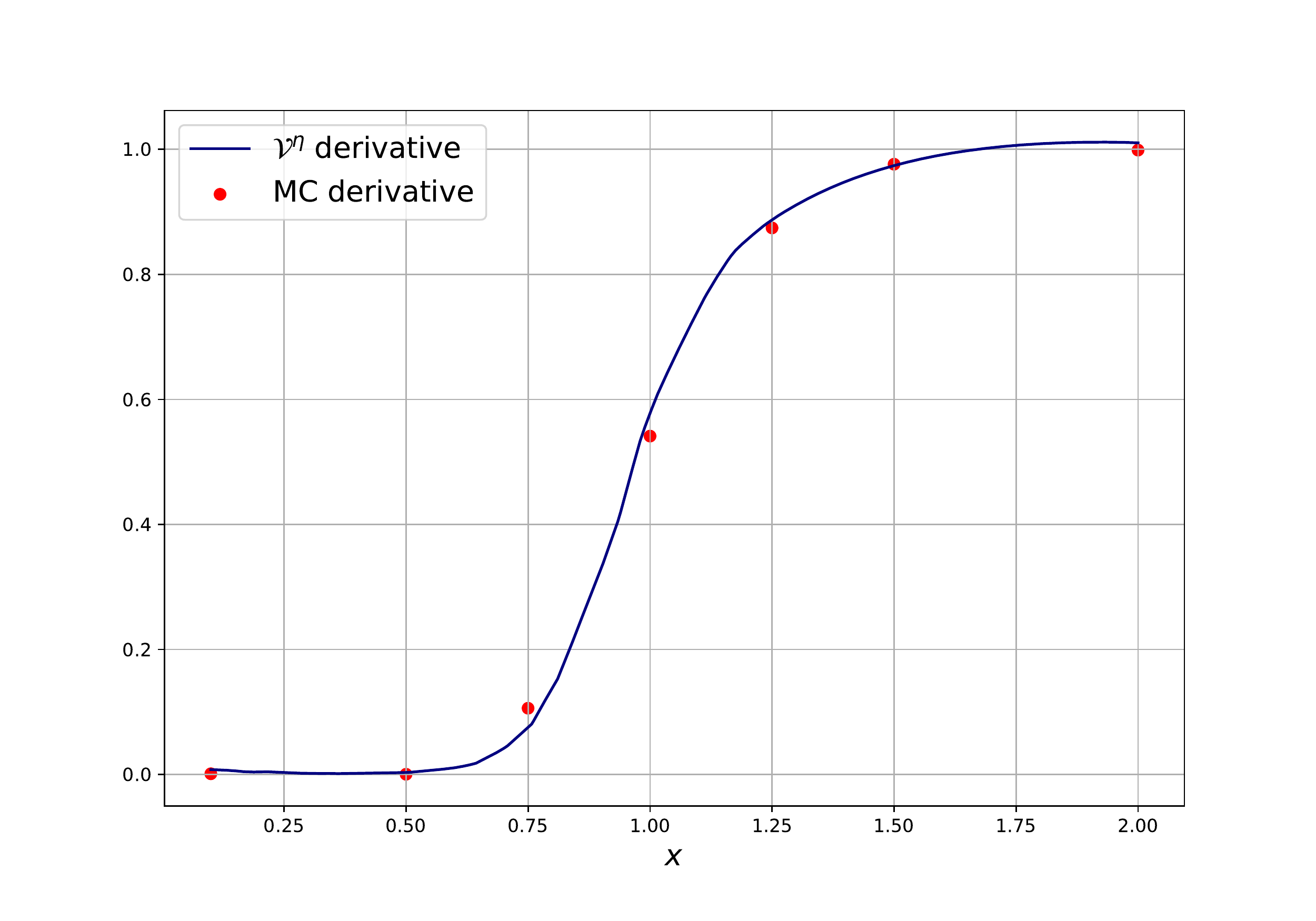}
        \caption[short]{$t=0.5$}
    \end{subfigure}
    \begin{subfigure}{.31\linewidth}
        \centering
        \includegraphics[height=3.67cm]{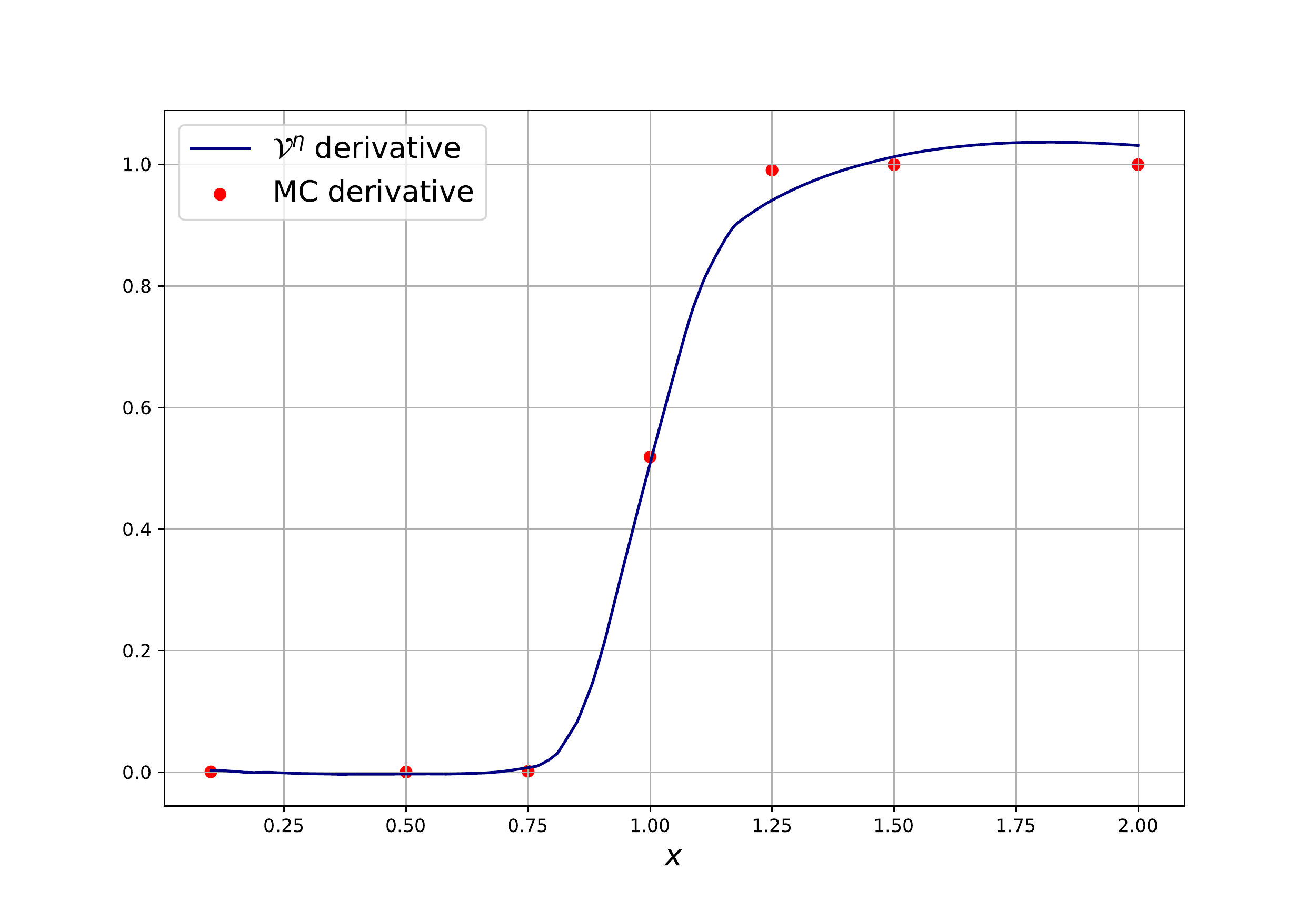}
        \caption[short]{$t=0.9$}
    \end{subfigure}
    \caption{
    \label{fig:value_differential_learning_call}
    \footnotesize{Value function $\vartheta^\eta$ and its derivative obtained by Differential Regression Learning (Algorithm \ref{algo:scheme_value_differential_learning}) for a call option  with strike $1$, with parameter $\sigma = 0.3$ and linear market impact factor $\lambda = 5e^{-3}$, plotted as functions of $x$, for fixed values of $t$.}
    }
\end{figure}

\begin{figure}[htp]
    \centering
    \begin{subfigure}{.31\linewidth}
        \centering
        \includegraphics[height=3.67cm]{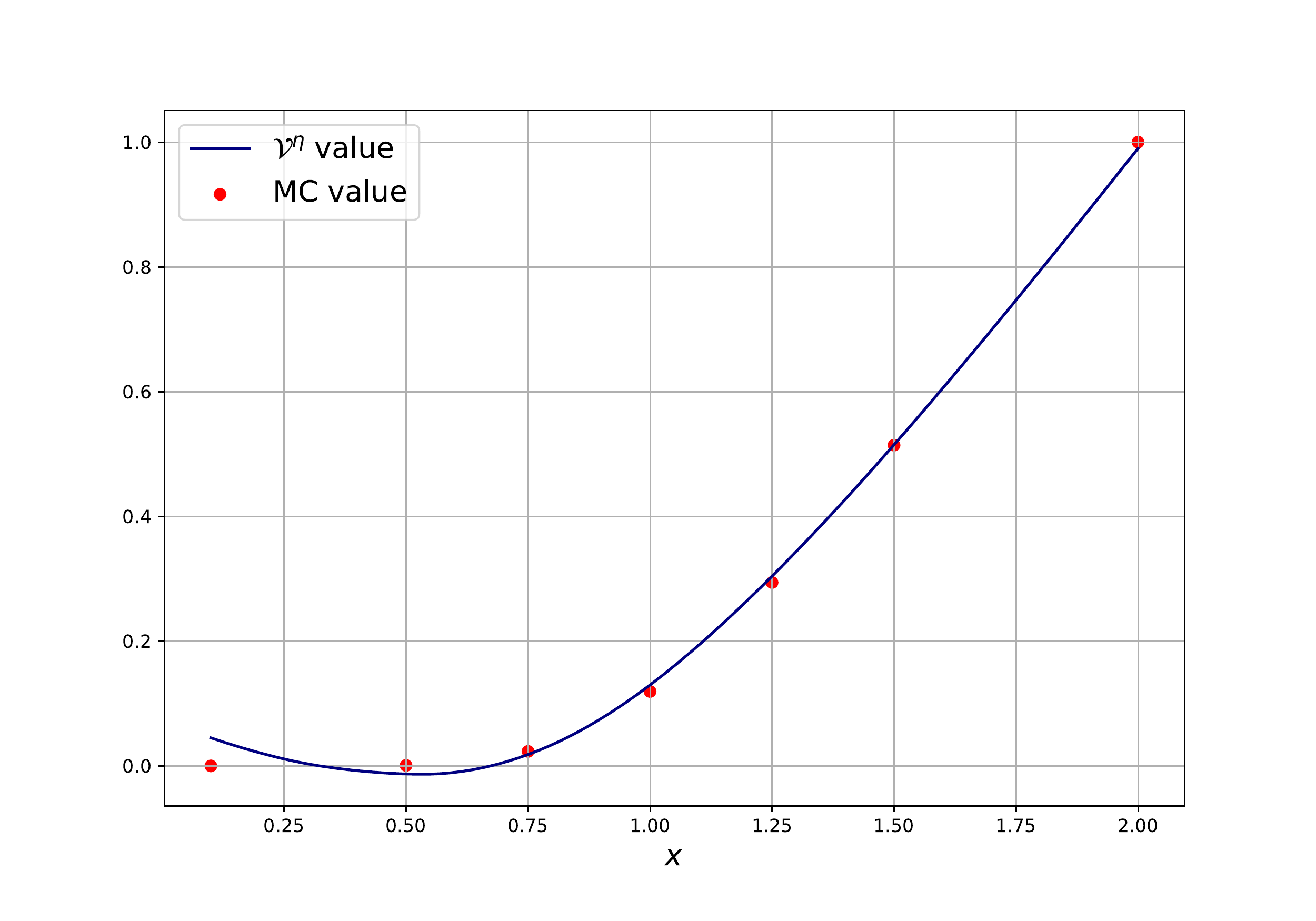} 
    \end{subfigure}
    \begin{subfigure}{.31\linewidth}
        \centering
        \includegraphics[height=3.67cm]{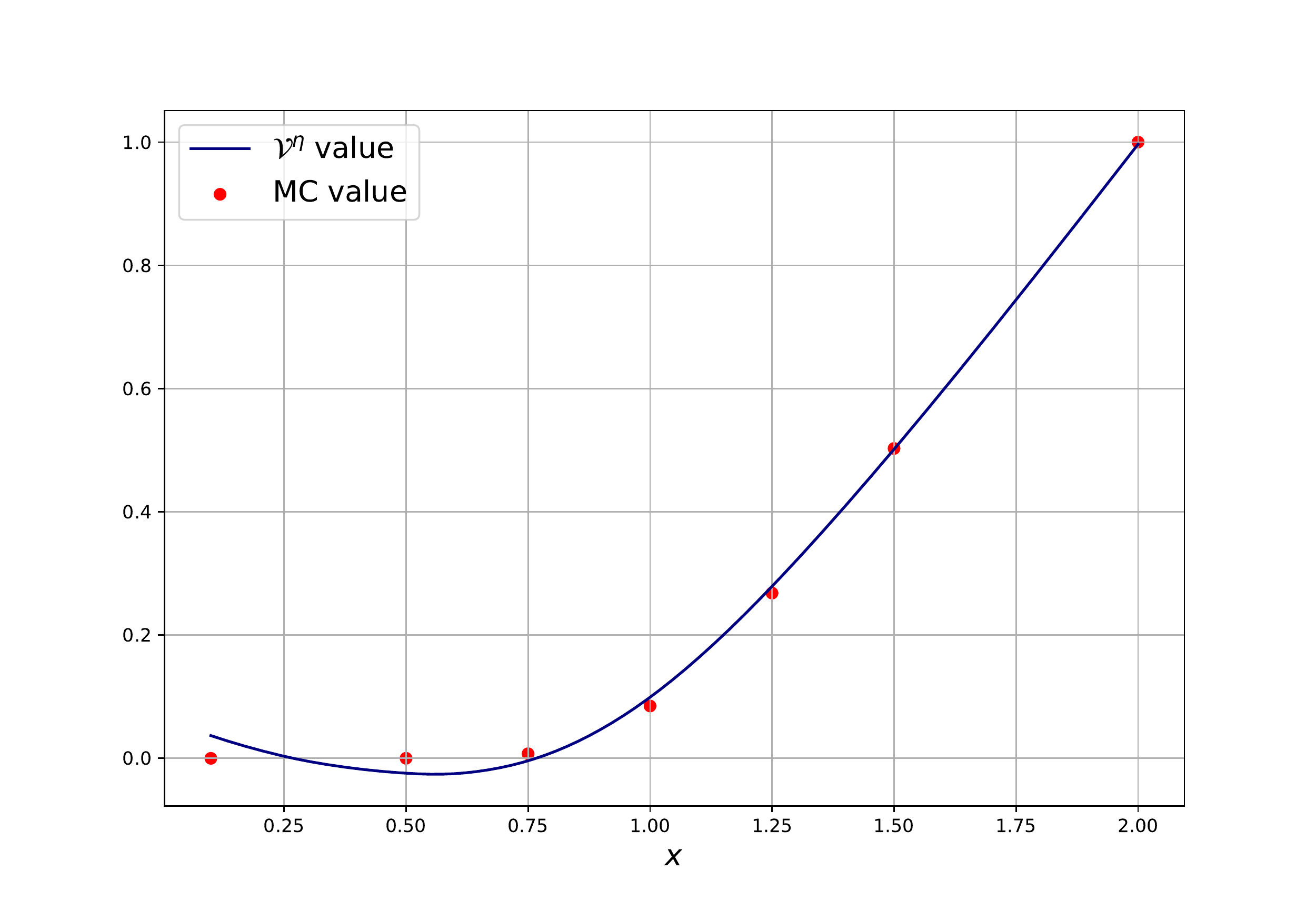}
    \end{subfigure}
    \begin{subfigure}{.31\linewidth}
        \centering
        \includegraphics[height=3.67cm]{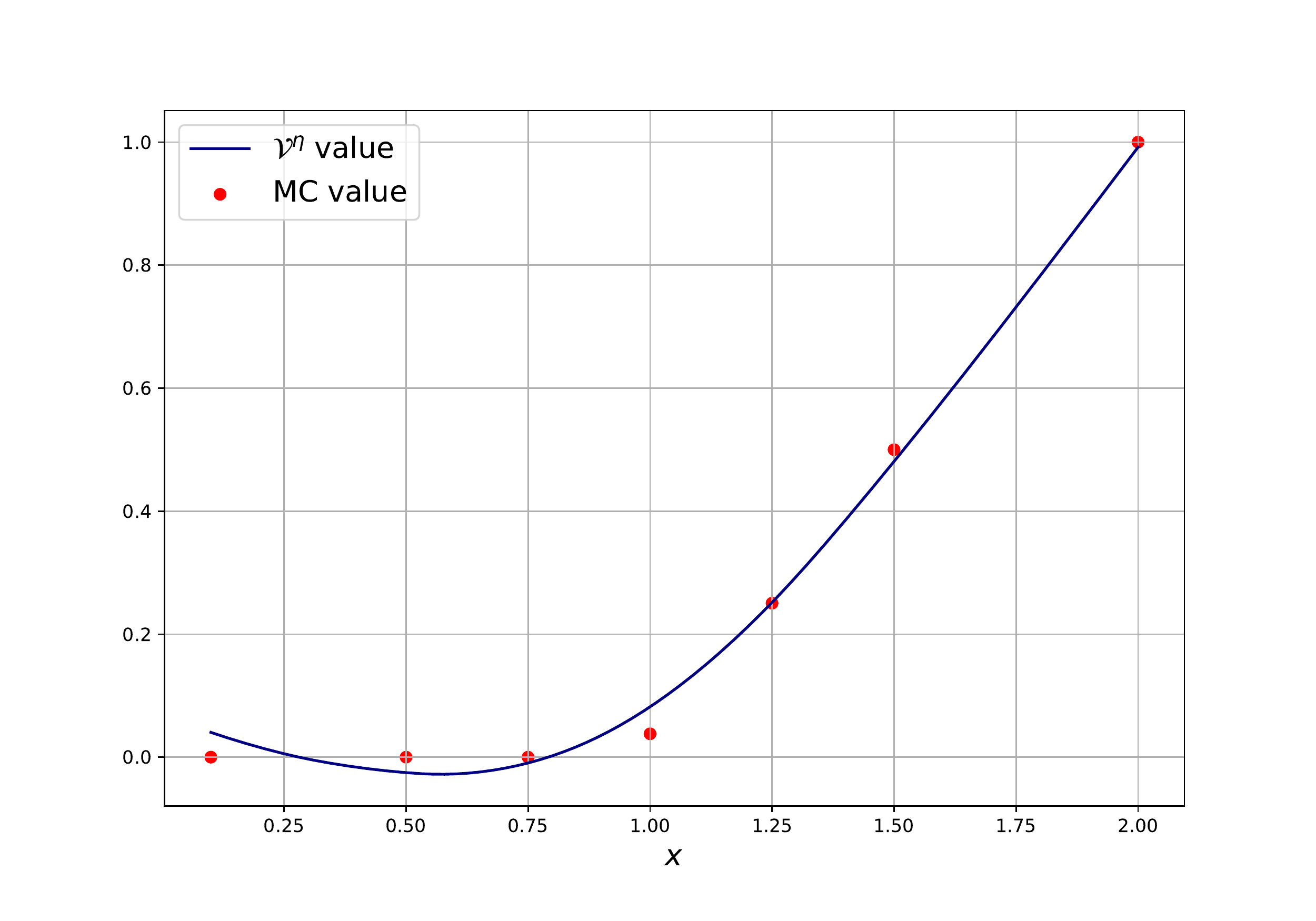}
    \end{subfigure}
    \begin{subfigure}{.31\linewidth}
        \centering
        \includegraphics[height=3.67cm]{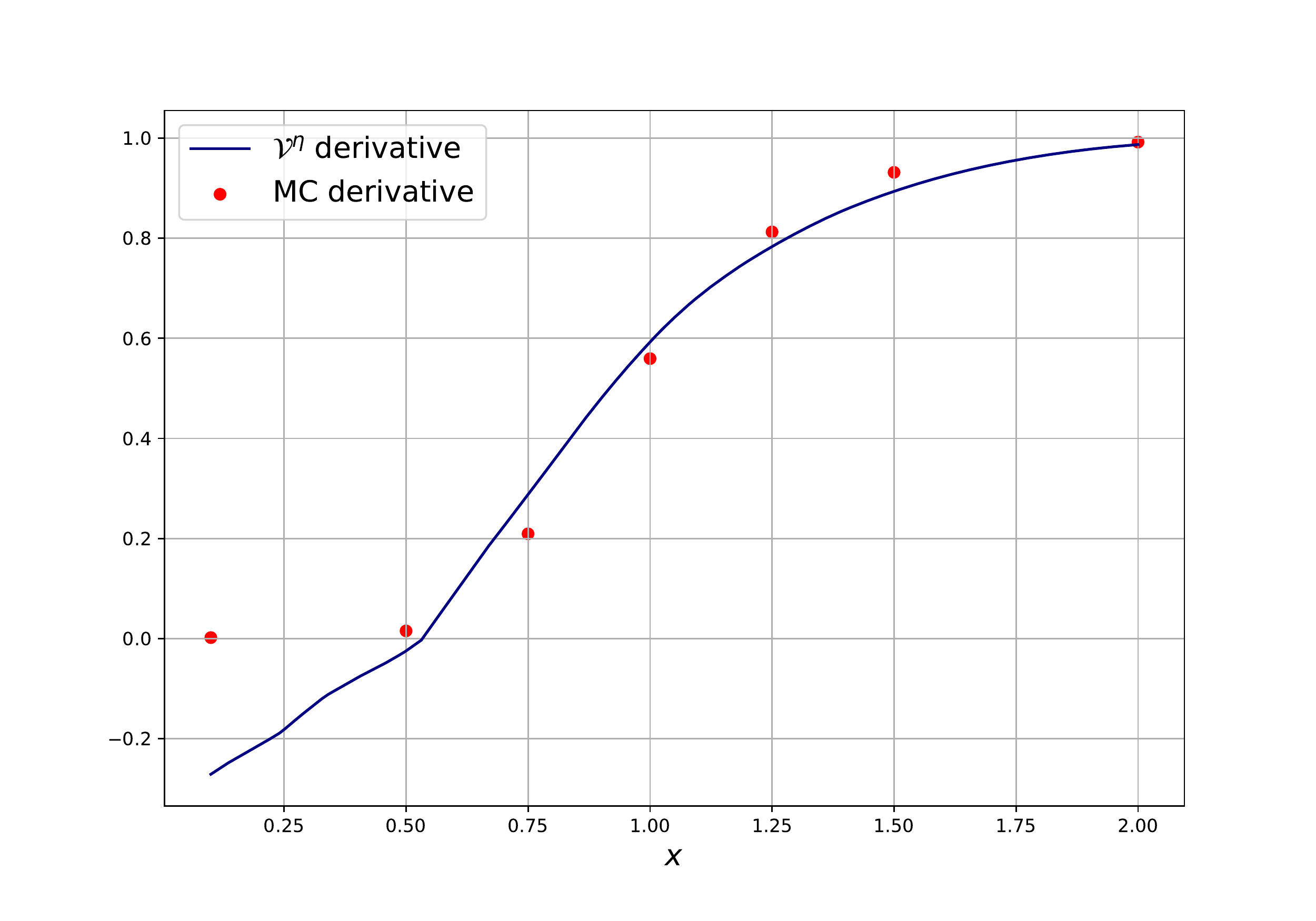} 
        \caption[short]{$t=0$}
    \end{subfigure}
    \begin{subfigure}{.31\linewidth}
        \centering
        \includegraphics[height=3.67cm]{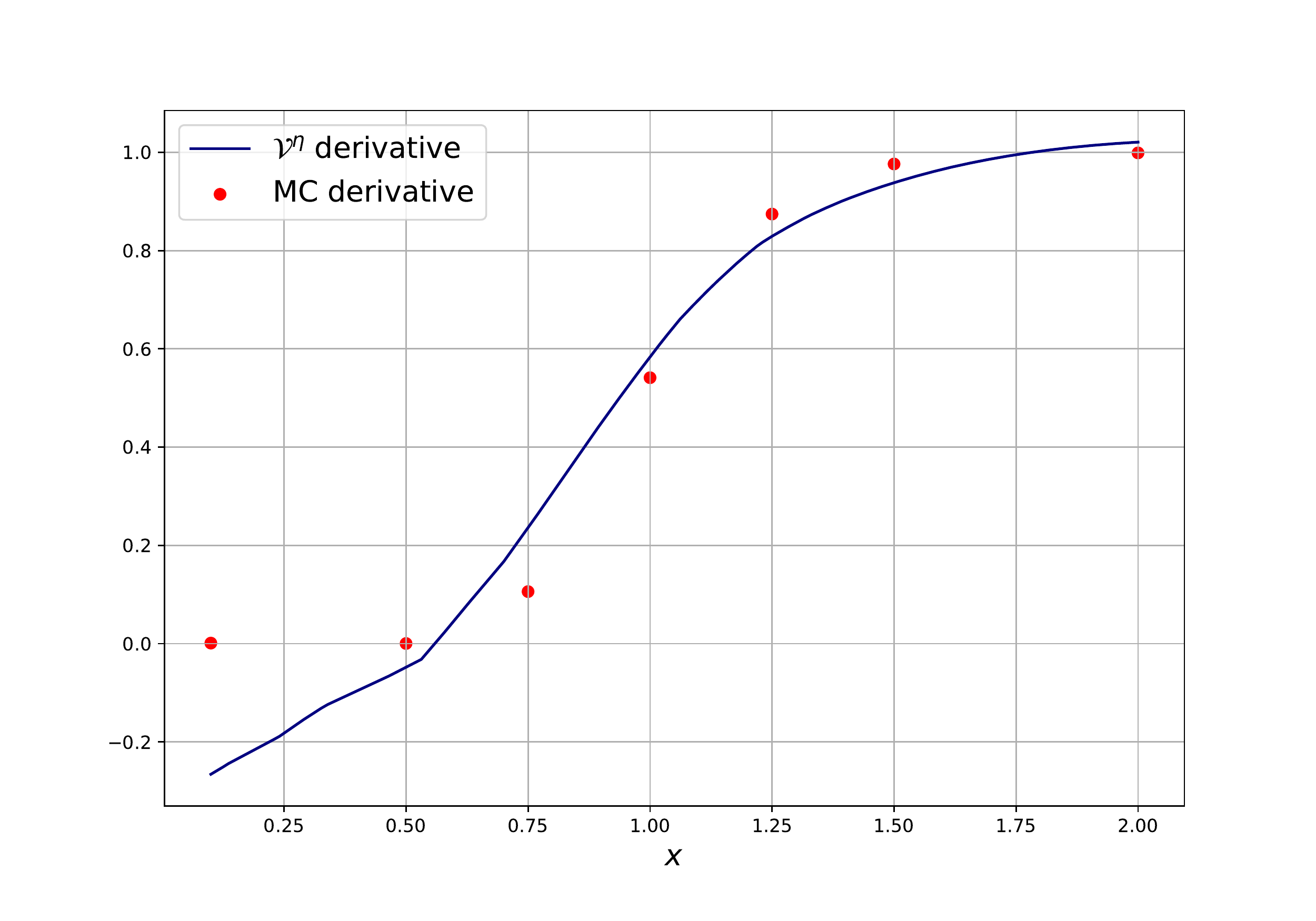}
        \caption[short]{$t=0.5$}
    \end{subfigure}
    \begin{subfigure}{.31\linewidth}
        \centering
        \includegraphics[height=3.67cm]{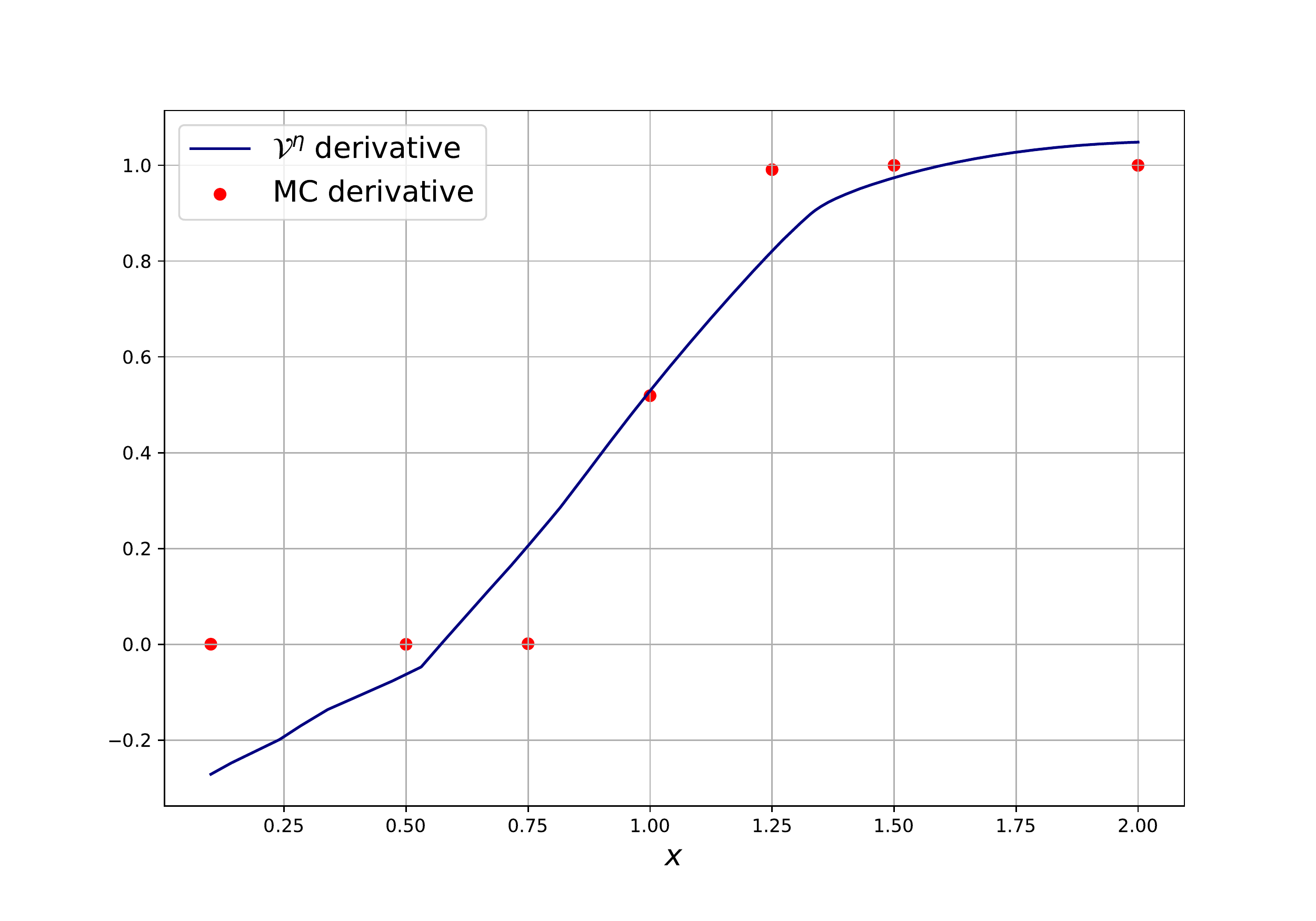}
        \caption[short]{$t=0.9$}
    \end{subfigure}
    \caption{
    \label{fig:value_pathwise_learning_call}
    \footnotesize{Value function $\vartheta^\eta$ and its derivative obtained by Pathwise Learning (Algorithm \ref{algo:scheme_value_pathwise_martingale_learning}) for a call option  with strike $1$, with parameter $\sigma = 0.3$ and linear market impact factor $\lambda = 5e^{-3}$, plotted as functions of $x$, for fixed values of $t$.}
    }
\end{figure}

\begin{figure}[htp]
    \centering
    \begin{subfigure}{.32\linewidth}
        \centering
        \includegraphics[height=3.75cm]{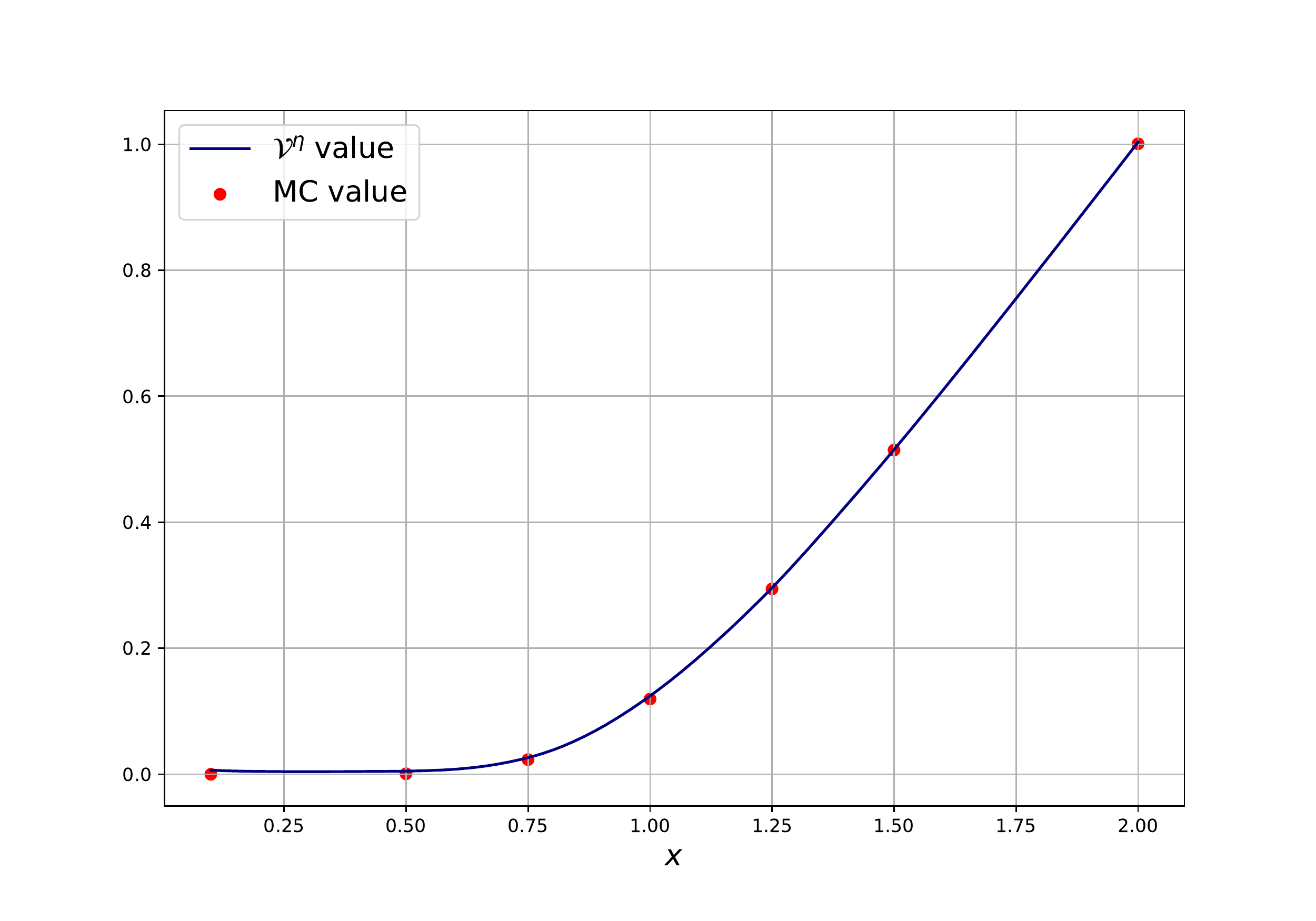}
    \end{subfigure}
    \begin{subfigure}{.32\linewidth}
        \centering
        \includegraphics[height=3.75cm]{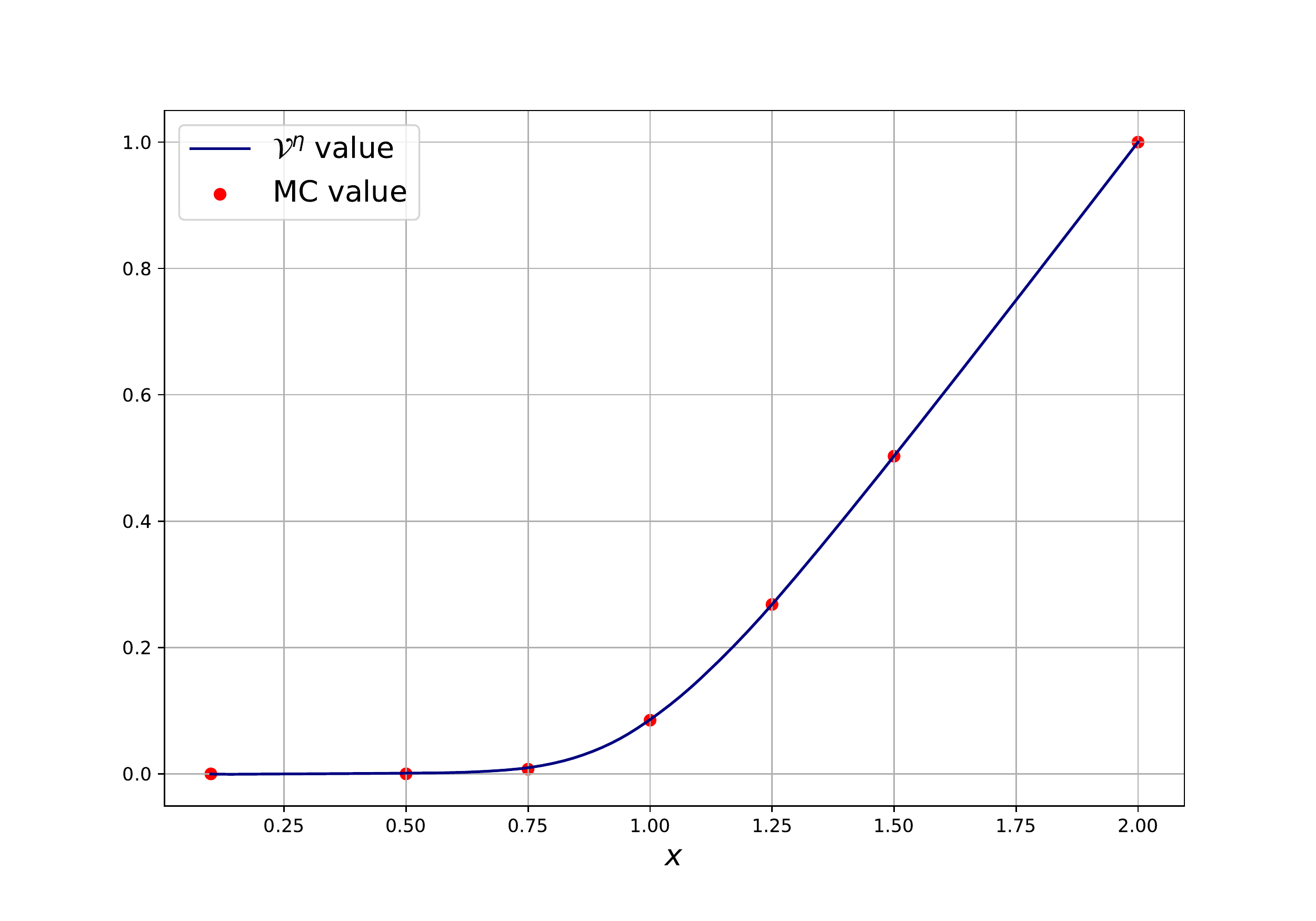}
    \end{subfigure}
    \begin{subfigure}{.32\linewidth}
        \centering
        \includegraphics[height=3.75cm]{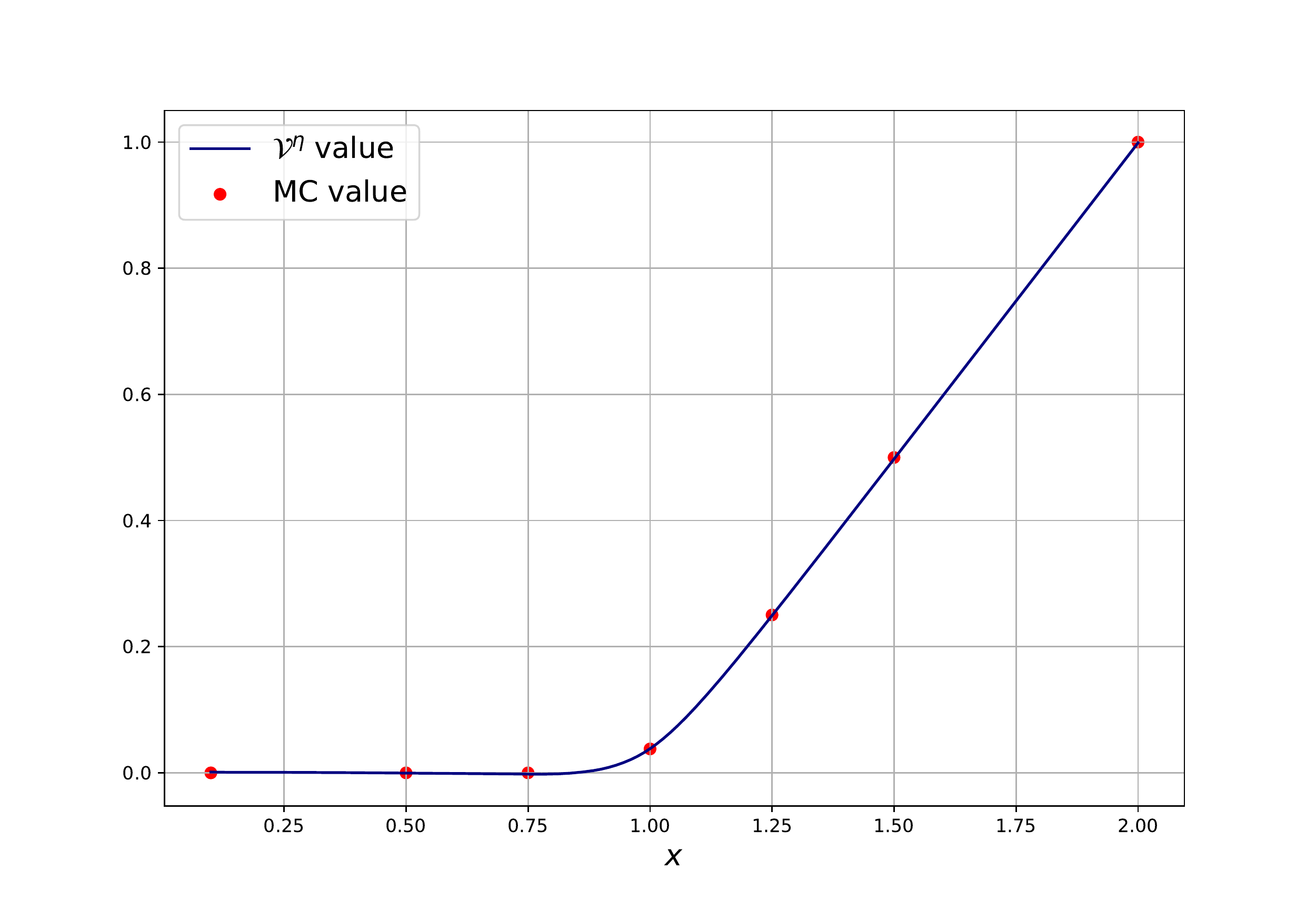}
    \end{subfigure}
    \begin{subfigure}{.32\linewidth}
        \centering
        \includegraphics[height=3.75cm]{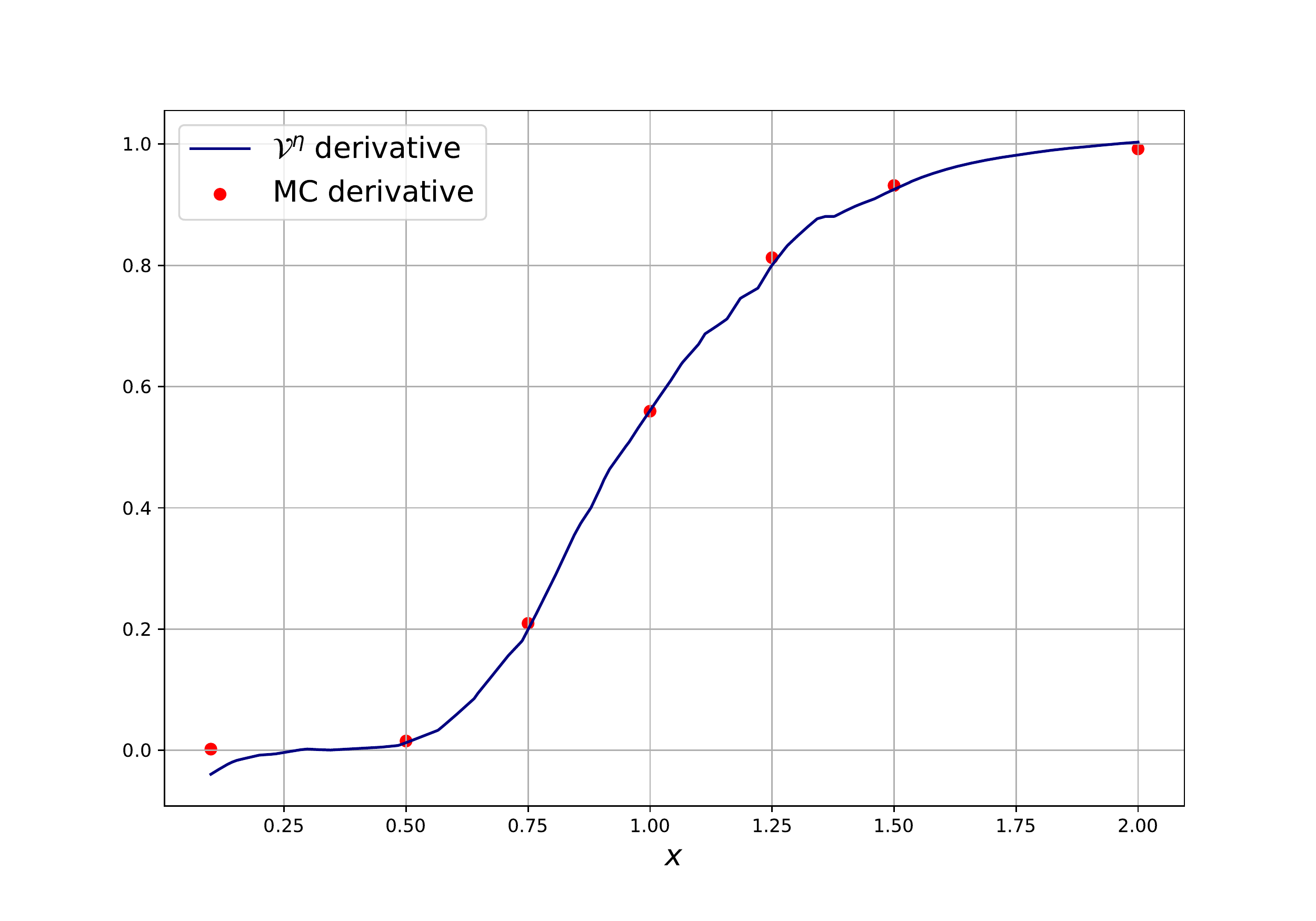} 
        \caption[short]{$t=0$}
    \end{subfigure}
    \begin{subfigure}{.32\linewidth}
        \centering
        \includegraphics[height=3.75cm]{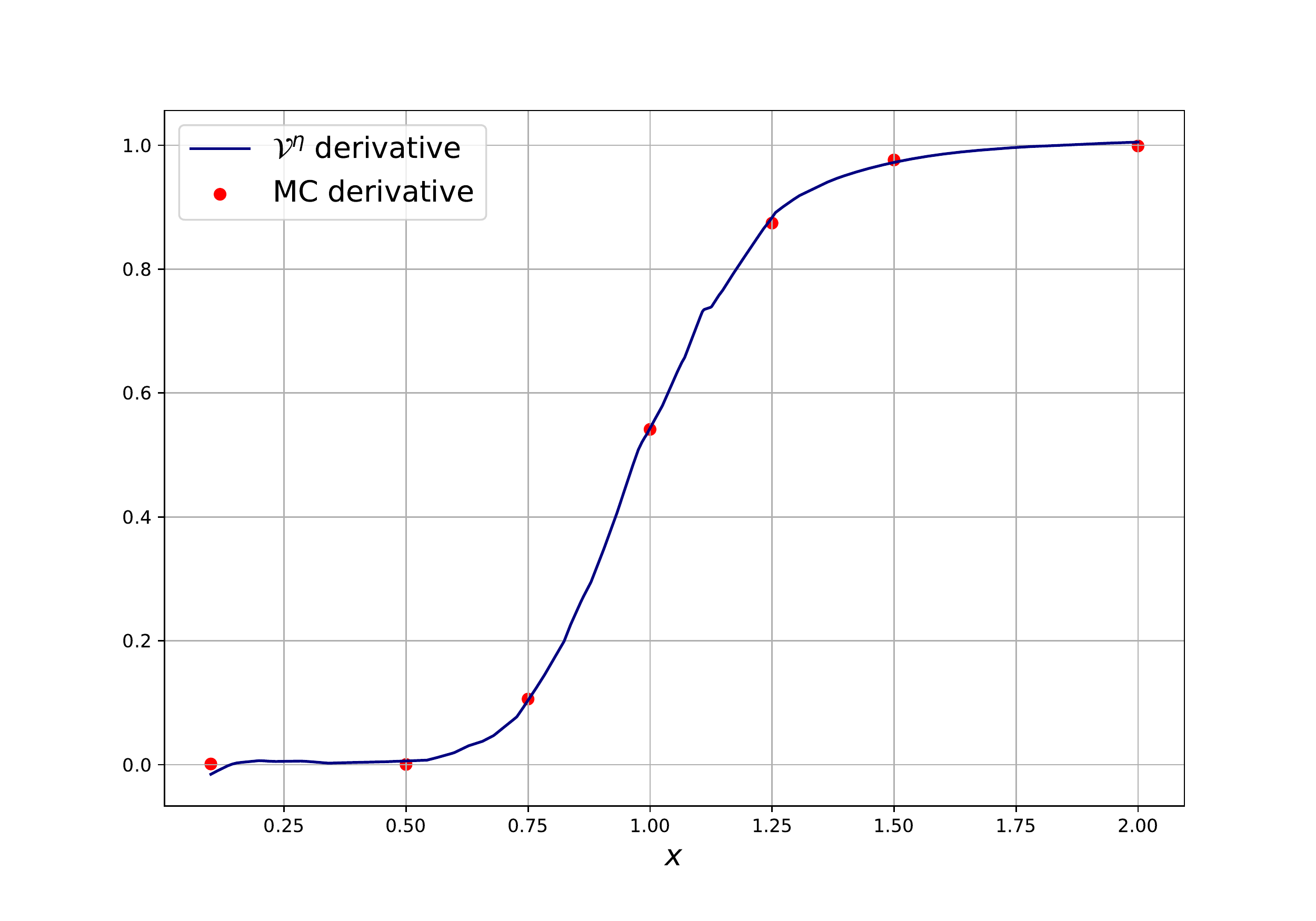}
        \caption[short]{$t=0.5$}
    \end{subfigure}
    \begin{subfigure}{.32\linewidth}
        \centering
        \includegraphics[height=3.75cm]{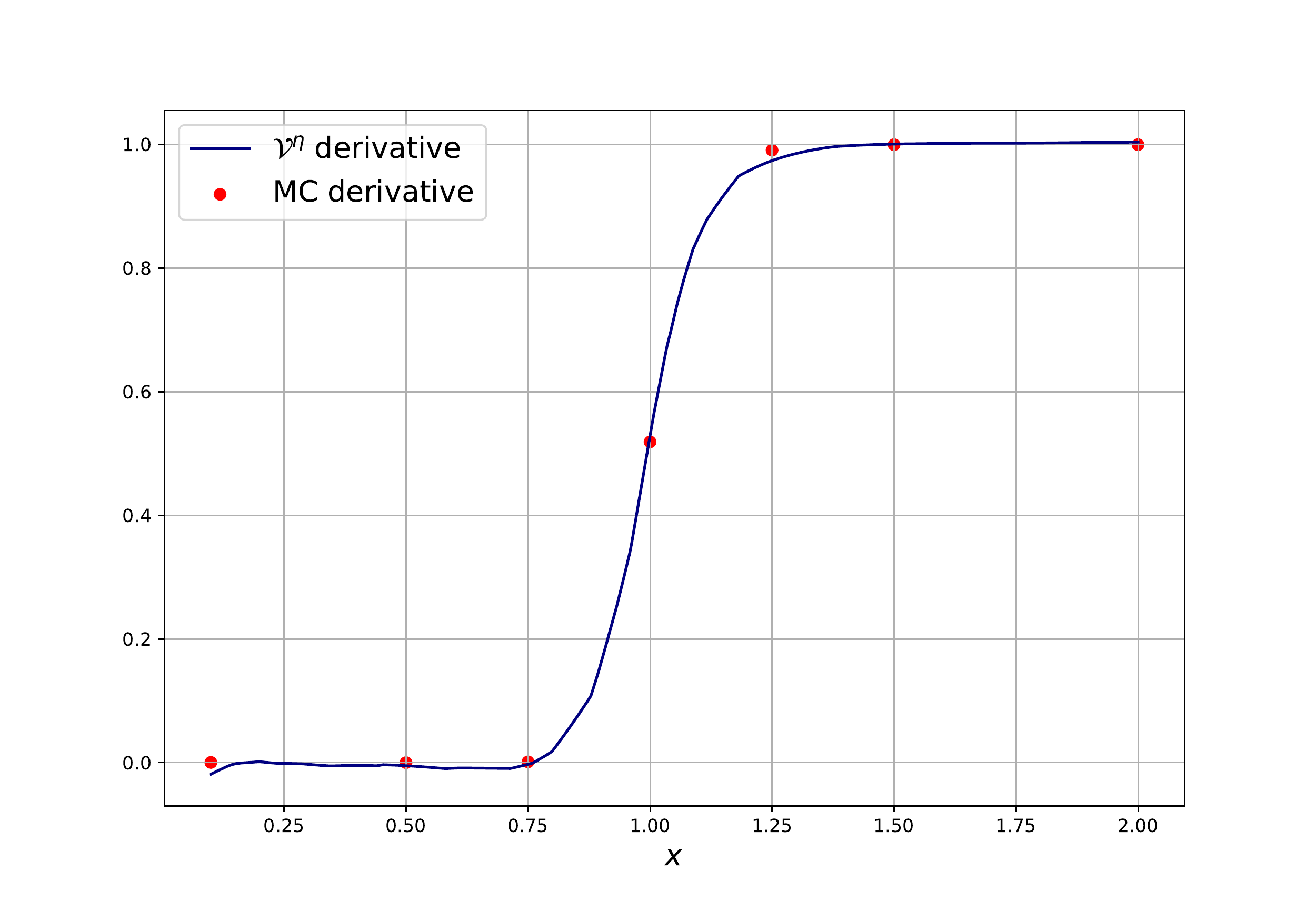}
        \caption[short]{$t=0.9$}
    \end{subfigure}
    \caption{
    \label{fig:value_pathwise_differential_learning_call}
    \footnotesize{Value function $\vartheta^\eta$ and its derivative obtained by Pathwise Differential Learning (Algorithm \ref{algo:scheme_value_pathwise_differential_learning}) for a call option  with strike $1$, with parameter $\sigma = 0.3$ and linear market impact factor $\lambda = 5e^{-3}$, plotted as functions of $x$, for fixed values of $t$.}
    }
\end{figure}

We plot again the value function $\vartheta^\eta(t,x)$ and its derivative $\partial_x \vartheta^\eta (t,x)$ for fixed values $t=0$, $t=0.5$, $t=0.9$, parameter $\sigma = 0.3$ and linear market impact factor $\lambda = 5e^{-3}$, and compare it with the Monte-Carlo estimation obtained. The Figure \ref{fig:value_differential_learning_call} corresponds to the Differential regression learning method, Figure \ref{fig:value_pathwise_learning_call} corresponds to the pathwise martingale learning  while Figure \ref{fig:value_pathwise_differential_learning_call} corresponds to the Pathwise differential learning method. Graphically, the results of the Differential regression learning and the Pathwise differential learning methods are very close to the points obtained by Monte Carlo estimation for the value and the first derivative. The Pathwise methods does not give a good approximation of the value and the derivatives for values of $x$ smaller than  1.
The difference of performance between the Pathwise and the Differential pathwise methods is analogous to the one observed between Differential regression learning and "simple" regression learning in Figure \ref{fig:comparison_diff_simple_learning_call}, demonstrating the interest of adding a regression term for the derivative of the neural network.

\section{Further step: resolution for parametric terminal functions}\label{sec:resolution_any_terminal_condition}
\subsection{Theory and network structure}


In line with the works in \cite{vidales2018unbiased}, \cite{glau2020deep}, \cite{remlinger2022robust}, our next goal  is to design a Machine Learning method allowing us to directly obtain a solution of problem \eqref{eq:var_form} for a parametric terminal condition $g_K$, for every value of the parameter $K\in \R^p$ in a compact set. In other words, we aim to learn the operator that maps the payoff function parameter $K$ to the solution of the PDE with terminal condition $g_K$. 
We want to be able to train the neural networks once and for all on a selection of parameter values and obtain a network which takes a couple $(t,x)$ and the parameter value $K$ and outputs the solution of the problem. 
In \cite{glau2020deep} the authors solve parametic PDEs by using a variant of highway networks \cite{srivastava2015training}, which are feedforward networks where each dense layer has an additional parameter called \textit{gate} which allows the layer to output a combination of the unmodified input and of the output of the affine and activation operations, alleviating the vanishing gradient problem and allowing to train deeper networks. Their network takes the time, space and PDE parameter as input and is trained in the spirit of the Deep Galerkin method. The PDE residuals is computed from the neural network on a random time, space and parameter grid and is minimized in order to approximate the PDE solution. In \cite{vidales2018unbiased} the authors give four methods to solve parametric PDEs using a fully connected network taking the time, space and PDE parameter as input and minimizing a loss averaged over random parameter values.
In this article we follow the methodology developped in \cite{remlinger2022robust} by relying on a class of neural networks, called  DeepONet, presented in \cite{lu2019deeponet}, and aiming to approximate functional operators. 
This method is based on the following universal approximation theorem for operator, due to Chen and Chen \cite{chen1995universal}.

\begin{theorem}
Suppose that $\sigma$ is a continuous non-polynomial function, $X$ is a Banach space, $K_1 \subset X$, $K_2 \subset \R^d$ are two compact sets in $X$ and $\R^d$, respectively, $V$ is a compact set in $C(K_1)$, $G$ is a nonlinear continuous operator, which maps $V$ into $C(K_2)$. Then for any $\epsilon >0$, there are positive integers $n$, $p$, $m$, constants $c_i^k$, $\xi_{ij}^k$, $\theta_i^k$, $\zeta_k\in\R$, $w_k\in\R^d$, $x_j\in K_1$, $i=1,...,n$, $k=1,...,p$, $j=1,...,m$, such that
\bes{
   \bigg\| G(u)(y) - \sum_{k=1}^p \underbrace{\sum_{i=1}^n c_i^k \sigma\bigg( \sum_{j=1}^m \xi_{ij}^k u(x_j) +\theta_i^k \bigg)}_{\text{branch net}} \underbrace{\sigma\left( w_k y + \zeta_k \right)}_{\text{trunk net}} \bigg\| < \epsilon,
}
holds for all $u\in V$ and $y\in K_2$.
\end{theorem}

\begin{figure}[h!]
    \centering
    \includegraphics[width=0.8\linewidth]{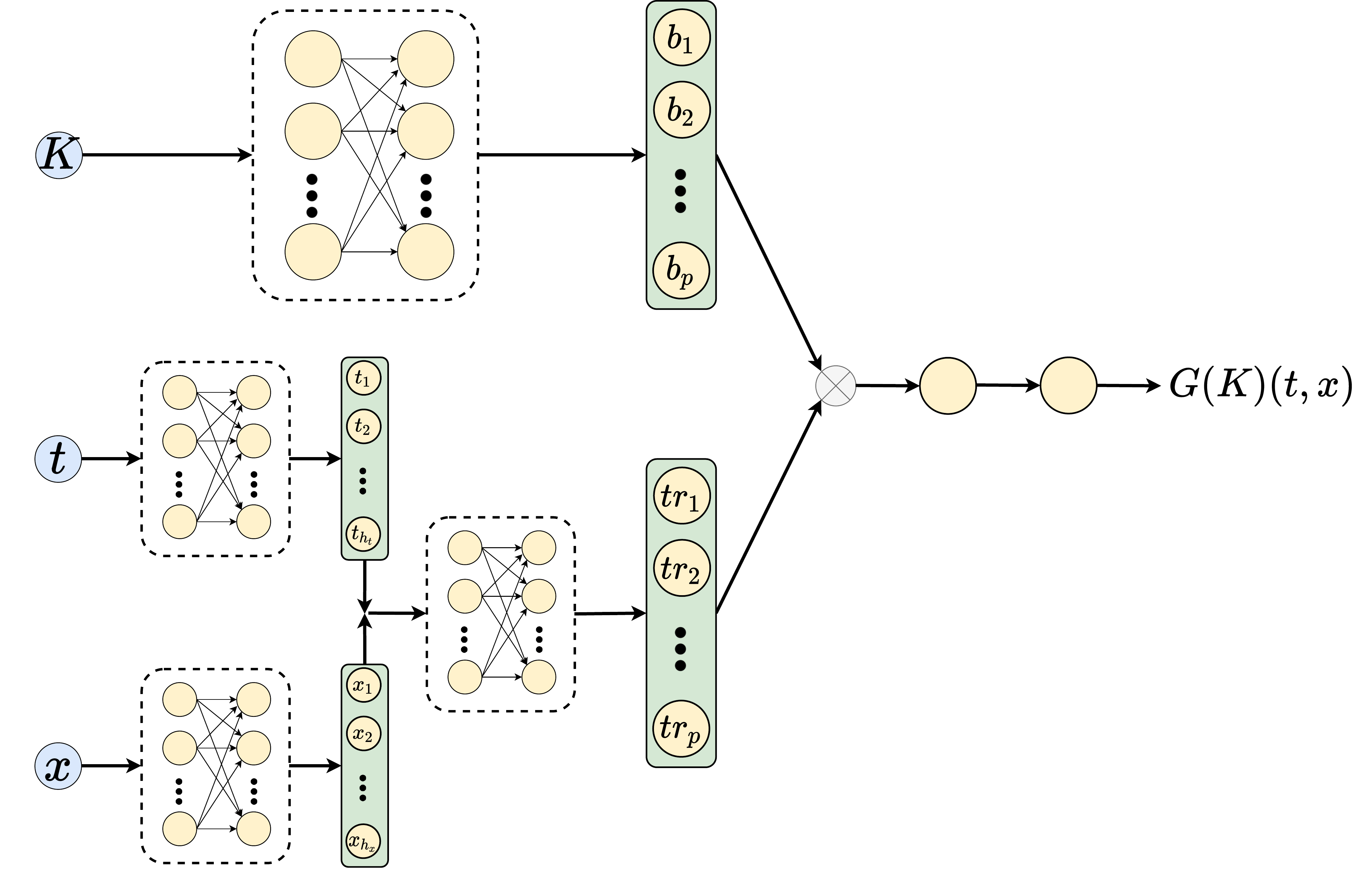}
\caption{Structure of the DeepONet network.}
\label{fig:structure_deeponet}
\end{figure}

The network used in \cite{lu2019deeponet} is composed of two sub networks, the \textit{branch net}, which takes the terminal function estimated on a fixed number of points called \textit{sensors} as input, and the trunk net, which takes the time and space coordinates as input. 
In our case, as in \cite{remlinger2022robust}, the \textit{branch net} takes the parametric terminal function estimated on a grid of \textit{sensors}, and will be trained for random values of the function's parameter. We represent the structure of this neural network in Figure \ref{fig:structure_deeponet}.

As our numerical methods proceed  in two stages, by first approximating the optimal control of problem \eqref{eq:var_form} with a neural network and then approximating the associated value function with another network, we again use a DeepONet to approximate the optimal control and then another one to approximate the value function $u$ by differential learning.


While the authors of \cite{lu2019deeponet} train a DeepONet in a supervised manner, 
we train  our network in an \textbf{unsupervised way}. 

As before, we start by training the control network $\mra_\theta$ in order to approximate the optimal control of problem. We use a batch of $M$ independent trajectories $\{x_{t_n}^{m,K_m,\theta}, t_n \in \Tc_N\}$, $m$ $=$ $1,\ldots,M$,  of $\{X_{t_n}^{K_m, \theta}, t_n \in \Tc_N\}$, where the $K$ superscript denotes that the trajectory is driven by a control with input parameter $K$, and $M$ random parameter values $K_m$ randomly sampled from a distribution $\mu_K$ with compact support in $\R^p$, and apply a stochastic gradient ascent method to the empirical gain function:
\bes{
J_M(\theta) &= \; \frac{1}{M} \sum_{m=1}^M \Big[ g(x_T^{m,K_m,\theta}, K_m) +  \sum_{n=0}^{N-1} f\big(x_{t_n}^{m,K_m,\theta},\mra_\theta(t_n,x_{t_n}^{m,K_m,\theta}, K_m)\big) \Delta t_n \Big].  
}
The pseudo-code is described in  Algorithm \ref{algo:scheme_control_global_method_deeponet}.  
The output of this algorithm yields a parameter $\theta^*$, and so 
an approximation of the optimal feedback control with  $\mra^*$ $=$ $\mra_{\theta^*}$, and of the associated optimal state process with $X^*$ $=$ $X^{\theta^*}$. 
 In the sequel, to alleviate notations, we shall omit the superscript $*$, and simply denote $\mra$ and $X$.

\begin{algorithm}
\scriptsize
\SetAlgoLined
\KwResult{A set of optimized parameters $\theta^*$;}
 Initialize the learning rate $l$ and the neural network $\mra_\theta$\; 
 Generate an $\R^{N+1}$-valued time grid $0=t_0 < t_1 < ... < t_N = T$ with time steps $(\Delta t_n)_{n=0,...,N-1}$\;
 Generate a batch of $M$ starting points $X_0$ $\sim$ $\mu_0$, Brownian increments $(\Delta W_{t_n})_{n=0,...,N-1}$ in $R^d$ and parameter values $K$ $\sim$ $\mu_K$\;
 \For{each batch element $m$}{
        Compute the trajectory $(x_{t_n}^{m, K_m, \theta})_{n=0,...,N}$ through the scheme
        \bes{
            x_{t_{n+1}}^{m, K_m, \theta} & = \;  x_{t_{n}}^{m, K_m, \theta}  +  \mrb(x_{t_{n}}^{m, K_m, \theta},\mra_\theta(t_n,x_{t_{n}}^{m, K_m, \theta}, K_m)) \Delta t_n  +  \sigma(x_{t_{n}}^{m, K_m, \theta},\mra_\theta(t_n,x_{t_{n}}^{m, K_m, \theta}, K_m)) \Delta w_{t_n}^m,
        }
        from the generated starting point $x_{t_0}^m$, Brownian increments $(\Delta w_{t_n}^m)_{n=0,...,N-1}$ and parameter $K_m$\; 
    } 
 \For{each epoch}{
    Compute the batch loss 
        \bes{
        J_M(\theta) &= \; \frac{1}{M} \sum_{m=1}^M \Big[ g(x_T^{m, K_m, \theta}, K_m) +  \sum_{n=0}^{N-1} f\big(x_{t_n}^{m, K_m, \theta},\mra_\theta(t_n,x_{t_n}^{m, K_m, \theta}, K_m)\big) \Delta t_n \Big]
        }
    Compute the gradients $\nabla_{\theta} J_M(\theta)$\;
    Update $\theta \leftarrow \theta - l \nabla_{\theta} J_M(\theta)$\;
}
\textbf{Return:} The set of  optimized parameters $\theta^*$\;
\caption{Deep learning scheme to solve the stochastic control problem \eqref{eq:var_form}}
\label{algo:scheme_control_global_method_deeponet}
\end{algorithm}

From this optimal control approximation, the value function is then approximated through the Differential regression learning algorithm presented in Section \ref{section:numerical_differential_learning} modified in order to train the network for different values of the terminal function parameter $K$. 
The target payoff and its derivative are then computed as 
\bes{
Y_T^{K_m, t_n} &= \; g(X_T^{K_m}, K_m) +  \sum_{q=n}^{N-1} f^{\mra^*}(t_q,X_{t_q}^{K_m}) \Delta t_q,  \quad n=0,\ldots,N,  \\
Z_T^{K_m, t_n} &= \;  \big(D_{X_{t_n}}X_T^{K_m} \big)^\top  D_x g(X_T^{K_m}, K_m)   \; + \;  \sum_{q=n}^{N-1}   \big(D_{X_{t_n}}X_{t_q}^{K_m}\big)^\top  D_x f^{\mra^*}(t_q,X_{t_q}^{K_m})  \Delta t_p,
}
with the convention that the above sum over $q$ is zero when $n$ $=$ $N$. 
For the training of the neural network $\vartheta^\eta$, we use a batch  of $M$ independent samples $(x_{t_n}^{m, K_m},y_T^{m, K_m, t_n}, z_T^{m, K_m, t_n})$, $m$ $=$ $1,\ldots,M$,  of $(X_{t_n}^{K_m},Y_T^{K_m, t_n},Z_T^{K_m, t_n})$, $n$ $=$ 
$0,\ldots,N$ and $M$ random parameter values $K^m$ randomly sampled from a distribution $\mu_K$ with compact support in $\R^p$, and apply stochastic gradient descent for the minimization of the mean squared error functions
\bes{\label{eq:MSE_differential_learning_deeponet}
MSE_{val}(\eta) &= \; \frac{1}{M} \sum_{m=1}^M \sum_{n=0}^{N-1} \big| y_T^{m, K_m, t_n} - \vartheta^\eta(t_n,x_{t_n}^{m, K_m}, K_m) \big|^2 \Delta t_n  \\
MSE_{der}(\eta) & = \; \frac{1}{M} \sum_{m=1}^M \sum_{n=0}^{N-1} \frac{1}{\| z_T^{K_m, t_n} \|^2}  \big| z_T^{m, K_m, t_n} - D_x \vartheta^\eta(t_n,x_{t_n}^{m, K_m}, K_m) \big|^2 \Delta t_n. 
}
The pseudo-code is described in Algorithm \ref{algo:scheme_value_differential_learning_deeponet}.

\begin{algorithm}[H] 
\scriptsize
\SetAlgoLined
\KwResult{A set of optimized parameters $\eta^*$;}
 Initialize the learning rate $l$, the neural networks $\vartheta^\eta$\; 
 Generate an $\R^{N+1}$-valued time grid $0=t_0 < t_1 < ... < t_N = T$ with time steps $(\Delta t_n)_{n=0,...,N-1}$\;
 Generate a batch of $M$ starting points $X_0$ $\sim$ $\mu_0$, Brownian increments $(\Delta W_{t_n})_{n=0,...,N}$ in $R^d$ and parameter values $K$ $\sim$ $\mu_K$\;
 \For{each batch element $m$}{
    Compute the trajectory $(x_{t_n}^{m, K_m})_{n=0,...,N}$ through the scheme
        \bes{
            x_{t_{n+1}}^{m, K_m} & = \;  x_{t_{n}}^{m, K_m}  +  \mrb^{\mra^*}(t_n,x_{t_{n}}^{m, K_m}) \Delta t_n  +  \sigma^{\mra^*}(t_n,x_{t_{n}}^{m, K_m}) \Delta w_{t_n}^m,
        }
        from the generated starting point $x_{t_0}^m$, Brownian increments $(\Delta w_{t_n}^m)_{n=0,...,N-1}$, parameter $K_m$ and previously trained control $a=a_{\theta^*}$\;
    Compute the value and derivative targets $(y_T^{m, K_m, t_n})_{n=0,...,N}$ and $(z_T^{m, K_m, t_n})_{n=0,...,N}$\;
}
 \For{each epoch}{
    \If{Epoch number is even}{
        Compute, for every batch element $m$, the integral $\sum_{n=0}^{N-1} \big| y_T^{m, K_m, t_n} - \vartheta^\eta(t_n,x_{t_n}^{m, K_m}, K_m) \big|^2 \Delta t_n$\;
        Compute the batch loss $MSE_{val}(\eta)$\;
        Compute the gradient $\nabla_{\eta} MSE_{val}(\eta)$\;
        Update $\eta \leftarrow \eta - l \nabla_{\eta} MSE_{val}(\eta)$\;
    }
    \Else{
        Compute, for every batch element $m$, the integral $\sum_{n=0}^{N-1} \big| z_T^{m, K_m, t_n} - D_x \vartheta^\eta(t_n,x_{t_n}^{m, K_m}, K_m) \big|^2 \Delta t_n$.\;
        Compute the batch loss $MSE_{der}(\eta)$\;
        Compute the gradient $\nabla_{\eta} MSE_{der}(\eta)$\;
        Update $\eta \leftarrow \eta - l \nabla_{\eta} MSE_{der}(\eta)$\;
    }
}
\textbf{Return:} The set of  optimized parameters  $\eta^*$\;
\caption{Deep learning scheme for Differential Regression learning}
\label{algo:scheme_value_differential_learning_deeponet}
\end{algorithm}

\subsection{Application to the Black-Scholes model with linear market impact}
\label{sec:BS_linear_market_impact_deeponet}

Using this DeepONet based algorithm, we revisit the resolution of the nonlinear Black Scholes equation presented in Section \ref{sec:BS_linear_market_impact}. In this Section, we use the Algorithms \ref{algo:scheme_control_global_method_deeponet} and \ref{algo:scheme_value_differential_learning_deeponet} in order to solve the PDE for a terminal function corresponding to a call option payoff $g(x, K) = \max(x-K, 0)$ with parameter (or \textit{strike}) $K$ $\in$ $\R_+^*$.

The \textit{branch net} of the \textbf{DeepONet used to approximate the control} as a function of the terminal function $g_K$ is a standard feed-forward network composed of two layers with 50 neurons and use the \textit{tanh} activation function. The \textit{trunk net} has the same structure as the network used in the previous sections and represented in Figure \ref{fig:structure_network}. It is composed of two sub-networks taking respectively the time $t$ and state $x$ as an input and each composed of two layers of 50 neurons using the \textit{tanh} activation function. The outputs of these two sub-networks are contatenated into a vector of $\R^{100}$ which passes through two additionnal layers of 50 neurons using \textit{tanh} activation. The output of the \textit{branch net} and the \textit{trunk net}, which have the same dimension, are then combined through a dot product whose output passes through a layer of one neuron using \textit{tanh} activation and a layer of one neuron using the \textit{Parametric ReLU} activation function ensuring, as explained in Section \ref{sec:numerical_approximation optimal control}, that the control obtained belongs to  the control space $A$.

The \textbf{DeepONet used to approximate the value function $u$} shares the same structure. The branch and trunk net have the same structures as the ones used in the control DeepONet with the same number of layers and neurons per layer and with \textit{Swish} activation function, defined as 
\bes{
Swish(x)=\frac{x}{1 + e^{-x}}.
}
After the dot product, the output also passes through a layer composed of one neuron using \textit{Swish} activation function and a last layer of one neuron using no activation function. 
\begin{remark}
For the control DeepONet, we used the \textit{tanh} activation function instead of the \textit{ELU} activation used in the previous sections as we encountered loss divergences during the training of the DeepONet with \textit{ELU} activation. Since the \textit{tanh} activation is bounded, the problem was resolved using this function.

For the value DeepONet, we used the \textit{Swish} activation function instead of the \textit{ELU} activation used in the previous sections as it empirically gave better results. As the \textit{ELU} function's second derivative is discontinuous, a kink was observed on the value function's derivative we obtained with the DeepONet. This effect was not present when performing the "simple" regression with a standard network in the previous sections, probably because the regression problem is simpler and the true value function fitted with a better accuracy. Since the \textit{Swish} activation is of class $C^\infty$, we obtained better results, without kinks, using this function.
\end{remark}

\begin{figure}[htp]
    \centering
    \begin{subfigure}{.49\linewidth}
        \centering
        \includegraphics[height=5.5cm]{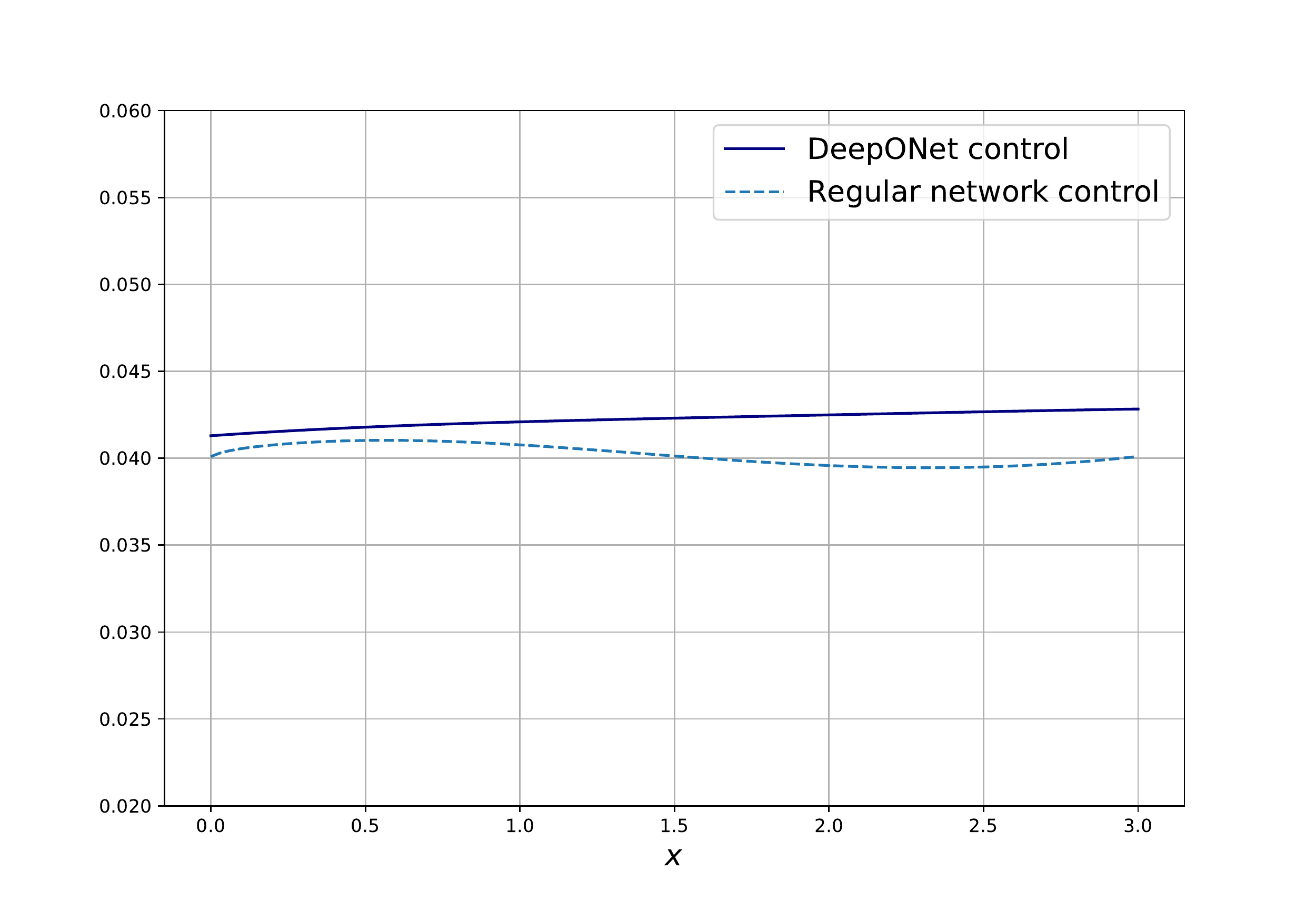}
        \caption[short]{$t=0$}
    \end{subfigure}
    \begin{subfigure}{.49\linewidth}
        \centering
        \includegraphics[height=5.5cm]{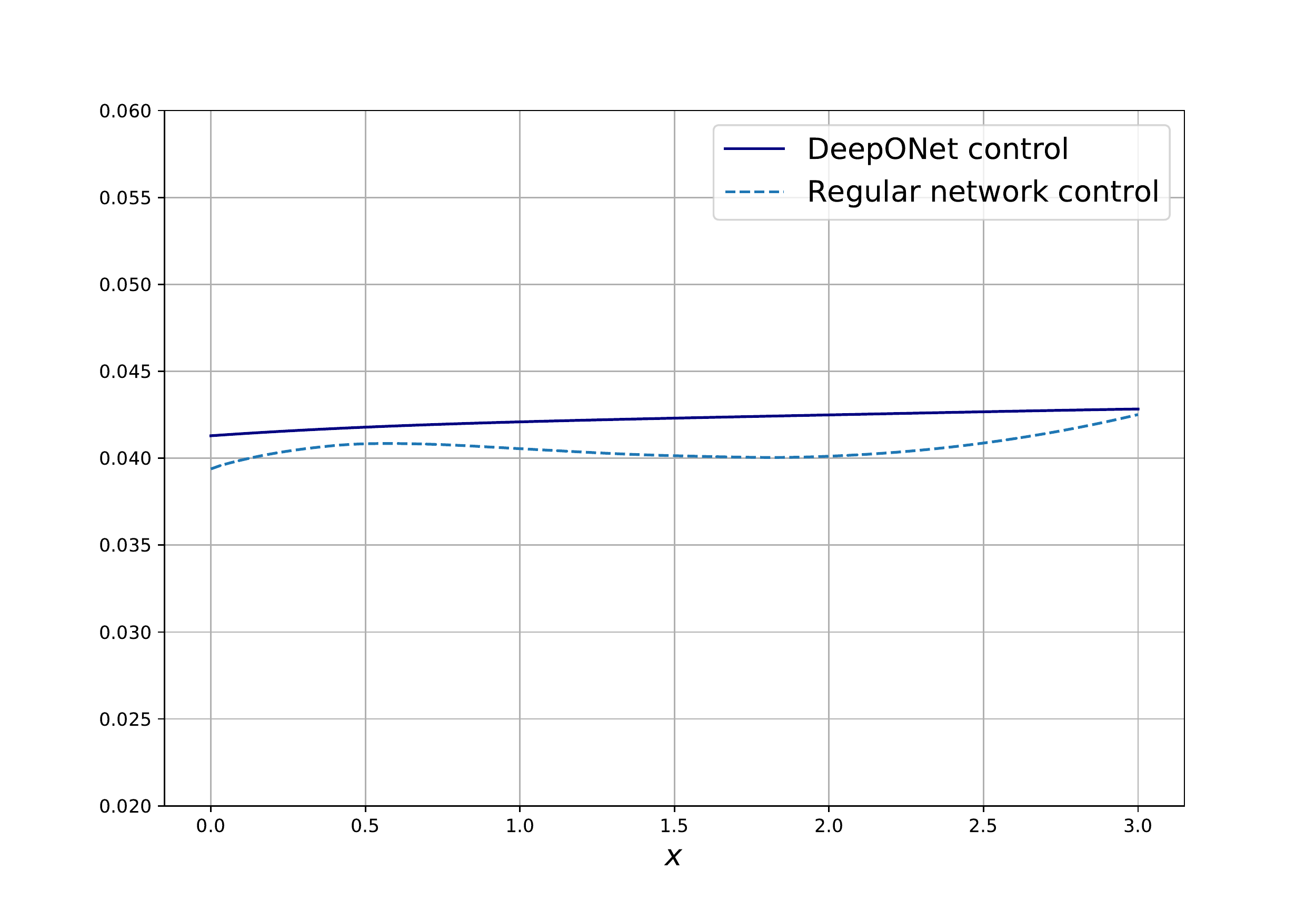}
        \caption[short]{$t=0.9$}
    \end{subfigure}
    \caption{
    \label{fig:control_call_deeponet}
    Control $a_\theta$ obtained by global method with DeepONet (Algorithm \ref{algo:scheme_control_global_method_deeponet}) for a call option  with strike $K=1$, with parameter $\sigma = 0.3$ and linear market impact factor $\lambda = 5e^{-3}$, plotted as functions of $x$, for fixed values of $t$.
    }
\end{figure}

For both trainings, we use the Adam optimizer with a learning rate equal to $1e^{-3}$ and train the network on 8192 random trajectories and strike values.
In order to test the generalization power of our method, we train the control and value neural networks on strikes randomly sampled from $\mathcal{U}([0.25, 0.75] \cup [1.5, 2])$ and test the network on strikes chosen in $[0.2, 2.1]$. We plot in figure \ref{fig:control_call_deeponet} below the optimal control approximation obtained along for $K=1$ along with the control approximation obtained by Algorithm \ref{algo:scheme_control_global_method} (denoted \textit{regular network control}) which serves as a reference.

We compute in Table \ref{table_deeponet_call}, the residual losses defined in  \eqref{residual} for the DeepONet $\vartheta^\eta$ by the Differential Regression Learning scheme (Algorithm \ref{algo:scheme_value_differential_learning_deeponet}). We compute the residual losses on a $102$x$102$ linearly spaced time and space grid with $t\in[0,0.9]$ and $x\in[0, 3]$ for a terminal call option payoff for strikes $K$ $\in$ $\{ 0.2,\ 0.5,\ 1,\ 1.75,\ 2,\ 2.1 \}$, with parameter $\sigma = 0.3$ and linear market impact factor $\lambda = 5e^{-3}$.  

\begin{scriptsize}
\begin{table}[h]
\centering
\begin{tabular}{|c|c|c|c|c|c|c|}
\hline  & $K=0.2$ & $K=0.5$ & $K=1$ & $K=1.75$ & $K=2$ & $K=2.1$ \\
\hline  Residual loss  & & & & & & \\
+ terminal loss & $ 1.918 e^{-3}$ & $ 1.461 e^{-4}$ & $ 1.700 e^{-3}$ & $1.162 e^{-3}$ & $2.002 e^{-3}$ & $2.175 e^{-3}$ \\
\hline
\end{tabular}
\caption{Residual and boundary losses computed on a $102$x$102$ time and space grid with $t\in[0,0.9]$ and $x\in[0, 3]$ for a terminal call option payoff $g(x)$ $=$ $\max(x-K,0)$ for $K$ $\in$ $\{ 0.2,\ 0.5,\ 1,\ 1.75,\ 2,\ 2.1 \}$, with parameter $\sigma = 0.3$ and linear market impact factor $\lambda = 5e^{-3}$.} 
\label{table_deeponet_call} 
\end{table}
\end{scriptsize}

We plot below the value function $\vartheta^\eta(t,x)$ and its derivative $\partial_x \vartheta^\eta (t,x)$ obtained by Differential Regression Learning with DeepONet networks, for fixed values $t=0$, $t=0.5$, $t=0.9$, parameter $\sigma = 0.3$ and linear market impact factor $\lambda = 5e^{-3}$, and compare it with the Monte-Carlo estimation obtained. We plot these value functions for strike values 
$K$ $\in$ $\{0.5, \; 2\}$ inside the training domain 
and $K$ $\in$ $\{ 0.2,\ 1,\ 2.1 \}$ outside the training domain. 

\begin{figure}[htp]
    \centering
    \begin{subfigure}{.32\linewidth}
        \centering
        \includegraphics[height=3.75cm]{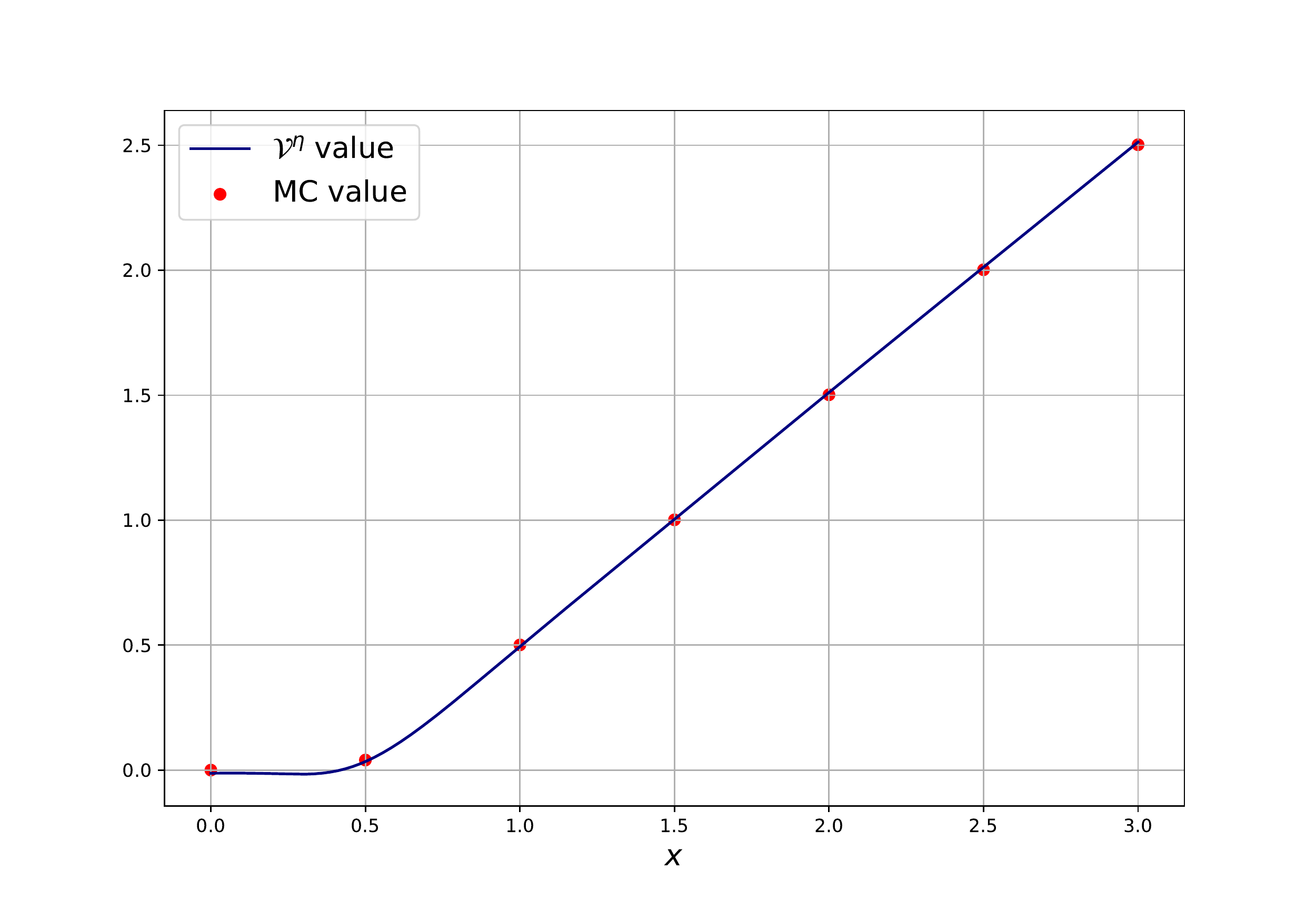} 
    \end{subfigure}
    \begin{subfigure}{.32\linewidth}
        \centering
        \includegraphics[height=3.75cm]{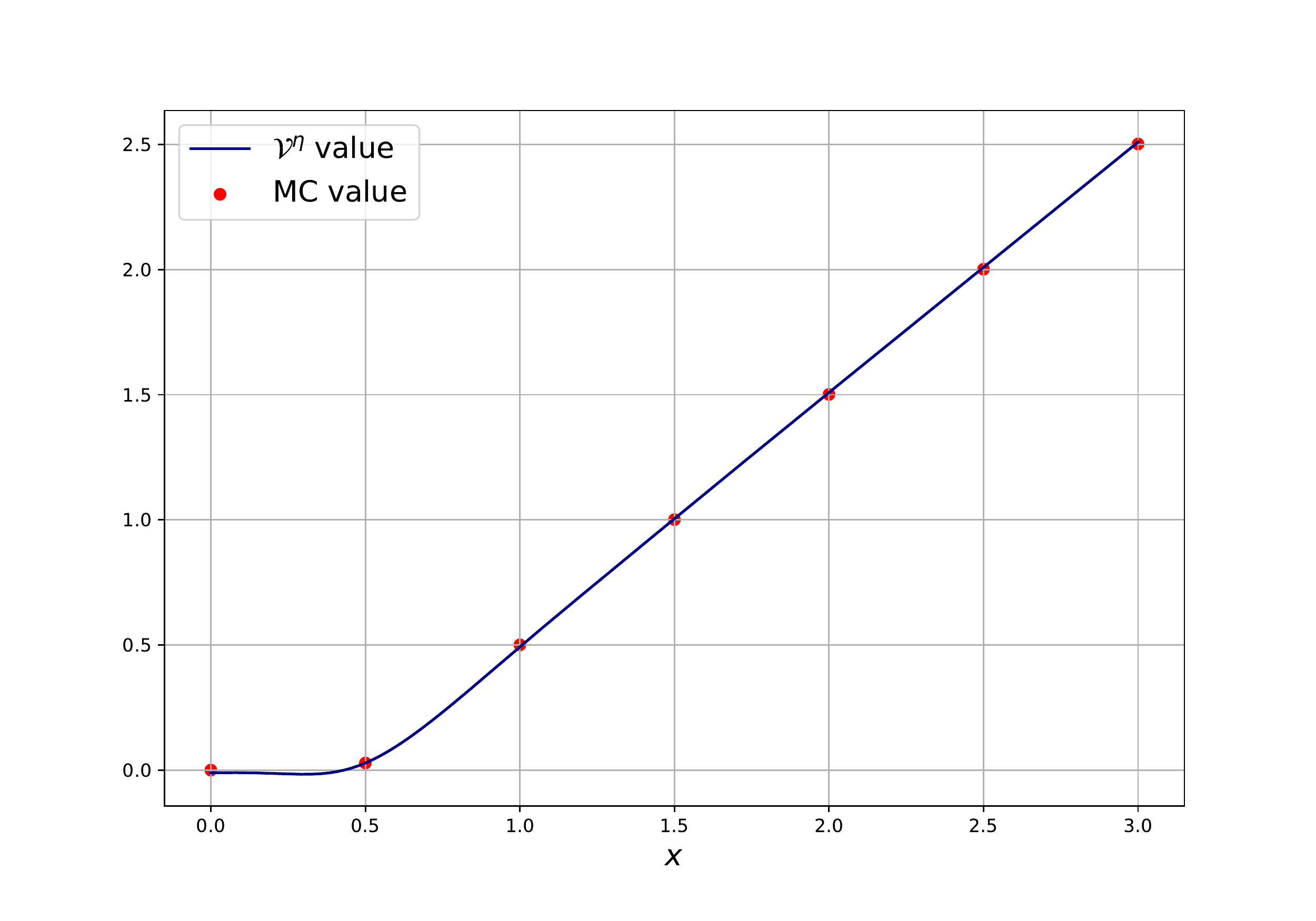}
    \end{subfigure}
    \begin{subfigure}{.32\linewidth}
        \centering
        \includegraphics[height=3.75cm]{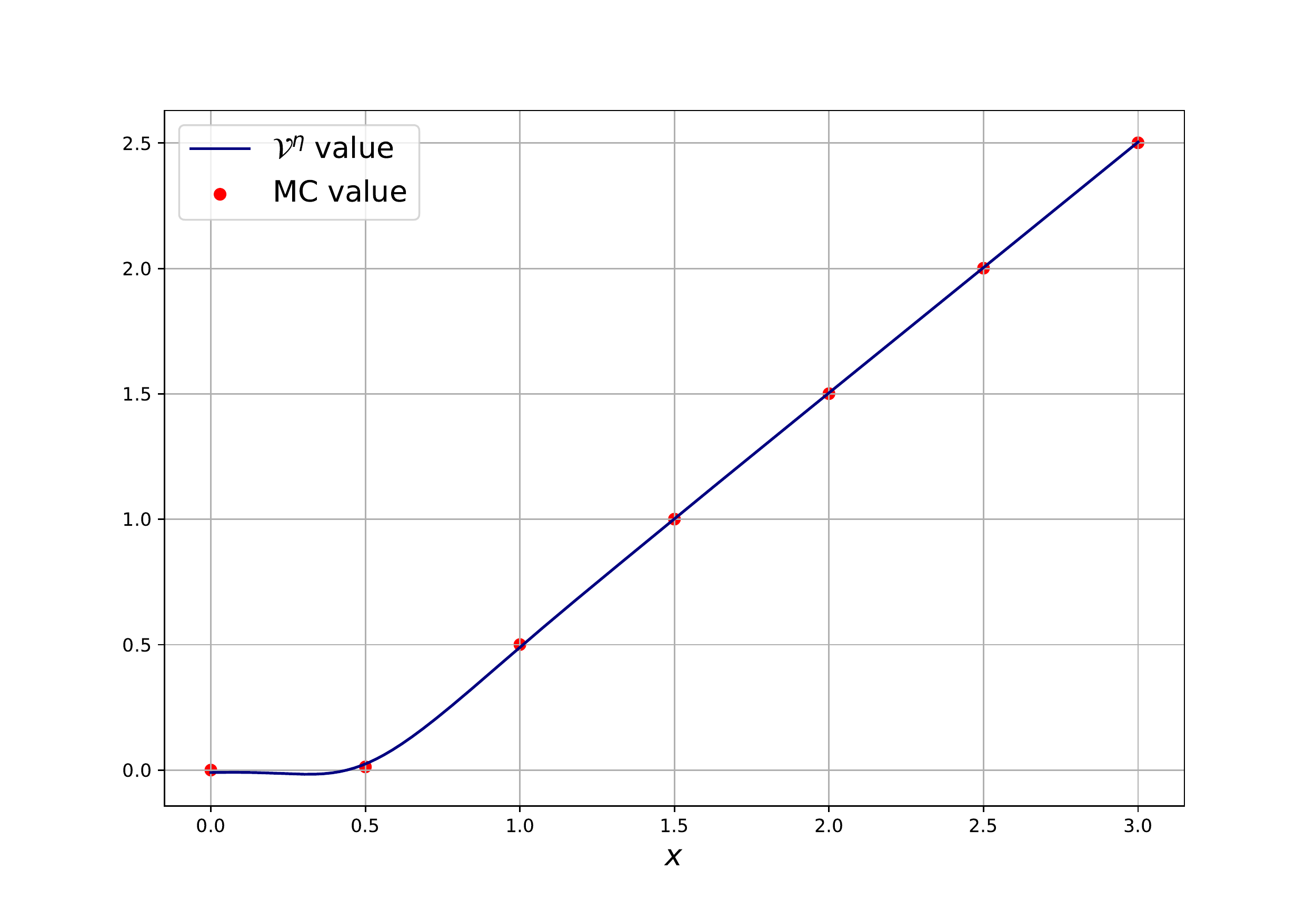}
    \end{subfigure}
    \begin{subfigure}{.32\linewidth}
        \centering
        \includegraphics[height=3.75cm]{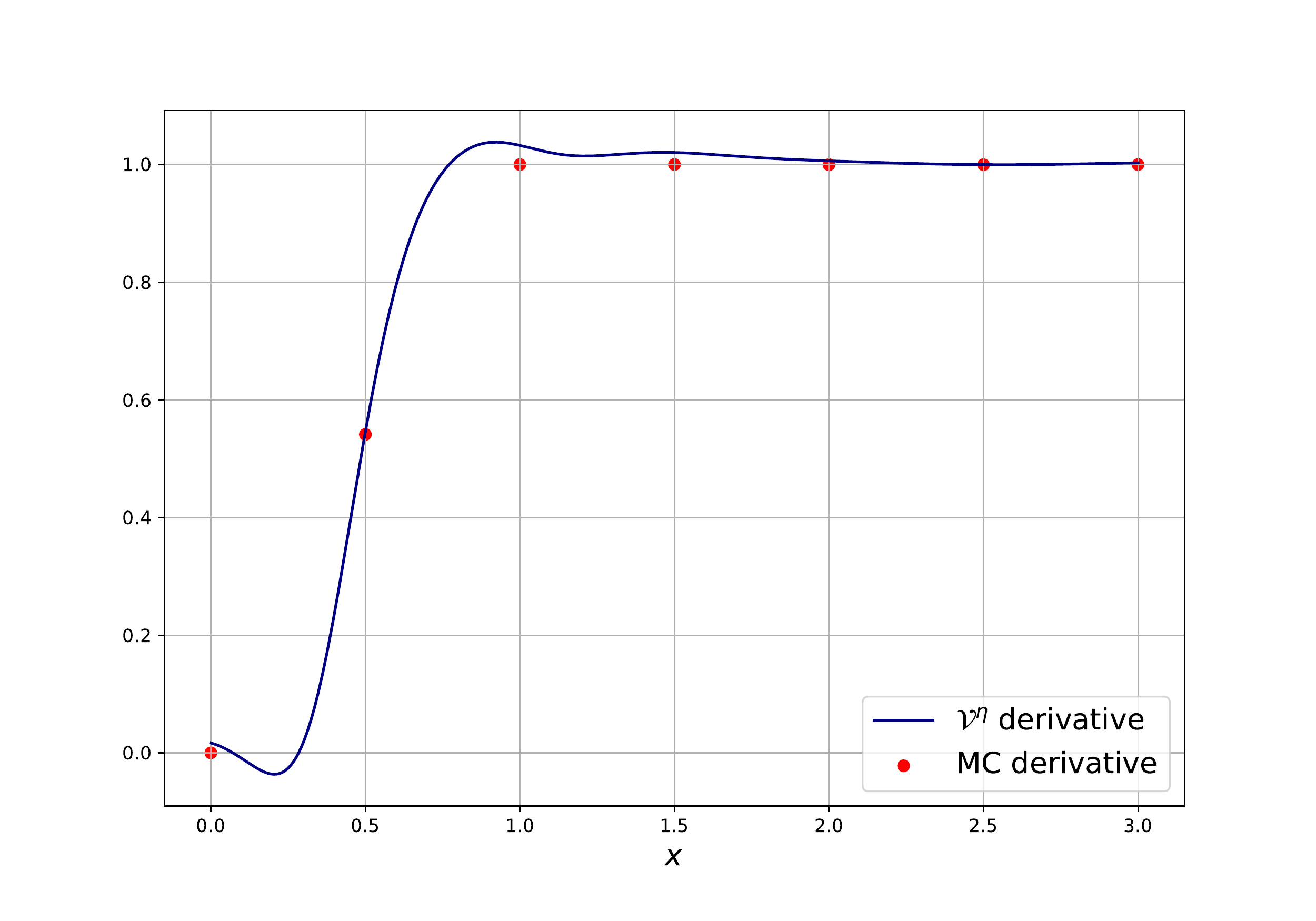} 
    \end{subfigure}
    \begin{subfigure}{.32\linewidth}
        \centering
        \includegraphics[height=3.75cm]{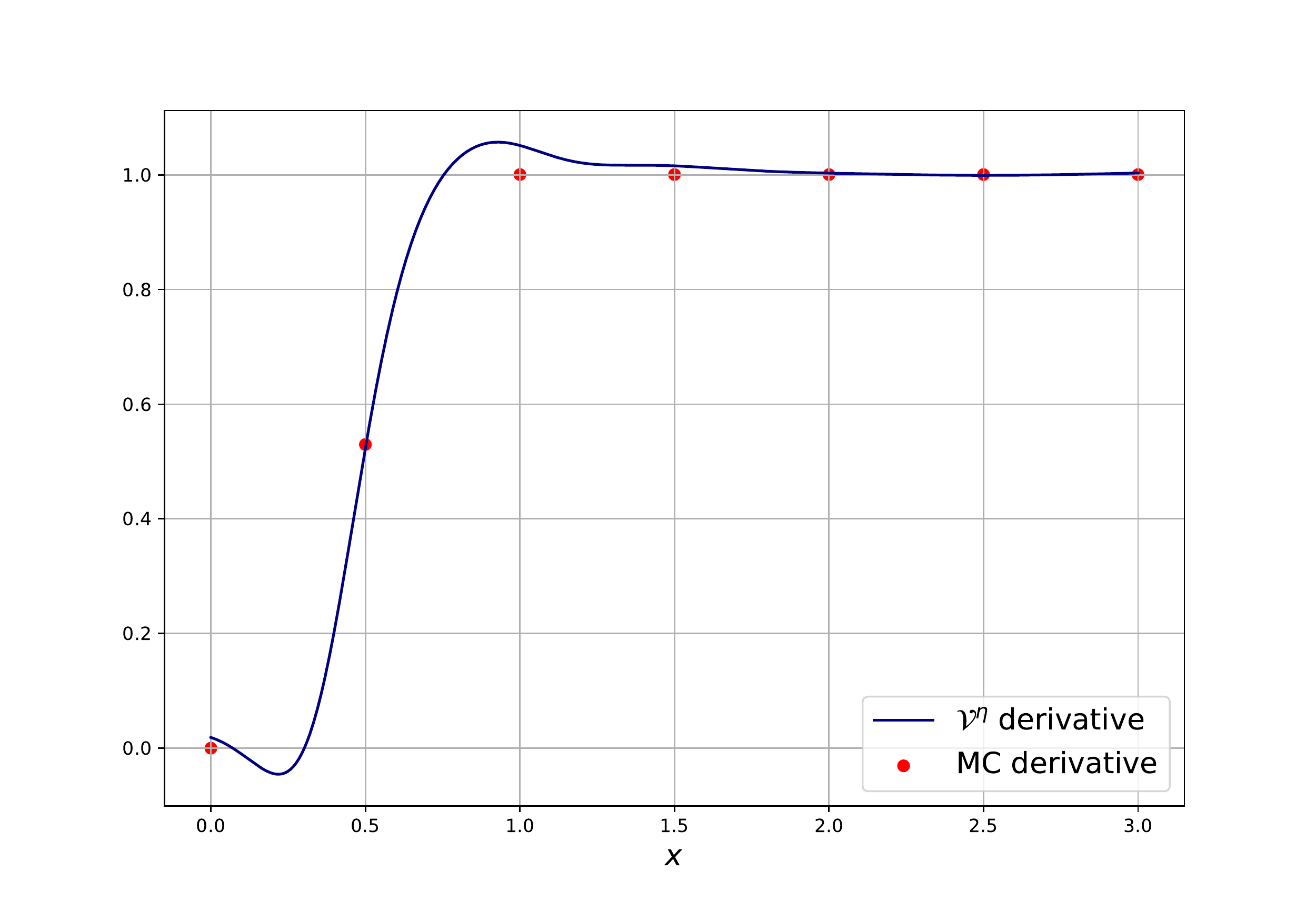}
    \end{subfigure}
    \begin{subfigure}{.32\linewidth}
        \centering
        \includegraphics[height=3.75cm]{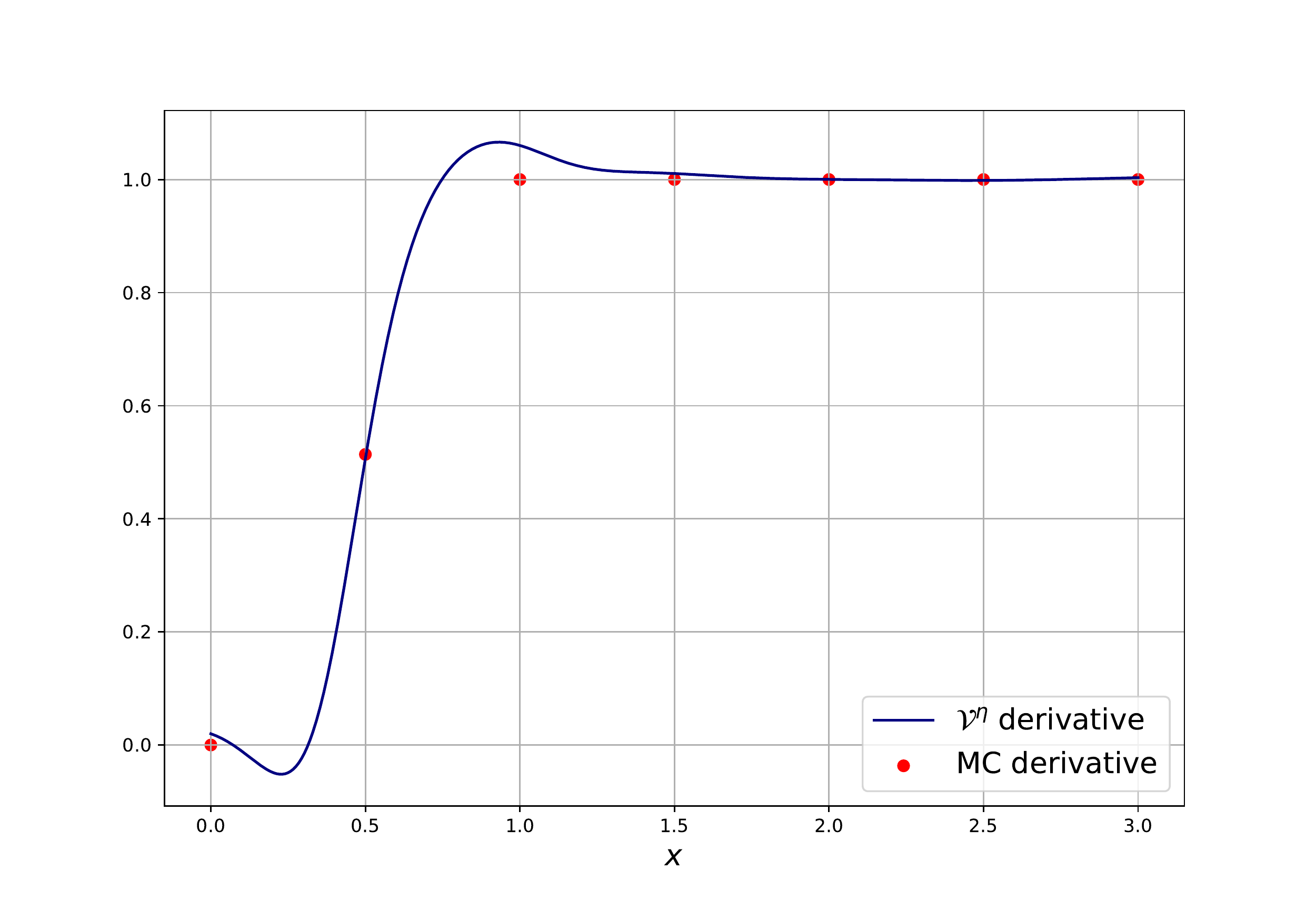}
    \end{subfigure}
    \caption{
    \label{fig:value_differential_learning_deeponet_call_K05}
    \footnotesize{Value function $\vartheta^\eta$ (first line) and its derivative (second line) obtained by Differential Regression Learning (Algorithm \ref{algo:scheme_value_differential_learning_deeponet}) for a terminal call option payoff with strike $K$ $=$ $0.5$, with parameter $\sigma = 0.3$ and linear market impact factor $\lambda = 5e^{-3}$, plotted as functions of $x$, for fixed values of $t$.}
    }
\end{figure}

\begin{figure}[htp]
    \centering
    \begin{subfigure}{.32\linewidth}
        \centering
        \includegraphics[height=3.75cm]{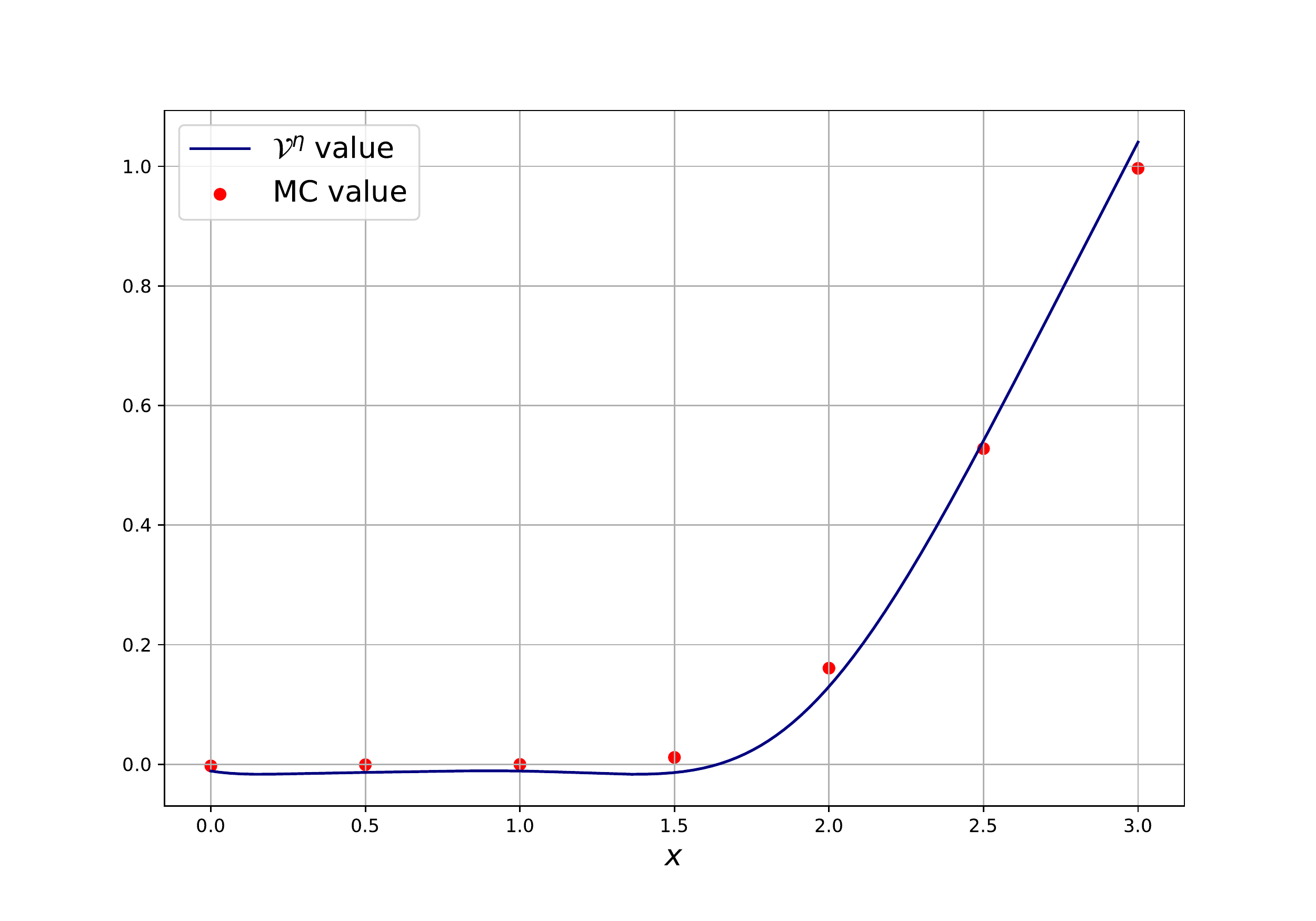} 
    \end{subfigure}
    \begin{subfigure}{.32\linewidth}
        \centering
        \includegraphics[height=3.75cm]{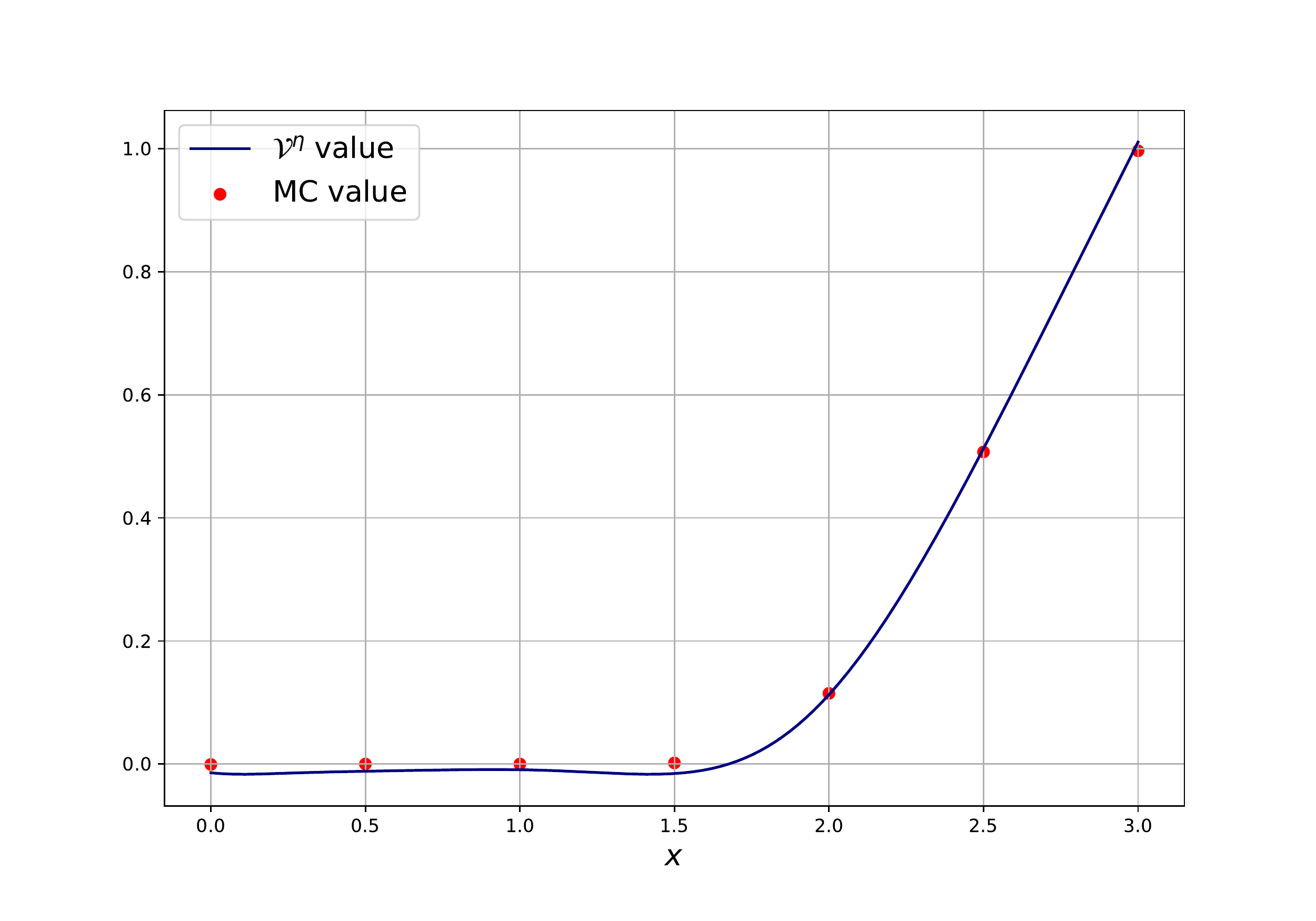}
    \end{subfigure}
    \begin{subfigure}{.32\linewidth}
        \centering
        \includegraphics[height=3.75cm]{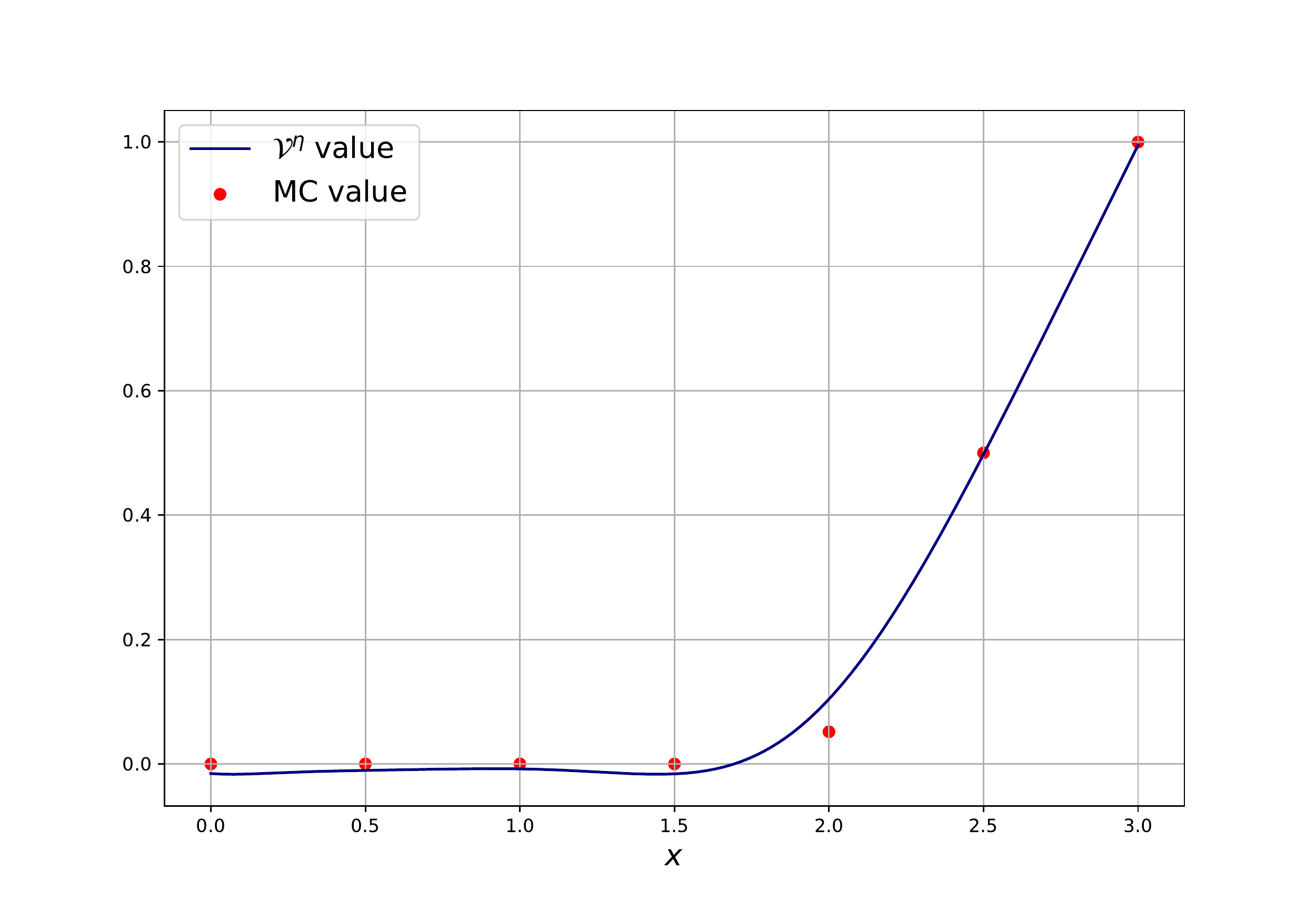}
    \end{subfigure}
    \begin{subfigure}{.32\linewidth}
        \centering
        \includegraphics[height=3.75cm]{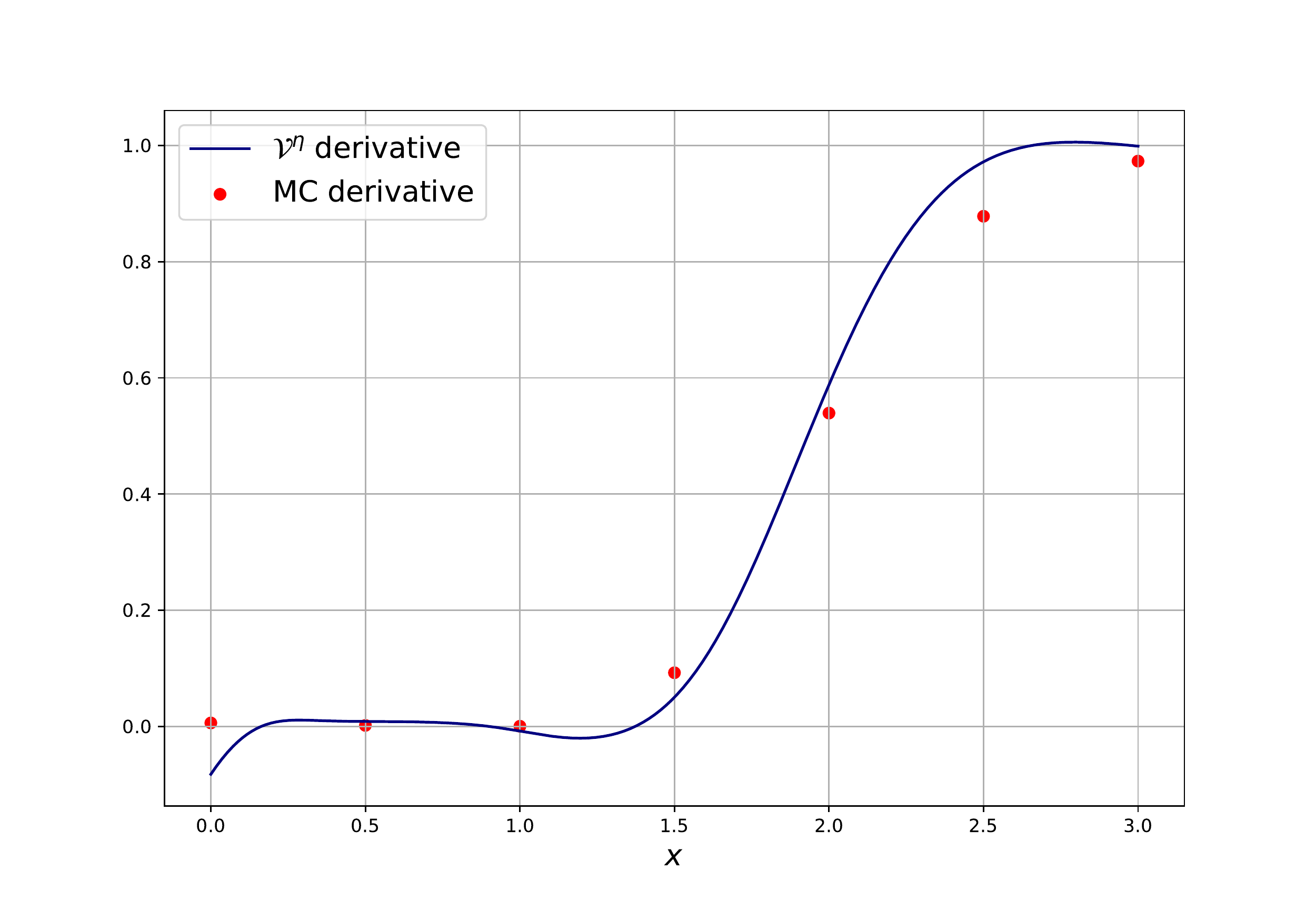} 
    \end{subfigure}
    \begin{subfigure}{.32\linewidth}
        \centering
        \includegraphics[height=3.75cm]{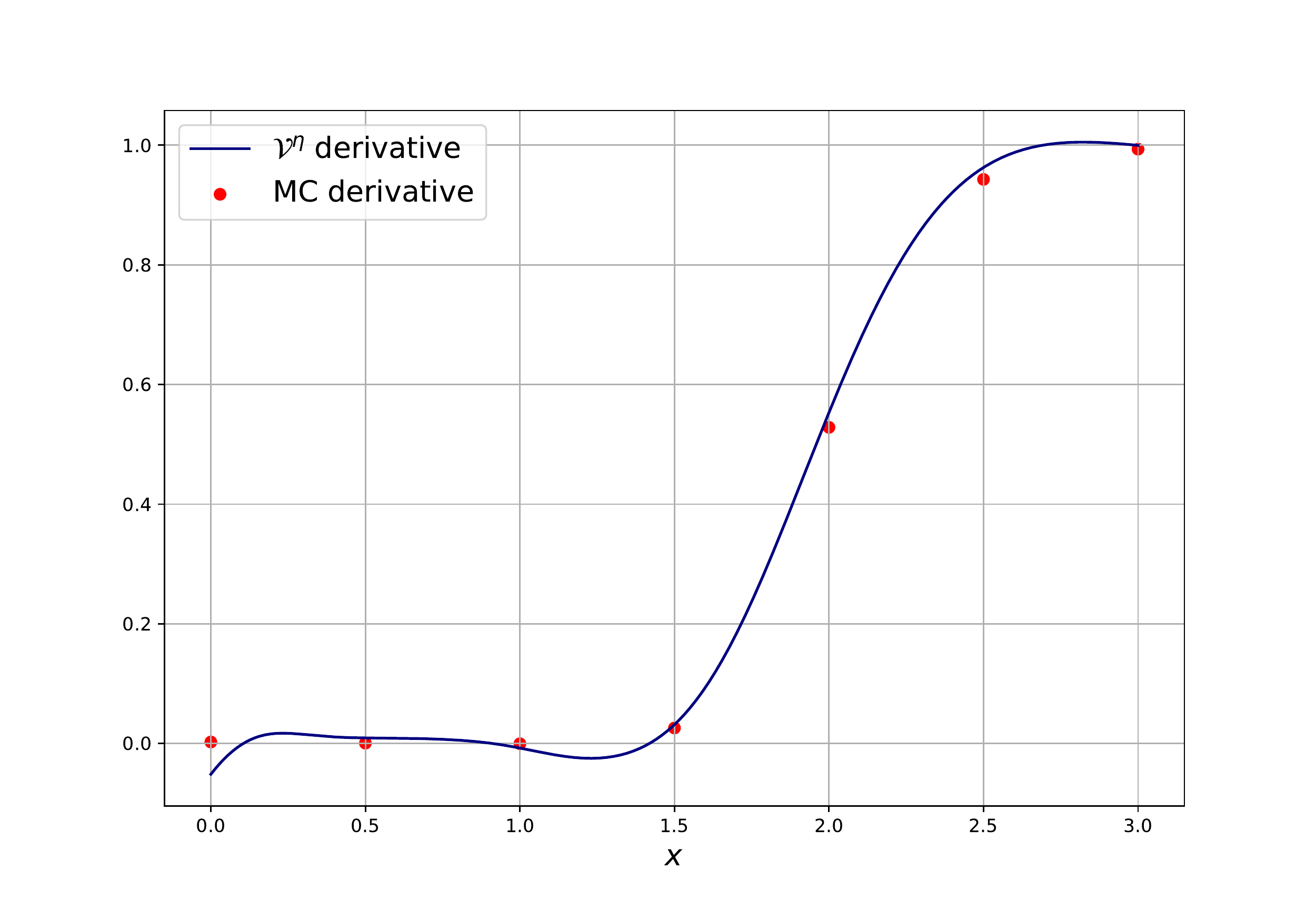}
    \end{subfigure}
    \begin{subfigure}{.32\linewidth}
        \centering
        \includegraphics[height=3.75cm]{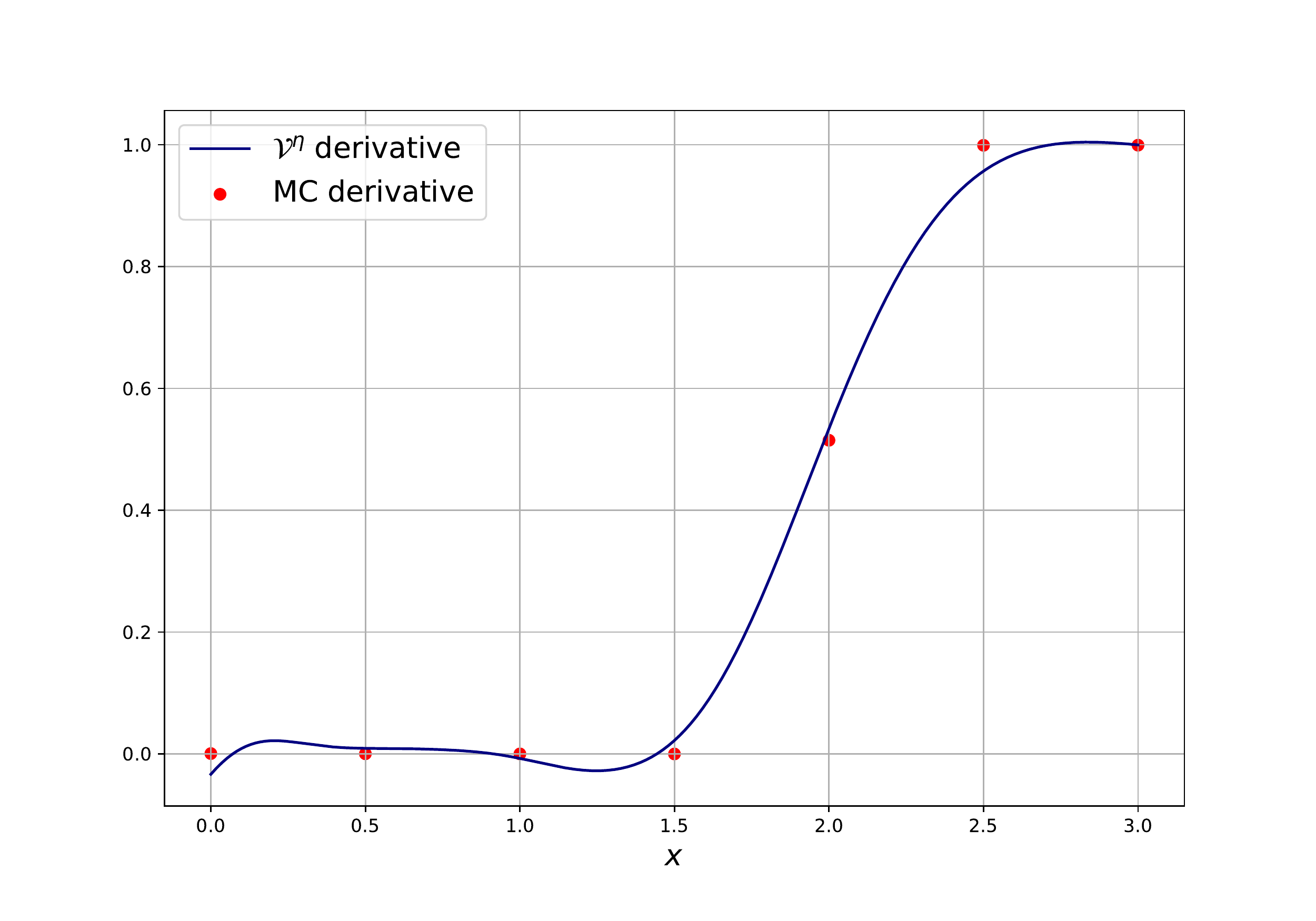}
    \end{subfigure}
    \caption{
    \label{fig:value_differential_learning_deeponet_call_K2}
    \footnotesize{Value function $\vartheta^\eta$ (first line) and its derivative (second line) obtained by Differential Regression Learning (Algorithm \ref{algo:scheme_value_differential_learning_deeponet}) for a terminal call option payoff with strike $K$ $=$ $2$, with parameter $\sigma = 0.3$ and linear market impact factor $\lambda = 5e^{-3}$, plotted as functions of $x$, for fixed values of $t$.}
    }
\end{figure}

\begin{figure}[htp]
    \centering
    \begin{subfigure}{.32\linewidth}
        \centering
        \includegraphics[height=3.75cm]{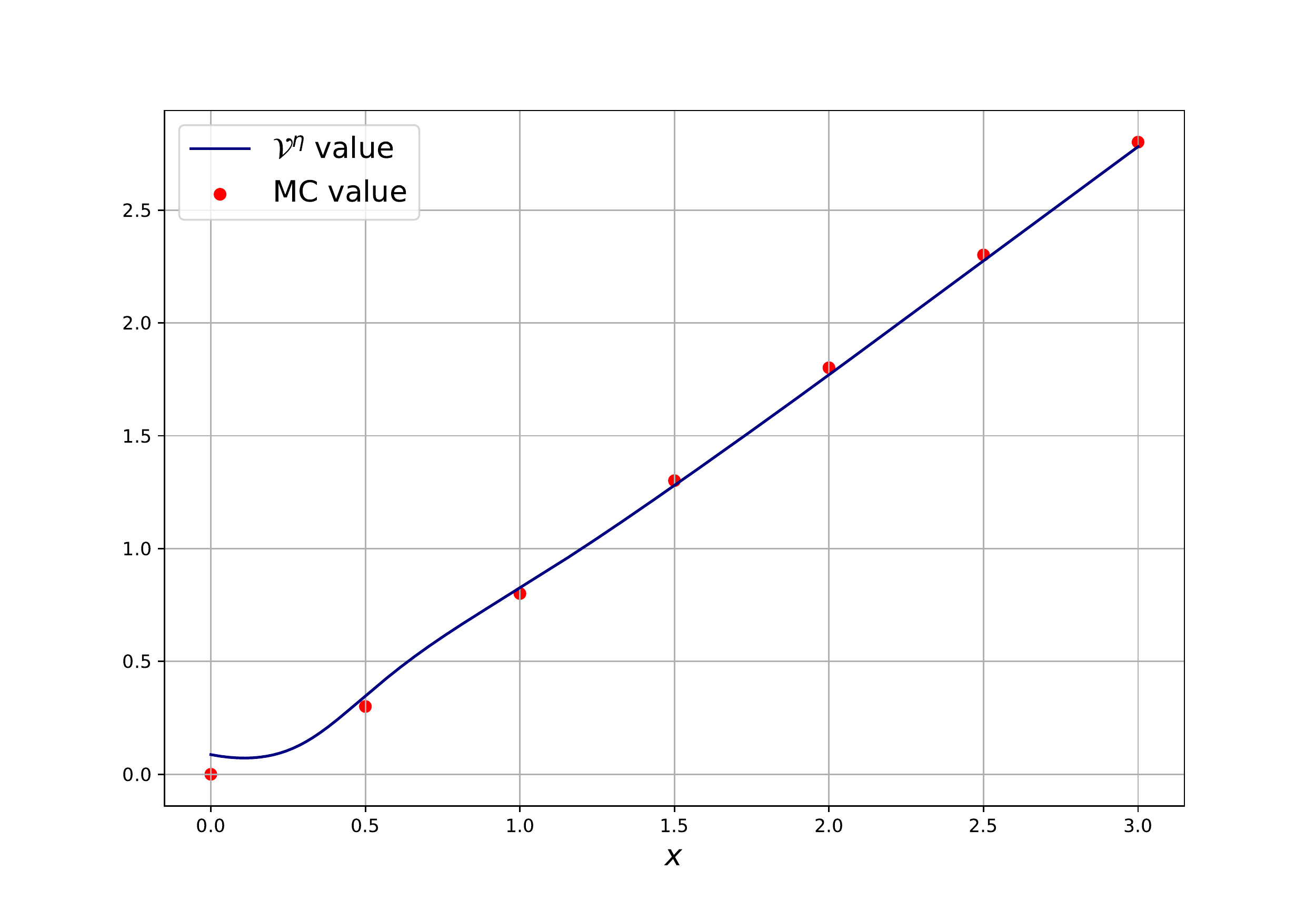} 
    \end{subfigure}
    \begin{subfigure}{.32\linewidth}
        \centering
        \includegraphics[height=3.75cm]{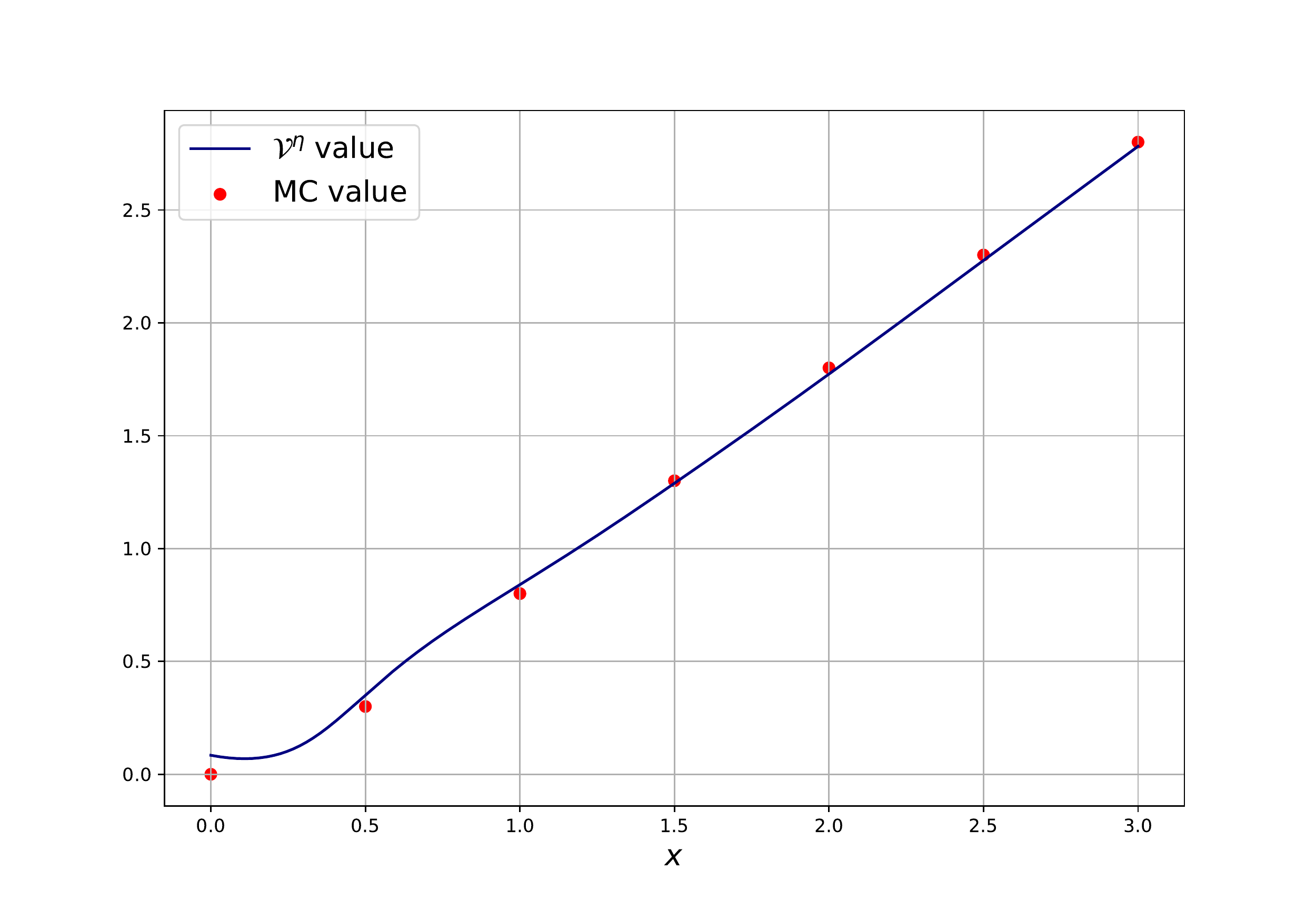}
    \end{subfigure}
    \begin{subfigure}{.32\linewidth}
        \centering
        \includegraphics[height=3.75cm]{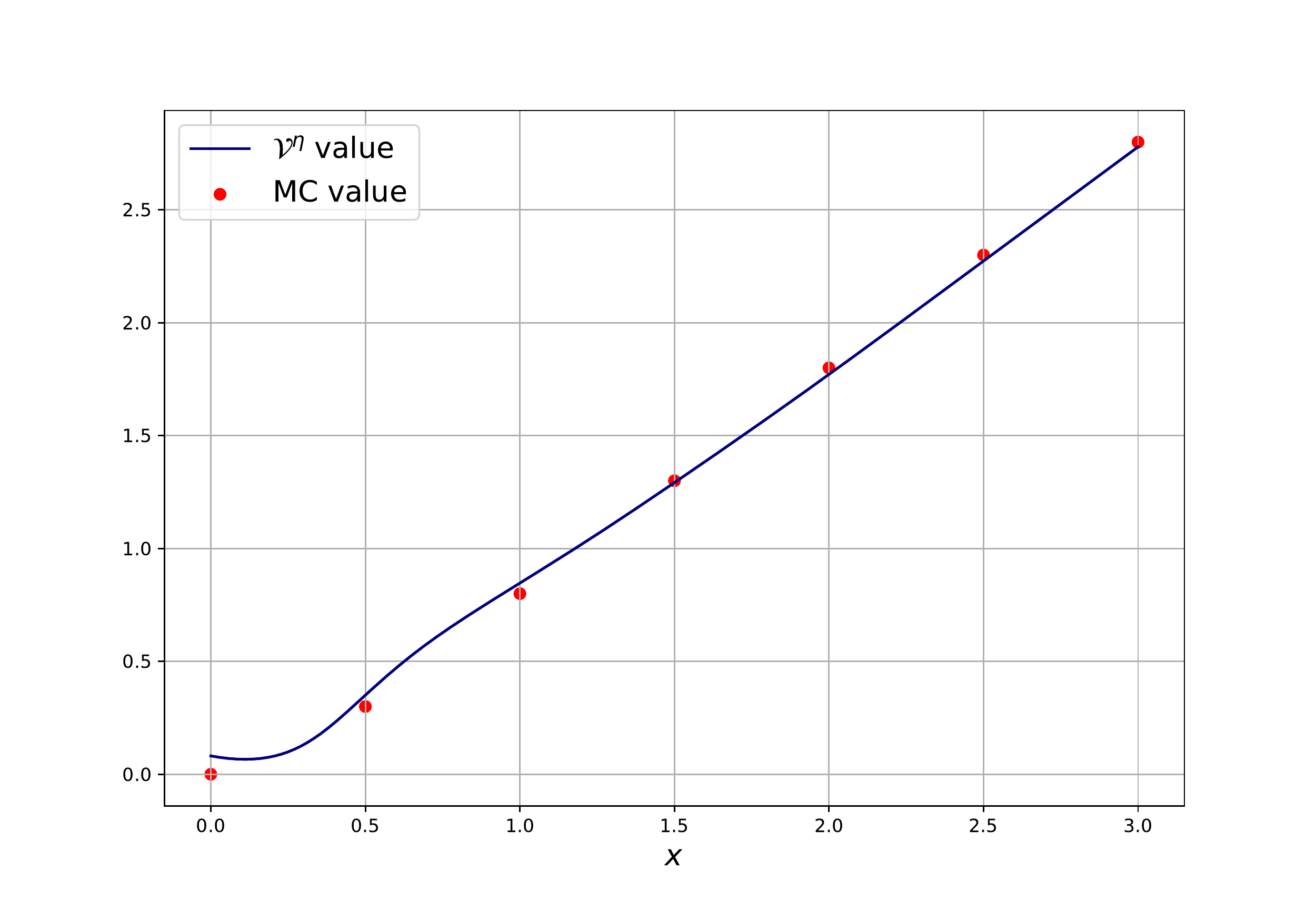}
    \end{subfigure}
    \begin{subfigure}{.32\linewidth}
        \centering
        \includegraphics[height=3.75cm]{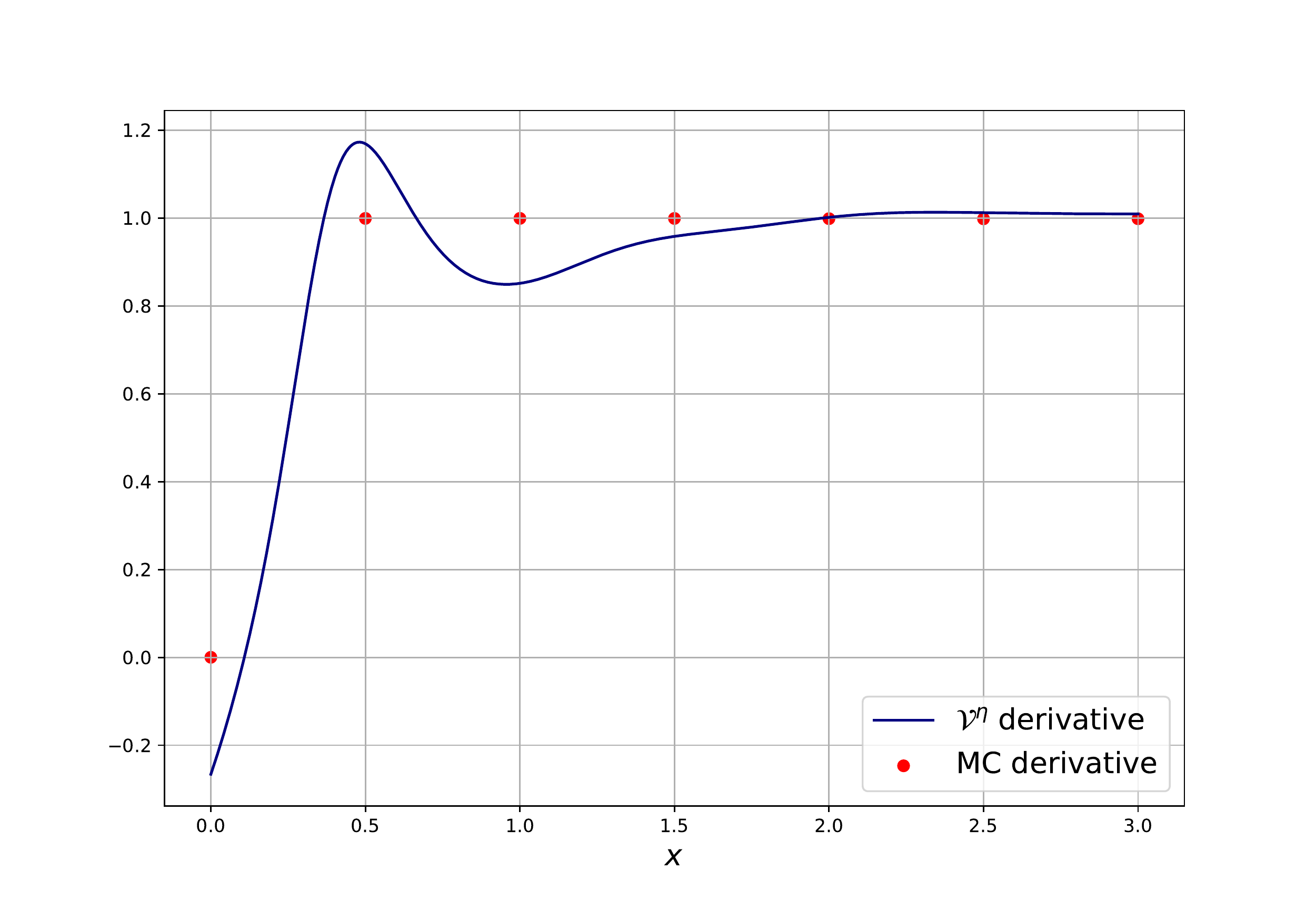} 
        \caption[short]{$t=0$}
    \end{subfigure}
    \begin{subfigure}{.32\linewidth}
        \centering
        \includegraphics[height=3.75cm]{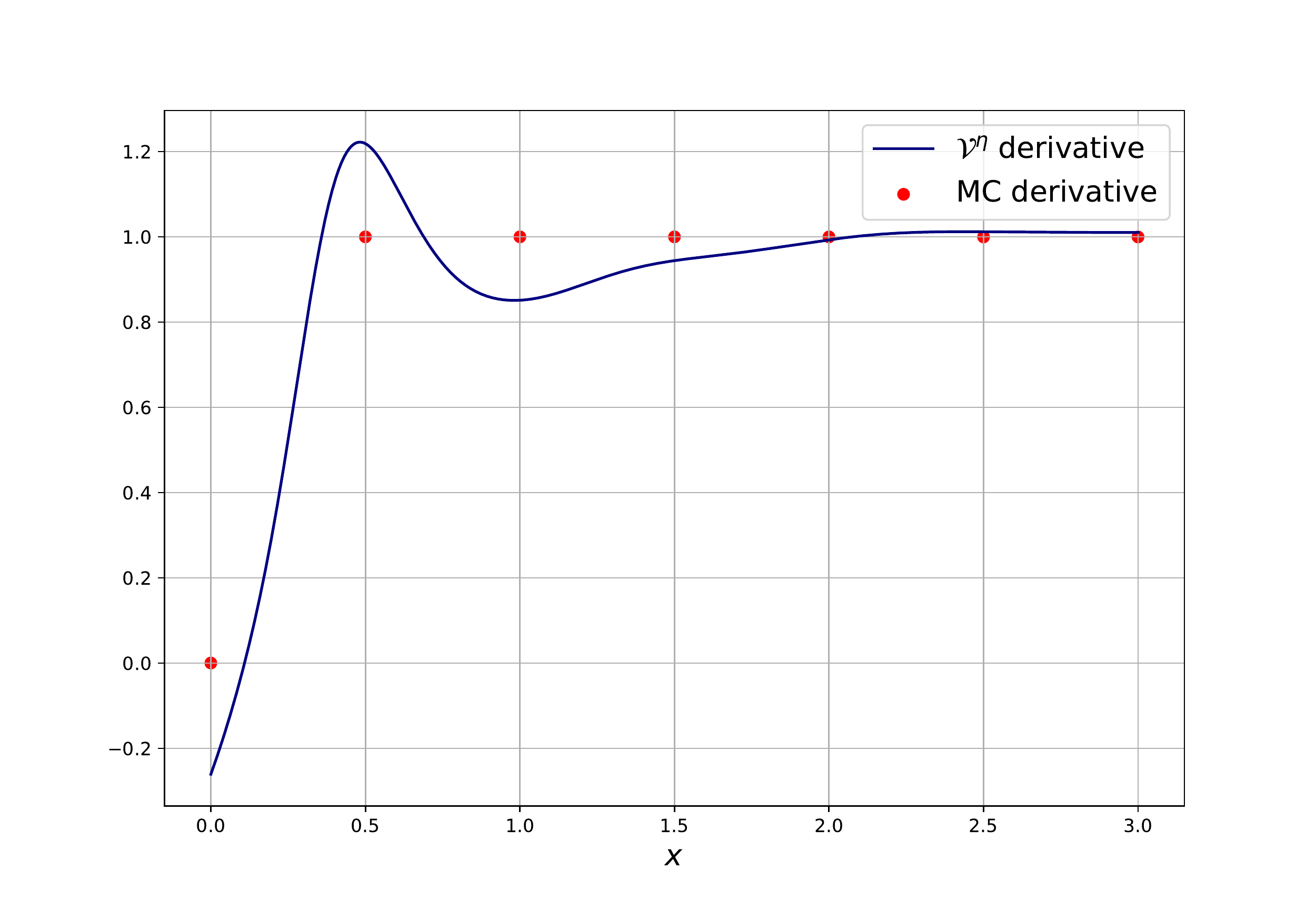}
        \caption[short]{$t=0.5$}
    \end{subfigure}
    \begin{subfigure}{.32\linewidth}
        \centering
        \includegraphics[height=3.75cm]{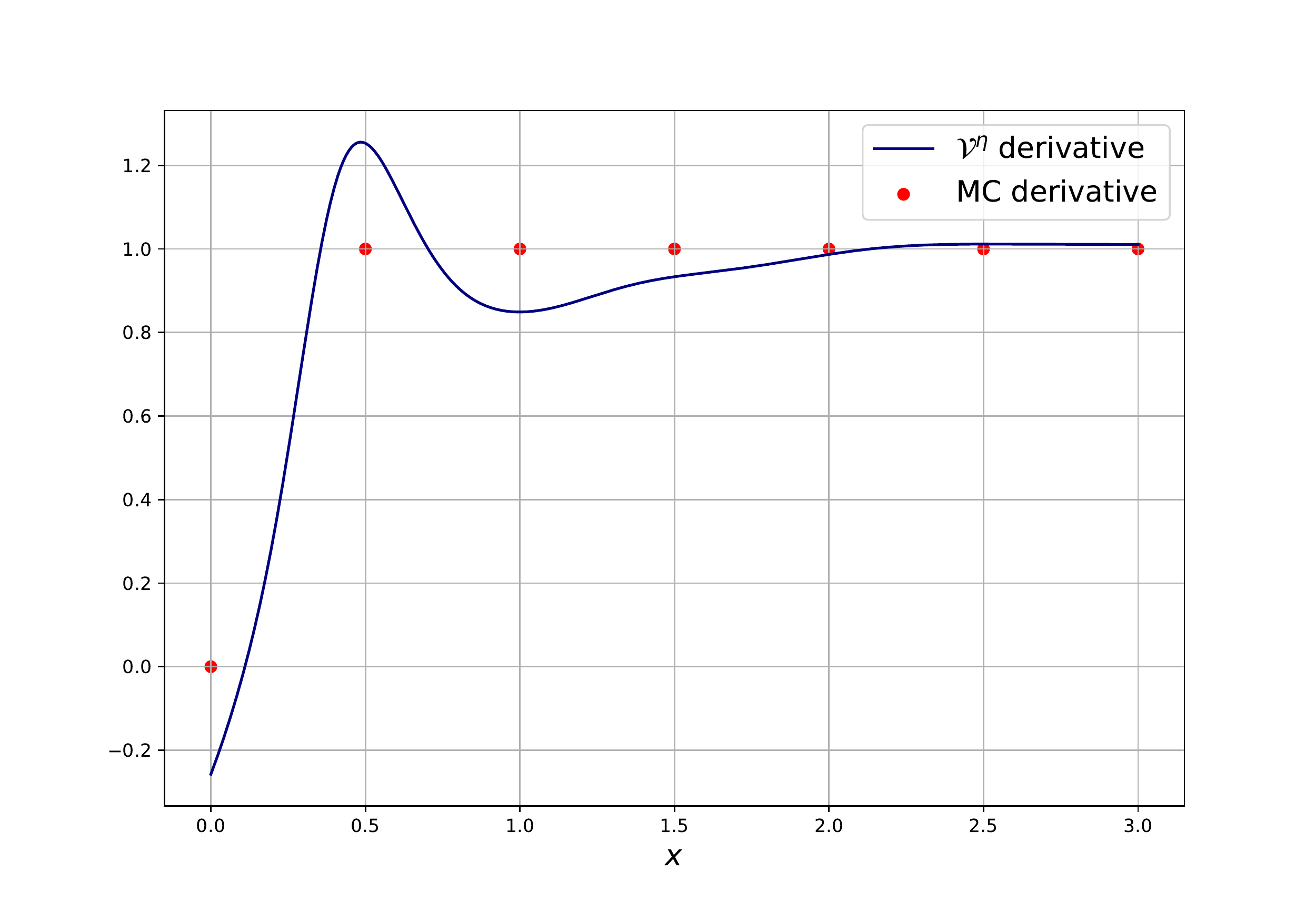}
        \caption[short]{$t=0.9$}
    \end{subfigure}
    \caption{
    \label{fig:value_differential_learning_deeponet_call_K02}
    \footnotesize{Value function $\vartheta^\eta$ (first line) and its derivative (second line) obtained by Differential Regression Learning (Algorithm \ref{algo:scheme_value_differential_learning_deeponet}) for a terminal call option payoff with strike $K$ $=$ $0.2$, with parameter $\sigma = 0.3$ and linear market impact factor $\lambda = 5e^{-3}$, plotted as functions of $x$, for fixed values of $t$.}
    }
\end{figure}

\begin{figure}[htp]
    \centering
    \begin{subfigure}{.32\linewidth}
        \centering
        \includegraphics[height=3.75cm]{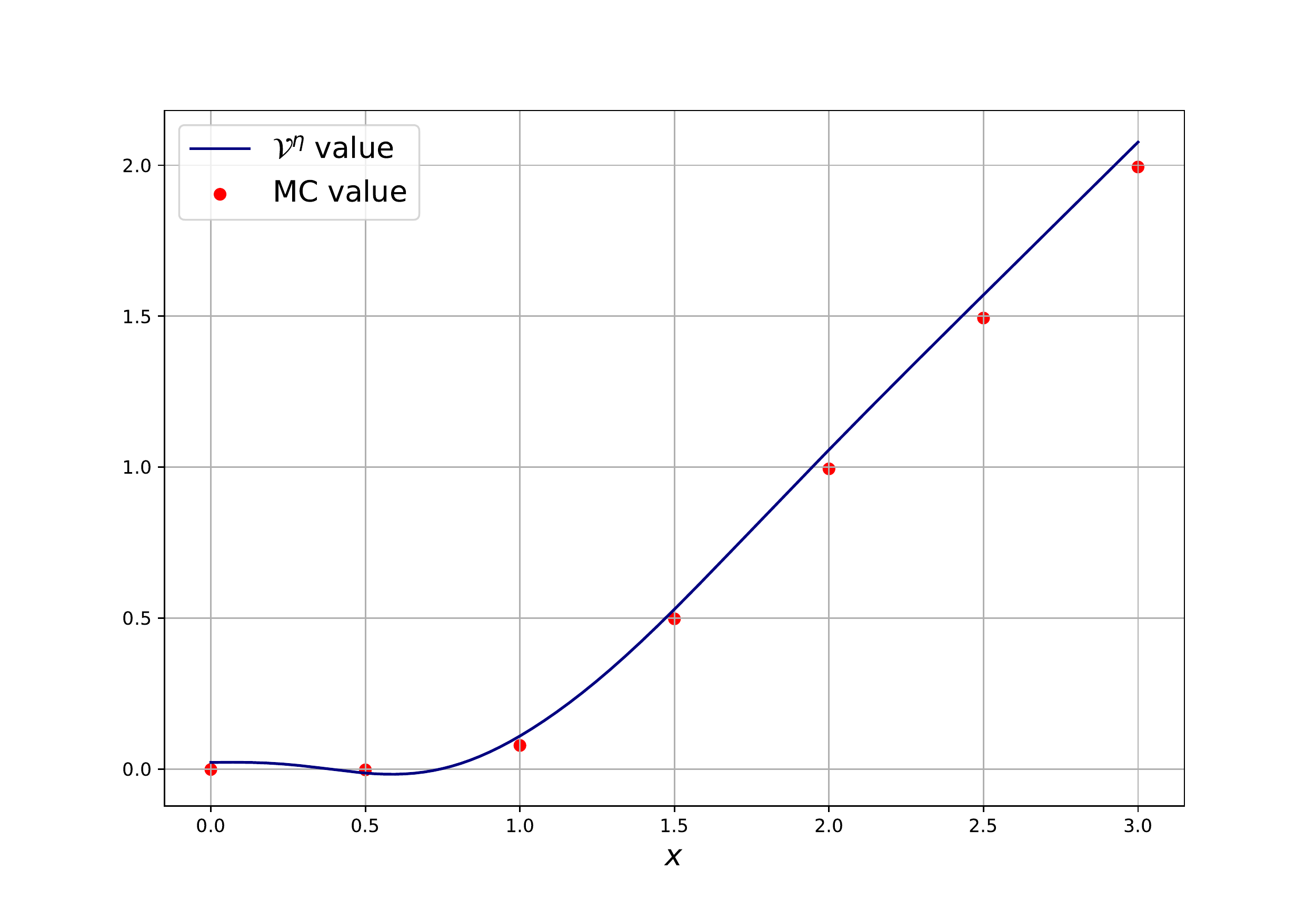} 
    \end{subfigure}
    \begin{subfigure}{.32\linewidth}
        \centering
        \includegraphics[height=3.75cm]{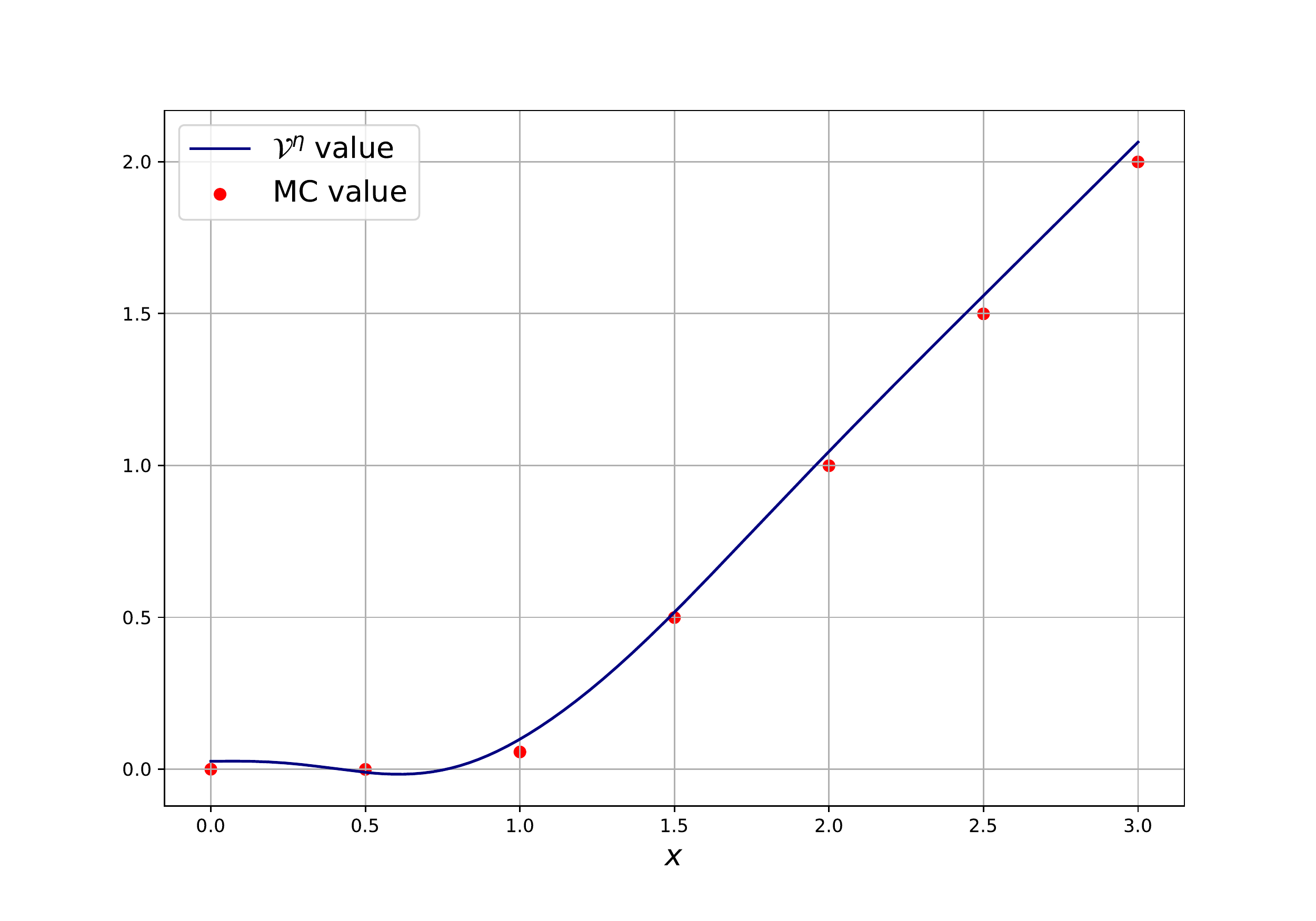}
    \end{subfigure}
    \begin{subfigure}{.32\linewidth}
        \centering
        \includegraphics[height=3.75cm]{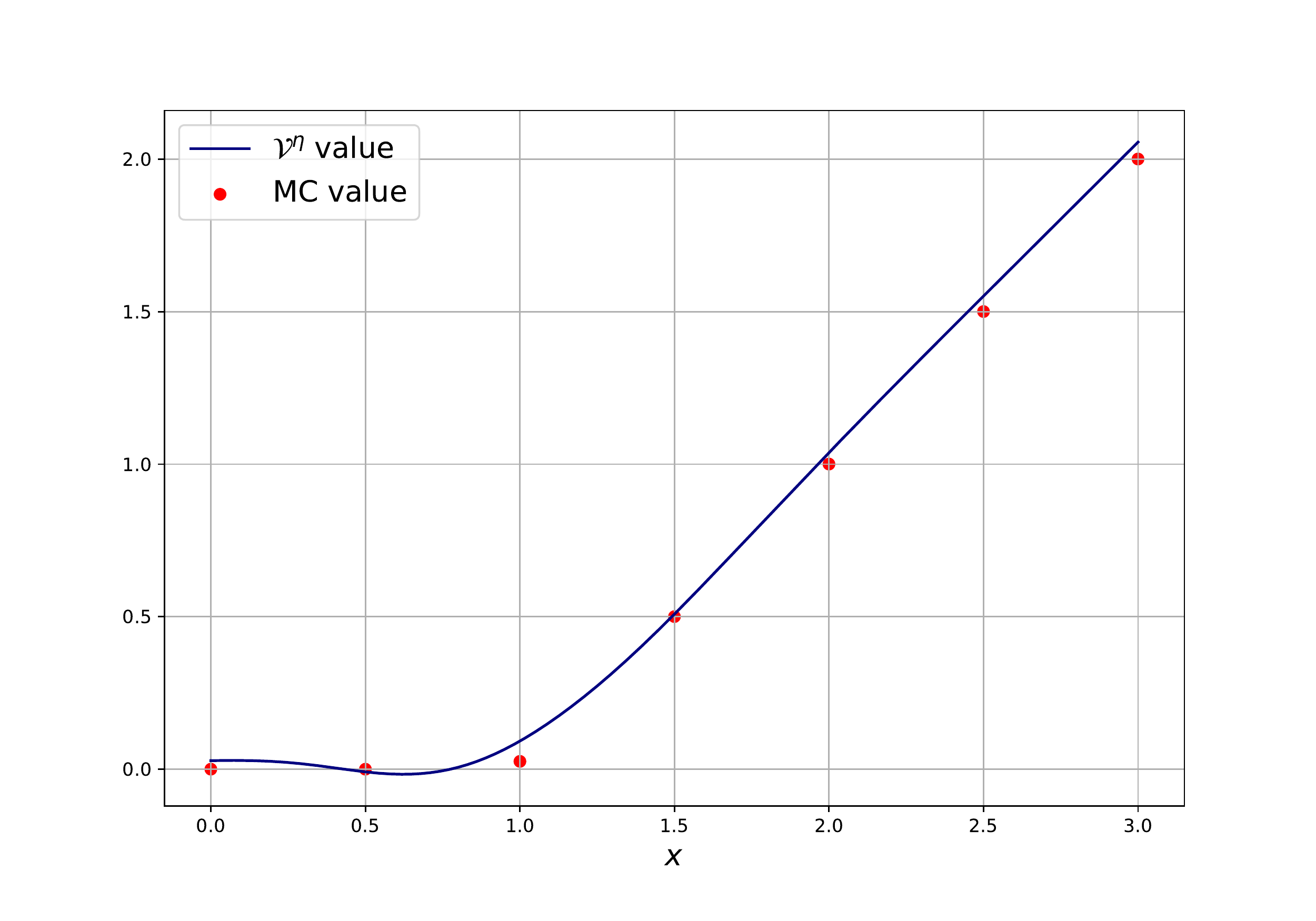}
    \end{subfigure}
    \begin{subfigure}{.32\linewidth}
        \centering
        \includegraphics[height=3.75cm]{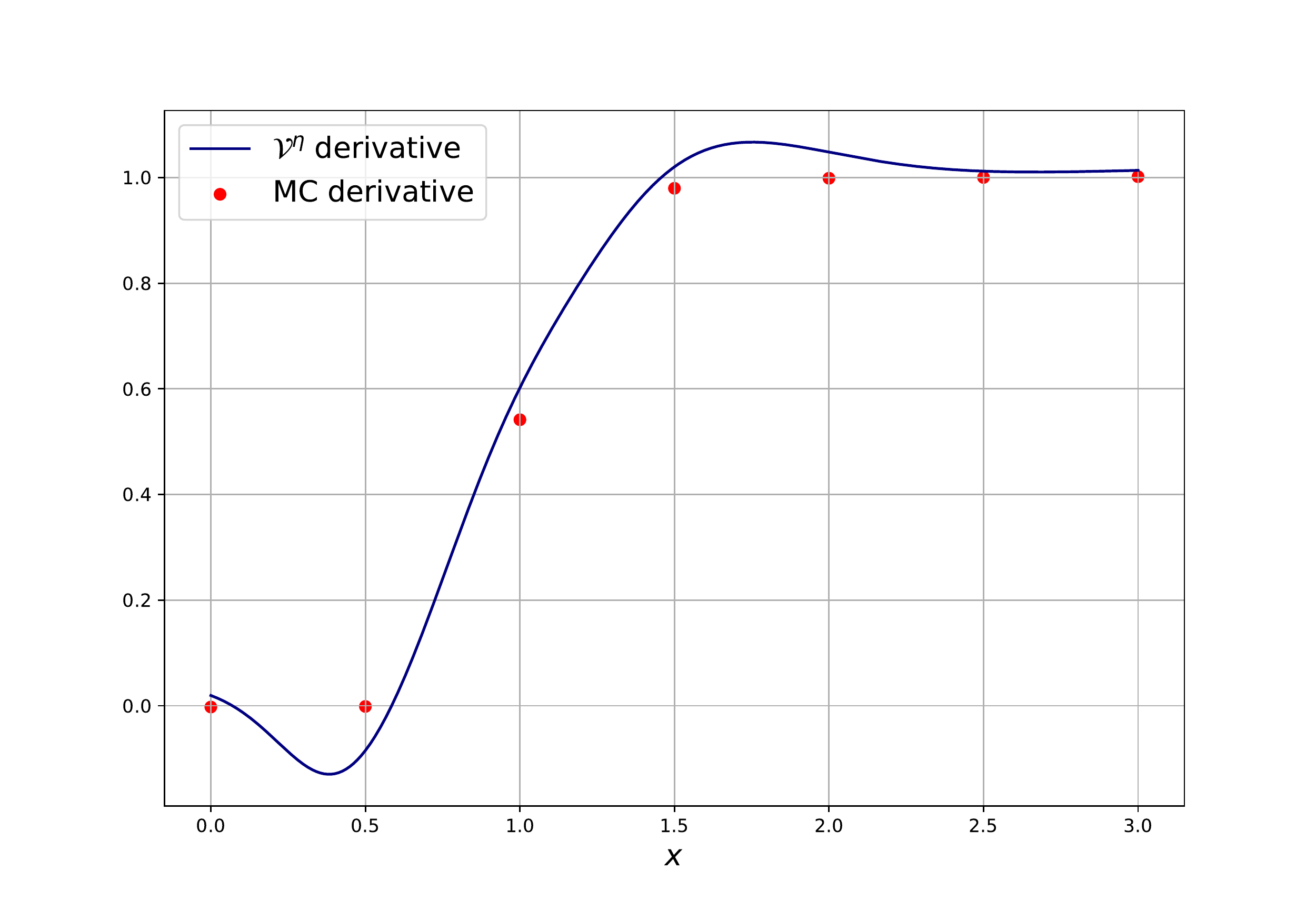} 
        \caption[short]{$t=0$}
    \end{subfigure}
    \begin{subfigure}{.32\linewidth}
        \centering
        \includegraphics[height=3.75cm]{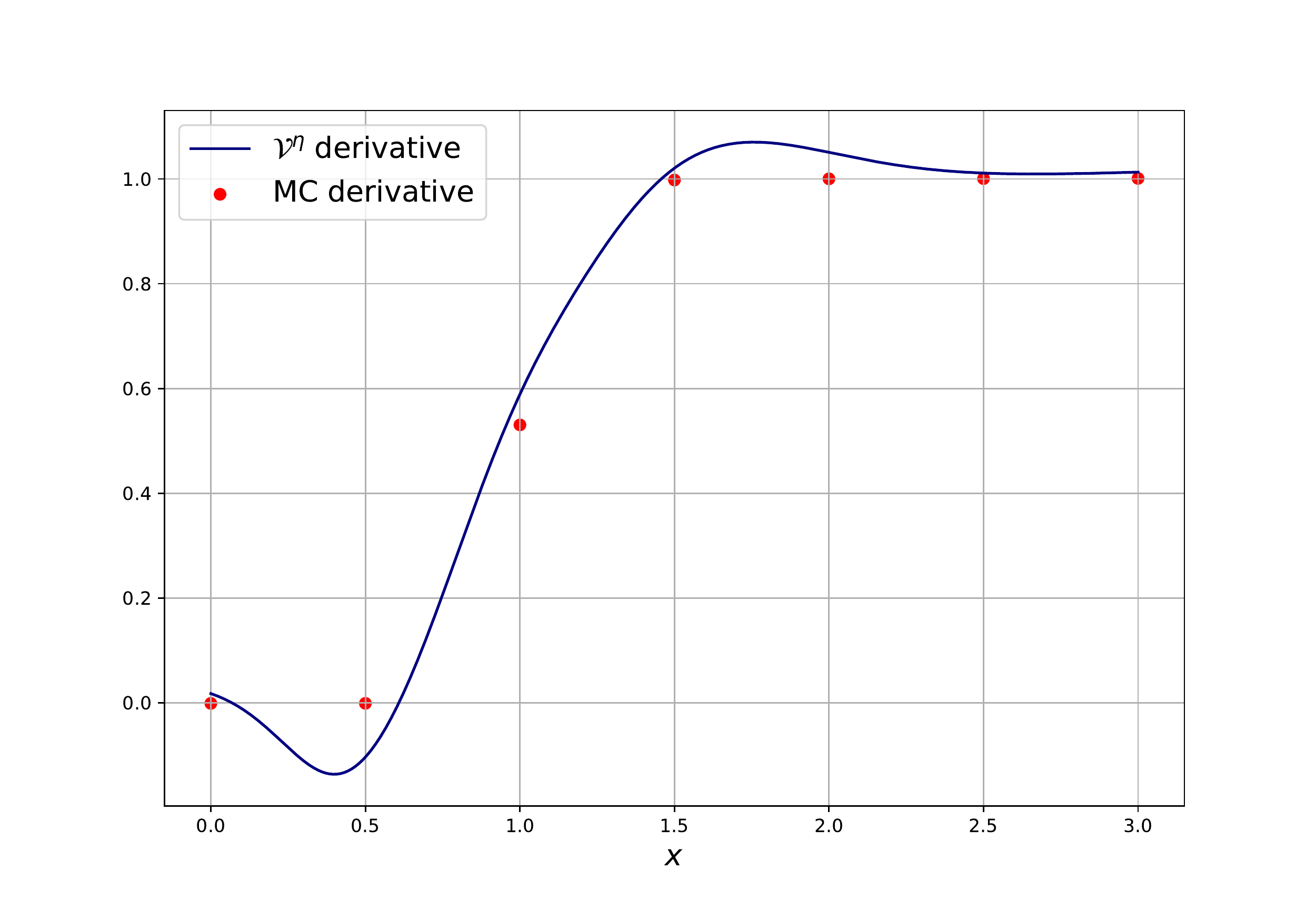}
        \caption[short]{$t=0.5$}
    \end{subfigure}
    \begin{subfigure}{.32\linewidth}
        \centering
        \includegraphics[height=3.75cm]{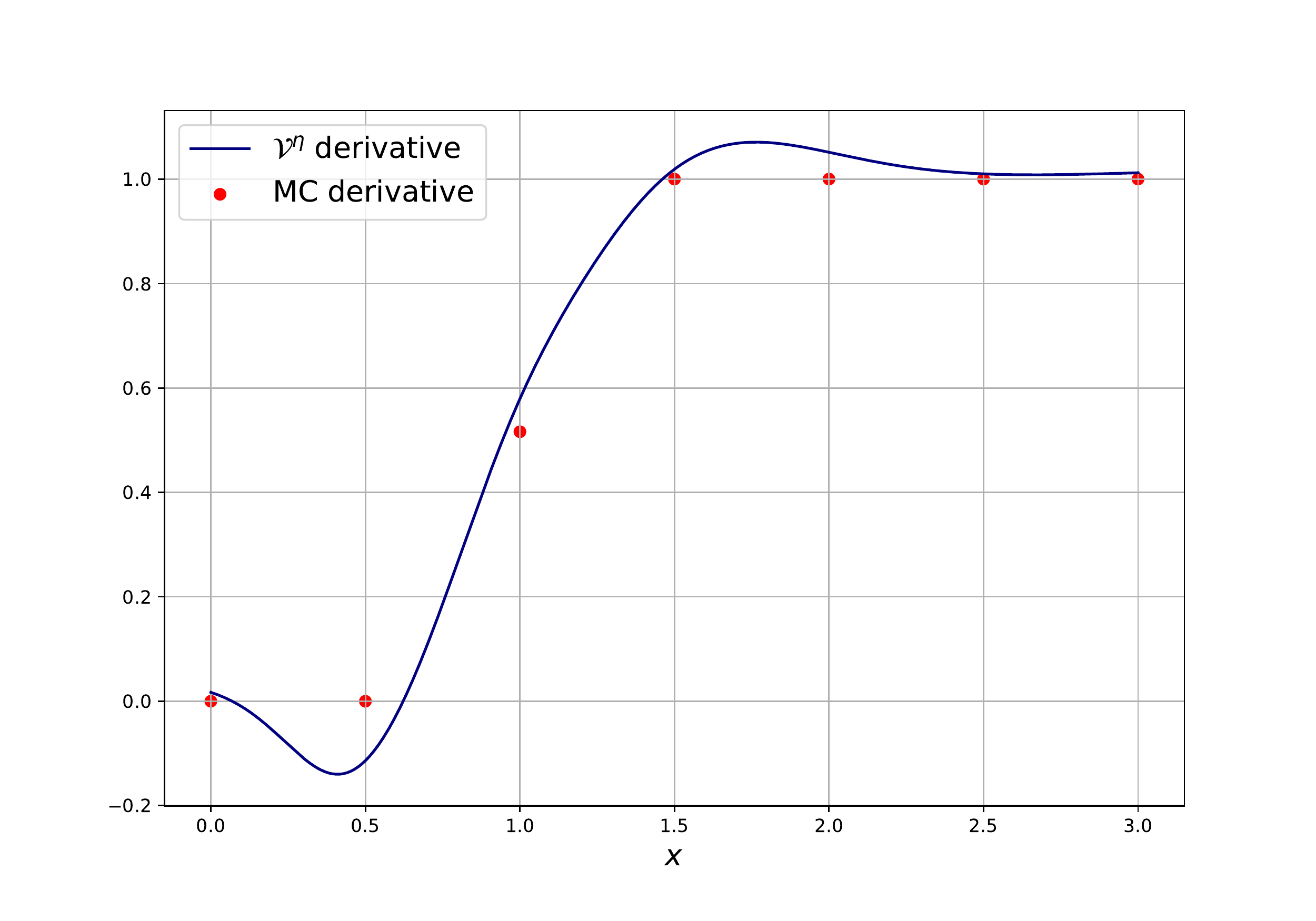}
        \caption[short]{$t=0.9$}
    \end{subfigure}
    \caption{
    \label{fig:value_differential_learning_deeponet_call_K1}
    \footnotesize{Value function $\vartheta^\eta$ (first line) and its derivative (second line) obtained by Differential Regression Learning (Algorithm \ref{algo:scheme_value_differential_learning_deeponet}) for a terminal call option payoff with strike $K$ $=$ $1$, with parameter $\sigma = 0.3$ and linear market impact factor $\lambda = 5e^{-3}$, plotted as functions of $x$, for fixed values of $t$.}
    }
\end{figure}

\begin{figure}[htp]
    \centering
    \begin{subfigure}{.32\linewidth}
        \centering
        \includegraphics[height=3.75cm]{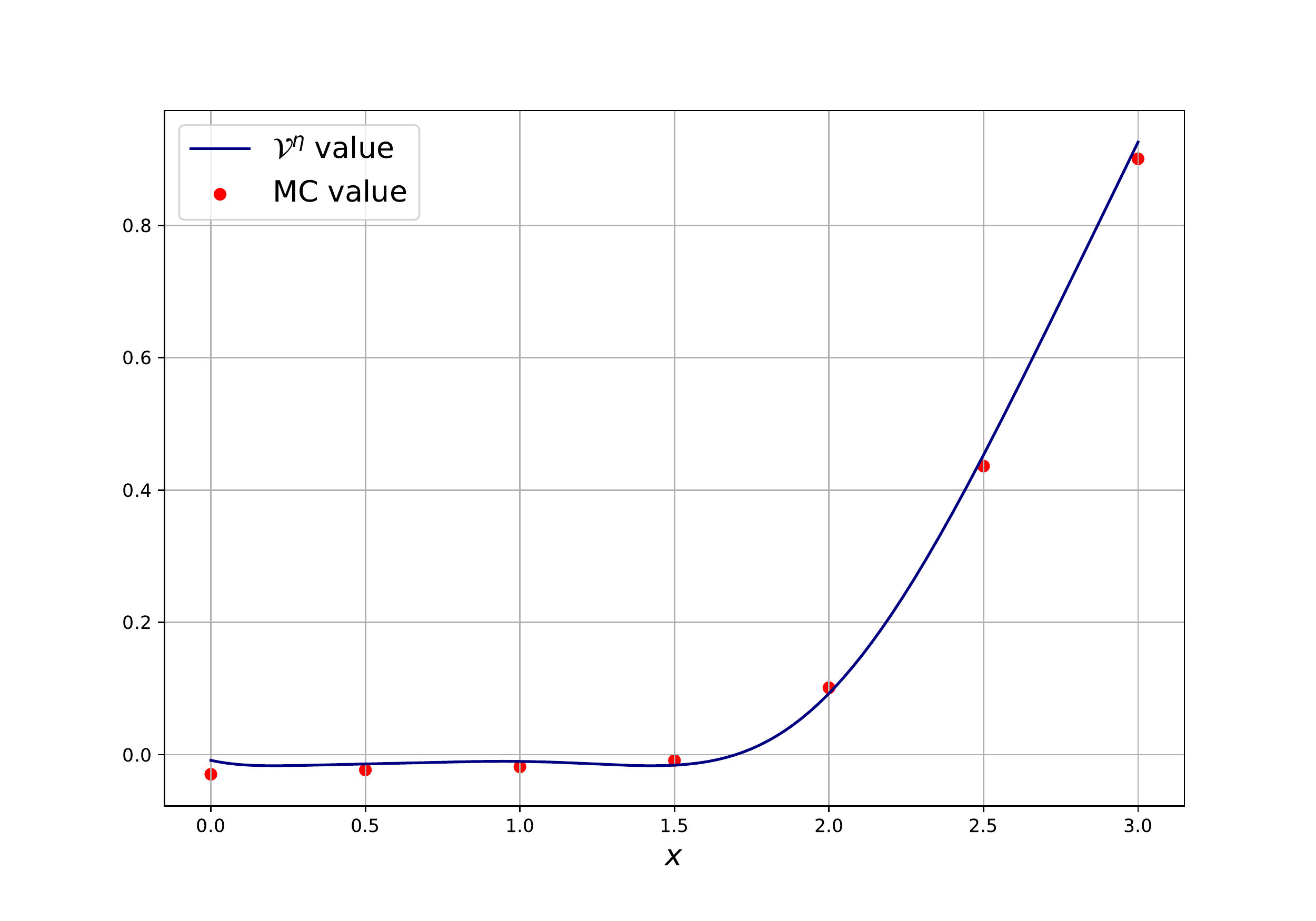} 
    \end{subfigure}
    \begin{subfigure}{.32\linewidth}
        \centering
        \includegraphics[height=3.75cm]{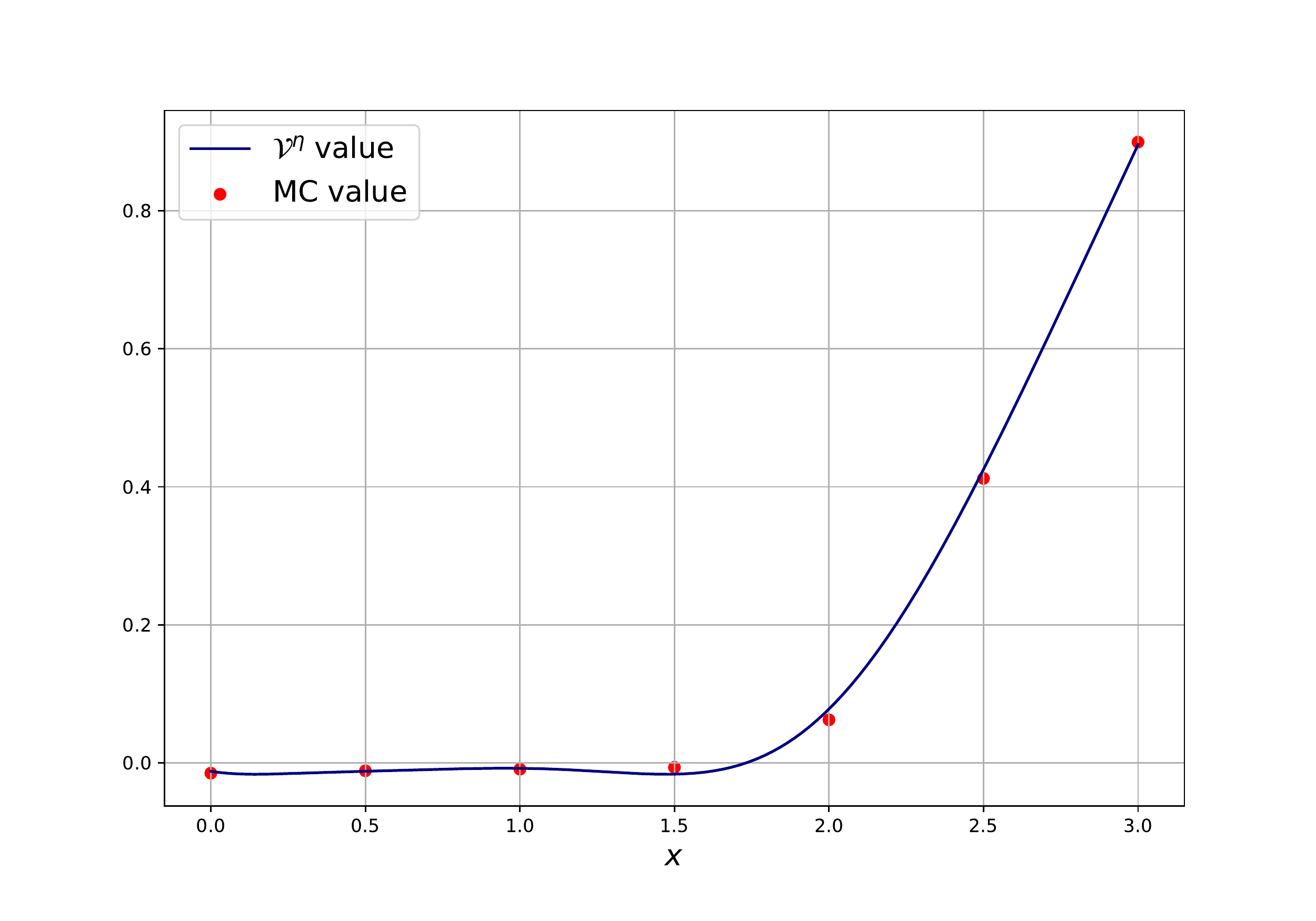}
    \end{subfigure}
    \begin{subfigure}{.32\linewidth}
        \centering
        \includegraphics[height=3.75cm]{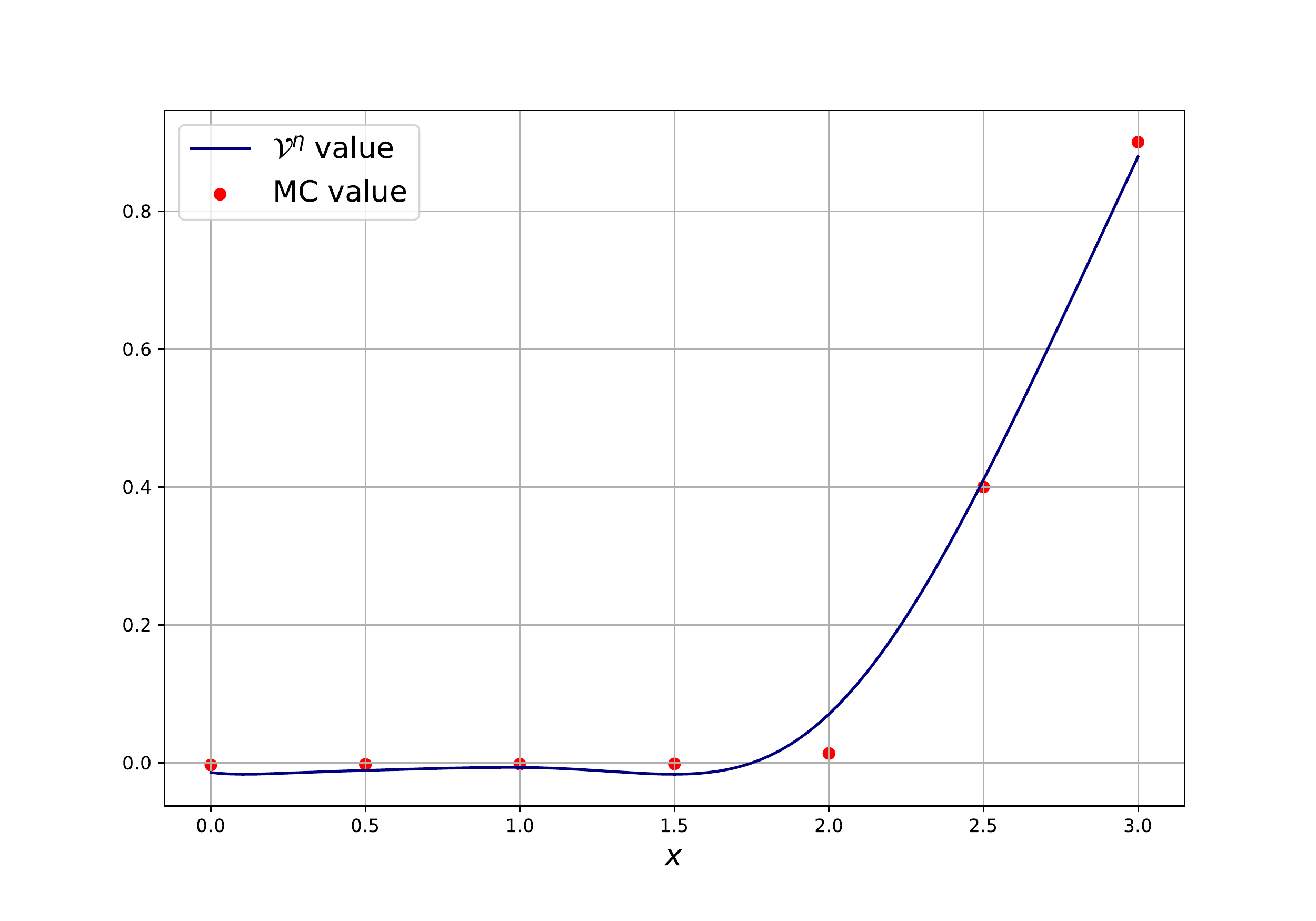}
    \end{subfigure}
    \begin{subfigure}{.32\linewidth}
        \centering
        \includegraphics[height=3.75cm]{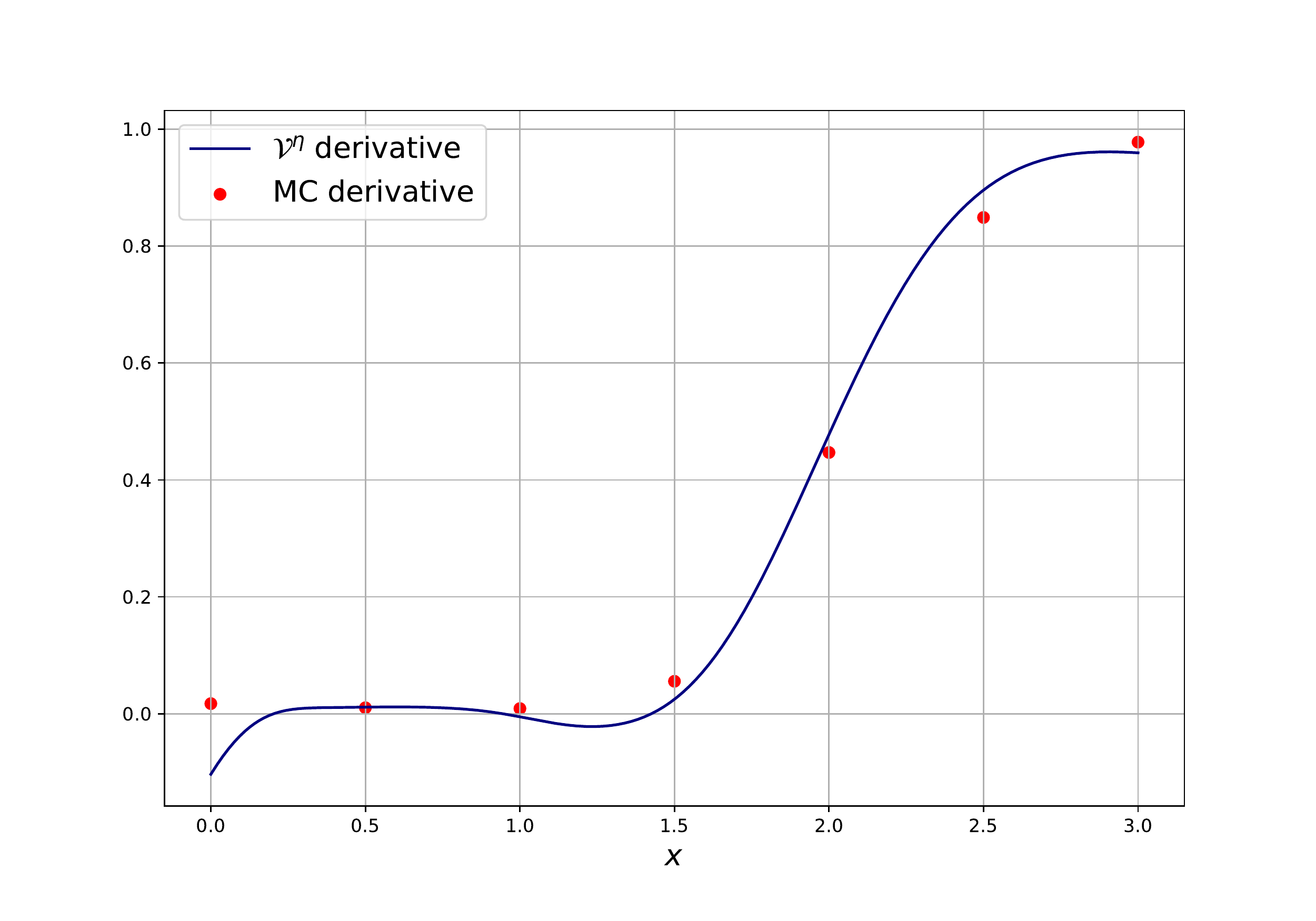} 
    \end{subfigure}
    \begin{subfigure}{.32\linewidth}
        \centering
        \includegraphics[height=3.75cm]{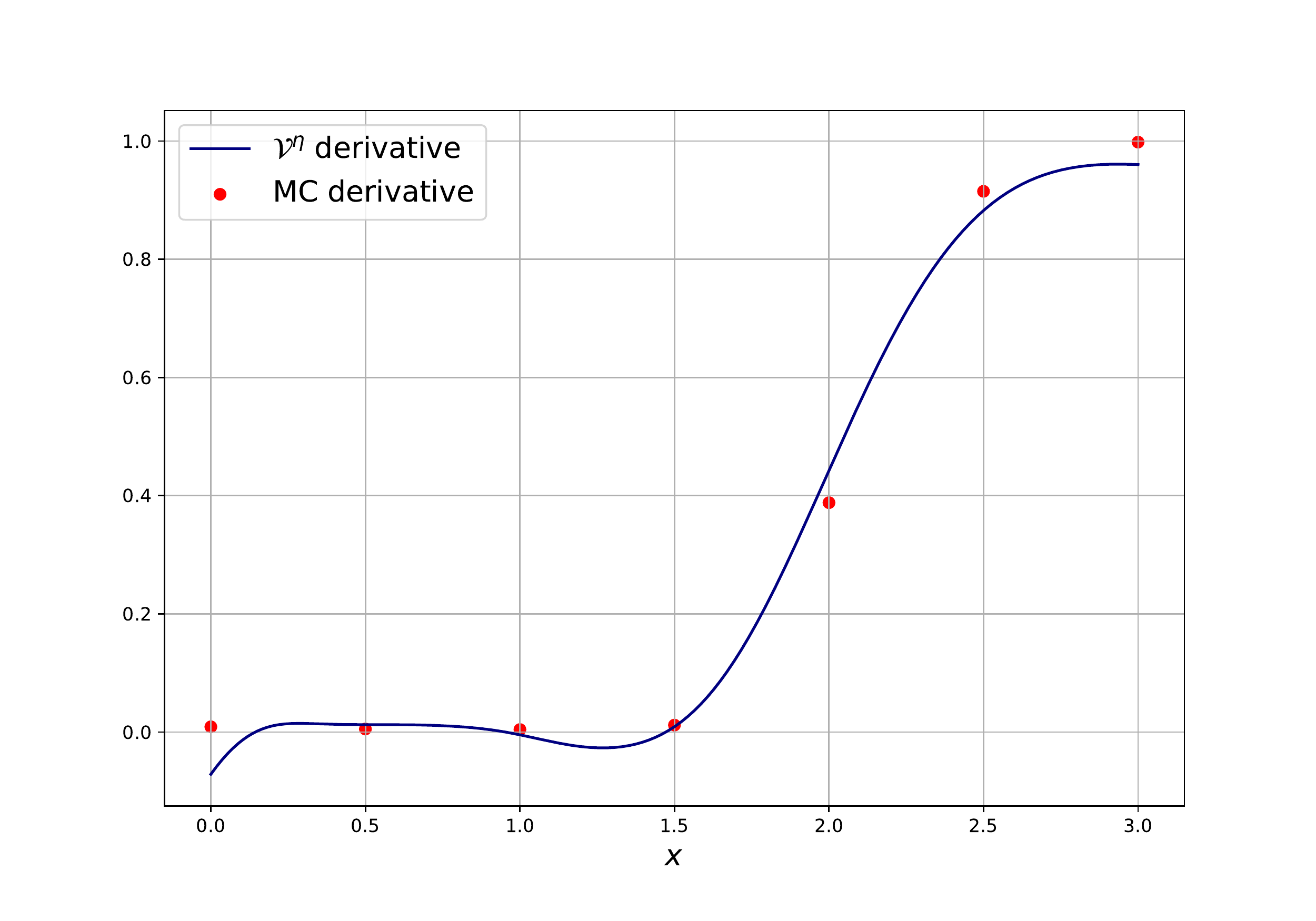}
    \end{subfigure}
    \begin{subfigure}{.32\linewidth}
        \centering
        \includegraphics[height=3.75cm]{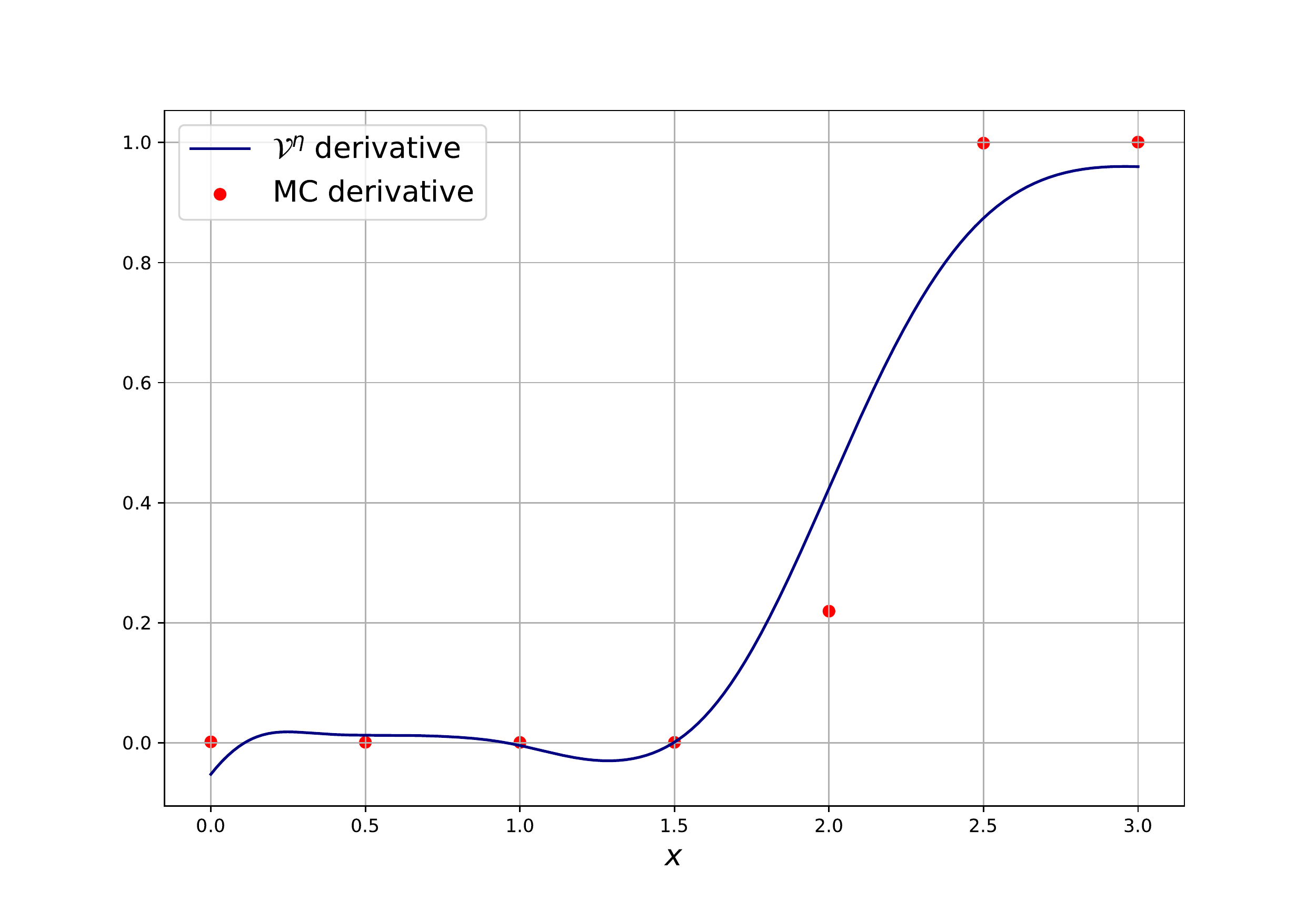}
    \end{subfigure}
    \caption{
    \label{fig:value_differential_learning_deeponet_call_K21}
    \footnotesize{Value function $\vartheta^\eta$ (first line) and its derivative (second line) obtained by Differential Regression Learning (Algorithm \ref{algo:scheme_value_differential_learning_deeponet}) for a terminal call option payoff with strike $K$ $=$ $2.1$, with parameter $\sigma = 0.3$ and linear market impact factor $\lambda = 5e^{-3}$, plotted as functions of $x$, for fixed values of $t$.}
    }
\end{figure}

On these graphs we see that the estimation of the value of the value function is good and consistent with the Monte Carlo estimator for every strike, inside or outside of the training domain. The estimation of the derivative of the value function is not as good and we can see on the graphs that we get the worst results for the value of the strike closest to zero, $K$ $=$ $0.2$ (Figure \ref{fig:value_differential_learning_deeponet_call_K02}), and for values of the strike out of the training domain, $K$ $=$ $1$ (Figure \ref{fig:value_differential_learning_deeponet_call_K1}) and $K$ $=$ $2.1$ (Figure \ref{fig:value_differential_learning_deeponet_call_K21}).

\appendix

\section{Alternative algorithms using multiple neural networks}\label{appendix:algorithms_multiple_NN}

We present below the Algorithm \ref{algo:scheme_value_pathwise_martingale_learningbis}, which is the version of Algorithm \ref{algo:scheme_value_pathwise_martingale_learning} using two neural networks, described in Section \ref{section:numerical_pathwise_learning}.

\vspace{1mm}

\begin{algorithm}[H] 
\scriptsize
\SetAlgoLined
\KwResult{A set of optimized parameters $\eta^*$, $\delta^*$;}
 Initialize the learning rate $l$, the neural networks $\vartheta^\eta$, $\Zcal^\delta$\; 
 Generate an $\R^{N+1}$-valued time grid $0=t_0 < t_1 < ... < t_N = T$ with time steps $(\Delta t_n)_{n=0,...,N-1}$\;
 Generate a batch of $M$ starting points $X_0$ $\sim$ $\mu_0$ and Brownian increments $(\Delta W_{t_n})_{n=0,...,N}$ in $R^d$\;
 \For{each batch element $m$}{
    Compute the trajectory $(x_{t_n}^{m})_{n=0,...,N}$ through the scheme
        \bes{
            x_{t_{n+1}}^{m} & = \;  x_{t_{n}}^{m}  +  \mrb^{\mra^*}(t_n,x_{t_{n}}^{m}) \Delta t_n  +  \sigma^{\mra^*}(t_n,x_{t_{n}}^{m}) \Delta w_{t_n}^m,
        }
        from the generated starting point $x_{t_0}^m$, Brownian increments $(\Delta w_{t_n}^m)_{n=0,...,N-1}$ and previously trained control $a=a_{\theta^*}$\;
    Compute the value target $(y_T^{m, t_n})_{n=0,...,N}$\;
}
 \For{each epoch}{
    Compute, for every batch element $m$, the integral $\sum_{n=0}^{N-1} \Big| y_T^{m,t_n} - \vartheta^\eta(t_n,x_{t_n}^m)  - \;  \sum_{p=n}^{N-1}  \big( \Zcal^\delta(t_p,x_{t_p}^m) \big)^\top \sigma^{\mra^*}(t_p,x_{t_p}^m) \Delta w_{t_p}^m   \Big|^2 \Delta t_n$\;
    Compute the batch loss $\tilde MSE_{mar}(\eta,\delta)$\;
    Compute the gradient $\nabla_{\eta} \tilde MSE_{mar}(\eta,\delta)$ and $\nabla_{\delta} \tilde MSE_{mar}(\eta,\delta)$ \;
    Update $\eta \leftarrow \eta - l \nabla_{\eta} \tilde MSE_{mar}(\eta,\delta)$, 
    $\delta \leftarrow \delta - l \nabla_{\delta} \tilde MSE_{mar}(\eta,\delta)$;
}
\textbf{Return:} The set of  optimized parameters  $\eta^*$, $\delta^*$\;
\caption{Deep learning scheme for Pathwise martingale learning with 2 NN}
\label{algo:scheme_value_pathwise_martingale_learningbis}
\end{algorithm}

\vspace{2mm}

In the same way, the Algorithm \ref{algo:scheme_value_pathwise_differential_learning3NN} below is the version of Algorithm \ref{algo:scheme_value_pathwise_differential_learning} using three neural networks described in Section \ref{section:numerical_pathwise_learning}.

\vspace{1mm}

\begin{algorithm}[H] 
\scriptsize
\SetAlgoLined
\KwResult{A set of optimized parameters $\eta^*$;}
 Initialize the learning rate $l$, the neural networks $\vartheta^\eta$\; 
 Generate an $\R^{N+1}$-valued time grid $0=t_0 < t_1 < ... < t_N = T$ with time steps $(\Delta t_n)_{n=0,...,N-1}$\;
 Generate a batch of $M$ starting points $X_0$ $\sim$ $\mu_0$ and Brownian increments $(\Delta W_{t_n})_{n=0,...,N}$ in $R^d$\;
 \For{each batch element $m$}{
    Compute the trajectory $(x_{t_n}^{m})_{n=0,...,N}$ through the scheme
        \bes{
            x_{t_{n+1}}^{m} & = \;  x_{t_{n}}^{m}  +  \mrb^{\mra^*}(t_n,x_{t_{n}}^{m}) \Delta t_n  +  \sigma^{\mra^*}(t_n,x_{t_{n}}^{m}) \Delta w_{t_n}^m,
        }
        from the generated starting point $x_{t_0}^m$, Brownian increments $(\Delta w_{t_n}^m)_{n=0,...,N-1}$ and previously trained control $a=a_{\theta^*}$\;
    Compute the value and derivative targets $(y_T^{m, t_n})_{n=0,...,N}$ and $(z_T^{m, t_n})_{n=0,...,N}$\;
}
\For{each epoch}{
    Compute, for every batch element $m$, the integral $\sum_{n=0}^{N-1} \Big| y_T^{m,t_n} - \vartheta^\eta(t_n,x_{t_n}^m) -  \sum_{p=n}^{N-1}  \big( D_x \vartheta^\eta(t_p,x_{t_p}^m) \big)^\top \sigma^{\mra^*}(t_p,x_{t_p}^m) \Delta w_{t_p}^m   \Big|^2 \Delta t_n$\;
    Compute the batch loss $MSE_{mar}(\eta)$\;
    Compute the gradient $\nabla_{\eta} MSE_{mar}(\eta)$\;
    Update $\eta \leftarrow \eta - l \nabla_{\eta} MSE_{mar}(\eta)$\;
    Compute, for every batch element $m$, the integral $\sum_{n=0}^{N-1} \Big| z_T^{m,t_n} -  D_x \vartheta^\eta(t_n,x_{t_n}^m) -  \sum_{p=n}^{N-1} \Big( \big[ D_x \sigma^{\mra^*}(t_p,x_{t_p}^m) \bullet_3 D_{x_{t_n}}  x_{t_p}^m  \big] \bullet_1 D_x \vartheta^\eta(t_p,x_{t_p}^m) $ \\
    $+ \sigma^{\mra^*}(t_p,x_{t_p}^m)^\top D_{xx} \vartheta^\eta(t_p,x_{t_p}^m) D_{x_{t_n}} x_{t_p}^m \Big)^\top \Delta w_{t_p}^m \Big|^2 \Delta t_n$\;
    Compute the batch loss $MSE_{dermar}(\eta)$\;
    Compute the gradient $\nabla_{\eta} MSE_{dermar}(\eta)$\;
    Update $\eta \leftarrow \eta - l \nabla_{\eta} MSE_{dermar}(\eta)$\;
}
\textbf{Return:} The set of  optimized parameters  $\eta^*$\;
\caption{Deep learning scheme for Pathwise differential learning with 3 NN}
\label{algo:scheme_value_pathwise_differential_learning3NN}
\end{algorithm}

\vspace{2mm}

\bibliographystyle{plain}

\bibliography{bibliography}

\begin{thebibliography}{10}

\bibitem{BBCJN19}
Chistian Beck, Sebastian Becker, Patrick Cheridito, Arnulf Jentzen, and Ariel
  Neufeld.
\newblock Deep splitting method for parabolic {PDE}s.
\newblock {\em SIAM Journal on Scientific Computing}, 43(5), 2021.

\bibitem{BEJ19}
Christian Beck, Weinan E, and Arnulf Jentzen.
\newblock Machine learning approximation algorithms for high-dimensional fully
  nonlinear partial differential equations and second-order backward stochastic
  differential equations.
\newblock {\em J. Nonlinear Sci.}, 29(4):1563--1619, 08 2019.

\bibitem{becetal20}
Christian Beck, Martin Hutzenthaler, Arnulf Jentzen, and Benno Kuckuck.
\newblock An overview on deep learning-based approximation methods for partial
  differential equations.
\newblock {\em arXiv preprint: 2012.12348}, 2020.

\bibitem{bergstra2012random}
James Bergstra and Yoshua Bengio.
\newblock Random search for hyper-parameter optimization.
\newblock {\em {Journal of Machine Learning Research}}, 13(2), 2012.

\bibitem{chen1995universal}
Tianping Chen and Hong Chen.
\newblock Universal approximation to nonlinear operators by neural networks
  with arbitrary activation functions and its application to dynamical systems.
\newblock {\em IEEE Transactions on Neural Networks}, 6(4):911--917, 1995.

\bibitem{Ehanjen17}
Weinan E., Jiequn Han, and Arnulf Jentzen.
\newblock Deep learning-based numerical methods for high dimensional parabolic
  partial differential equations and backward stochastic differential
  equations.
\newblock {\em Commun. Math. Stat.}, 5(4):349--380, 2017.

\bibitem{elkquepen97}
Nicole El~Karoui, Marie-Claire Quenez, and Shige Peng.
\newblock Backward stochastic differential applications in finance.
\newblock {\em Mathematical Finance}, 7(1):1--71, 1997.

\bibitem{gerphawar21}
Maximilien Germain, Huy{\^e}n Pham, and Xavier Warin.
\newblock Neural networks-based algorithms for stochastic control and {PDEs} in
  finance.
\newblock {\em to appear in Machine learning for financial markets: a guide to
  contemporary practices}, 2021.

\bibitem{glasserman2013monte}
Paul Glasserman.
\newblock {\em Monte Carlo methods in financial engineering}, volume~53.
\newblock Springer Science \& Business Media, 2013.

\bibitem{glau2020deep}
Kathrin Glau and Linus Wunderlich.
\newblock The deep parametric {PDE} method: application to option pricing.
\newblock {\em arXiv preprint arXiv:2012.06211}, 2020.

\bibitem{gobmun05}
Emmanuel Gobet and R{\'e}mi Munos.
\newblock Sensitivity analysis using {I}t\^o-{M}alliavin calculus and
  martingales, and application to stochastic optimal control.
\newblock {\em SIAM J. Control Optim.}, 43(5):1676--1713, 2005.

\bibitem{HanE16}
Jiequn Han and Weinan E.
\newblock Deep learning approximation for stochastic control problems.
\newblock {\em Deep Reinforcement Learning Workshop, NIPS, arXiv preprint:
  1611.07422}, 2016.

\bibitem{HJE17}
Jiequn Han, Arnulf Jentzen, and Weinan E.
\newblock Solving high-dimensional partial differential equations using deep
  learning.
\newblock {\em Proc. Natl. Acad. Sci. USA}, 115, 2017.

\bibitem{huge2020differential}
Brian~Norsk Huge and Antoine Savine.
\newblock Differential machine learning.
\newblock {\em Available at SSRN 3591734}, 2020.

\bibitem{hure2021deep}
C{\^o}me Hur{\'e}, Huy{\^e}n Pham, Achref Bachouch, and Nicolas Langren{\'e}.
\newblock Deep neural networks algorithms for stochastic control problems on
  finite horizon: convergence analysis.
\newblock {\em SIAM Journal on Numerical Analysis}, 59(1):525--557, 2021.

\bibitem{hure2020deep}
C{\^o}me Hur{\'e}, Huy{\^e}n Pham, and Xavier Warin.
\newblock Deep backward schemes for high-dimensional nonlinear {PDEs}.
\newblock {\em Mathematics of Computation}, 89(324):1547--1579, 2020.

\bibitem{penetal20}
Shaolin Ji, Shige Peng, Ying Peng, and Xichuan Zhang.
\newblock Three algorithms for solving high-dimensional fully coupled {FBSDE}
  through deep learning.
\newblock {\em {IEEE Intelligent Systems}}, 35(3):71--84, 2020.

\bibitem{loeper2018option}
Gr\'egoire Loeper.
\newblock Option pricing with linear market impact and nonlinear
  {Black--Scholes} equations.
\newblock {\em Annals of Applied Probability}, 28(5):2664--2726, 2018.

\bibitem{longstaff2001valuing}
Francis~A Longstaff and Eduardo~S Schwartz.
\newblock Valuing american options by simulation: a simple least-squares
  approach.
\newblock {\em The review of financial studies}, 14(1):113--147, 2001.

\bibitem{lu2019deeponet}
Lu~Lu, Pengzhan Jin, and George~Em Karniadakis.
\newblock Deeponet: Learning nonlinear operators for identifying differential
  equations based on the universal approximation theorem of operators.
\newblock {\em arXiv preprint arXiv:1910.03193}, 2019.

\bibitem{mazha02}
Jin Ma and Jianfeng Zhang.
\newblock Representation theorems for backward stochastic differential
  equations.
\newblock {\em Annals of Applied Probability}, 12(4):1390--1418, 2002.

\bibitem{negyesi21}
Balint Negyesi, Kristoffer Andersson, and Cornelis Oosterlee.
\newblock The one step {Malliavin} scheme: new discretization of bsdes
  implemented with deep learning regressions.
\newblock {\em arXiv preprint arXiv:2110.05421}, 2021.

\bibitem{nua95}
David Nualart.
\newblock {\em The Malliavin calculus and related topics}.
\newblock Springer-Verlag, Berlin, 1995.

\bibitem{nusric21}
Nikolas N\"uskens and Lorenz Richter.
\newblock Interpolating between {BSDEs and PINNs}: deep learning for elliptic
  and parabolic boundary value problems.
\newblock {\em arXiv:2112.03749}, 2021.

\bibitem{pham2021neural}
Huy\^en Pham, Xavier Warin, and Maximilien Germain.
\newblock Neural networks-based backward scheme for fully nonlinear {PDEs}.
\newblock {\em SN Partial Differential Equations and Applications}, 2(1):1--24,
  2021.

\bibitem{potters2001hedged}
Marc Potters, Jean-Philippe Bouchaud, and Dragan Sestovic.
\newblock Hedged {Monte-Carlo}: low variance derivative pricing with objective
  probabilities.
\newblock {\em Physica A: Statistical Mechanics and its Applications},
  289(3-4):517--525, 2001.

\bibitem{protter2005stochastic}
Philip~E Protter.
\newblock Stochastic differential equations.
\newblock In {\em Stochastic integration and differential equations}, pages
  249--361. Springer, 2005.

\bibitem{raissi2019physics}
Maziar Raissi, Paris Perdikaris, and George~E Karniadakis.
\newblock Physics-informed neural networks: A deep learning framework for
  solving forward and inverse problems involving nonlinear partial differential
  equations.
\newblock {\em Journal of Computational Physics}, 378:686--707, 2019.

\bibitem{remlinger2022robust}
Carl Remlinger, Joseph Mikael, and Romuald Elie.
\newblock Robust operator learning to solve {PDE}.
\newblock 2022.

\bibitem{sirignano2018dgm}
Justin Sirignano and Konstantinos Spiliopoulos.
\newblock Dgm: A deep learning algorithm for solving partial differential
  equations.
\newblock {\em Journal of Computational Physics}, 375:1339--1364, 2018.

\bibitem{soner2013dual}
H~Mete Soner, Nizar Touzi, and Jianfeng Zhang.
\newblock Dual formulation of second order target problems.
\newblock {\em The Annals of Applied Probability}, 23(1):308--347, 2013.

\bibitem{srivastava2015training}
Rupesh~K Srivastava, Klaus Greff, and J{\"u}rgen Schmidhuber.
\newblock Training very deep networks.
\newblock {\em Advances in neural information processing systems}, 28, 2015.

\bibitem{van2021optimally}
Remco van~der Meer, Cornelis~W Oosterlee, and Anastasia Borovykh.
\newblock Optimally weighted loss functions for solving {PDEs} with neural
  networks.
\newblock {\em Journal of Computational and Applied Mathematics}, page 113887,
  2021.

\bibitem{vidales2018unbiased}
Marc~Sabate Vidales, David Siska, and Lukasz Szpruch.
\newblock Unbiased deep solvers for parametric {PDEs}.
\newblock {\em arXiv preprint arXiv:1810.05094}, 2018.

\end{thebibliography}

\end{document}